\documentclass[a4paper,11pt]{article}
\pdfoutput=1

\usepackage{jcappub} 
\usepackage{natbib}
\setcitestyle{square,comma,numbers,sort&compress}
\usepackage{enumerate}
\usepackage{tabularx}
\usepackage[left = 1.0in, top=1.0in,right=1.0in,bottom=1.0in,a4paper]{geometry}
\usepackage[dvipsnames]{xcolor}
\usepackage{comment}
\usepackage{cancel}

\usepackage[section]{placeins}
\usepackage{listings}
\usepackage{amsbsy}
\usepackage{slashed}
\usepackage{textcomp}
\usepackage{slashed}
\newcolumntype{Y}{>{\centering\arraybackslash}X}

\definecolor{myRED}{rgb}{0.8, 0.25, 0.33}
\usepackage[normalem]{ulem}
\usepackage{dsfont}
\usepackage{pbox}
\usepackage{graphicx}
\usepackage{multirow}
\usepackage{tikz}
\usepackage{nicematrix}
\usetikzlibrary{backgrounds}
\usetikzlibrary{decorations.pathreplacing,calc}
\usetikzlibrary{arrows}
\usepackage{enumitem} 
\usepackage{ulem}
\usepackage{soul}
\usepackage{lipsum} 
\usepackage{tikz-cd}
\usepackage{tikz-feynman}
\usepackage{xcolor,pifont}
\usepackage{adjustbox} 
\usepackage{subcaption}
\usepackage{float}
\usepackage{placeins} 
\usepackage{siunitx}
\RequirePackage{luatex85}
\usepackage[toc,page]{appendix}

\title{\boldmath\Huge  
TeV Scale Quark-Lepton Unification}

\author[a]{K.S. Babu,}
\author[a]{Sumit Biswas,}
\author[b]{Shaikh Saad}

\emailAdd{babu@okstate.edu, sumit.biswas@okstate.edu, shaikh.saad@ijs.si}

\affiliation[a]{Department of Physics, Oklahoma State University, Stillwater, OK 74078, USA}

\affiliation[b]{Jožef Stefan Institute, Jamova 39, P.\ O.\ Box 3000, SI-1001 Ljubljana, Slovenia}

\abstract{
We propose a quark-lepton symmetric Pati-Salam (PS)  model based on the gauge group $SU(2)_L \times SU(2)_R \times SU(4)_C$ with an $E_6$-inspired particle spectrum which naturally accommodates a multi-TeV leptoquark gauge boson $X_\mu(3,1,\frac{2}{3})$. A softly broken $Z_2$ symmetry plays a crucial role in realizing this scenario, under which the Standard Model (SM) fermions are even, while new vector-like fermions present in the model are odd. A notable feature of this model is that the PS gauge boson $X_\mu$  itself is $Z_2$-odd, causing it to couple exclusively between SM fermions and the vector-like fermions, except in the right-handed down-quark sector, where mixing is induced by the soft breaking of $Z_2$. This structure leads to helicity suppression of tree-level meson decays mediated by $X_\mu$,  with $Z_2$-unsuppressed contributions arising via one-loop diagrams. We show that $X_\mu$ can be as light as 1.1\, TeV while being compatible with all flavor-violating constraints. However, mass relations among the $(W'^\pm_\mu, Z'_\mu, X_\mu)$ gauge bosons push the $X_\mu$ mass limit up to 4.3\, TeV from the stringent LHC bounds on the $Z^\prime$ mass.  This comparatively low PS-breaking scale opens up promising collider opportunities for probing the leptoquark gauge boson, as well as the distinctive signature of vector-like down-type quarks carrying an unusual baryon number of $2/3$. The model can be further tested via lepton flavor-violating processes induced by the leptoquark gauge boson, such as $\mu \to e \gamma$, $\mu \to e e e$, and $\mu$-$e$ conversion in nuclei. Neutrino masses arise in the model through dimension-seven operators at tree-level, as well as from dimension-five operators via one-loop diagrams.
}

\makeatletter
\gdef\@fpheader{}
\makeatother
\begin{document}
\maketitle
\flushbottom

\section{Introduction}
The first successful attempt to unify quarks and leptons is the Pati–Salam~\cite{Pati:1973rp,Pati:1974yy} (PS) model, where they transform as common multiplets of the gauge symmetry $SU(2)_L \times SU(2)_R \times SU(4)_C$. Lepton number is treated as the fourth color of the enhanced color group $SU(4)_C$. Beyond quark–lepton unification, the PS model has several attractive features: its non-Abelian nature implies that electric charge is quantized; the presence of right-handed neutrinos dictated by the gauge structure paves the way for light neutrino mass generation via the seesaw mechanism; and the matter--antimatter asymmetry of the universe can be explained via leptogenesis arising from the decays of the right-handed neutrinos. Furthermore, unlike Grand Unified Theories (GUTs) based on simple gauge groups, the new leptoquark gauge boson in the PS model, $X_\mu (3,1,\tfrac{2}{3})$, where the quantum numbers refer to the Standard Model symmetry $SU(3)_C \times SU(2)_L \times U(1)_Y$, does not mediate proton decay, allowing the PS-breaking scale to be much lower than the traditional GUT scale of order $10^{16}$ GeV. There is, however, a strong lower limit on the PS-breaking scale from meson decay to leptons. In particular  $K^0_L \rightarrow \mu^{\pm} e^{\mp}$ constraint requires $M_X \gtrsim \mathcal{O}(2500)$ TeV~\cite{Kuznetsov:1994tt,Valencia:1994cj,Smirnov:2007hv}. This scale is still well beyond the reach of any foreseeable collider experiments.

The goal of this paper is to realize Pati-Salam symmetry at the multi-TeV scale, which can be probed directly at the Large Hadron Collider (LHC) and other near-future collider experiments. Our proposal makes use of an $E_6$-inspired particle spectrum, which enables us to evade the stringent lower limits on $X_\mu$ mass from meson decay constraints. In $E_6$ GUTs~\cite{Gursey:1975ki}, all fermions of a generation are unified in a $27_F$-dimensional representation.  Apart from the Standard Model (SM) fermions, this contains new vector-like fermions -- an iso-singlet down-type quark, a vector-like lepton doublet, and two singlet neutrinos $\nu^c$ and $N$. An additional vector-like fermion belonging to $(1,1,6)$ of the Pati-Salam gauge symmetry is also introduced for realizing quark-lepton symmetry in the multi-TeV range. We are agnostic about the origin of this multiplet, but we note that it has no gauge anomaly and it can arise in specific compactification schemes that break $E_6$ down to the PS gauge group.

Our proposed model incorporates a softly broken $Z_2$ symmetry, under which all SM fermions are even, while the new vector-like fermions are odd. An exception to this assignment is that a vectorlike quark with baryon number of $+2/3$, $\textcolor{red} {D'^c}$, is odd under $Z_2$. An important feature of the model is that the PS gauge boson $X_\mu$ is also $Z_2$-odd, causing its coupling to involve one SM fermion and one vector-like fermion.  The $Z_2$ symmetry is softly broken, which leads to mixing of the $d_R$ quarks of the SM with $D_R$ vector-like quarks.  Meson decays arising at tree-level through the $d_R-D_R$ mixings and mediated by the $X_\mu$ gauge boson are helicity suppressed, with $Z_2$-unsuppressed contributions arising only via loop diagrams.   Consequently, the $X_\mu$ gauge boson can be remarkably as light as a TeV, while being consistent with all flavor-violating constraints. Although the model introduced in this work is new, the idea of suppressing meson decays through the inclusion of vector-like fermions has been explored previously in the context of the PS framework in Refs.~\cite{Kuznetsov:1994tt,Valencia:1994cj,Kuznetsov:1994wa,Volkas:1995yn,Foot:1997pb,Foot:1999wv,Smirnov:2007hv,Dolan:2020doe,Calibbi:2017qbu}. What is genuinely novel in our construction, however, is that the specific set of exotic fermionic multiplets we introduce allows the softly broken $Z_2$ symmetry to play a significant role in the suppression mechanism and admits a consistent baryon-number assignment for all fields, thereby ensuring absolute proton stability. This combination of features---PS structure, vector-like fermions with a softly broken $Z_2$ symmetry, and exact baryon-number conservation---has not been realized in any of the prior models.

Our detailed analysis shows that the mass of the leptoquark gauge boson $X_\mu$ can be as low as 1.1\,TeV while being fully consistent with all flavor-violating constraints. Such a low bound on $M_X$ is obtained by eliminating the kaon decay at tree level by a suitable choice of flavor structure. The most stringent flavor bound then arises from $B_s \to \mu e$ and $B_s \to \mu\mu$ decays in this scenario.
However, current LHC experiments set a lower limit of  5.6\,TeV on the $Z'$ mass, from the production and decay $Z'\to \ell\overline{\ell}$ process. Taking account of the theoretical relations among the masses of $Z'$ and $X_\mu$, this translates into an indirect collider limit of $M_X \gtrsim 4.3$~TeV. Such a low PS-breaking scale opens up promising opportunities for collider probes.  While direct searches for $W'$ and $X$ gauge bosons are less sensitive, we find that the high-luminosity LHC (HL-LHC) can probe the $Z'$ mass up to 6.6\,TeV, whereas the FCC-hh has the potential to reach a sensitivity of order $\mathcal{O}(40)$~TeV~\cite{Helsens:2642473}. Owing to the mass relations among the gauge bosons, this reach would also probe indirectly $X_\mu$ mass of the same order. The model predicts the existence of vector-like down-type quarks ${\color{red} D'}$ with an unusual baryon number of $2/3$. Pair production of ${\color{red} D'}$ would lead to multi-jet final states, which already place a lower bound of $\mathcal{O}(1.5)$~TeV on ${\color{red} D'}$ mass, which can be further tested in upcoming LHC upgrades and at future colliders. 
The leptoquark gauge boson also mediates lepton flavor violation (LFV), with the most stringent constraints arising from $\mu\to e\gamma$, $\mu\to eee$, and $\mu-e$ conversion in nuclei. If we assume a CKM-like mixing structure for the mixing matrix involving $X_\mu$, these processes would yield a lower bound of roughly $M_X \gtrsim 1.64$~TeV. The model proposed, therefore, can be tested indirectly by the observation of these LFV processes.

Within this setup, we find that there are two comparable sources of neutrino masses. The first one arises through mixing of $\nu$ with the other neutral leptons $\nu^c$ and $N$ present in the model, leading to an effective dimension-7 operator $LL HH H^\dagger H$ (with $L$ and $H$ denoting the lepton and SM Higgs doublet)~\cite{Babu:2009aq,Bambhaniya:2013yca}.   The second contribution arises from one-loop diagrams analogous to neutrino mass generation in the scotogenic model~\cite{Ma:2006km,Tao:1996vb}. Neutrino masses arising from either of these sources are sufficiently small and allow for the PS breaking scale to be in the multi-TeV range. Depending on the choice of parameters of the model, one or the other can dominate, or both can be comparable.  We also point out that, due to low-scale PS breaking, monopoles produced at a temperature of TeV are cosmologically harmless~\cite{Preskill:1979zi,Volkas:1995yn,Jeannerot:2000sv}. In sum,  TeV-scale PS-breaking is not only theoretically motivated and consistent, but also ensures that the model remains fully accessible to experimental probes.

Extended versions of low scale Pati-Salam models have been studied in the context of explaining certain $B$-meson decay anomalies~\cite{Barbieri:2017tuq,Blanke:2018sro,Heeck:2018ntp,Fuentes-Martin:2020pww,Greljo:2018tuh,FernandezNavarro:2022gst}. In contrast to these extended models, the model proposed here is more traditional. 
There have been several studies on realizing Pati-Salam theory in four dimensions from string compactification, see for example, Ref.~\cite{Kobayashi:2004ya,Cvetic:2004ui,Assel:2010wj,Chamseddine:2013rta,Hewett:1988xc}. For discussions on various other aspects of the Pati-Salam model, such as its origin from $SO(10)$ unified theory, baryon number violation, solution to the gauge hierarchy problem, and symmetry breaking sector, see Ref.~\cite{Mohapatra:1980qe,Aulakh:2002zr,Arkani-Hamed:1998jmv,deAdelhartToorop:2010vtu,King:2014iia,Saad:2017pqj}.  A modified realization of the PS gauge group, \( SU(4)\times SU(2)_L \times U(1)_R \), has also been investigated in the literature; see, for example, Refs.~\cite{FileviezPerez:2013zmv,Butterworth:2025ttw}.

The layout of the paper is as follows. In Sec.~\ref{sec:2}, we introduce the model and provide all necessary details. These include the fermion, gauge boson and Higgs boson spectra, interconnections among the gauge boson masses, and gauge interactions in the physical basis of fermions. Collider implications are discussed in Sec.~\ref{sec:3}. A detailed analysis of flavor-violating processes, including meson decays, meson--antimeson oscillations, and lepton flavor violation (LFV), is presented in Sec.~\ref{sec:4}. This analysis includes tree-level induced processes as well as loop-induced processes. Finally, we conclude in Sec.~\ref{sec:5}. Some computational details are relegated to Appendices~\ref{app:B}, \ref{app:C}, and \ref{app:D}.

\section{$\boldmath{E_6}$-inspired Pati-Salam  Model}\label{sec:2}
In this section, we introduce the proposed Pati-Salam model with an $E_6$-inspired fermion spectrum and provide the relevant details necessary to develop strategies to test the model.

\subsection{Fermionic content}

The proposed model is based on the Pati-Salam gauge symmetry  $\mathcal{G}_\mathrm{PS}\equiv SU(2)_L\times SU(2)_R\times SU(4)_C$. The fermionic content of the theory is inspired by $E_6$ GUT.
To see this, first consider the decomposition of the fundamental representation 
$27$ of $E_6$ under its maximal subgroup $SO(10) \times U(1)$ as well as its subgroup $\mathcal{G}_{\mathrm{PS}} \times U(1)$:
\begin{align}
    27 &=16_1\oplus 10_{-2}\oplus 1_4~~\left[E_6 \supset SO(10) \times U(1)\right] \notag \\ 
    &= (2,1,4)_1\oplus (1,2,\overline{4})_1\oplus(2,2,1)_{-2}\oplus(1,1,6)_{-2}\oplus(1,1,1)_{4} ~~\left[E_6 \supset \mathcal{G}_\mathrm{PS}  \times U(1)  \right].
\end{align}
Our model contains three families of these fermion fields arising from $27$.  In addition, we introduce the multiplet $\Omega\equiv (1,1,6)_F$ of fermions, one per family, which may have its origin
from a pair of vector-like fermions $27_F + \overline{27}_F$, or from compactification of a higher dimensional theory. The $\Omega$ field is crucial for defining baryon number consistently. The full fermionic content of the theory is given below, where all fields are taken to be left-handed.
\begin{align}
& \psi=(2,1,4)=\left(\begin{array}{llll}
u_{1} & u_{2} & u_{3} & \nu_{E}\\
d_{1} & d_{2} & d_{3} & E
\end{array}\right),  \label{eq:fermion:01}\\
& \psi^{c}=(1,2, \overline{4})=\left(\begin{array}{llll}
D_{1}^{c} & D_{2}^{c} & D_{3}^{c} & e^{c} \\
u_{1}^{c} & u_{2}^{c} & u_{3}^{c} & \nu^{c}
\end{array}\right), \\
& \chi=(2,2,1)=\left(\begin{array}{ll}
E^{c} & \nu \\
\nu_{E}^{c} & e
\end{array}\right), \\
& \omega=(1,1,6)=\frac{1}{\sqrt{2}}\left(\begin{array}{cccc}
0 & d_{3}^{c} & -d_{2}^{c} & \textcolor{red}{D_{1}'} \\
-d_{3}^{c} & 0 & d_{1}^{c} & \textcolor{red}{D_{2}'} \\
d_{2}^{c} & -d_{1}^{c} & 0 & \textcolor{red}{D_{3}'}\\
-\textcolor{red}{D_{1}'} & -\textcolor{red}{D_{2}'} & -\textcolor{red}{D_{3}'} & 0
\end{array}\right),\\
& N=(1,1,1),\\
&\Omega=(1,1,6)=\frac{1}{\sqrt{2}}\left(\begin{array}{cccc}
0 & \textcolor{red}{D_{3}'^{c}} & -\textcolor{red}{D_{2}'^{c}} & D_{1} \\
-\textcolor{red}{D_{3}'^{c}} & 0 & \textcolor{red}{D_{1}'^{c}} & D_{2} \\
\textcolor{red}{D_{2}'^{c}} & -\textcolor{red}{D_{1}'^{c}} & 0 & D_{3}\\
-D_{1} & -D_{2} & -D_{3} & 0
\end{array}\right).   \label{eq:fermion:02}
\end{align}
A feature worth noting about this assignment is the flipping of $d^c \leftrightarrow D^c$ and 
$(\nu, e) \leftrightarrow (\nu_E, E)$, compared to the standard PS 
assignment, which does not include the $\chi, \omega, N,$ and $\Omega$ fields. This flipping plays an important role in realizing the PS symmetry at the multi-TeV scale, as we shall see later. The new fermions ${\color{red} D',D'^{c}}$ are denoted in red color throughout, to easily distinguish them from the usual quarks and the $D$ vector-like quarks, as they carry an unusual baryon number of $\pm 2/3$.

\subsection{Scalar content}

The scalar sector of our theory is rather minimal and consists of the following Higgs fields:
\begin{align}
& \widetilde{\psi}=(2,1,4)=\left(\begin{array}{llll}
\widetilde{u}_{1}& \widetilde{u}_{2} & \widetilde{u}_{3} & \widetilde{\nu}_{E} \\
\widetilde{d}_{1} & \widetilde{d}_{2} & \widetilde{d}_{3} & \widetilde{E}
\end{array}\right),  \label{eq:scalar:01} \\
& \widetilde{\psi}^{c}=(1,2, \overline{4})=\left(\begin{array}{llll}
\widetilde{D}_{1}^{c} & \widetilde{D}_{2}^{c} & \widetilde{D}_{3}^{c} & \widetilde{e}^{c} \\
\widetilde{u}_{1}^{c} & \widetilde{u}_{2}^{c} & \widetilde{u}_{3}^{c} & \widetilde{\nu}^{c}
\end{array}\right),\\
& \widetilde{\chi}=(2,2,1)=\left(\begin{array}{ll}
\widetilde{E}^{c} & \widetilde{\nu} \\
\widetilde{\nu}_{E}^{c} & \widetilde{e}
\end{array}\right). \label{eq:scalar:02}
\end{align}
Here we have used common symbols for the scalars and the fermions with the same quantum numbers, but with a tilde added for the scalars.
Note again that these states can arise from 
$27_H \supset (2,1,4) \oplus (1,2,\bar{4}) \oplus (2,2,1)$, 
with the remaining submultiplets of $27_H$ assumed to be decoupled 
and residing at the high scale.

The full particle content of the theory, as given in Eqs.~\eqref{eq:fermion:01}--\eqref{eq:fermion:02} 
and Eqs.~\eqref{eq:scalar:01}--\eqref{eq:scalar:02}, allows the imposition of a discrete $Z_2$ symmetry. 
Table~\ref{tab:Z_2} shows the relevant transformations of the particle content under this $Z_2$.  It is apparent that all SM are even and exotic fermions (except $\textcolor{red}{D'^c}$), including the right-handed neutrino $\nu^c$, are odd under this symmetry. Note that the $Z_2$ symmetry does not commute with the gauge symmetry. Such a discrete $Z_2$ has been known to be present in $E_6$-inspired left-right symmetric models~\cite{Ma:1986we,Babu:1987kp} where the symmetry may remain unbroken.  In the context of low scale Pati-Salam model, the $Z_2$ has to be softly broken, otherwise baryon number violation would occur through the exchange of scalar leptoquarks, or some of the vector-like quarks  would be massless. As will be discussed in Sec.~\ref {subsec:softly_broken_Z2}, the soft breaking of $Z_2$ arises through a $\omega-\Omega$ mass mixing, which is necessary to give {\color{red} $D'$} a mass. 

The neutral components of the Higgs fields $\widetilde{\psi}, \widetilde{\psi}^{c}$, and $\widetilde{\chi}$ all acquire vacuum expectation values (VEV) which  are denoted as: 
\begin{align}
    \langle \widetilde{\nu}_{E} \rangle=\kappa_L,~~~~~\langle \widetilde{\nu}^{c} \rangle=\kappa_R,~~~~~\langle \widetilde{\nu}_{E}^{c} \rangle=v,~~~~~\langle \widetilde{\nu} \rangle=\textcolor{magenta}{w}.
    \label{eq:VEVs}
\end{align}
While the VEVs of $\tilde{\nu}_E$, $\tilde{\nu^c}$ and $\tilde{\nu_E^c}$ do not break the $Z_2$ symmetry, the VEV of $\tilde{\nu}$ breaks the $Z_2$ and is thus shown in magenta color.  The soft breaking of $Z_2$ through the fermionic $\omega-\Omega$ mixing implies that there is a linear term in $\tilde{\nu}$ in the Higgs potential that needs to be shifted away, leading to a nonzero ${\color{magenta} w}$ with a hierarchy ${\color{magenta} w} \ll \kappa_L, v$. 
The spontaneous symmetry breaking proceeds in two stages as follows:
\begin{align}
SU(2)_L\times SU(2)_R\times SU(4)_C\xrightarrow[]{\langle \widetilde \psi^c\rangle} SU(3)_C \times SU(2)_L\times U(1)_Y   \xrightarrow[]{\langle \widetilde \psi, \widetilde \chi\rangle} SU(3)_C \times U(1)_\mathrm{em} \,. \end{align}
We are interested in a scenario with $\kappa_R \sim \mathcal{O}(\mathrm{few~TeV}) > \kappa_L, v \sim \mathcal{O}(100~\mathrm{GeV}) >  \textcolor{magenta}{w} \sim \mathcal{O}(100~\mathrm{MeV})$. 
\begin{table}[t]
\centering
\renewcommand{\arraystretch}{1.3}
\begin{tabular}{|c|c|c||c|c|c|}
\hline
\multicolumn{3}{|c||}{\textbf{$Z_2$ Even}} & \multicolumn{3}{c|}{\textbf{$Z_2$ Odd}} \\
\hline
\textbf{Field} & \textbf{Type} & ~~$B$~~ &
\textbf{Field} & \textbf{Type} & $B$ \\
\hline
$u,d$ & F & $\tfrac13$ & $D, D^c$ & F & $(\tfrac13,-\tfrac13)$ \\
$u^c,d^c$ & F & $-\tfrac13$ & $\textcolor{red}{D'}$ & F & $-\tfrac23$ \\
$e,\nu$ & F & $0$ & $\nu^c$ & F & $0$ \\
$e^c$ & F & $0$ & $\nu_E,E$ & F & $0$ \\
$\textcolor{red}{D'^c}$ & F & $\tfrac23$ & $E^c,\nu_E^c$ & F & $0$ \\
$\widetilde{\nu}^c$ & S & $0$ & $N$ & F & $0$ \\
$\widetilde{\nu}_E$ & S & $0$ & $\widetilde{u},\widetilde{d}$ & S & $\tfrac13$ \\
$\widetilde{\nu}^c_E$ & S & $0$ & $\widetilde{u}^c$ & S & $-\tfrac13$ \\
$\widetilde{D}^c$ & S & $-\tfrac13$ & $\widetilde{\nu},\widetilde{e}$ & S & $0$ \\
$\widetilde{E}$ & S & $0$ & $\widetilde{e}^c$ & S & $0$ \\
$\widetilde{E}^c$ & S & $0$ & $X_\mu^{+2/3}$ & GB & $\tfrac13$ \\
$Z^{\prime}_\mu$ & GB & $0$ & $W_\mu^{\prime\pm}$ & GB & $0$ \\
\hline
\end{tabular}
\caption{$Z_2$-parity and baryon number $B$ for all fermions (F), scalars (S), and gauge bosons (GB) of the theory.  }\label{tab:Z_2}
\end{table}

\subsection{Yukawa interactions and charged fermion masses}
\label{subsec:Yukawa_charged_fermion}
With the particle spectrum specified, we can write down the most general Yukawa interactions involving the fermions and the scalars of the model.  In doing so, we impose baryon number as well as the $Z_2$ symmetry given in Table~\ref{tab:Z_2}, but we allow for a soft breaking of the $Z_2$ symmetry. The gauge–invariant Yukawa Lagrangian for the charged fermions can be written as
\begin{align}
    \mathcal{L}_{\mathrm{Yuk}}&=Y_u\psi_{\alpha i } (\psi^c)^i_{\dot{\alpha}}\widetilde{\chi}_{\beta \dot{\beta}}\epsilon^{\alpha \beta} \epsilon ^{\dot{\alpha}\dot{\beta}}+Y_E\psi_{\alpha i } (\widetilde{\psi}^c)^i_{\dot{\alpha}}\chi_{\beta \dot{\beta}}\epsilon^{\alpha \beta} \epsilon ^{\dot{\alpha}\dot{\beta}}+Y_e\widetilde{\psi}_{\alpha i } (\psi^c)^i_{\dot{\alpha}}\chi_{\beta \dot{\beta}}\epsilon^{\alpha \beta} \epsilon ^{\dot{\alpha}\dot{\beta}}\notag\\
    &+ \frac{Y_d}{\sqrt{2}} \psi_{\alpha i}\omega_{k l}\widetilde{\psi}_{\beta j}\epsilon^{\alpha \beta} \epsilon^{ijkl}+\sqrt{2}Y_D (\psi^c)_{\dot{\alpha}}^i \Omega_{ij}(\widetilde{\psi}^c)_{\dot{\beta}}^j \epsilon^{\dot{\alpha}\dot{\beta}} + \frac{M_\omega}{2} \omega_{ij} \Omega_{kl} \epsilon^{ijkl} + \mathrm{h.c.}.
    \label{eqn:L-Yukawa-0}
    \end{align}
where $\alpha,\beta$ denotes $SU(2)_L$ indices, $\dot{\alpha},\dot{\beta}$ denotes $SU(2)_R$ indices, and $i,j,k,l$ presents $SU(4)_C$ indices. The last term in Eq. (\ref{eqn:L-Yukawa-0}) breaks $Z_2$ softly.  In component form \(\mathcal{L}_{\mathrm{Yuk}}\) reads as
\begin{align}
 \mathcal{L}_{\mathrm{Yuk}}&=Y_u\left((uD^c+\nu_Ee^c)\widetilde{e}-(uu^c+\nu_E\nu^c)\widetilde{\nu}_E^c-(dD^c+E e^c)\widetilde{\nu}+(du^c+E \nu^c)\widetilde{E}^c\right)\notag\\
    &+Y_{E}\left((u\widetilde{D}^c+\nu_E\widetilde{e}^c)e-(u \widetilde{u}^c+\nu_E\widetilde{\nu}^c)\nu_E^c-(d\widetilde{D}^c+E \widetilde{e}^c)\nu+(d\widetilde{u}^c+E \widetilde{\nu}^c)E^c\right)\notag\\
    &+Y_e\left((\widetilde{u}D^c+\widetilde{\nu}_Ee^c)e-(\widetilde{u}u^c+\widetilde{\nu}_E\nu^c)\nu_E^c-(\widetilde{d}D^c+\widetilde{E} e^c)\nu+(\widetilde{d}u^c+\widetilde{E}\nu^c)E^c\right)\notag\\
    &+Y_d\left(d^c(u\widetilde{E}-d\widetilde{\nu}_E)+d^c(E \widetilde{u}-\nu_E\widetilde{d})+\textcolor{red}{D'}_i u_j \widetilde{d}_k \epsilon^{ijk}+\textcolor{red}{D'}_i d_j \widetilde{u}_k \epsilon^{ijk}\right)\notag\\
    &+Y_D\left((\textcolor{red}{D'^c})_i(D^c)_j(\widetilde{u}^c)_k\epsilon^{ijk}-(\textcolor{red}{D'^c})_i(u^c)_j(\widetilde{D}^c)_k\epsilon^{ijk}+DD^c\widetilde{\nu}^c-De^c\widetilde{u}^c\right.\nonumber\\
 &\left.+D\nu^c\widetilde{D}^c-Du^c\widetilde{e}^c\right) + M_{\omega}\left(d^cD+\textcolor{red}{D'D'^c}\right)+\mathrm{h.c.}
 \label{eqn:L-Yukawa}
\end{align}
The soft breaking of $Z_2$ symmetry by the $M_\omega$ term allows us to generate a mass term for $\textcolor{red}{D'D'^c}$ which forms a Dirac fermion. The Yukawa interactions of the singlet fermion $N$ field will be discussed in Sec.~\ref{subsec:Yukawa_for_N}, where we also discuss the neutrino mass spectrum.

The charged fermion mass matrices obtained from Eq. (\ref{eqn:L-Yukawa}), with the VEVs defined in Eq. (\ref{eq:VEVs}), have the following form in a basis $f_i^T M_{ij} f^c_j$ (for $f=u,d,e,E, D,\textcolor{red}{D'})$:
\begin{eqnarray}
M_u^{3 \times 3} = Y_u \,v,~ M_{\textcolor{red}{D'}}^{3 \times 3} = M_\omega,~ {\cal M}_E^{6 \times 6} = \left( \begin{matrix} Y_e^T \kappa_L & 0 \cr -Y_u \textcolor{magenta}{w} & ~Y_E \kappa_R  \end{matrix} \right), ~{\cal M}_D^{6 \times 6} = \left( \begin{matrix} -Y_d^T \kappa_L & -Y_u \textcolor{magenta}{w} \cr M_\omega^T & Y_D \kappa_R \end{matrix} \right).
\label{eq:massmatrices}
\end{eqnarray}
Here, the basis chosen for the $2\times 2$ block matrices is $e-E$ and $d-D$.  We define, for these matrices,
\begin{equation}
\hat{M}_D = Y_D\, \kappa_R,~~~\hat{M}_E = Y_E^T\, \kappa_R
\label{eqn:MDhat}
\end{equation}
and proceed to diagonalize them perturbatively. 
The elements of the blocks in the first row of ${\cal M}_D$ are much smaller than those in the second row, due to the hierarchy $w \ll \kappa_L \ll \kappa_R$ realized in the model.  Similarly, the first column of ${\cal M}_E$ is much smaller than the nonzero second column. A general block matrix of the form
\begin{eqnarray}
{\cal M} = \left( \begin{matrix} m_0 & m \cr M' & M\end{matrix} \right)
\end{eqnarray}
with $m_0, m \ll M', M$ can be block-diagonalized to second order in $m, m_0$ by the following transformation:
\begin{equation}
U_L^T {\cal M} U_R = M^{\rm Block},
\label{eq:Block}
\end{equation}
where $U_R$ and $U_L$ are unitary matrices given, to linear order linear in $m_0, m$, as
\begin{eqnarray}
U_R = \left(\begin{matrix} (\mathbb{I} + x^\dagger x)^{-1/2} & x^\dagger M^\dagger \left(M (\mathbb{I} + x x^\dagger) M^{\dagger}  \right)^{-1/2} \cr -x(\mathbb{I}+x^\dagger x)^{-1/2} &  M^\dagger \left(M (\mathbb{I} + x x^\dagger) M^{\dagger}  \right)^{-1/2} \end{matrix}   \right),~~~U_L = \left( \begin{matrix}\mathbb{I} & \rho^* \cr -\rho^T & \mathbb{I}  \end{matrix} \right).
\end{eqnarray}
Here we have defined the $3 \times 3$ matrices $x$ and $\rho$ as
\begin{eqnarray}
x \equiv  M^{-1} M',~~~ \rho = (m_0\, x^\dagger + m)(\mathbb{I}+x x^\dagger)^{-1} M^{-1}~.
\end{eqnarray}
The block diagonal mass matrix has the form
\begin{eqnarray}
M^{\rm Block} = \left(\begin{matrix}  (m_0-m\,x)(1+x^\dagger x)^{-1/2} & \mathbb{O} \cr \mathbb{O} & (M M^\dagger + M' M'^\dagger)^{1/2} \end{matrix}  \right)~.
\end{eqnarray}
Here we use the symbols $\mathbb{I}$ to denote the $3 \times 3$ identity matrix and $\mathbb{O}$ for the $3 \times 3$ zero matrix with all elements equal to zero.

Each of the block diagonal matrix, written as $f'^T M^{\rm Block} f'^c$, where
\begin{equation}
f=U_L\, f',~~f^c = U_R \,f'^c,
\label{eq:prime}
\end{equation}
is further diagonalized by the transformation
\begin{equation}
f' = V_f f_0,~~f'^c = V_{f^c} f^c_0
\end{equation}
with the $f_0$ and $f^c_0$ denoting the mass eigenstates (for $f=u,d,e,E, D,\textcolor{red}{D'}$). These unitary matrices $V_f, V_{f^c}$ diagonalize the block mass matrix:
\begin{equation}
V_f^T M^{\rm Block} V_{f^c} = M^{\rm diagonal}~.
\label{eq:diag}
\end{equation}
The transformation from the gauge eigenstates to the mass eigenstates can then be obtained from the combined effects of Eq. (\ref{eq:Block}), Eq. (\ref{eq:prime}) and Eq. (\ref{eq:diag}).

These general results can be applied to block diagonalize ${\cal M}_D$ of Eq. (\ref{eq:massmatrices}).  We identify 
\begin{equation}
M = \hat{M}_D,~~x=\hat{M}_D^{-1} M_\omega^T,~~\rho = -(Y_d^T \kappa_L x^\dagger +Y_u \textcolor{magenta}{w})(\mathbb{I}+x x^\dagger)^{-1} \hat{M}_D^{-1}~. 
\label{eq:rho}
\end{equation}
We can now write down the transformations that take $(d,D)$ and $(d^c, D^c)$ into their respective mass eigenstates as
\begin{eqnarray}
\left( \begin{matrix} d \cr D\end{matrix} \right) = U_L \left( \begin{matrix} V_d &  \mathbb{O} \cr \mathbb{O} & V_D \end{matrix} \right) \left( \begin{matrix} d_0 \cr D_0 \end{matrix} \right),~~~
\left( \begin{matrix} d^c \cr D^c\end{matrix} \right) = U_R \left( \begin{matrix} V_{d^c} &  \mathbb{O} \cr \mathbb{O} & V_{D^c} \end{matrix} \right) \left( \begin{matrix} d^c_0 \cr D^c_0 \end{matrix} \right)~.
\end{eqnarray}
More explicitly, we have
\begin{eqnarray}
d &=& V_d \, d_0 + \rho^* V_D\, D_0 \nonumber \\
D &=& V_D \, D_0 -\rho^T V_d \, d_0 \nonumber \\
d^c &=& (\mathbb{I} + x^\dagger x)^{-1/2} \, V_{d^c} \, d^c_0 + x^\dagger M^\dagger \left(M(\mathbb{I} + x x^\dagger)M^\dagger\right)^{-1/2} V_{D^c} D_0^c \nonumber \\
D^c &=& -x\,(\mathbb{I} +x^\dagger x)^{-1/2}\,  V_{d^c} \,d^c_0 + M^\dagger \left(M(\mathbb{I} + x x^\dagger) M^\dagger \right)^{-1/2} V_{D^c}D_0^c~
\end{eqnarray}
with $M$, $x$ and $\rho$ defined in Eq. (\ref{eq:rho}).

Similarly, the $6 \times 6$ matrix ${\cal M}_E^T$ arising from Eq. (\ref{eq:massmatrices}) can be block diagonalized with the identification
\begin{equation}
x = 0,~~\rho_\ell  = -(Y_u^T \textcolor{magenta}{w}) \,\hat{M}_E^{-1}~.
\label{eqn:rhol}
\end{equation}
The block-diagonal matrices are then diagonalized following the definitions of Eq. (\ref{eq:diag}). We find explicitly the following transformation equations (with $U_R$ being an identity matrix):
\begin{eqnarray}
e &=& V_e\, e_0 \nonumber\\
E &=& V_E\, E_0 \nonumber \\
e^c &=& V_{e^c}\, e^c_0 + \rho_\ell ^* V_{E^c} \,E^c_0 \nonumber \\
E^c &=& -\rho_\ell ^T V_{e^c}\, e^c_0 + V_{E^c} \,E^c_0~.
\end{eqnarray}

As an application, the gauge interactions of $X_\mu$ in the original basis given as $\overline{D^c} \, \gamma_\mu\,  e^c X^\mu$ will contain light fermion fields as
\begin{equation}
\overline{D^c} \, \gamma_\mu\, e^c X^\mu \supset   \overline{d^c_0}\, \left[-V_{d^c}^\dagger \, (\mathbb{I}+x^\dagger x)^{-1/2} x^\dagger \, V_{e^c}\right]  \, \gamma_\mu \, e^c_0  \,X^\mu~.
\end{equation}
We shall present the complete interactions of the gauge bosons with the mass eigenstate fermions later in Sec.~\ref{sec:gauge} using these results. 
\subsection{Softly broken \texorpdfstring{$Z_2$}{Z2} symmetry}
\label{subsec:softly_broken_Z2}
\begin{figure}[!h]
    \centering
    \includegraphics[scale=0.2]{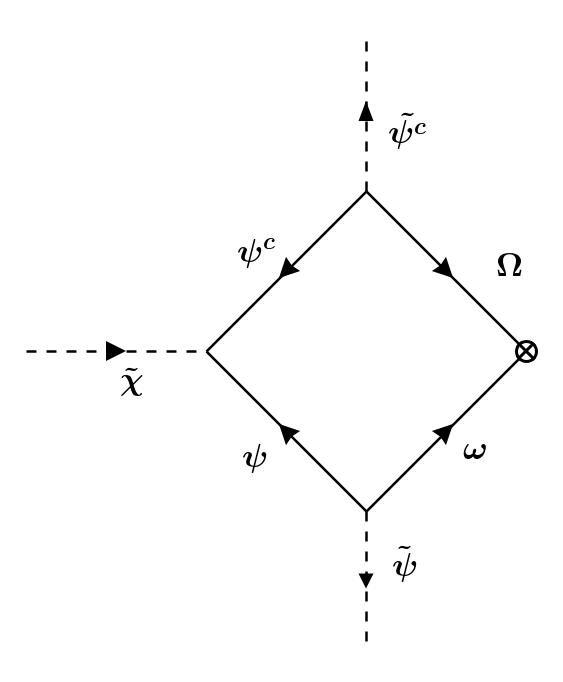}
    \caption{Induced cubic scalar coupling $\tilde{\psi}\tilde{\psi}^c \tilde{\chi}^*$ through the  softly broken $Z_2$ mass term $M_{\omega} \Omega\,\omega$.}
    \label{fig:induced vev}
\end{figure} 
Within the model, baryon number $B$ of each particle can be uniquely determined from its gauge interactions, which are listed in Table~\ref{tab:Z_2}. We shall assume that  $B$ is an exact symmetry of the Lagrangian. All quarks carry $B=1/3$, except for $\textcolor{red}{D'}$ which has $B = -2/3$. Note also that scalars with identical quantum numbers carry the same $B$-charge as the corresponding fermions. 

The couplings $\psi^{c}\omega\,\widetilde{\psi}^{\,c}$, $\omega\omega$, and $\Omega\Omega$ are absent in $\mathcal{L}_{\mathrm{Yuk}}$ (see Eq.~\eqref{eqn:L-Yukawa-0}), since they do not conserve baryon number. But the last term in Eq.~\eqref{eqn:L-Yukawa-0}, \(\frac{M_\omega}{2}\,\omega_{ij}\,\Omega_{kl}\,\epsilon^{ijkl},
\)
does conserve baryon number and is necessary to generate a Dirac mass for the down-type state $\textcolor{red}{D'}$ with $B$ charge of $-2/3$. This term breaks the $Z_2$ symmetry softly, and will induce mixing between $d^{c}$ and $D^{c}$ anti-quark fields. Such mixings are proportional to the matrix $x$ defined in Eq. (\ref{eq:rho}), viz.,
\begin{equation}
x_{ij} =(\hat{M}_D^{-1} M_\omega^T)_{ij} \approx 0.2~ \forall~ i,j.
\label{eq:xij}
\end{equation}
Here $\hat{M}_D = Y_D \kappa_R$. 
This numerical choice is motivated by the desire to realize the lowest mass leptoquark gauge boson $X_\mu$ in the model.  The logic behind this is as follows. The $x_{ij}$ parameters control the $d^c-D^c$ mixing angles, which lead to tree-level meson decays mediated by $X_\mu$. Only right-handed quarks take part in these decays.  To suppress these tree-level decays, we wish to have $x_{ij}$ as small as possible.  Reducing $M_\omega$ would help, but collider bounds on the vector–like quark $\textcolor{red}{D'}$ require $M_\omega \gtrsim 1.3~\text{TeV}$ (see Sec.~\ref{sec:Collider Implications}). Increasing $\kappa_R$ would also suppress $x_{ij}$, but at the price of raising the leptoquark gauge boson mass $M_X$ (see Eq.~\eqref{eqn:M_X}). We therefore go with the smallest $\kappa_R$ allowed by the $M_{Z'}$ mass limit, and require that all elements of $x_{ij}$ are of the same order.\footnote{Flavor effects may further lower the $d^c-D^c$ mixing parameters. If we consider the presence of $\Phi(1,1,15)$ Higgs with $\langle \Phi \rangle= v_{\Phi}/(2\sqrt{6})\,\mathrm{diag}\left(1,1,1,-3\right)$, the coupling $\Omega^{ab}\Phi^c_{b}\epsilon_{acdf}\omega^{df}$ is allowed, as well as a bare mass term, and the  $\textcolor{red}{D'D'^c}$ mass becomes $\left(M_{\omega}+v_{\Phi}/\sqrt{6}\right)$, and $d^c-D$ entry will be now $\left(M_{\omega}-v_{\Phi}/\sqrt{6}\right)$. In this case, $d^c-D^c$ mixing may even vanish.}
We assume $Y_D \sim \mathcal{O}(1)$ (consistent with perturbative unitarity~\cite{Allwicher:2021rtd}), $M_\omega \simeq 1.3~\text{TeV}$, and $\kappa_R \!\geq\! 6.3~\text{TeV}$, the lowest value allowed by the limit $M_{Z'} \!\geq\! 5.6~\text{TeV}$, and arrive at $x_{ij} \sim 0.2$. A benchmark point for the matrix elements $x_{ij}$ is given in Eq.~\eqref{eqn:x_bench2}.
In principle, all soft $Z_2$ breaking terms that conserve baryon number should be allowed in the Lagrangian. These include the terms
\begin{align}
    \mathcal{L}_{\text{soft}}
    &\supset 
    \, M_\chi \, \chi \chi+ M_{\tilde{\chi}}^2 \, \tilde{\chi} \tilde{\chi}+\mu \tilde{\psi}\tilde{\psi}^c \tilde{\chi}^*+\mathrm{h.c.}
    \label{eq:soft}
\end{align}
Among these three, only the third term $\tilde{\psi}\tilde{\psi}^c \tilde{\chi}^*$ will get induced in the presence of the term soft $Z_2$ breaking  $M_{\omega} \Omega\,\omega$. This can be seen by the induced trilinear term shown in Fig.~\ref{fig:induced vev} which is log-divergent. Therefore, we must allow this term in the Lagrangian. On the other hand, the terms $M_\chi \, \chi \chi$ and $ M_{\tilde{\chi}}^2 \, \tilde{\chi} \tilde{\chi}$ will not be induced through  $M_{\omega} \Omega\,\omega$ coupling. We can think of $M_{\omega}$ as arising from the VEV of a spurion field $(1,1,15)_H$ through the coupling $\Omega\,\omega\,(1,1,15)_H$. But this putative Higgs cannot couple to $\chi \chi$ or $\tilde{\chi} \tilde{\chi}$. Thus, it is consistent to consider only the soft breaking term \(\mu \tilde{\psi}\tilde{\psi}^c \tilde{\chi}^*\) in the Lagrangian, along with the  $M_{\omega} \Omega\,\omega$ mass term. To summarize, the soft $Z_2$ breaking Lagrangian is taken to be
\begin{equation}
{\cal L}_{{\rm soft}~\cancel{Z_2}} = M_\omega \Omega \omega + \mu \tilde{\psi}\tilde{\psi}^c \tilde{\chi}^* + \mathrm{h.c.}
\label{eq:Z2final}
\end{equation}

To get an estimate of the order of magnitude for the coefficient $\mu$ in Eq. (\ref{eq:Z2final}) we evaluate the log-divergent correction from Fig.~\ref{fig:induced vev}, which is found to be
\begin{equation}
    \mu \approx \dfrac{3}{16\pi^2}\,\mathrm{Tr}\left(Y_uY_D^{\dagger}M_{\omega}Y_d^{\dagger}\right)\left(\mathrm{ln}\dfrac{\Lambda^2}{M_D^2}\right),
    \label{eq:estimate}
\end{equation}
where 3 is the color factor. With the assumption that the cut-off scale for the theory is \(\Lambda\sim 10^{11}\,\mathrm{GeV}\), $Y_{\mathrm{top}}\sim Y_{D}\sim 1, Y_{\mathrm{bottom}}\sim 1.6 \times 10^{-2}, M_D\sim 6.3 \,\mathrm{TeV},~M_{\omega}\sim 1.3\,\mathrm{TeV}$, Eq. (\ref{eq:estimate}) yields
\begin{equation}
\mu \sim 13\, \mathrm{GeV}.
\label{eq:mu}
\end{equation}
We may translate this naturalness bound on \(\mu\) into a constraint on the induced vacuum expectation value $\langle \widetilde{\nu} \rangle=\textcolor{magenta}{w}$ using
\(\mu_{\tilde{\chi}}\sim 10~\mathrm{TeV}\)  for the mass of the corresponding scalar:
\begin{equation}
    \textcolor{magenta}{w}=\frac{\mu \kappa_L\kappa_R}{\mu_{\tilde{\chi}}^2}\sim 100\,\mathrm{MeV}.
\end{equation}
Here, we have used $\kappa_R \geq 6.3$ TeV using the limit $M_{Z'}\geq 5.6$ TeV, see Fig.~\ref{fig:LHC:ZR}.

\subsection{Neutrino mass generation}
\label{subsec:Yukawa_for_N}

With the choice  of \(N\) being \(Z_2\) odd (see Table~\ref{tab:Z_2}), we can write down the Yukawa couplings involving \(N\):
\begin{align}
  \mathcal{L}_{\text{Yuk}}^N=&Y_{\nu_E}\psi_{\alpha i}\widetilde{\psi}^{*\alpha i}N+Y_{\nu^c}(\psi^c)^i_{\dot{\alpha}} (\widetilde{\psi}^c)_i^{*\dot{\alpha}}N+Y_{\nu}\chi_{\alpha\dot{\alpha}}(\tau_2\widetilde{\chi}^*\tau_2)_{\beta\dot{\beta}}N\epsilon^{\alpha\beta}\epsilon^{\dot{\alpha}\dot{\beta}}+\frac{1}{2}M_NN^2+\mathrm{h.c.}
\end{align}
Expanding in component form, this becomes
\begin{align}  \mathcal{L}_{\text{Yuk}}^N=&Y_{\nu_E}(u\widetilde{u}^*+\nu_E\widetilde{\nu}_E^*+d\widetilde{d}^*+E\widetilde{E}^*)N+Y_{\nu^c}(u^c\widetilde{u}^{c*}+\nu^c\widetilde{\nu}^{c*}+D^c\widetilde{D}^{c*}+e^c\widetilde{e}^{c*})N\notag\\
&+Y_{\nu}(E^c\widetilde{E}^{c*}+\nu_E^c\widetilde{\nu}_E^{c*}+\nu\widetilde{\nu}^*+e\widetilde{e}^*)N+\frac{1}{2}M_NN^2+\mathrm{h.c.}
\label{eq:Yuk-N}
\end{align}
The term \(\chi_{\alpha\dot{\alpha}}\widetilde{\chi}_{\beta\dot{\beta}}N\epsilon^{\alpha\beta}\epsilon^{\dot{\alpha}\dot{\beta}}\) is absent here because it would be a hard breaking of \(Z_2\) by a dimension four operator. 
\begin{figure}[!h]
    \centering
    \includegraphics[scale=.25]{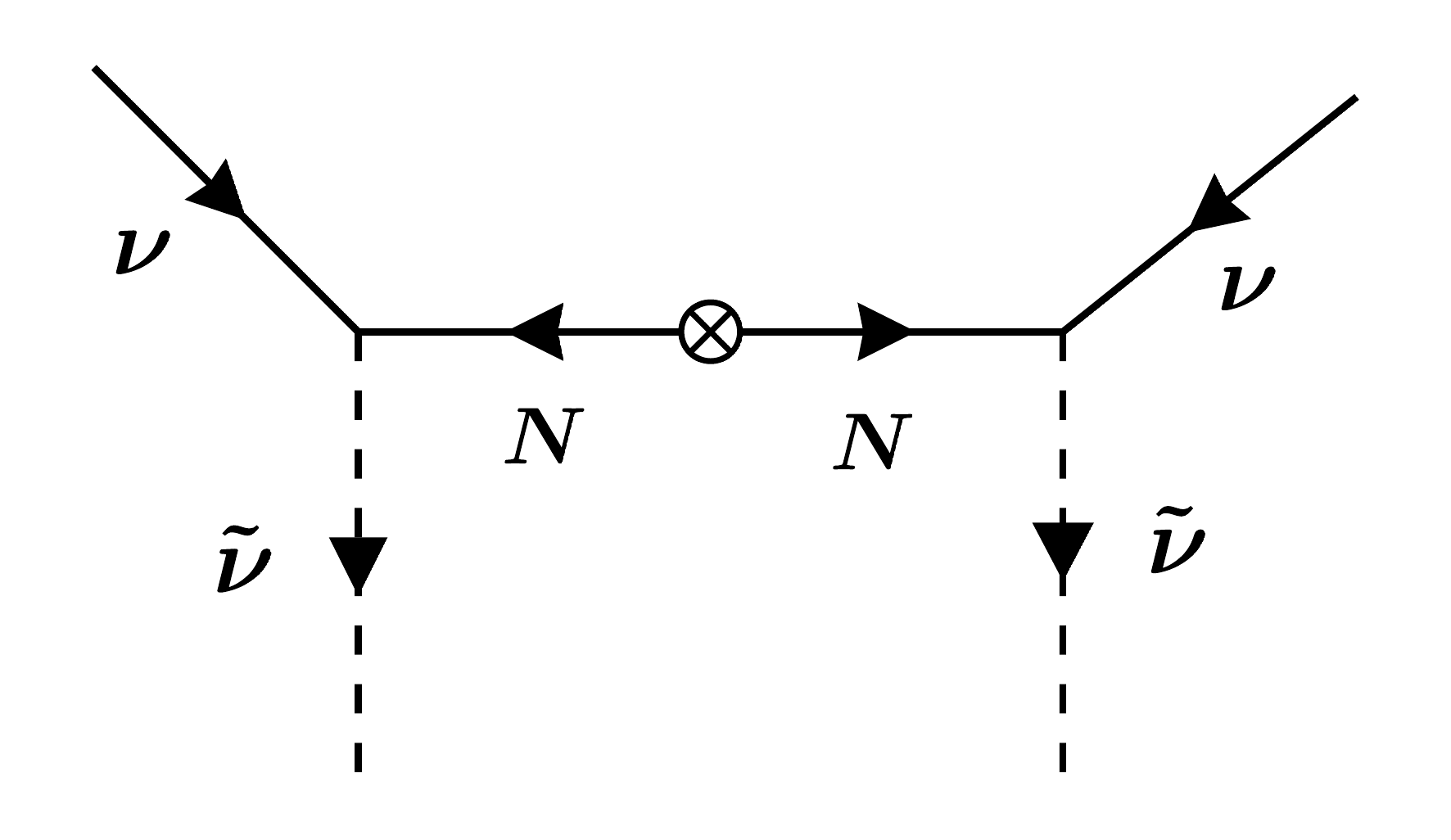}
    \includegraphics[scale=.25]{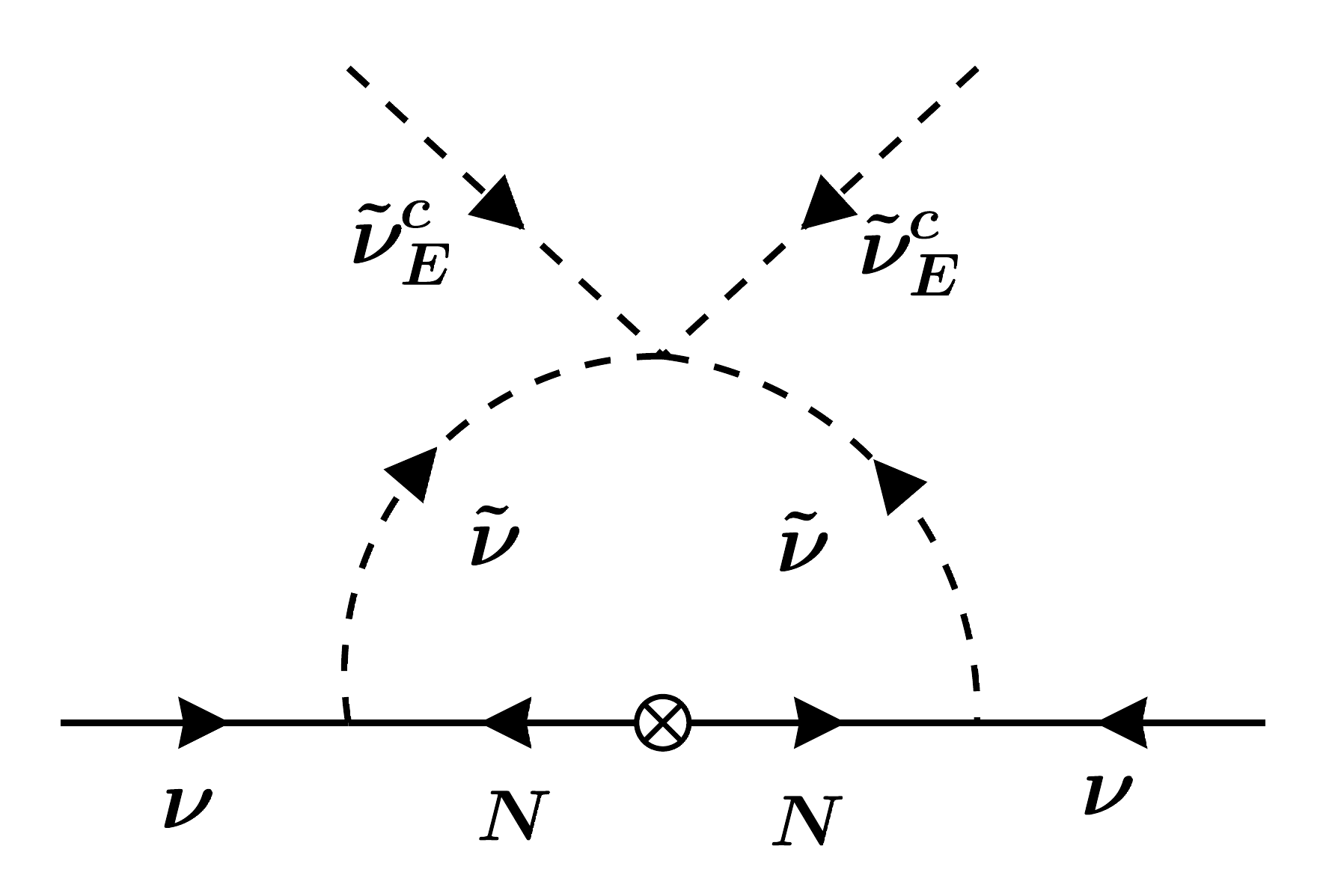} \\
 (a)~~~~~~~~~~~~~~~~~~~~~~~~~~~~~~~~~~~~~~~~~~~~~~~~~~~~~~(b) \\
    \caption{(a) Tree-level neutrino mass through seesaw diagram, (b) One-loop neutrino mass through Scotogenic-type diagram.}
    \label{fig:neutrino_mass}
\end{figure} 
Collecting terms from Eq. (\ref{eqn:L-Yukawa}) and Eq. (\ref{eq:Yuk-N}), we can write down the neutral lepton mass matrix in the basis $(\nu,\nu_E, \nu_E^c, \nu^c, N)$ as:
\begin{equation}
M_\nu=\begin{pmatrix}
0 & \mathcal B\\[2pt]
\mathcal B^{T} & \mathcal{H}
\end{pmatrix}
\end{equation}
with 
\begin{equation}
\mathcal B=\big(0~0~0~Y_\nu \textcolor{magenta}{w^*}\big),\qquad \mathcal{H}=\begin{pmatrix}
0 & -Y_E\kappa_R & -Y_uv & Y_{\nu_E}\kappa_L\\
-\,Y_E^{T}\kappa_R & 0 & -Y_e^{T}\kappa_L & Y_\nu v^*\\
-Y_u^Tv& -Y_e^{T}\kappa_L & 0 & Y_{\nu^c}\kappa_R\\
Y_{\nu_E}^{T}\kappa_L & Y_\nu^{T}v^* & Y_{\nu^c}^{T}\kappa_R & M_N
\end{pmatrix}.
\label{eq:H-matrix}
\end{equation}
Integrating out the heavy block $\mathcal{H}$~\cite{Mohapatra:1979ia,Grimus:2000vj} one would get the tree-level light neutrino mass matrix as 
\begin{equation}
m_\nu^{\rm tree}=-\mathcal B\mathcal{H}^{-1} \mathcal B^{T}
=-\big(Y_\nu \textcolor{magenta}{w^*}\big)\big[\mathcal{H}^{-1}\big]_{NN}\big(Y_\nu \textcolor{magenta}{w^*}\big)^{T}.
\label{eq:type-I neutrino mass}
\end{equation}
To calculate the closed form  expression for\(\big[\mathcal{H}^{-1}\big]_{NN}\), let's partition \(\mathcal H\) into \(2\times 2\) blocks by grouping \((\nu_E,\nu_E^c)\) and \((\nu^c,N)\): 
\begin{equation}
 \mathcal H=
\begin{pmatrix}
\underline{A} & \underline{C}\\ \underline{C}^{T} & \underline{D}
\end{pmatrix}.
\end{equation}
With the assumption that \(Y_E\kappa_R\) is invertible, the Schur complement can be written as\footnote {For the inverse of a 2$\times$2 block matrix, assuming that the  (1,1) block is non-singular, we can implement the following recipe~\cite{lu2002inverses}: 
\begin{align*}
    \left( \begin{matrix}
        A & B \\
        C & D 
    \end{matrix}
    \right)^{-1}=\begin{pmatrix}
 A^{-1}+A^{-1}B\,S^{-1}C\,A^{-1} & -\,A^{-1}B\,S^{-1}\\[2pt]
-\,S^{-1}C\,A^{-1} & \ \ S^{-1}
        \end{pmatrix}
\end{align*}
where the Schur complement $S=D-CA^{-1}B$ of $A$ is invertible.}
\begin{equation}
\underline{S}\ \equiv\ \underline{D}- \underline{C}^{T}\underline{A}^{-1}\underline{C} \ =\ \begin{pmatrix}s_{11}&s_{12}\\ s_{21}&s_{22}\end{pmatrix} 
\end{equation}
which gives 
\begin{equation}
    \big[\mathcal H^{-1}\big]_{NN}=(\underline{S}^{-1})_{22}
=\big(s_{22}-s_{21}\,s_{11}^{-1}\,s_{12}\big)^{-1}~.
\label{eq:HNN-block}
\end{equation}
With the definition \(\underline{E}\equiv Y_E\kappa_R\), the block multiplication leads to the relations 
\begin{align}
s_{11}&= v\kappa_L\Big(Y_u^{T}(\underline{E}^{-1})^TY_e^{T}+Y_e\underline{E}^{-1}Y_u\Big),\label{eq:s11}\\
s_{12}&=Y_{\nu^c}\kappa_R\ +|v|^{2}Y_u^{T}(\underline{E}^{-1})^TY_\nu\ +\kappa_L^{2}Y_e\underline{E}^{-1}Y_{\nu_E},\\
s_{21}&=s_{12}^{T},\\
s_{22}&=M_N+\kappa_L v^*\Big(Y_{\nu_E}^{T}(\underline{E}^{-1})^TY_\nu+Y_\nu^{T}\underline{E}^{-1}Y_{\nu_E}\Big).
\label{eq:s22}
\end{align}
In the hierarchical limit \(\kappa_R\gg \kappa_L, v\), the \((\nu_E,\nu_E^c)\) block dominates the heavy sector. Let's define,
\begin{equation}
   \underline{X}\equiv Y_eY_E^{-1}Y_u,\qquad \underline{S_0}\equiv \underline{X}+\underline{X}^{T}.
\end{equation}
Then from Eq.~\eqref{eq:s11}, we see that \(s_{11}=\dfrac{v\kappa_L}{\kappa_R}\,\underline{S}_0\), while  \(s_{12}\approx Y_{\nu^c}\kappa_R\) and \(s_{22}=M_N+\mathcal O\!\big(\kappa_Lv^*/\kappa_R\big)\). Keeping the leading term in \(v/\kappa_R,\kappa_L/\kappa_R\) we find from Eq.~\eqref{eq:HNN-block}
\begin{equation}
\big[\mathcal H^{-1}\big]_{NN}\simeq\
\frac{v\kappa_L}{\kappa_R^{3}}
\Big(Y_{\nu^c}^{T}\underline{S}_0^{-1}Y_{\nu^c}\Big)^{-1}
\label{eq:HNN-leading}
\end{equation}
Here, the explicit \(M_N\) dependence is sub-leading provided that \(M_N \ll \kappa_R^{3}/(v\,\kappa_L)\) after the Yukawa factors have been accounted for. In this controlled limit, we see a ``triple seesaw" structure for neutrino mass:
\begin{equation}
    \ m_\nu^{\rm tree}\ \simeq\
\frac{v\,\kappa_L}{\kappa_R^{3}}\;
(Y_\nu \textcolor{magenta}{w^*})\ \Big(Y_{\nu^c}^{T}\,\underline{S}_0^{-1}\,Y_{\nu^c}\Big)^{-1}\ (Y_\nu \textcolor{magenta}{w^*})^{T}~.
\end{equation}
The effective operator that generates this mass is of dimension-seven, given as $(LL HH) (H^\dagger H)$~\cite{Babu:2009aq}. 
Thus, we can write down the light neutrino mass matrix in terms of the original matrices explicitly as
\begin{equation}
m_\nu^{\rm tree}\simeq
\frac{v\kappa_L}{\kappa_R^{3}}
(Y_\nu \textcolor{magenta}{w^*})Y_{\nu^c}^{-1}\Big[Y_eY_E^{-1}Y_u+Y_u^T(Y_E^{-1})^TY_e^T\Big](Y_{\nu^c}^{T})^{-1}(Y_\nu \textcolor{magenta}{w^*})^{T}.
\label{eq:mnuleading}
\end{equation}

We note in passing that in the opposite limit where  \(M_N \gg \kappa_R^{3}/(v\,\kappa_L)\), the neutrino mass matrix takes the form
 \(m_{\nu}^{\rm tree}\approx-\big(Y_\nu \textcolor{magenta}{w^*}\big)M_N^{-1}\big(Y_\nu \textcolor{magenta}{w^*}\big)^{T}\). This is a dimension-five operator shown in Fig.~\ref{fig:neutrino_mass} (a).
 However, in this limit, the mass of the right-handed neutrino states $\nu^c$ will be well below the electroweak scale, which we wish to avoid in our setup.

We now present a benchmark point that gives the right order of magnitude of the neutrino masses from $m_\nu^{\rm tree}$ given in Eq. (\ref{eq:mnuleading}):
\[ {\rm Input:}~~
\kappa_R = 7~\text{TeV},\qquad 
\kappa_L = v = 123~\text{GeV},\qquad 
\textcolor{magenta}{w} = 0.1~\text{GeV},
\]
\[
Y_\nu = 0.24,\qquad 
Y_e = 10^{-2},\quad Y_E^{-1}= 1,\quad Y_u= 1,\qquad
Y_{\nu^c} = 0.10,\]
\[
{\rm Output:}~~~    m_{\nu}^{\mathrm{tree}}\approx 5\times10^{-2}~\mathrm{eV}.
\]
\paragraph{Radiative (scotogenic-type) neutrino mass:}
The tree-level contribution to neutrino mass shown in Eq. (\ref{eq:mnuleading}) is proportional to ${\color{magenta} w^2}$, where ${\color{magenta} w} \sim 100$ MeV.  There are also one-loop corrections to the neutrino mass from the diagram shown in Fig.~\ref{fig:neutrino_mass}(b).  These loop corrections are independent of the small VEV {\color{magenta} $w$} and can be comparable to or exceed the tree-level mass. Here we evaluate these loop-induced neutrino masses in the limit of neglecting ${\color{magenta} w}$.  In this limit, the discrete $Z_2$ symmetry is unbroken, and the diagram shown in Fig.~\ref{fig:neutrino_mass} (b) is analogous to the radiative neutrino mass generation in the Scotogenic model~\cite{Ma:2006km,Tao:1996vb}. 

The scalar field that couples with the external \(\nu\) line in Fig.~\ref{fig:neutrino_mass}(b) is $\widetilde{\nu}$ which is $Z_2$–odd (see Table~\ref{tab:Z_2}), which does not mix with the other neutral scalars in the limit of setting ${\color{magenta} w}$ to zero.  The real and imaginary components of the $\widetilde{\nu}$ will acquire slightly different masses from the quartic couplings in the Higgs potential. We define their mass splitting as
\begin{equation}
m_{R,a}^2=m_a^2+\tfrac12\Delta m_a^2,\qquad
m_{I,a}^2=m_a^2-\tfrac12\Delta m_a^2.
\label{eqn:Deltam2}
\end{equation}
$\Delta m_a^2$ is small since the splitting is proportional to $v^2$, while the $\widetilde{\nu}$ is in the multi-TeV range.  
To evaluate the loop diagrams we choose a basis where the mass matrix for the $N$ fields, $M_N$, is real and diagonal. The radiatively induced neutrino mass is found to be
\begin{equation}
(m_\nu^{\mathrm {loop}})_{\alpha\beta}\;\simeq\;\frac{1}{16\pi^2}\sum_k (Y_\nu)_{\alpha k}(Y_\nu)_{\beta k}\;
M_{N_k}\,\frac{\Delta m^2}{m^2-M_{N_k}^2}\!
\left[1-\frac{M_{N_k}^2}{m^2-M_{N_k}^2}\ln\!\frac{m^2}{M_{N_k}^2}\right].
\end{equation}
with \(m=m_{\tilde{\nu}}\), and \(\Delta m^2=\Delta m_{\tilde{\nu}}^2\) is defined in Eq.~\eqref{eqn:Deltam2}.

We present a benchmark point where the right order of magnitude for the neutrino mass arises from the loop corrections.
\[ {\rm Input:}~~
\kappa_R=7~\text{TeV},\quad \kappa_L=123~\text{GeV},\quad v=123~\text{GeV},\quad \textcolor{magenta}{w}=0~\text{GeV},
\]
\[
m=1.5~\text{TeV},\quad M_N=1.0~\text{TeV},\quad Y_\nu=0.1,\quad   (Y_\nu V_N)_{\alpha k}(Y_\nu V_N)_{\beta k}\sim 0.01.
\]
\begin{equation}
{\rm Output:}~~  m_\nu^{\mathrm {loop}}\approx 5\times 10^{-2}~\mathrm{eV},
\end{equation}
with a modest CP-even/odd splitting of \(\Delta m=0.05~\mathrm{GeV}\). 

This analysis shows that naturally small neutrino masses can be generated in the model even with the PS breaking scale being in the multi-TeV range.  The mass generation mechanism may be either tree-level $d=7$ operator, or loop-induced $d=5$ operator, or a combination of the two.
\subsection{Gauge Sector}
\label{sec:gauge}
Now we turn to the masses of the various gauge bosons in the model.

\subsubsection{Charged Gauge Boson Masses}
The kinetic terms for each scalar field are given by:
\begin{equation}    
\mathcal{L}_{\text{kin}}^S=\mathrm{Tr}\left[(D_{\mu} \widetilde{\psi})^{\dagger}(D^{\mu} \widetilde{\psi})+(D_{\mu} \widetilde{\psi}^{c})^{\dagger}(D^{\mu} \widetilde{\psi}^{c})+(D_{\mu} \widetilde{\chi})^{\dagger}(D^{\mu} \widetilde{\chi})\right].
\end{equation}
The covariant derivatives are defined as:
\begin{align}
    D_{\mu} \widetilde{\psi}^{\alpha i}&= \partial_{\mu}\widetilde{\psi}^{\alpha i}-ig_{L}(W_{L\mu}^b\frac{\sigma^b}{2})^{\alpha}_{\beta}\widetilde{\psi}^{\beta i}-ig_4 (G_{\mu}^d\frac{T_4^d}{2})^{i}_{\gamma}\widetilde{\psi}^{\alpha \gamma}\notag\\
    D_{\mu} (\widetilde{\psi}^{c})^{\dot{\alpha}}_{i}&= \partial_{\mu}(\widetilde{\psi}^{c})^{\dot{\alpha}}_{i}-ig_{R}(W_{R\mu}^b\frac{\sigma^b}{2})^{\dot{\alpha}}_{\dot{\beta}}(\widetilde{\psi}^{c})^{\dot{\beta}}_{i}+ig_4 (G_{\mu}^d\frac{T_4^d}{2})^{\gamma}_{i}(\widetilde{\psi}^{c})^{\dot{\alpha}}_{\gamma}\notag \\
    D_{\mu} \widetilde{\chi}^{\alpha \dot{\alpha}}&= \partial_{\mu}\widetilde{\chi}^{\alpha \dot{\alpha}}-ig_{L}(W_{L\mu}^b\frac{\sigma^b}{2})^{\alpha}_{\beta}\widetilde{\chi}^{\beta \dot{\alpha}}-ig_{R}(W_{R\mu}^b\frac{\sigma^b}{2})^{\dot{\alpha}}_{\dot{\beta}}\widetilde{\chi}^{\alpha \dot{\beta}}~.
    \label{eqn: Covariant-derivative}
\end{align}
Here, $b=1-3$ runs over  $SU(2)_{L/R}$ generators, while $d=1-15$  runs over $SU(4)_C$ generators. The indices \(\alpha\), \(\dot{\alpha}\), and \(i\), respectively, denote the indices for \(SU(2)_L\), \(SU(2)_R\), and \(SU(4)_C\).
 The charged gauge bosons $W_{L\mu}^\pm$ and $W_{R\mu}^\pm$ mix, with the mixing angle being proportional to the soft $Z_2$  breaking term, and their mass matrix is given by:
\begin{equation}
M_{W-W'}^2=\left(
\begin{array}{cc}
 \frac{1}{2} g_L^2(\kappa_L^2+v^2+\textcolor{magenta}{w^2}) & g_L\,g_R\,v\,\textcolor{magenta}{w} \\
\bullet & \frac{1}{2}g_R^2(\kappa_R^2+v^2+\textcolor{magenta}{w^2}) \\
\end{array}
\right)~.
\end{equation}
We can define the physical states as,
\begin{equation}
W_1^{\mu \pm} = \cos \zeta\, W^{\mu \pm} - \sin\zeta\, W'^{\mu \pm}, \quad W_2 = \sin\zeta\,W^{\mu \pm} + \cos\zeta\,W'^{\mu \pm}, \nonumber \\
\end{equation}
with the mixing angle $\zeta$ given by
\begin{equation}
\mathrm{tan}\,2\zeta= \frac{4\,g_L g_R\, v \,\textcolor{magenta}{w}}{
g_R^{2}\!\left(k_R^{2}+v^{2}+\textcolor{magenta}{w^2}\right)
- g_L^{2}\!\left(k_L^{2}+v^{2}+\textcolor{magenta}{w^2}\right)}.
\end{equation}
Their mass eigenvalues are given by
\begin{align}
M_{W_1^{\mu \pm}}^2 &\simeq \frac{1}{2} g_L^2(\kappa_L^2+v^2+\textcolor{magenta}{w^2}),\\
M_{ W_2^{\mu \pm}}^2 &\simeq \frac{1}{2}g_R^2(\kappa_R^2+v^2+\textcolor{magenta}{w^2}).
\label{eqn:WL-WR}
\end{align}
With \(\textcolor{magenta}{w}=0\), $W_1^{\mu \pm} $ is identified as the $W$ boson. 

The mass of the leptoquark gauge boson $X_\mu(3,1,\frac{2}{3})$ is given by
\begin{equation}
    M_{X^{\mu}}^2=\frac{1}{2} g_4^2(\kappa_L^2+\kappa_R^2).
    \label{eqn:M_X}
\end{equation}
\subsubsection{Neutral Gauge Boson Masses}
\label{Neutral Gauge Boson Masses}
The neutral gauge bosons associated with the diagonal generators of $SU(2)_L$, $SU(2)_R$,  and $SU(4)_C$ are denoted as $W_{\mu L}^0$, $W_{\mu R}^0$,  and $G_\mu^{15} $, respectively.  These gauge fields will mix and produce the massless photon field $A_{\mu}$, while the other two orthogonal states $Z_{\mu L}$ and $Z_{\mu R}$ will further mix. The corresponding mass-squared matrix in the basis $(W_{\mu L}^0, W_{\mu R}^0, G_\mu^{15})$ reads as
\begin{equation}
M_0^2=
\begin{pmatrix}
        \frac{1}{2}g_L^2(\kappa_L^2+v^2+\textcolor{magenta}{w^2}) & -\frac{1}{2}g_Lg_R(v^2+\textcolor{magenta}{w^2}) & -\sqrt{\frac{3}{8}}g_4g_L\kappa_L^2 \\
        \bullet & \frac{1}{2}g_R^2(\kappa_R^2+v^2+\textcolor{magenta}{w^2}) & -\sqrt{\frac{3}{8}}g_4g_R\kappa_R^2 \\
        \bullet & \bullet & \frac{3}{4}g_4^2(\kappa_L^2+\kappa_R^2)
\end{pmatrix}.
\end{equation}
where `$\bullet$' represents the symmetric elements. The compositions of these fields, expressed in a suitable basis, take the form:
\begin{align}
    A_{\mu}&=\frac{\sqrt{3}g_4 g_R W_{\mu L}^0+\sqrt{3} g_4 g_L W_{\mu R}^0+\sqrt{2} g_L g_R G_\mu^{15}}{\sqrt{2 g_L^2 g_R^2 + 3 g_4^2 (g_L^2+g_R^2)}},\\
    Z_{\mu}&=\frac{g_L(3 g_4^2+2 g_R^2) W_{\mu L}^0 - 3 g_4^2 g_R W_{\mu R}^0 - \sqrt{6} g_R^2 g_4 G_\mu^{15}}{\sqrt{(3 g_4^2+2 g_R^2)(2 g_L^2 g_R^2 + 3 g_4^2 (g_L^2 + g_R^2))}},
    \label{eq:ZL basis}\\
    Z'_{\mu}&= \frac{-\sqrt{2} g_R W_{\mu R}^0 +\sqrt{3} g_4  G_\mu^{15}}{\sqrt{3 g_4^2+2 g_R^2}}
    \label{eq:ZR basis}.
\end{align}
In this basis, the photon field decouples from other gauge fields, leading to a $2\times 2$ mass-mixing matrix $B^2$ for $Z-Z'$ which reads as:
\begin{equation}
M_{Z-Z'}^2=\frac{1}{2} \left(
\begin{array}{cc}
 \left(g_L^2+g_Y^2\right) \left(\kappa _L^2+v^2+\textcolor{magenta}{w^2}\right) & \quad \sqrt{\dfrac{g_L^2+g_Y^2}{g_R^2-g_Y^2}} \left(\left(v^2+\textcolor{magenta}{w^2}\right)\left(g_R^2-g_Y^2\right)-g_Y^2 \kappa _L^2\right)\\
 
 \bullet & \dfrac{\left(v^2+\textcolor{magenta}{w^2}\right) \left(g_R^2-g_Y^2\right){}^2+g_Y^4 \kappa _L^2+g_R^4 \kappa _R^2}{g_R^2-g_Y^2} \\
\end{array}
\right)~.
\end{equation}
Here we have traded $g_4$ in favor of $g_Y$ using the following coupling matching condition
\begin{equation}
\dfrac{1}{g_Y^2} = \dfrac{1}{g_R^2} + \dfrac{2}{3}\dfrac{1}{g_4^2}.
\end{equation}
The physical states can be written as
\begin{equation}
   Z= \cos \xi\, Z_L - \sin\xi\, Z_R, \quad Z' = \sin\xi\, Z_L + \cos\xi \,Z_R. 
\end{equation}
with the small mixing angle $\xi$ given approximately by
\begin{equation}
\xi \simeq \frac{\left(\left(g_R^2-g_Y^2\right)\left(v^2+\textcolor{magenta}{w^2}\right) -g_Y^2\kappa _L^2\right)}{g_R^4 \kappa _R^2}\sqrt{\left(g_L^2+g_Y^2\right) \left(g_R^2-g_Y^2\right)}~.
\end{equation}
The mass eigenvalues of $Z$ and $Z'$ are given by 
\begin{align}
    M_{Z}^2 &\simeq \frac{1}{2} (g_L^2+g_Y^2)\, (\kappa_L^2+v^2+\textcolor{magenta}{w^2}),\\
    M_{Z'}^2&\simeq
\frac{1}{2}\dfrac{\left(v^2+\textcolor{magenta}{w^2}\right)\left(g_R^2-g_Y^2\right){}^2+g_Y^4 \kappa _L^2+g_R^4 \kappa _R^2}{g_R^2-g_Y^2}.
\label{eqn:M_Z2}
\end{align}
With small \(\xi\), $Z$ is identified as the SM $Z$ boson.
\begin{figure}[h!]
\centering
\includegraphics[scale=0.3]{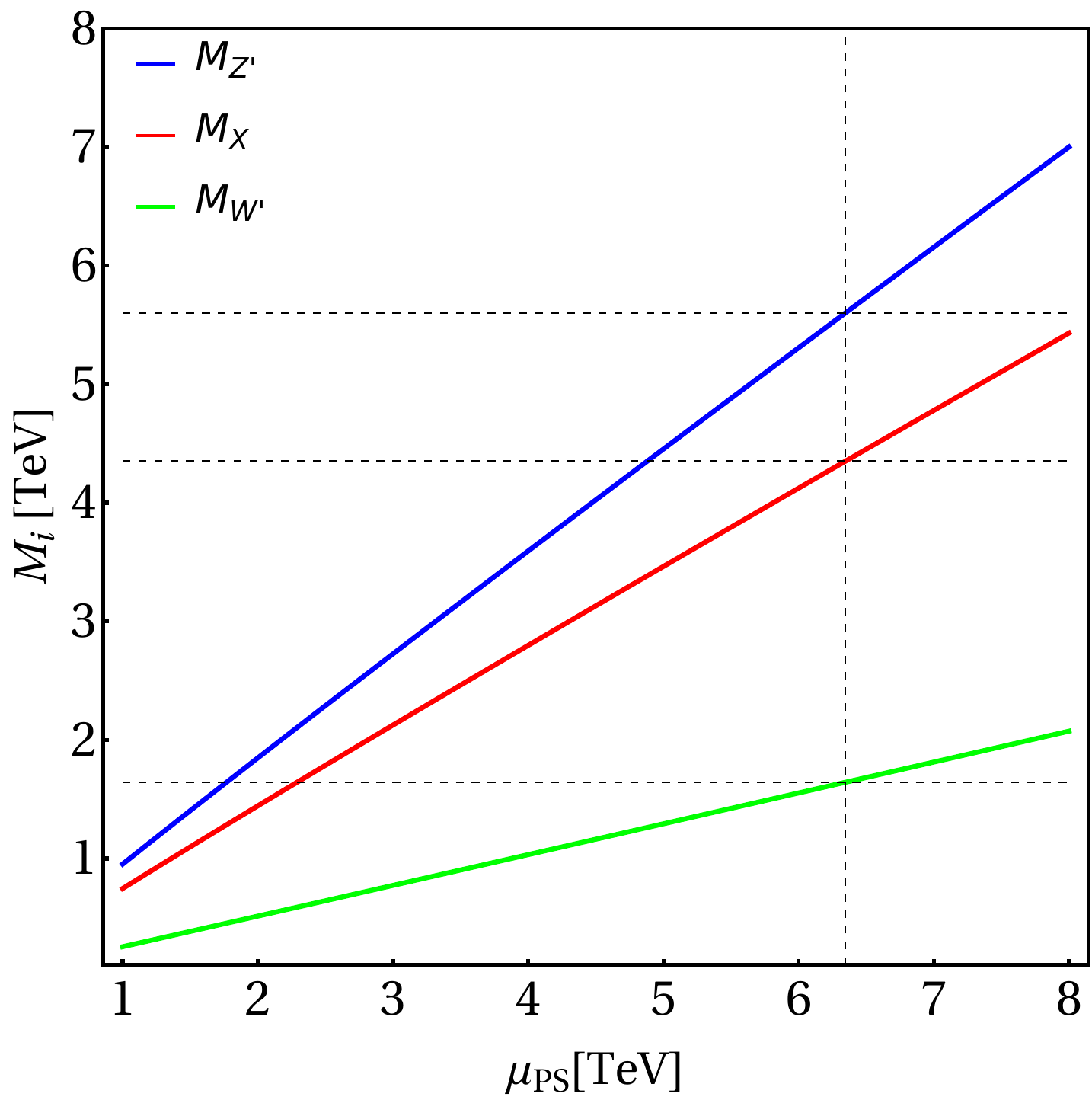}
\caption{Masses of the heavy gauge bosons for different choices of the Pati--Salam breaking scale \(\mu_{PS}\). The curves for \(M_{W'}\), \(M_X\), and \(M_{Z'}\) are obtained numerically from Eqs.~\eqref{eqn:WL-WR}, \eqref{eqn:M_X}, and \eqref{eqn:M_Z2} respectively, using gauge couplings fixed by the matching conditions in Eq.~\eqref{eqn:matching_PS} and by running the Standard Model gauge couplings up to \(\mu_{PS}\) with their renormalization group equations as discussed in Sec.~\ref{Neutral Gauge Boson Masses}. The horizontal dashed line at 5.6 TeV corresponds to the lower limit on the $Z^\prime$ mass from LHC constraints on the $Z^\prime\to \ell\overline \ell$ process (see Sec.~\ref{sec:3}). This collider constraint on $M_{Z^\prime}$ indirectly sets the lower bounds on the masses of $X_\mu$ and $W^\prime_{\mu}$.  }
\label{fig:gauge_boson_masses}
\end{figure}

To get the relations among the gauge boson masses, we solve the two-loop renormalization group equations for the gauge couplings within the SM:
\begin{equation}
    \mu\frac{dg_{i}}{d\mu}=\frac{1}{(4\pi)^2}\beta_{g_i}^{(1)}+\frac{1}{(4\pi)^4}\,\beta^{(2)}_{g_i}.
    \label{eqn:beta_function}
\end{equation}
Since the new particles present in the model lie around or above the PS breaking scale, they do not contribute to the running of the gauge couplings below the scale of PS symmetry breaking. Thus, we include the two-loop beta functions in the SM given as
\begin{align}
    \beta_{g_i}^{(1)}&=b_i[g_i(\mu)]^3,
    \label{eqn:beta1}\\
    \beta^{(2)}_{g_i}
  &= g_i^3(\mu)\left[
       \sum_{j} b_{ij}\,g_j^2(\mu)-C_{iu}\,\mathrm{Tr}\big(Y_uY_u^\dagger\big)
-C_{id}\,\mathrm{Tr}\big(Y_dY_d^\dagger\big)
   -C_{ie}\,\mathrm{Tr}\big(Y_eY_e^\dagger\big)
     \right].
     \label{eqn:beta2}
\end{align}
with $b_{i},\,b_{ij},\,C_{i(u,d,e)}$ are the SM two-loop coefficients~(summarized in  appendix~\ref{app:B}) and $Y_u,Y_d,Y_e$ are the SM Yukawa matrices. $g_1$ has the usual \(SU(5)\) normalization $g_1^2=(5/3) g_Y^2$. The PS gauge couplings are related to the SM gauge couplings at the PS breaking scale (\(\mu_{\mathrm{PS}}\)) by the following matching conditions:
\begin{equation}
g_L^2(\mu_{\mathrm{PS}})=g_2^2(\mu_{\mathrm{PS}}),\quad g_R^2(\mu_{\mathrm{PS}})=\dfrac{3g_Y^2(\mu_{\mathrm{PS}})g_3^2(\mu_{\mathrm{PS}})}{3g_3^2(\mu_{\mathrm{PS}})-2g_Y^2(\mu_{\mathrm{PS}})},\quad g_4^2(\mu_{\mathrm{PS}})= g_3^2(\mu_{\mathrm{PS}}).
\label{eqn:matching_PS}
\end{equation}
With these considerations, we show in Fig.~\ref{fig:gauge_boson_masses} the interplay of the heavy gauge boson masses due to the presence of a single  $(1,2,\bar{4})$ Higgs VEV, $\kappa_R$. The neutral gauge boson, $Z'$, is heavier ($ M_{Z'}>M_{X}>M_{W'})$ than the charged ones due to the additional mixing effects occurring in the neutral sector. If we take the limit on the mass of $Z'$ from LHC for our setup (see Fig.~\ref{fig:LHC:ZR}), which is $M_{Z'}\approx 5.6$ TeV, that immediately puts a lower limit on $M_X$ and $M_{W'}$ given by  $M_X \geq 4.3$ TeV and $M_{W'} \geq 1.8$ TeV. Later, we will see that this LHC limit on \(M_{Z'}\) is the most stringent bound on the mass of the leptoquark gauge boson \(X_{\mu}\) in the model.

\subsection{Higgs Sector}
\label{sec:Higgs}
Now we turn to the symmetry-breaking sector of the theory. One of the goals here is to establish the consistency of the two-step symmetry breaking. 
The most general renormalizable potential involving the ($\widetilde{\psi},\widetilde{\psi}^c,\widetilde{\chi}$) fields can be written as: 
\begin{align}
V=&-\mu_{L}^2\widetilde{\psi}_{\alpha i}\widetilde{\psi}^{\alpha i}-\mu_{R}^2(\widetilde{\psi}^c)_{\dot{\alpha}}^i(\widetilde{\psi}^c)^{\dot{\alpha}}_i-\mu_{\tilde{\chi}}^2\widetilde{\chi}_{\alpha \dot{\alpha}}\widetilde{\chi}^{\alpha \dot{\alpha}}+\lambda_{L1}(\widetilde{\psi}_{\alpha i}\widetilde{\psi}^{\alpha i})^2+\lambda_{R1}((\widetilde{\psi}^c)_{\dot{\alpha}}^i(\widetilde{\psi}^c)^{\dot{\alpha}}_i)^2 \nonumber \\
&+\lambda_{\chi_1}(\widetilde{\chi}_{\alpha \dot{\alpha}}\widetilde{\chi}^{\alpha \dot{\alpha}})^2+\lambda_{L2}\widetilde{\psi}_{\alpha i}\widetilde{\psi}^{\alpha j}\widetilde{\psi}_{\beta j}\widetilde{\psi}^{\beta i}+\lambda_{R2}(\widetilde{\psi}^c)_{\dot{\alpha}}^i(\widetilde{\psi}^c)^{\dot{\alpha}}_j(\widetilde{\psi}^c)_{\dot{\beta}}^j(\widetilde{\psi}^c)^{\dot{\beta}}_i \nonumber \\
&+\lambda_{LR1}\widetilde{\psi}_{\alpha i}\widetilde{\psi}^{\alpha j}(\widetilde{\psi}^c)^{\dot{\alpha}}_j(\widetilde{\psi}^c)_{\dot{\alpha}}^i
+\lambda_{LR2}\widetilde{\psi}_{\alpha i}\widetilde{\psi}^{\alpha i}(\widetilde{\psi}^c)_{\dot{\alpha}}^j(\widetilde{\psi}^c)^{\dot{\alpha}}_j+
\lambda_{L\chi_1}\widetilde{\chi}_{\alpha \dot{\alpha}}\widetilde{\chi}^{\alpha \dot{\alpha}}\widetilde{\psi}_{\beta i}\widetilde{\psi}^{\beta i}\nonumber \\
&+\lambda_{R\chi_1}\widetilde{\chi}_{\alpha \dot{\alpha}}\widetilde{\chi}^{\alpha \dot{\alpha}}(\widetilde{\psi}^c)_{\dot{\beta}}^i(\widetilde{\psi}^c)^{\dot{\beta}}_i+\lambda_{L\chi_2}\widetilde{\psi}^{\alpha i}\widetilde{\psi}_{\beta i}\widetilde{\chi}^{\beta \dot{\alpha}}\widetilde{\chi}_{\alpha \dot{\alpha}}+\lambda_{R\chi_2}(\widetilde{\psi}^c)^{\dot{\alpha}}_i(\widetilde{\psi}^c)_{\dot{\beta}}^i\widetilde{\chi}^{\alpha \dot{\beta}}\widetilde{\chi}_{\alpha \dot{\alpha}}\nonumber \\
&+\lambda_{\chi_2}((\widetilde{\chi}_{\alpha \dot{\alpha}}\epsilon^{\alpha \beta}\widetilde{\chi}_{\beta \dot{\beta}}\epsilon^{\dot{\beta} \dot{\alpha}})^2+\mathrm{h.c.})+\lambda_{\chi_3} \widetilde{\chi}_{\alpha \dot{\alpha}}\epsilon^{\alpha\beta}\widetilde{\chi}_{\beta\dot{\beta}}\epsilon^{\dot{\beta} \dot{\alpha}}\epsilon_{\gamma\delta}\widetilde{\chi}^{\delta \dot{\delta}}\epsilon_{\dot{\delta} \dot{\gamma}}\widetilde{\chi}^{\gamma \dot{\gamma}}\nonumber \\
&+(\rho~\widetilde{\psi}^{\alpha i}\epsilon_{\alpha \beta}\widetilde{\chi}^{\beta \dot{\beta}}\epsilon_{\dot{\beta}\dot{\alpha} }(\widetilde{\psi}^c)^{\dot{\alpha}}_i+\mathrm{h.c.})+\textcolor{red}{(\mu\widetilde{\psi}^{\alpha i}\widetilde{\chi}_{\alpha \dot{\alpha}}(\widetilde{\psi}^c)^{\dot{\alpha}}_i+\mathrm{h.c.})}
\end{align}
Here, we have used the notation $\widetilde{\psi}^{\alpha i}=(\widetilde{\psi}^{*})_{\alpha i}; (\widetilde{\psi}^c)^{\dot{\alpha}}_i=((\widetilde{\psi}^c)^*)_{\dot{\alpha}}^{i}$ and $\widetilde{\chi}^{\alpha \dot{\alpha}}=(\widetilde{\chi}^*)_{\alpha \dot{\alpha}}$. The parameters $\rho$ and $\mu$ are complex in general. One of these, which we take to be $\rho$, can be made real by field redefinitions, in which case the other ($\mu$) will remain complex. Since this is a small parameter that breaks $Z_2$ symmetry softly with its value estimated in Eq.~\eqref{eq:mu}, its complexity will have little effect on the Higgs boson spectrum. Therefore, for simplicity, we take $\mu$ and $\rho$ to be real. 

Among the 40 real degrees of freedom -- 16 come from the $SU(2)_L$ doublet \(\widetilde{\psi}\), 16 from the $SU(2)_R$ doublet \(\widetilde{\psi}^c\), and 8 from the bi-doublet \(\widetilde{\chi}\)-- 28 fields get mass, while the remaining 12 fields are Nambu-Goldstone modes absorbed by the 12 massive gauge bosons. To determine the Higgs mass spectrum, it is necessary to consider small oscillations around the true vacuum. We choose to eliminate $\mu_L^2, \mu_R^2, \mu_{\tilde{\chi}}^2$ and $\rho$ in favor of the VEVs given in Eq. (\ref{eq:VEVs}) through the extrema conditions:
\begin{align}
\mu_L^2 = &2(\lambda_{L1} + \lambda_{L2}) \kappa_{L}^2
+ \dfrac{v^2}{\kappa_{L}^2}\left(\lambda_{R\chi_2}\kappa_{R}^2 - 4(2 \lambda_{\chi_2} - \lambda_{\chi_3})(v^2-\textcolor{magenta}{w^2})\right)+\dfrac{\mu}{\textcolor{magenta}{w}}\dfrac{\kappa_{R}}{\kappa_{L}}(v^2 + \textcolor{magenta}{w^2})\nonumber\\
&\quad+(\lambda_{LR1} + \lambda_{LR2}) \kappa_{R}^2
+ (\lambda_{L\chi_1} + \lambda_{L\chi_2})(v^2+\textcolor{magenta}{w^2}),\\
\mu_R^2= &2(\lambda_{R1} + \lambda_{R2}) \kappa_{R}^2
+ \dfrac{v^2}{\kappa_{R}^2}\left(\lambda_{L\chi_2}\kappa_{L}^2 - 4(2 \lambda_{\chi_2} - \lambda_{\chi_3})(v^2-\textcolor{magenta}{w^2})\right)+\dfrac{\mu}{\textcolor{magenta}{w}}\dfrac{\kappa_{L}}{\kappa_{R}}(v^2 + \textcolor{magenta}{w^2})\nonumber\\
&\quad+(\lambda_{LR1} + \lambda_{LR2}) \kappa_{L}^2
+ (\lambda_{R\chi_1} + \lambda_{R\chi_2})(v^2+\textcolor{magenta}{w^2}),\\
\mu_{\tilde{\chi}}^2=&(\lambda_{L\chi_1} + \lambda_{L\chi_2}) \kappa_{L}^2
+ (\lambda_{R\chi_1} + \lambda_{R\chi_2}) \kappa_{R}^2
+ 2\lambda_{\chi_1}(v^2 + \textcolor{magenta}{w^2})- 4 (2\lambda_{\chi_2}-\lambda_{\chi_3})v^2 +\frac{\mu}{\textcolor{magenta}{w}}\kappa_{L} \kappa_{R},\\
\rho =& \left(\lambda_{L\chi_2}\frac{\kappa_{L}}{\kappa_{R}}
\;+\; \lambda_{R\chi_2}\frac{\kappa_{R}}{\kappa_{L}}
\;+\; \frac{\mu}{\textcolor{magenta}{w}}\right)v-4(2\lambda_{\chi_2}-\lambda_{\chi_3}) \frac{v}{\kappa_{L}}\frac{1}{\kappa_{R}}(v^{2}-\textcolor{magenta}{w^2}) .
\end{align}

With these conditions, we derive the mass spectrum of all scalars by taking the second derivatives of the potential with respect to the fields. Our results are summarized below, sector by sector.

\vspace*{0.1in}
\noindent \textbf{Leptoquark scalars with \(\pm2/3\) charge:}
The $SU(3)$ fundamental and $Z_2$ odd up-type charged leptoquark fields, $\widetilde{u}(\widetilde{u}^c)$, coming from $\widetilde{\psi}(\widetilde{\psi^c})$, have the following mixing matrix in the $(\widetilde{u},(\widetilde{u}^c)^*)$ basis:
\begin{equation}
M_{\tilde{u}}^2 = A_{\tilde{u}} 
\begin{pmatrix}
1 & -\dfrac{\kappa_L}{\kappa_R} \\[6pt]
 -\dfrac{\kappa_L}{\kappa_R}& \dfrac{\kappa_L^2}{\kappa_R^2}
\end{pmatrix}.
\end{equation}
Here we have defined 
\begin{equation}
A_{\tilde{u}}= \frac{v^{2}}{\kappa_{L}^{2}}\bigl(4(2\lambda_{\chi_2}-\lambda_{\chi_3})(v^{2}-\textcolor{magenta}{w^2})-\lambda_{R\chi_2}\kappa_{R}^{2}\bigr)
- \frac{\mu}{\textcolor{magenta}{w}}\frac{\kappa_{R}}{\kappa_{L}}(v^{2}+\textcolor{magenta}{w^2})
- \lambda_{LR1}\kappa_{R}^{2}
- \lambda_{L\chi_2}v^{2}.
\end{equation}
The physical state and its mass is given by 
\begin{equation}
   \phi_X=\frac{-\kappa_R \widetilde{u} + \kappa_L (\widetilde{u}^c)^*}{\sqrt{\kappa_L^2 + \kappa_R^2}}, \qquad m_{\phi_X}^2 =A_{\tilde{u}} \left(1+\kappa_L^2/\kappa_R^2\right)
\end{equation} 
while the orthogonal state, given by 
\begin{equation}
    G_X = \frac{\kappa_L \widetilde{u} + \kappa_R(\widetilde{u}^c)^*}{\sqrt{\kappa_L^2 + \kappa_R^2}}
    \label{eqn:G_x}
\end{equation}
corresponds to the Goldstone mode, which gets absorbed and becomes the longitudinal component of leptoquark gauge bosons $X^\mu$.

\vspace*{0.1in}
\noindent\textbf{Leptoquark scalars with \(\pm1/3\) charge:}
The colored \(\pm1/3\) charged Higgs field \(\widetilde{d}(\widetilde{D}^c)\), coming from $\widetilde{\psi}(\widetilde{\psi^c})$, has the following mixing matrix on the basis $(\widetilde{d},(\widetilde{D}^c)^*)$:
\begin{equation}
M_{\tilde{d}}^2 =  
\begin{pmatrix}
A_{\tilde{d}} & B_{\tilde{d}} \\[6pt]
 B_{\tilde{d}}& C_{\tilde{d}}
\end{pmatrix}.
\end{equation}
where 
\begin{align}
A_{\tilde{d}}&= \frac{v^{2}}{\kappa_{L}^{2}}\bigl(4(2\lambda_{\chi_2}-\lambda_{\chi_3})(v^{2}-\textcolor{magenta}{w^2})-\lambda_{R\chi_2}\kappa_{R}^{2}\bigr)
- \frac{\mu}{\textcolor{magenta}{w}}\frac{\kappa_{R}}{\kappa_{L}}(v^{2}+\textcolor{magenta}{w^2})-2\lambda_{L2}\kappa_L^2
- \lambda_{LR1}\kappa_{R}^{2}
- \lambda_{L\chi_2}\textcolor{magenta}{w^2}.\\
B_{\tilde{d}}&= -\frac{\textcolor{magenta}{w}\,v}{\kappa_{L}\kappa_{R}}\bigl(4(2\lambda_{\chi_2}-\lambda_{\chi_3})(v^{2}-\textcolor{magenta}{w^2})-\lambda_{R\chi_2}\kappa_{R}^{2}\bigr)
+2\mu v+\dfrac{\kappa_{L}}{\kappa_{R}}\lambda_{L\chi_2}\textcolor{magenta}{w}\,v\\
C_{\tilde{d}}&= \frac{v^{2}}{\kappa_{R}^{2}}\bigl(4(2\lambda_{\chi_2}-\lambda_{\chi_3})(v^{2}-\textcolor{magenta}{w^2})-\lambda_{R\chi_2}\left(\dfrac{\textcolor{magenta}{w^2}}{v^2}\right)\kappa_{R}^{2}\bigr)
- \frac{\mu}{\textcolor{magenta}{w}}\frac{\kappa_{L}}{\kappa_{R}}(v^{2}+\textcolor{magenta}{w^2})
-2\lambda_{R2}\kappa_R^2\nonumber\\
&\quad -\lambda_{LR1}\kappa_{L}^{2}- \lambda_{L\chi_2}\left(\dfrac{\kappa_L^2}{\kappa_R^2}\right)v^{2}.
\end{align}
Here $\widetilde{d}$ is $Z_2$ odd and $\widetilde{D}^c$ is $Z_2$ even, and the elements of the off-diagonal block depend on $\langle \nu \rangle=\textcolor{magenta}{w}$ as well as the parameter $\mu$ of the $Z_2$ soft-breaking Lagrangian of Eq. (\ref{eq:soft}). Note that there is no Goldstone mode appearing from this sector.

\vspace*{0.1in}
\noindent\textbf{Color singlet scalars with \(\pm 1\) charge:}
The $Z_2$ odd electron-type charged Higgs fields $\widetilde{e}(\widetilde{e}^c)$ originating from $\widetilde{\chi}(\widetilde{\psi^c})$ and the $Z_2$ even electron-type charged Higgs fields $\widetilde{E}(\widetilde{E}^c)$ originating from $\widetilde{\psi}(\widetilde{\chi})$ mix, with the mass matrix is given in the $(\widetilde{e},(\widetilde{e}^c)^*,\widetilde{E},(\widetilde{E}^c)^*)$ as:
\begin{equation}
M_{\tilde{e}\widetilde{E}}^{2}=\left(\begin{array}{c|c}
\left(m_{\tilde{e}}^{2}\right)_{2 \times 2} & \hspace{10pt} \left(m_{\tilde{e}\widetilde{E}}^{2}\right)_{2 \times 2} \\[.3 cm]
\hline
\hspace{10pt} \left(m_{\tilde{e}\widetilde{E}}^{2}\right)_{2 \times 2} & \left(m_{\tilde{E}}^{2}\right)_{2 \times 2}
\end{array}\right).
\end{equation}
The off-diagonal $\left(m_{\tilde{e}\widetilde{E}}^{2}\right)_{2 \times 2}$ block arises from the mixing of the $Z_2$ even and odd sectors due to its soft breaking, and the various blocks have the following structures:
\begin{itemize}

\item The upper-half block:
\begin{equation}
m_{\tilde{e}}^2 =  
\begin{pmatrix}
A_{\tilde{e}} & B_{\tilde{e}} \\[6pt]
 B_{\tilde{e}}& C_{\tilde{e}}
\end{pmatrix}.
\end{equation}
where 
\begin{align}
A_{\tilde{e}}&= 4 v^2 \bigl(2 \lambda_{\chi_2} - \lambda_{\chi_3}\bigr)
-  \lambda_{L\chi_2} \kappa_L^2 -\dfrac{\mu}{\textcolor{magenta}{w}}\kappa_L\kappa_R,\\
B_{\tilde{e}}&= \frac{v}{\kappa_{R}}\bigl(4(2\lambda_{\chi_2}-\lambda_{\chi_3})(v^{2}-\textcolor{magenta}{w^2})-\lambda_{L\chi_2}\kappa_{L}^{2}\bigr)
-\dfrac{\mu}{\textcolor{magenta}{w}}\kappa_Lv,\\
C_{\tilde{e}}&= \frac{v^{2}}{\kappa_{R}^{2}}\bigl(4(2\lambda_{\chi_2}-\lambda_{\chi_3})(v^{2}-\textcolor{magenta}{w^2})-\lambda_{L\chi_2}\kappa_{L}^{2}\bigr)
- \frac{\mu}{\textcolor{magenta}{w}}\frac{\kappa_{L}}{\kappa_{R}}(v^{2}+\textcolor{magenta}{w^2})- \lambda_{R\chi_2}\textcolor{magenta}{w^2}.
\end{align}

\item The non-diagonal block:
\begin{equation}
m_{\tilde{e}\tilde{E}}^2 =  
\begin{pmatrix}
A_{\tilde{e}\tilde{E}} & B_{\tilde{e}\tilde{E}} \\[6pt]
C_{\tilde{e}\tilde{E}}& D_{\tilde{e}\tilde{E}}
\end{pmatrix}.
\end{equation}
where 
\begin{align}
A_{\tilde{e}\tilde{E}}&= \lambda_{L\chi_2} \textcolor{magenta}{w}\,\kappa_L + \mu\,\kappa_R,\\
B_{\tilde{e}\tilde{E}}&= 4\,v\,\textcolor{magenta}{w}(2\lambda_{\chi_2}-\lambda_{\chi_3}),\\
C_{\tilde{e}\tilde{E}}&= -\dfrac{\textcolor{magenta}{w}\,v}{\kappa_{L}\kappa_{R}}\bigl(4(2\lambda_{\chi_2}-\lambda_{\chi_3})(v^{2}-\textcolor{magenta}{w^2})-\lambda_{L\chi_2}\kappa_{L}^{2}-\lambda_{R\chi_2}\kappa_{R}^{2}\bigr)+2\mu\,v,\\
D_{\tilde{e}\tilde{E}}&= \lambda_{R\chi_2} \textcolor{magenta}{w}\,\kappa_R + \mu\,\kappa_L.
\end{align}

\item The lower-half block:
\begin{equation}
m_{\tilde{E}}^2 =  
\begin{pmatrix}
A_{\tilde{E}} & B_{\tilde{E}} \\[6pt]
 B_{\tilde{E}}& C_{\tilde{E}}
\end{pmatrix}.
\end{equation}
where 
\begin{align}
A_{\tilde{E}}&=\frac{v^{2}}{\kappa_{L}^{2}}\bigl(4(2\lambda_{\chi_2}-\lambda_{\chi_3})(v^{2}-\textcolor{magenta}{w^2})-\lambda_{R\chi_2}\kappa_{R}^{2}\bigr)
- \frac{\mu}{\textcolor{magenta}{w}}\frac{\kappa_{R}}{\kappa_{L}}(v^{2}+\textcolor{magenta}{w^2})- \lambda_{L\chi_2}\textcolor{magenta}{w^2},\\
B_{\tilde{E}}&= \frac{v}{\kappa_{L}}\bigl(4(2\lambda_{\chi_2}-\lambda_{\chi_3})(v^{2}-\textcolor{magenta}{w^2})-\lambda_{R\chi_2}\kappa_{R}^{2}\bigr)
-\dfrac{\mu}{\textcolor{magenta}{w}}\kappa_R\,v,\\
C_{\tilde{E}}&= 4 v^2 \bigl(2 \lambda_{\chi_2} - \lambda_{\chi_3}\bigr)
-  \lambda_{R\chi_2} \kappa_R^2 -\dfrac{\mu}{\textcolor{magenta}{w}}\kappa_L\kappa_R.
\end{align}
\end{itemize}
The combinations of orthogonal massless states, denoted by \( G_{W}\) and \( G_{W'} \) are given by:
\begin{equation}
 G_{W}= \left(-\textcolor{magenta}{w}(v^2 - \textcolor{magenta}{w^2} + \kappa_R^2)\, \widetilde{e} - 2v\,\textcolor{magenta}{w}\,\kappa_R(\widetilde{e}^c)^* - \kappa_L(v^2 + \textcolor{magenta}{w^2} + \kappa_R^2)\, \widetilde{E} + v(v^2 - \textcolor{magenta}{w^2} + \kappa_R^2)(\widetilde{E}^c)^*\right)/N_1,
\end{equation}
where the normalization factor \( N_1 \) is given by
\begin{equation}
N_1 = \sqrt{(v^2 + \textcolor{magenta}{w^2} + \kappa_R^2) \left((v^2 - \textcolor{magenta}{w^2})^2 + (v^2 + \textcolor{magenta}{w^2})\kappa_R^2 + (v^2 + \textcolor{magenta}{w^2}+ \kappa_R^2)\kappa_L^2\right)},
\end{equation}
and 
\begin{equation}
G_{W'} = \left(-v\, \widetilde{e} + \kappa_R (\widetilde{e}^c)^* + \textcolor{magenta}{w}(\widetilde{E}^c)^*\right)/N_2, \quad \text{with} \quad N_2 = \sqrt{\kappa_R^2 + v^2 + \textcolor{magenta}{w^2}}.
\end{equation}
These states are eaten by the charged gauge bosons  $W_\mu^\pm$, and $W_\mu^{'\pm}$  by the Higgs mechanism. Later, we will take the limit \( \textcolor{magenta}{w} \approx 0\) in our calculations of flavor observables, and the definition of \( G_{W}/G_{W'}\) becomes identical to \( G_{W_L}/G_{W_R}\):
\begin{align}
    G_{W_L} = \frac{-\kappa_L\tilde{E} + v(\tilde{E}^c)^*}{\sqrt{\kappa_L^2 + v^2}},\qquad 
    G_{W_R}= \frac{-v\,\tilde{e} + \kappa_R (\tilde{e}^c)^*}{\sqrt{\kappa_R^2 + v^2}}.
    \label{eqn:W-LW-R}
\end{align}

\vspace*{0.1in}
\noindent\textbf{Neutral Higgs:}
Among the four complex neutral Higgs fields, $\widetilde{\nu}_E$, $\widetilde{\nu}^c$, $\widetilde{\nu}_E^c$, and $\widetilde{\nu}$ three are $Z_2$-even,  namely $\widetilde{\nu}_E$, $\widetilde{\nu}^c$, and $\widetilde{\nu}_E^c$,  and naturally acquire vacuum expectation values. On the other hand, although $\widetilde{\nu}$ is $ Z_2$-odd, due to the soft $Z_2$ breaking, it will acquire a small VEV $\textcolor{magenta}{w}$. The absence of scalar-pseudo-scalar mixing allows us to express the field as, for example, $\widetilde{\nu} = \dfrac{1}{\sqrt{2}}(\mathrm{Re}[\widetilde{\nu}] + i~ \mathrm{Im}[\widetilde{\nu}])$, and consider the real and imaginary components separately.\par

The mass matrix for the neutral Higgs fields in the basis $\left(\mathrm{Im}[\widetilde{\nu}_E], \mathrm{Im}[\widetilde{\nu}^c], \mathrm{Im}[\widetilde{\nu}], \mathrm{Im}[\widetilde{\nu}_E^c]\right)$ is given by:
\begin{equation}
M_{\mathrm{Im}[\tilde{N}]}^2 =
\begin{pmatrix}
A_{\mathrm{Im}[\tilde{N}]}& -\dfrac{\kappa_L}{\kappa_R} A_{\mathrm{Im}[\tilde{N}]} & \mu \kappa_R & B_{\mathrm{Im}[\tilde{N}]} \\[6pt]
\bullet & \dfrac{\kappa_L^2}{\kappa_R^2} A_{\mathrm{Im}[\tilde{N}]} & -\mu \kappa_L & -\dfrac{\kappa_L}{\kappa_R} B_{\mathrm{Im}[\tilde{N}]} \\[6pt]
\bullet & \bullet & 16\lambda_{\chi_2}v^2 - \frac{\mu}{\textcolor{magenta}{w}}\kappa_L\kappa_R & -16\lambda_{\chi_2}v\,\textcolor{magenta}{w} \\[6pt]
\bullet & \bullet & \bullet & C_{\mathrm{Im}[\tilde{N}]}
\label{eq:neutral Higgs mass I} 
\end{pmatrix}.
\end{equation}
where
\begin{align}
A_{\mathrm{Im}[\tilde{N}]} &= \frac{v^2}{\kappa_L^2} 
\bigl(4(2\lambda_{\chi_2} - \lambda_{\chi_3})(v^2 - \textcolor{magenta}{w^2}) - \lambda_{R\chi_2}\kappa_R^2\bigr) 
- \frac{\mu}{\textcolor{magenta}{w}} \frac{\kappa_R}{\kappa_L}(v^2 + \textcolor{magenta}{w^2}) 
- \lambda_{L\chi_2} v^2, \\[6pt]
B_{\mathrm{Im}[\tilde{N}]} &= -\frac{v}{\kappa_L} 
\bigl(4(2\lambda_{\chi_2} - \lambda_{\chi_3})(v^2 - \textcolor{magenta}{w^2}) - \lambda_{R\chi_2}\kappa_R^2\bigr) 
+\frac{\mu}{\textcolor{magenta}{w}} v \kappa_R
+\lambda_{L\chi_2} \kappa_L v,  \\[6pt]
C_{\mathrm{Im}[\tilde{N}]} &= 4 v^2 \bigl(2 \lambda_{\chi_2}-\lambda_{\chi_3}\bigr) 
+ 4 \textcolor{magenta}{w^2} \bigl(2 \lambda_{\chi_2} + \lambda_{\chi_3}\bigr)- \dfrac{\mu} {\textcolor{magenta}{w}}\kappa_L \kappa_R -\lambda_{L\chi_2} \kappa_L^2 
- \lambda_{R\chi_2} \kappa_R^2.
\end{align}
Among these four states, two orthogonal states,
\begin{equation}
G_{Z} = \left(\kappa_L\, \mathrm{Im}[\widetilde{\nu}_E] + \textcolor{magenta}{w}\, \mathrm{Im}[\widetilde{\nu}] + v\, \mathrm{Im}[\widetilde{\nu}_E^c]\right)/N_4, \quad \text{with} \quad N_4 = \sqrt{\kappa_L^2 + v^2 + \textcolor{magenta}{w^2}},
\end{equation}
and
\begin{equation}
G_{Z}' = \left(\kappa_L(v^2 + \textcolor{magenta}{w^2})\, \mathrm{Im}[\widetilde{\nu}_E] + \kappa_R N_5\, \mathrm{Im}[\widetilde{\nu}^c] - \kappa_L^2 \textcolor{magenta}{w}\, \mathrm{Im}[\widetilde{\nu}] - \kappa_L^2 v\, \mathrm{Im}[\widetilde{\nu}_E^c] \right)/N_5,
\end{equation}
where the normalization factor \( N_5 \) is given by:
\begin{equation}
N_5 = \sqrt{(\kappa_L^2 + v^2 + \textcolor{magenta}{w^2})\left(\kappa_R^2(v^2 + \textcolor{magenta}{w^2}) + \kappa_L^2(\kappa_R^2 + v^2 + \textcolor{magenta}{w^2})\right)},
\end{equation}
are massless and are absorbed by $Z$ and $Z'$, respectively.

The mass matrix for the real components of the neutral Higgs fields in the basis\\ $\left(\mathrm{Re}[\widetilde{\nu}_E], \mathrm{Re}[\widetilde{\nu}^c],\mathrm{Re}[\widetilde{\nu}],  \mathrm{Re}[\widetilde{\nu}_E^c]\right)$ can be expressed as:
\begin{equation}
M_{\mathrm{Re}[\tilde{N}]}^2 =
\begin{pmatrix}
A_{\mathrm{Re}[\tilde{N}]} & B_{\mathrm{Re}[\tilde{N}]} & 2(\lambda_{L\chi_1} + \lambda_{L\chi_2})\textcolor{magenta}{w}\kappa_L + \mu\kappa_R & C_{\mathrm{Re}[\tilde{N}]} \\[6pt]
\bullet & D_{\mathrm{Re}[\tilde{N}]} &  2(\lambda_{R\chi_1} + \lambda_{R\chi_2})\textcolor{magenta}{w}\kappa_R+\mu\kappa_L & E_{\mathrm{Re}[\tilde{N}]} \\[6pt]
\bullet & \bullet & 4\lambda_{\chi_1}\textcolor{magenta}{w^2} - \frac{\mu}{\textcolor{magenta}{w}}\kappa_L\kappa_R & 4 \bigl(\lambda_{\chi_1}-4 \lambda_{\chi_2} + 2 \lambda_{\chi_3}\bigr) v \textcolor{magenta}{w} \\[6pt]
\bullet & \bullet & \bullet & F_{\mathrm{Re}[\tilde{N}]}
\label{eq:neutral Higgs mass R} 
\end{pmatrix}.
\end{equation}
where 

\begin{align}
A_{\mathrm{Re}[\tilde{N}]} &= \frac{v^2}{\kappa_L^2} 
\bigl(4(2\lambda_{\chi_2} - \lambda_{\chi_3})(v^2 - \textcolor{magenta}{w^2}) - \lambda_{L\chi_2}\kappa_L^2 - \lambda_{R\chi_2}\kappa_R^2\bigr) 
- \frac{\mu}{\textcolor{magenta}{w}} \frac{\kappa_R}{\kappa_L}(v^2 + \textcolor{magenta}{w^2})\notag\\
&\hspace{10cm}+ 4 (\lambda_{L1} + \lambda_{L2})\kappa_L^2 , \\[6pt]
B_{\mathrm{Re}[\tilde{N}]} &=-\frac{\kappa_L}{\kappa_R}\bigl( \frac{v^2}{\kappa_L^2} 
\bigl(4(2\lambda_{\chi_2} - \lambda_{\chi_3})(v^2 -\textcolor{magenta}{w^2}) - \lambda_{L\chi_2}\kappa_L^2 - \lambda_{R\chi_2}\kappa_R^2\bigr) 
- \frac{\mu}{\textcolor{magenta}{w}} \frac{\kappa_R}{\kappa_L}(v^2 + \textcolor{magenta}{w^2})\bigr.\notag\\
&\bigl.\hspace{10cm}-2(\lambda_{L1} + \lambda_{L2})\kappa_R^2\bigr)\\[6pt]
C_{\mathrm{Re}[\tilde{N}]} &= -\frac{v}{\kappa_L} 
\biggl(4 \bigl(2\lambda_{\chi_2} - \lambda_{\chi_3}\bigr)(v^2 - \textcolor{magenta}{w^2})- \lambda_{R\chi_2}\kappa_R^2
- \frac{\mu}{\textcolor{magenta}{w}}\kappa_L \kappa_R-\bigl(2\lambda_{L\chi_1} + \lambda_{L\chi_2}\bigr)\kappa_L^2 
\biggr), \\[6pt]
D_{\mathrm{Re}[\tilde{N}]} &= \frac{v^2}{\kappa_R^2} 
\bigl(4(2\lambda_{\chi_2} - \lambda_{\chi_3})(v^2 - \textcolor{magenta}{w^2}) - \lambda_{L\chi_2}\kappa_L^2 - \lambda_{R\chi_2}\kappa_R^2\bigr) 
- \frac{\mu}{\textcolor{magenta}{w}} \frac{\kappa_L}{\kappa_R}(v^2 + \textcolor{magenta}{w^2})\notag\\
&\hspace{9cm}+4 (\lambda_{R1} + \lambda_{R2})\kappa_R^2 , \\[6pt]
&=A_{\mathrm{Re}[\tilde{N}]}(L\leftrightarrow R),\\[6pt]
E_{\mathrm{Re}[\tilde{N}]} &= -\frac{v}{\kappa_R} 
\biggl(4 \bigl(2\lambda_{\chi_2} - \lambda_{\chi_3}\bigr)(v^2 - \textcolor{magenta}{w^2})- \lambda_{L\chi_2}\kappa_L^2
- \frac{\mu}{\textcolor{magenta}{w}}\kappa_L \kappa_R-\bigl(2\lambda_{R\chi_1} + \lambda_{R\chi_2}\bigr)\kappa_R^2 
\biggr), \\[6pt]
&=C_{\mathrm{Re}[\tilde{N}]}(L\leftrightarrow R),\\[6pt]
F_{\mathrm{Re}[\tilde{N}]} &=  4(2\lambda_{\chi_2} - \lambda_{\chi_3})(v^2 - \textcolor{magenta}{w^2})-\lambda_{L\chi_2}\kappa_L^2- \lambda_{R\chi_2} \kappa_R^2 - \frac{\mu}{\textcolor{magenta}{w}}\kappa_L\kappa_R 
+ 4\lambda_{\chi_1}v^2.
\end{align}

All the mass eigenvalues of physical scalars of the model are positive for a wide range of parameter space of the model, which shows consistency with symmetry breaking.

In summary, this subsection establishes a consistent Higgs spectrum that is compatible with the spontaneous symmetry breaking pattern. We have identified the physical scalar states and the Goldstone modes eaten by the gauge bosons, and expressed the neutral and charged Higgs sectors in such a form that will be used. In particular, the neutral Higgs analysis has already entered our discussion of neutrino mass generation (see Sec.~\ref{subsec:Yukawa_for_N}), and the identification of the Goldstone modes will be needed when we compute flavor–violating processes in the 't Hooft–Feynman gauge in subsequent sections.

\subsection{Interactions of the gauge bosons 
with light and heavy fermions}
\label{subsec:Interactions of the gauge bosons 
with light and heavy fermions}
The gauge bosons couple to fermions that are charged under the corresponding gauge groups through their kinetic terms:
\begin{equation}
    \mathcal{L}_\mathrm{kin}= i \overline{\psi}\slashed{D}\psi+ i \overline{\psi^c}\slashed{D}\psi^c+i
    \overline{\chi}\slashed{D}\chi+i
    \overline{\omega}\slashed{D}\omega+i \overline{\Omega}\slashed{D}\Omega.
    \label{eq:kinetic terms}
\end{equation}
For fields in the $(1,1,6)$ representation, say \(\omega\) (same for \(\Omega\) as well), we can define the covariant derivative as follows:
\begin{align}
     D_{\mu}\omega_{ij}&=\partial_{\mu}\omega_{ij}-ig_4(G_{\mu}^d\frac{T_4^d}{2})_i^k\omega_{kj}-ig_4(G_{\mu}^d\frac{T_4^d}{2})_j^k\omega_{ik},
\end{align}
where $d$ runs over the $SU(4)_C$ generators and
$i,j,k$ are $SU(4)_C$ indices. The definitions of the other fields charged under the gauge group are given in Eq.~\eqref{eqn: Covariant-derivative}.

\paragraph{Couplings of leptoquark gauge boson, \(X_{\mu}\):}
We can expand the covariant derivative to get the explicit couplings of $X_{\mu}$ to fermions in the gauge eigenstates:
\begin{equation}
   \mathcal{L}_{\mathrm{kin}}^{X_{\mu}}=\frac{g_4}{\sqrt{2}} \Bigl[
(\overline{u}\gamma^\mu \nu_{E} \;+\; \overline{d}\gamma^\mu E) X_{\mu} 
\;+\; (\overline{D^c}\gamma^\mu e^{c} \;+\; \overline{u^c}\gamma^\mu \nu^{c}) X_{\mu}^{\ast} 
\;+\epsilon^{ijk}(\overline{d^c_j}\gamma^\mu \textcolor{red}{D'_k}\;+\;\overline{\textcolor{red}{D'^c_j}}\gamma^{\mu}D_k) X_{i\mu}
\Bigr]+\mathrm{h.c}. \label{eq:X}
\end{equation}
Now we rotate the gauge eigenstates to the mass eigenstates following the definitions used in Sec.~\ref{subsec:Yukawa_charged_fermion} for charged fermions and Sec.~\ref{subsec:Yukawa_for_N} for neutral lepton, and write down the light-light~(L-L) fermion and light-heavy~(L-H) fermion interactions with \(X_{\mu}\) (the heavy-heavy interactions are irrelevant for our analysis):  
\begin{align}
    \mathcal{L}_{\mathrm{L\text{--}L}}^{X_{\mu}}&=\frac{g_4}{\sqrt{2}}\overline{d^c_0}\,  K_{d^ce^c} \, \gamma^\mu \, e^c_0  \,X_\mu^*+\mathrm{h.c.}\label{eqn:L-L-X}\\
  \mathcal{L}_{\mathrm{L\text{--}H}}^{X_{\mu}}&= \frac{g_4}{\sqrt{2}} 
\Bigl[\overline{u_0}V_{uE}\gamma^\mu\nu_{E0}X_{\mu}+ \overline{d_0}V_{dE}\gamma^\mu E_0X_{\mu}+\overline{d^c_0}\,K_{d^cE^c}\gamma^{\mu}E^c_0X_{\mu}^{\ast}+\overline{D^c_0}K_{D^ce^c}\gamma^\mu e^{c}_0 X_{\mu}^{\ast}\Bigr.\notag\\
&\Bigl.+\overline{u^c_0} V_{u^c\nu^c}\gamma^\mu \nu^{c}_0 X_{\mu}^{\ast}+\epsilon^{ijk}\overline{(d^c_j)}_0 K_{d^c\textcolor{red}{D'}} \gamma^\mu\textcolor{red}{D'_{k0}}X_{i\mu}+\epsilon^{ijk}\overline{\textcolor{red}{(D'^c_j)_0}} K_{\textcolor{red}{D'^c}d}\gamma^{\mu}d_{k0} X_{i\mu}\Bigr]+\mathrm{h.c.}
    \label{eqn:H-L-X}
\end{align} 
where we have defined
\begin{align}
   K_{d^ce^c}&=-V_{d^c}^\dagger \, (\mathbb{I}+x^\dagger x)^{-1/2} x^\dagger \, V_{e^c},
   \label{eqn:Kdcec}\\
   V_{f_1f_2}&=V_{f_1}^{\dagger}V_{f_2},\\
   K_{d^cE^c}&=-V_{d^c}^\dagger \, (\mathbb{I}+x^\dagger x)^{-1/2} x^\dagger \, \rho_\ell ^* V_{E^c},\\
   K_{D^ce^c}&=V_{D^c}^{\dagger}\left(\hat{M}_D(\mathbb{I} + x x^\dagger) \hat{M}_D^\dagger \right)^{-1/2}\hat{M}_D\,V_{e^c},
   \label{eqn:kDcec}\\
   K_{d^c\textcolor{red}{D'}}&=V_{d^c}^{\dagger}(\mathbb{I} + x^\dagger x)^{-1/2} V_{\textcolor{red}{D'}},\\
   K_{\textcolor{red}{D'^c}d}&=-V_{\textcolor{red}{D'^c}}^{\dagger}\rho^T V_d.
\end{align}
Here the definition for \(\hat{M}_D\) can be found in Eq.~\eqref{eqn:MDhat}, \(x\) and \(\rho\) are given in Eq.~\eqref{eq:rho} and \(\rho_{\ell}\) in Eq.~\eqref{eqn:rhol}.
Later, for the flavor-violating processes, we would follow ’t Hooft-Feynman gauge, and we would require the couplings of the unphysical mode of \(X_{\mu}\)~(defined as \(G_X\) in Eq.~\eqref{eqn:G_x}) with the fermions. One such example is given by
\begin{align}
    \mathcal{L}_{\mathrm{L\text{--}L}}^{G_X}&=\frac{g_4}{\sqrt{2}M_X}\left[ (d_0^c)^{\dagger} K_{d^ce^c}M_e^{\mathrm{diag}}e_0^*-d_0^TM_d^{\mathrm{diag}}K_{d^ce^c}e_0^c\right]G_X^{*}+\mathrm{h.c.}\label{eqn:L-L-GX}\\
    \mathcal{L}_{\mathrm{L\text{--}H}}^{G_X}& \supset \frac{g_4}{\sqrt{2}M_X}\Bigl[-u_0^{\dagger}V_{uE}M_E^{\mathrm{diag}}(\nu_{E_0}^c)^*G_X+d_0^{\dagger}V_{dE}M_E^{\mathrm{diag}}(E_0^c)^*G_X-D_0^TM_D^{\mathrm{diag}}K_{D^ce^c}e^cG_X^*\Bigr.\notag\\
    &\Bigl.+(u_0^c)^{\dagger}V_{u^c\nu^c}m_{N0}^{\mathrm{diag}}(\nu^c_0)^*G_X^*+\epsilon^{ijk}(d^c_j)_0^{\dagger}K_{d^c\textcolor{red}{D'}}M_{\omega}^{\mathrm{diag}}(\textcolor{red}{D_{k0}^{'c}})^*(G_X)_i\Bigr]+\mathrm{h.c.}
    \label{eqn:L-H-GX}
\end{align}
In the \(\mathcal{L}_{\mathrm{L\text{--}H}}^{G_X}\) equation, we have not shown the couplings proportional to the light quark mass and the couplings involving the terms \(\rho, \rho_{\ell}\) proportional to soft breaking, but they can be arrived at similarly.
\paragraph{Couplings of \(SU(2)_R\) gauge boson \(W'_{\mu}\):}
Ignoring the mixing between \(W\text{--}W'\), we can expand the covariant derivative to get the couplings of $W'_{\mu}$ to fermions:
\begin{equation}
   \mathcal{L}_{\mathrm{kin}}^{W'_{\mu}}=\frac{g_R}{\sqrt{2}} \Bigl[
(\overline{D^c}\gamma^\mu u^c \;+\; \overline{e^c}\gamma^\mu \nu^{c})W_{\mu}^{'+} 
+ (\overline{E^c}\gamma^\mu \nu \;+\; \overline{\nu_E^{c}}\gamma^\mu e)W_{\mu}^{'+}\Bigr]+\mathrm{h.c}., \label{eq:Wprime}
\end{equation}
Here, the generation indices and color indices are implicit. Rotating the gauge eigenstates to the mass eigenstates allows us to write down the interactions with light-light~(L-L) fermion and light-heavy~(L-H) fermions as:  
\begin{align}
    \mathcal{L}_{\mathrm{L\text{--}L}}^{W'_{\mu}}&=\frac{g_R}{\sqrt{2}}\Big[\overline{d^c_0}\,  K_{d^cu^c}\gamma^\mu \, u^c_0  \,W_{\mu}^{'+}+\overline{e^c_0}\,  K_{e^c\nu}\gamma^\mu \, \nu_0  \,W_{\mu}^{'+}\Big]+\mathrm{h.c.}\label{eqn:L-L-WR}\\
  \mathcal{L}_{\mathrm{L\text{--}H}}^{W'_{\mu}}&= \frac{g_R}{\sqrt{2}} 
\Bigl[\overline{D^c_0}K_{D^cu}\gamma^\mu u^c_{0}W_{\mu}^{'+}+ \overline{E^c_0}V_{E^c\nu}\gamma^\mu \nu_0W_{\mu}^{'+}+\overline{\nu^c_{E0}}V_{\nu_E^c e}\gamma^\mu e_0W_{\mu}^{'+}\Bigr]+\mathrm{h.c.}
    \label{eqn:H-L-WR}
\end{align} 
where we have defined
\begin{align}
   K_{d^cu^c}&=-V_{d^c}^\dagger \, (\mathbb{I}+x^\dagger x)^{-1/2} x^\dagger \, V_{u^c},
   \label{eqn:Kdcuc}\\
   K_{e^c\nu}&=-V_{e^c}^\dagger \rho_\ell ^*,\\
   K_{D^cu}&=V_{D^c}^{\dagger}\left(\hat{M}_D(\mathbb{I} + x x^\dagger) \hat{M}_D^\dagger \right)^{-1/2}\hat{M}_D\,V_{u^c}.
\end{align}
The couplings with the associated Goldstone, \(G'_W\) (see Eq.~\eqref{eqn:W-LW-R} for definition) can be understood similarly as discussed in Eq.~\eqref{eqn:L-L-GX}-\eqref{eqn:L-H-GX}.
\paragraph{Couplings of SM neutral gauge boson \(Z_{\mu}\):}
Ignoring the mixing between \(Z\text{--}Z'\), we can expand the covariant derivative in Eq.~\eqref{eq:kinetic terms} to get the couplings of $Z_{\mu}$~(see Eq.~\eqref{eq:ZL basis} for the definition of \(Z_{\mu}\) basis) with fermions in their gauge eigenstates:
\begin{align}
\mathcal{L}^{Z_{\mu}}_{\mathrm{kin}} = \frac{g_L}{\mathrm{cos}\theta_W} 
\left[
\bar{f} 
\left\{ T_{3L}^f - Q_f \mathrm{sin}^2\theta_W \right\} 
\gamma^\mu f 
+ 
\overline{f^c} 
\left( -Q_f \mathrm{sin}^2\theta_W \right) 
\gamma^\mu f^c
\right] Z_{\mu}.
\end{align}
Here, $\mathrm{tan}\theta_W=g_Y/g_L$ and $T_{3L}^f$ is the \(SU(2)_L\) generator (third component of weak isospin) acting on \(f\). $T_{3L}^f$ charge assignment for $\chi(2,2,1)$ is 
\begin{equation}
T_{3L}^{E^c} = \frac{1}{2}, \quad 
T_{3L}^{\nu_E^c} = -\frac{1}{2}, \quad 
T_{3L}^{\nu} = \frac{1}{2}, \quad 
T_{3L}^e = -\frac{1}{2}.
\end{equation}
The interactions of light fermions in their mass eigenbasis with \(Z_{\mu}\) become:
\begin{align}
      \mathcal{L}_{\mathrm{L\text{--}L}}^{Z_{\mu}}&=\frac{g_L}{\mathrm{cos}\theta_W}\Bigl[\overline{d_0}L^d\gamma^{\mu}d_0+\overline{d^c_0}L^{d^c}\gamma^{\mu}d^c_0+\overline{u_0}L^u\gamma^{\mu}u_0+\overline{u^c_0}L^{u^c}\gamma^{\mu}u^c_0+\Bigr.\notag \\
   &\Bigl.\overline{e_0}L^e\gamma^{\mu}e_0+\overline{e^c_0}L^{e^c}\gamma^{\mu}e^c_0+\overline{\nu_0}L^{\nu}\gamma^{\mu}\nu_0\Bigr]Z_{\mu}.
   \label{eq:L-L-ZL}
\end{align} 
where we have defined
\begin{align}
   L^d&=-\frac{1}{2}\,\mathbb{I}+\frac{1}{3}\mathrm{sin}^2\theta_W\left(\mathbb{I}+V_d^{\dagger}\rho^*\rho^TV_d\right),
    \label{eq:ZFCNC}\\
   L^{d^c}&=-\frac{1}{3}\mathrm{sin}^2\theta_W\,\mathbb{I},\\
   L^{u}&=\frac{1}{2}\,\mathbb{I}-\frac{2}{3}\mathrm{sin}^2\theta_W\,\mathbb{I},\\
   L^{u^c}&=\frac{2}{3}\mathrm{sin}^2\theta_W\,\mathbb{I},\\
   L^{e}&=-\frac{1}{2}\,\mathbb{I}+\mathrm{sin}^2\theta_W\,\mathbb{I},\\
   L^{e^c}&=-\mathrm{sin}^2\theta_W\left(\mathbb{I}+V_{e^c}^{\dagger}\rho_{\ell}^*\rho_{\ell}^TV_{e^c}\right),\\
   L^{\nu}&=\frac{1}{2}\mathbb{I}.
\end{align}

\paragraph{Couplings of heavy neutral gauge boson \(Z'_{\mu}\):}
Similarly, we can get the  couplings of $Z'_{\mu}$~(see Eq.~\eqref{eq:ZR basis} for the definition of \(Z'_{\mu}\) basis) with fermions in their gauge eigenstates:
\begin{equation}
\mathcal{L}_{\mathrm{kin}}^{Z'_{\mu}} =-\frac{g_R^2}{\sqrt{g_R^2 - g_Y^2}} 
\Bigl[\overline{f} \gamma^\mu 
\left(
T_{3R}^f - \frac{Y_f}{2} \frac{g_Y^2}{g_R^2}
\right)
f+\overline{f^c} \gamma^\mu 
\left(
T_{3R}^{f^c} - \frac{Y_{f^c}}{2} \frac{g_Y^2}{g_R^2}
\right)
f^c \Bigr]Z'^\mu. \label{eq:Zprime}
\end{equation}
Here $d,u, u^c$ have the usual hypercharge with the normalization $Y_{e^c}=2$. Thus, $Y_d=-2/3,Y_u=4/3, Y_{u^c}=- 4/3$. For $D^c$ coming from $(1,2,4)$ also has the usual assignment $Y_{D^c}=2/3$ follows. For (1,1,6), we identify $d^c$ as SM right-handed down quark with the usual $Y_{d^c}=2/3$, then the gauge interactions fix $Y_D, Y_{\textcolor{red}{D'}}=-2/3,\,Y_{\textcolor{red}{D'^c}}= 2/3$ and \(Y_{E^c}=1\). $T_{3R}^f$ is the \(SU(2)_R\) generator (third component of weak isospin) acting on \(f\), its charge assignments for the components of $\chi(2,2,1)$ are
\begin{equation}
T_{3R}^{E^c} = \frac{1}{2}, \quad  
T_{3R}^{\nu} = -\frac{1}{2}, \quad 
T_{3R}^{\nu_E^c} = \frac{1}{2}, \quad
T_{3R}^e = -\frac{1}{2}.
\end{equation}
We can now write down the interactions of light fermions in their mass eigenbasis with \(Z'_{\mu}\) as:
\begin{align}
   \mathcal{L}_{\mathrm{L\text{--}L}}^{Z'_{\mu}}&=-\frac{g_R^2}{\sqrt{g_R^2 - g_Y^2}}\Bigl[\overline{d_0}P^d\gamma^{\mu}d_0+\overline{d^c_0}P^{d^c}\gamma^{\mu}d^c_0+\overline{u_0}P^u\gamma^{\mu}u_0+\overline{u^c_0}P^{u^c}\gamma^{\mu}u^c_0+\Bigr.\notag \\
   &\Bigl.\overline{e_0}P^e\gamma^{\mu}e_0+\overline{e^c_0}P^{e^c}\gamma^{\mu}e^c_0+\overline{\nu_0}P^{\nu}\gamma^{\mu}\nu_0\Bigr]Z'_{\mu}.
   \label{eq:L-L-ZR}
\end{align}
Here, we have defined 
\begin{align}
P^d&=\frac{1}{3}\frac{g_Y^2}{g_R^2}\left(-\frac{1}{2}\,\mathbb{I}+V_d^{\dagger}\rho^*\rho^TV_d\right),\\
P^{d^c}&=V_{d^c}^{\dagger}(\mathbb{I} + x^\dagger x)^{-1/2}\left(\frac{1}{2}x^\dagger x-\frac{1}{3} (\mathbb{I} + x^\dagger x)\frac{g_Y^2}{g_R^2}\right) (\mathbb{I} + x^\dagger x)^{-1/2} V_{d^c},
\label{eqn:Pdc}\\
P^u&=-\frac{1}{6}\frac{g_Y^2}{g_R^2}\,\mathbb{I},\\
P^{u^c}&=-\frac{1}{2}\,\mathbb{I}+\frac{2}{3}\frac{g_Y^2}{g_R^2}\,\mathbb{I},\\
P^e&=-\frac{1}{2}\,\mathbb{I}+\frac{1}{2}\frac{g_Y^2}{g_R^2}\,\mathbb{I},\\
P^{e^c}&=\frac{1}{2}\left(\mathbb{I}+V_{e^c}^{\dagger}\rho_{\ell}^*\rho_{\ell}^TV_{e^c}\right)-\frac{g_Y^2}{g_R^2}\left(\mathbb{I}+\frac{1}{2}V_{e^c}^{\dagger}\rho_{\ell}^*\rho_{\ell}^TV_{e^c}\right),\\
P^{\nu}&=-\frac{1}{2}\,\mathbb{I}+\frac{1}{2}\frac{g_Y^2}{g_R^2}\,\mathbb{I}.
\end{align}
From now on, for notational simplicity, we drop the subscript ``0'' on
mass–eigenstate fields, thus \(d^c\) should be understood as
\(d^c_0\), \(e^c\) as \(e^c_0\), and so on. 

To summarize, in this subsection, we have collected the couplings of the light–light and light–heavy fermion combinations with the gauge bosons \(X_\mu\), \(W'_\mu\), \(Z_\mu\), and \(Z'_\mu\). These couplings form the basic ingredients for the rest of the paper. They will enter the collider analysis to get constraints on the new gauge boson masses and production channels, and all subsequent calculations of flavor–violating observables.

\section{Collider Implications}\label{sec:3}
\label{sec:Collider Implications}

Realizing TeV-scale Pati-Salam symmetry in our model, together with the presence 
of new vector-like fermions and a softly broken $Z_2$ symmetry, opens up unique 
opportunities for collider probes. The relatively light leptoquark gauge boson $X_\mu$ as well as  $Z^\prime$ and $W^\prime$ gauge bosons, and the vector-like quarks carrying unusual baryon number $-2/3$, lead to distinct experimental signatures. In this section, we explore the current bounds from the LHC and discuss the prospects of discovery at the high-luminosity LHC.  The limits derived from direct searches will then be used in the subsequent discussion of flavor violation mediated by these particles.

\begin{figure}[h!]
\centering
\begin{tabular}{cc}
\begin{adjustbox}{valign=m}
  \resizebox{0.35\textwidth}{!}{%
\includegraphics[width=0.5\textwidth]{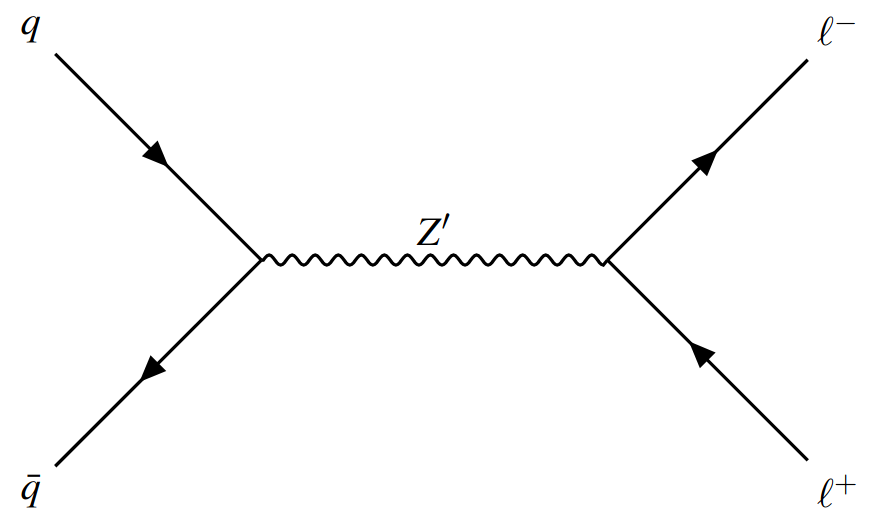}
}%
\end{adjustbox}
&
\begin{adjustbox}{valign=m}
  \includegraphics[width=0.5\textwidth]{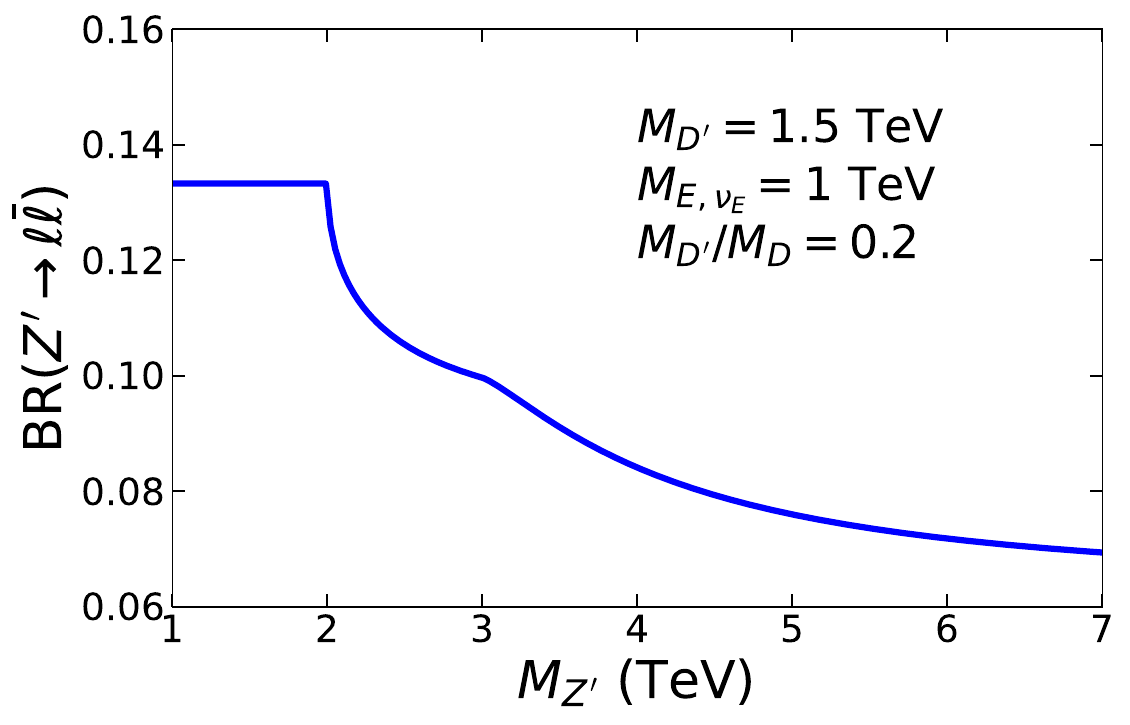}
\end{adjustbox}
\end{tabular}
\caption{Left panel: Production and decay of $Z^\prime$ gauge boson at the LHC. Right panel: Branching ratio of $Z^\prime \to \ell\overline{\ell}$ ($\ell = e, \mu$) in our theory. Here, we have assumed the masses of all three generations of $\textcolor{red}{D'}$ ($E, \nu_E$) to $M_{\textcolor{red}{D'}}=1.5$ TeV ($M_{E,\nu_E}=1$ TeV).  Moreover, $Z^\prime$ decays to scalars is not considered, which can be arranged by taking their masses somewhat heavy, see text for details.   }
\label{fig:LHC:production-ZR}
\end{figure}

\begin{figure}[b!]
\centering
\includegraphics[scale=0.4]{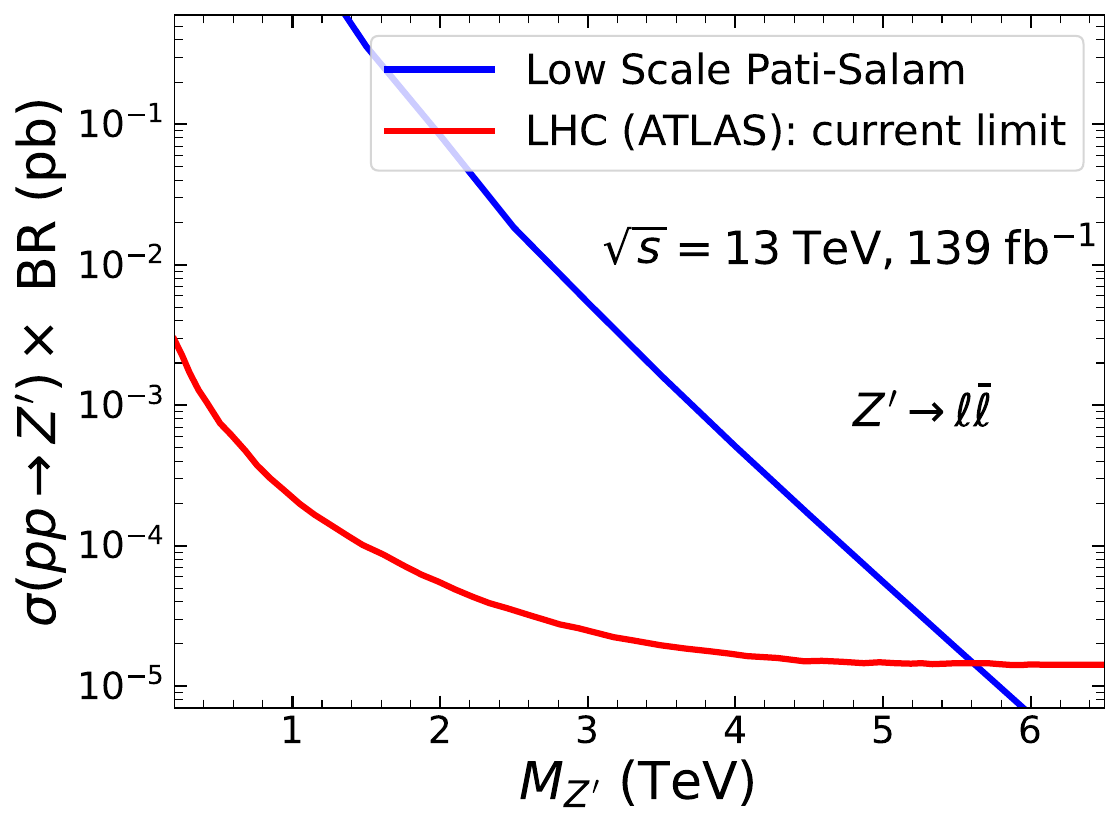}
\includegraphics[scale=0.4]{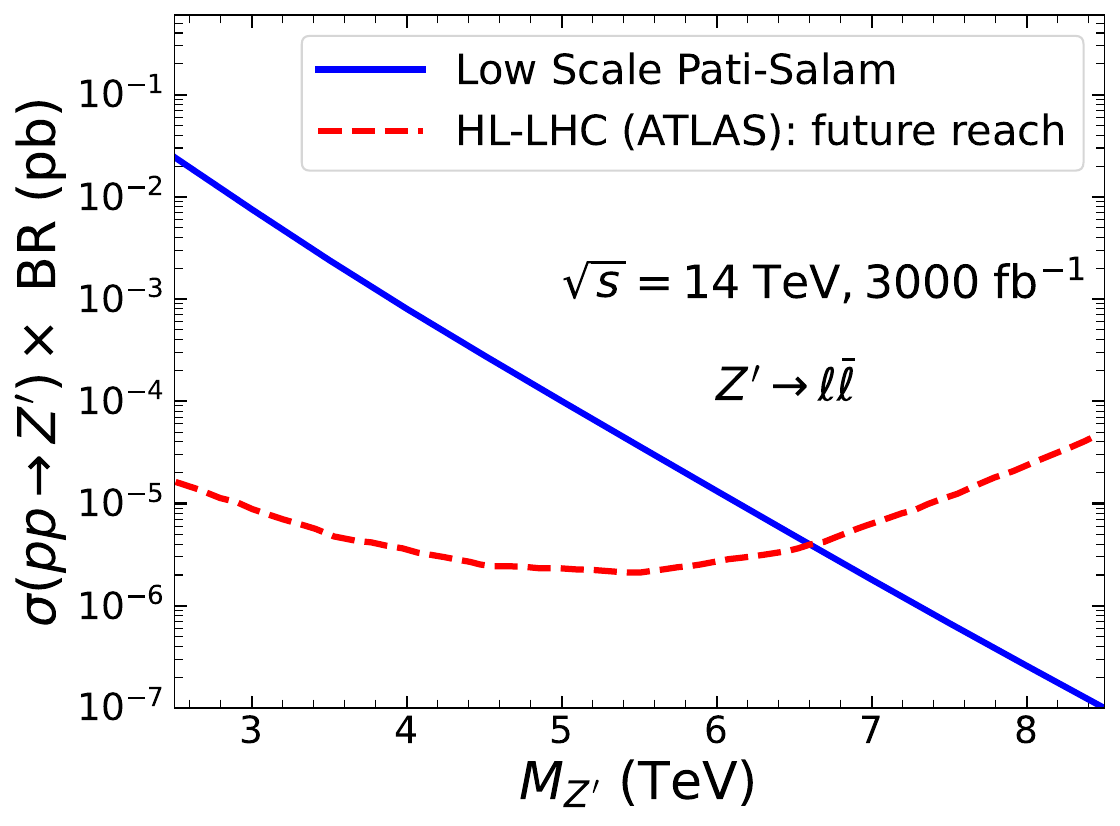}
\caption{Left panel: branching ratio of $Z^\prime\to \ell\overline \ell$ ($\ell=e, \mu$).  Right left panel: current LHC limit on $Z^\prime$ mass from $pp\to Z^\prime\to \ell\overline \ell$ searches. Lower right panel:  future reach for the same process at the HL-LHC.
}
\label{fig:LHC:ZR}
\end{figure}

\textbf{Phenomenology of $Z^\prime$:}
First, we investigate the production and the leptonic decay of  $Z^\prime$ at the LHC, which proceeds through the Drell-Yan (DY) processes, as shown in Fig.~\ref{fig:LHC:production-ZR}.  The relevant interactions of $Z^\prime$ are provided in Eq.~\eqref{eq:Zprime}, which we implemented in \texttt{FeynRules}~\cite{Alloul:2013bka} (similarly, other processes, to be discussed below, are also implemented in \texttt{FeynRules} for the purpose of computing the production cross sections).  The associated \texttt{Universal FeynRules Output} (\texttt{UFO})~\cite{Degrande:2011ua} file is then fed to \texttt{MadGraph5 aMC@NLO} (\texttt{MG5aMC})~\cite{Alwall:2014hca} to compute the production cross section. For this numerical analysis, we use gauge coupling values at 1 TeV, which are  $\left(g_1, g_R, g_2, g_C \right)=\left(0.467, 0.377, 0.638, 1.060\right)$.  The decay branching ratios of $Z^\prime$, on the other hand, are dependent on the mass spectrum of the BSM states. Following the aforementioned discussion on flavor constraints, here we fix the mass ratio $M_{\textcolor{red}{D'}}/M_D\sim 0.2$  as a benchmark value and take $M_{\textcolor{red}{D'}}=1.5$ TeV, $M_{E,\nu_E}=1$ TeV. Furthermore, $Z^\prime$ decays to scalars are not considered, which can be arranged by taking their masses somewhat heavy. The corresponding relevant branching ratio, namely, $Z^\prime\to \ell\overline\ell$ ($\ell=e, \mu$) is presented in Fig.~\ref{fig:LHC:production-ZR} (left panel). With these, we obtain a very strong constraint of $M_{Z^\prime}\gtrsim 5.6$ TeV from the current LHC searches~\cite{ATLAS:2019erb}, as shown in the left panel of Fig.~\ref{fig:LHC:ZR}. By appropriately assigning the couplings, the procedure adopted correctly reproduces the sequential SM $Z^\prime$~\cite{Altarelli:1989ff} mass limit from LHC, which is an independent check of our method.  Moreover, the future sensitivity~\cite{ATLAS:2018tvr} of  HL-LHC is also depicted in Fig.~\ref{fig:LHC:ZR} (right panel). This figure shows that, if not discovered, the LHC operating at high luminosity can rule out a $Z^\prime$-boson mass up to $M_{Z^\prime}\gtrsim 6.6$ TeV. Intriguingly, if the FCC-hh is built, it could potentially reach a sensitivity 
of order $M_{Z^\prime}\gtrsim \mathcal{O}(40)$~TeV~\cite{Helsens:2642473}.

\begin{figure}[t!]
\centering
\begin{tabular}{cc}
\begin{adjustbox}{valign=m}
  \resizebox{0.35\textwidth}{!}{%
\includegraphics[width=0.5\textwidth]{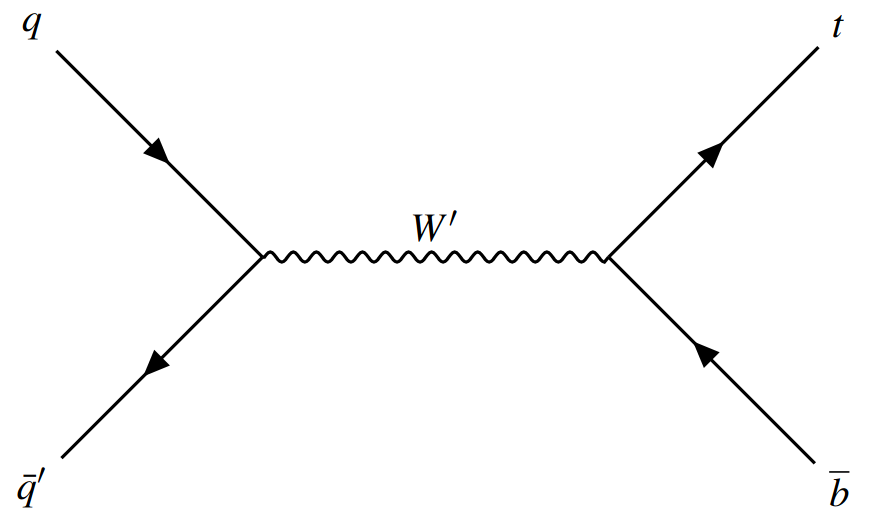}
  }%
\end{adjustbox}
&
\begin{adjustbox}{valign=m}
\includegraphics[scale=0.4]{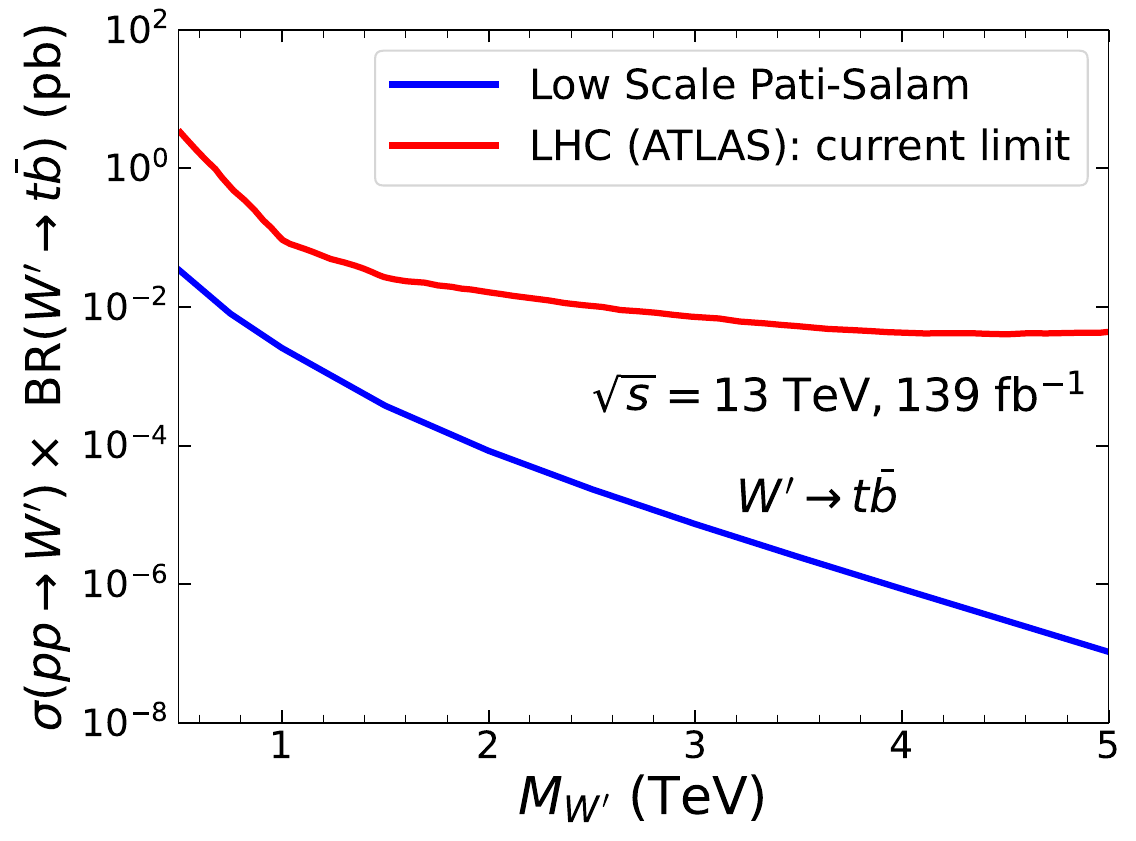}
\end{adjustbox}
\end{tabular}
\caption{Current LHC limit on $W^\prime$ mass from $pp\to W^\prime\to t\overline b$ searches.  The $W^\prime$ production  is $d^c-D^c$ mixing suppressed, and as before, we have assumed $M_{\textcolor{red}{D'}}/M_D\sim 0.2$ and $2 M_{\nu^c}= M_{E,\nu_E} = 1$ TeV. }
\label{fig:LHC:WR}
\end{figure}

\begin{figure}[b!]
\centering
\begin{tabular}{cc}
\begin{adjustbox}{valign=m}
  \resizebox{0.4\textwidth}{!}{%
\includegraphics[scale=0.15]{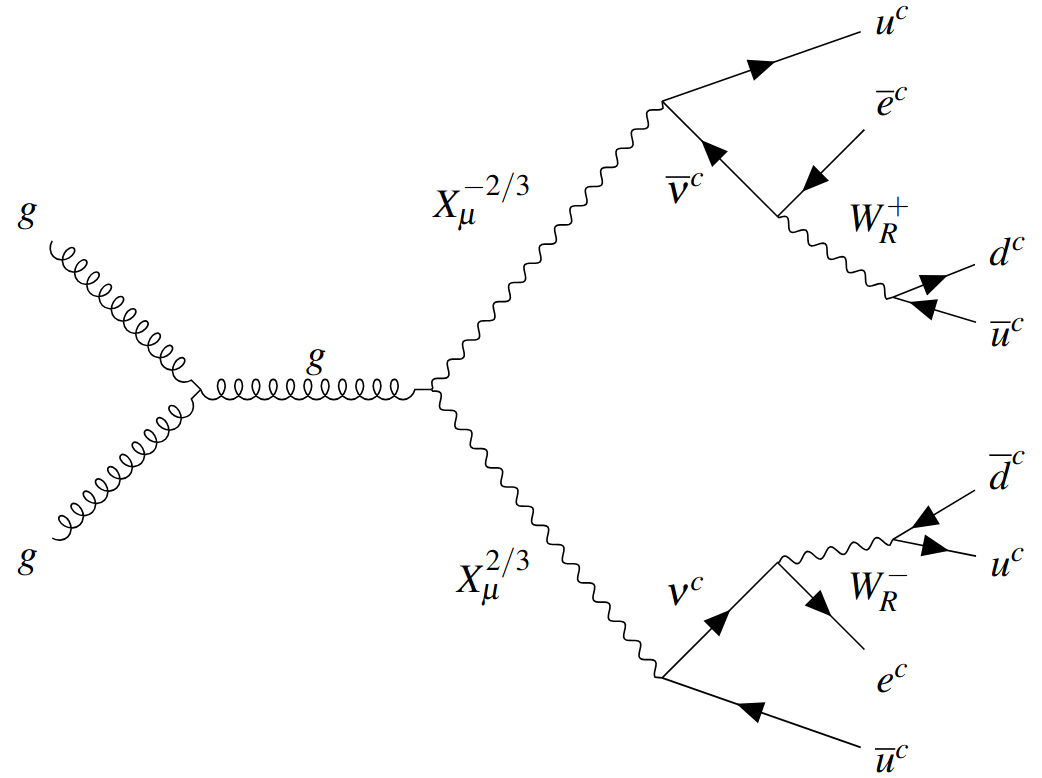}
  }%
\end{adjustbox}
&
\begin{adjustbox}{valign=m}
\includegraphics[scale=0.4]{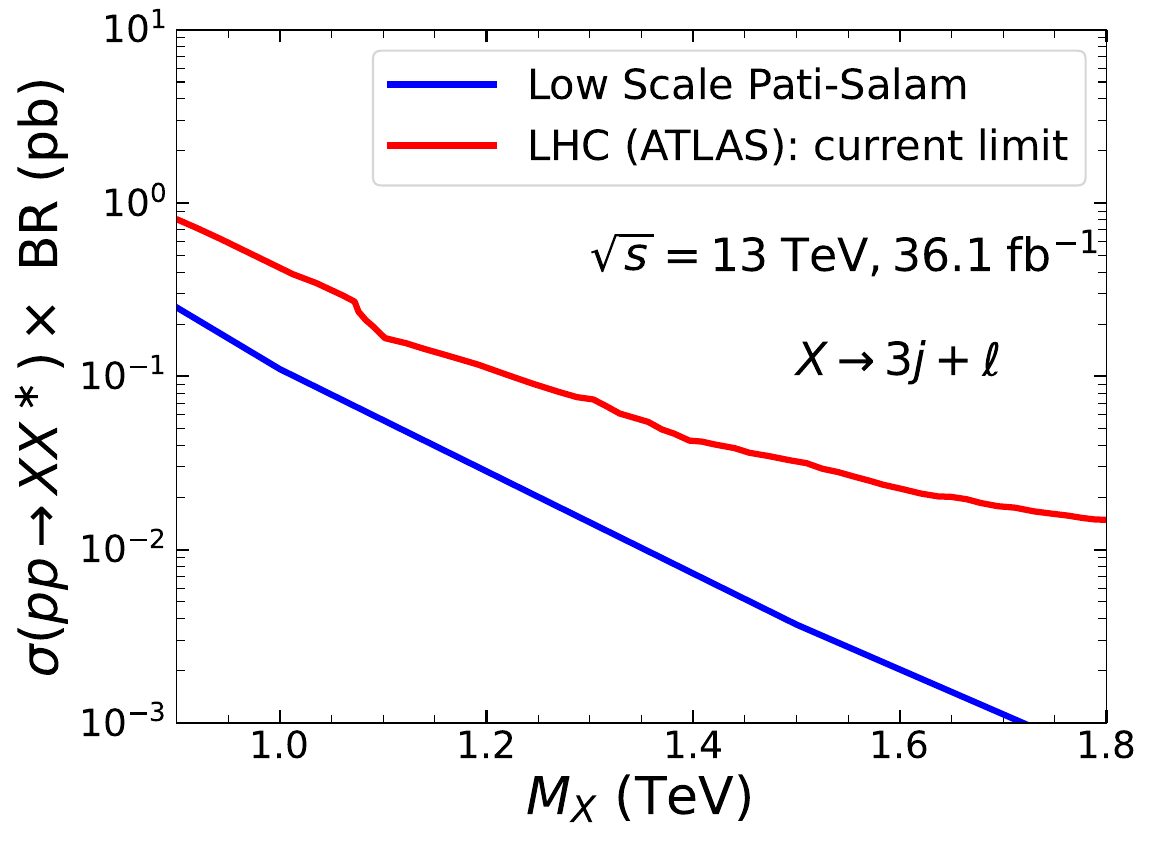}
\end{adjustbox}
\end{tabular}
\caption{  Pair-production of Pati-Salam gauge boson at the LHC $pp\to XX^*$. The left panel shows the production cross section $pp\to XX^*$ (for simplicity, here, the branching ratio is set to unity) along with the current LHC limit on $6j$ final state.  }
\label{fig:LHC:X:production}
\end{figure}

\textbf{Phenomenology of $W^\prime$:}
Although, in most of the models, $W^\prime$ gauge boson mass is also strongly constrained from LHC searches, within our model, its production, see Fig.~\ref{fig:LHC:WR}, is highly suppressed. It is because $W^\prime$, in the flavor basis, couples to $u^cD^c$, see Eq.~\eqref{eq:Wprime}. Therefore, the production $pp\to W^\prime$ as well as the decay $W^\prime\to t\overline b$ are $d^c-D^c$ mixing suppressed. Consequently, current LHC searches cannot provide a meaningful bound to its mass. However,  for completeness, in Fig.~\ref{fig:LHC:WR}, we show the corresponding production cross section times the branching ratio along with the current LHC bounds~\cite{ATLAS:2023ibb} as a function of its mass.  For producing this plot, as before, we take $M_{\textcolor{red}{D'}}/M_D\sim 0.2$ and to incorporate the rest of its decays, we assumed $2 M_{\nu^c}= M_{E,\nu_E} = 1$ TeV.

\textbf{Phenomenology of $X_\mu$:}
Since the Pati-Salam gauge boson, $X_\mu$, carries color charge, they can be pair produced at the LHC through gluon fusion, as shown in Fig.~\ref{fig:LHC:X:production}. Each gauge boson then decays to three quarks and a lepton, leading to, for example,  $pp\to XX^*\to 6j+\ell\overline \ell$. To estimate an approximate collider bound on the mass of this exotic gauge boson, we employ the LHC limit~\cite{ATLAS:2018umm} on $6j$ final state. Ref.~\cite{ATLAS:2018umm} considered R-parity violating (RPV) supersymmetric (SUSY) scenario, where gluinos are pair produced, and subsequently, each gluino decays to three quarks utilizing RVP coupling, already present within the minimal SUSY extension of the SM.    As can be seen in Fig.~\ref{fig:LHC:X:production}, this, however, for our scenario,  does not provide a meaningful bound on its mass.

\begin{figure}[t!]
\centering
\begin{tabular}{cc}
\begin{adjustbox}{valign=m}
  \resizebox{0.5\textwidth}{!}{%
\includegraphics[scale=0.3]{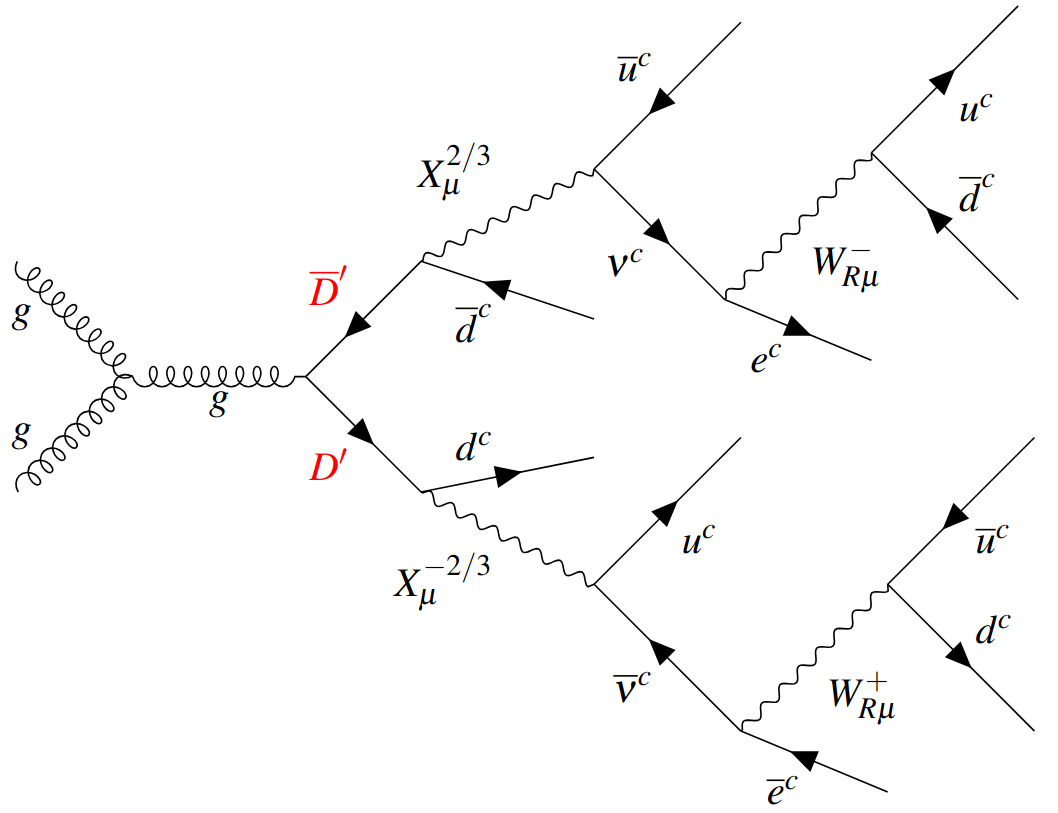}
  }%
\end{adjustbox}
&
\begin{adjustbox}{valign=m}
\includegraphics[scale=0.22]{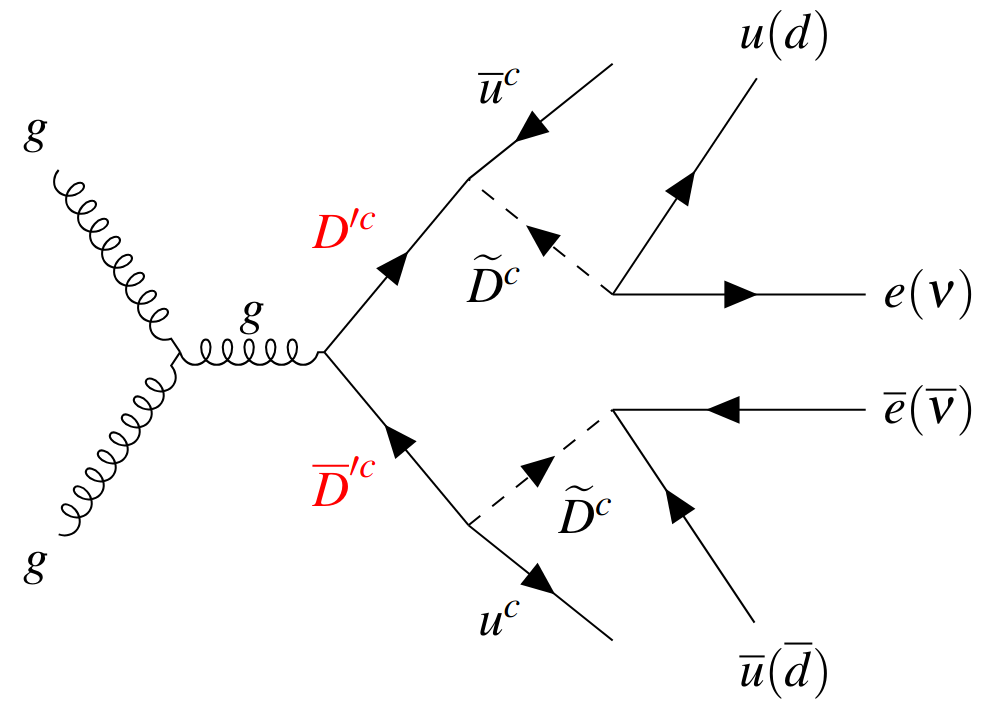}
\end{adjustbox}
\end{tabular}
\caption{ The VLQ can be pair produced at the LHC via gluon fusion $pp\to  \textcolor{red}{D'} \overline {\textcolor{red}{D'}}$. Through cascade decay, the process on the left leads to $pp\to \textcolor{red}{D'} \overline {\textcolor{red}{D'}} \to 8j+\ell\overline\ell$ final states. Whereas, the process on the right can give rise to  4 jets+opposite sign leptons, or 4 jets+lepton and missing energy,  4 jets+missing energy.  }
\label{fig:LHC:Dprime:production}
\end{figure}

\textbf{Phenomenology of $\textcolor{red}{D'}$ VLQ:}
To realize multi-TeV scale Pati-Salam theory, one of the main predictions of our model is that the vector-like quark $\textcolor{red}{D'}$, having a baryon number of $-2/3$, must be as light as possible. However, LHC searches restrict its mass to lie below the TeV scale. In our model, the pair-produced VLQs, through cascade decays, lead to multi-jet final states along with two leptons. In finding collider limit on its mass,  relevant decay chains are shown in Fig.~\ref{fig:LHC:Dprime:production}.

\begin{figure}[!]
\centering
\includegraphics[scale=0.4]{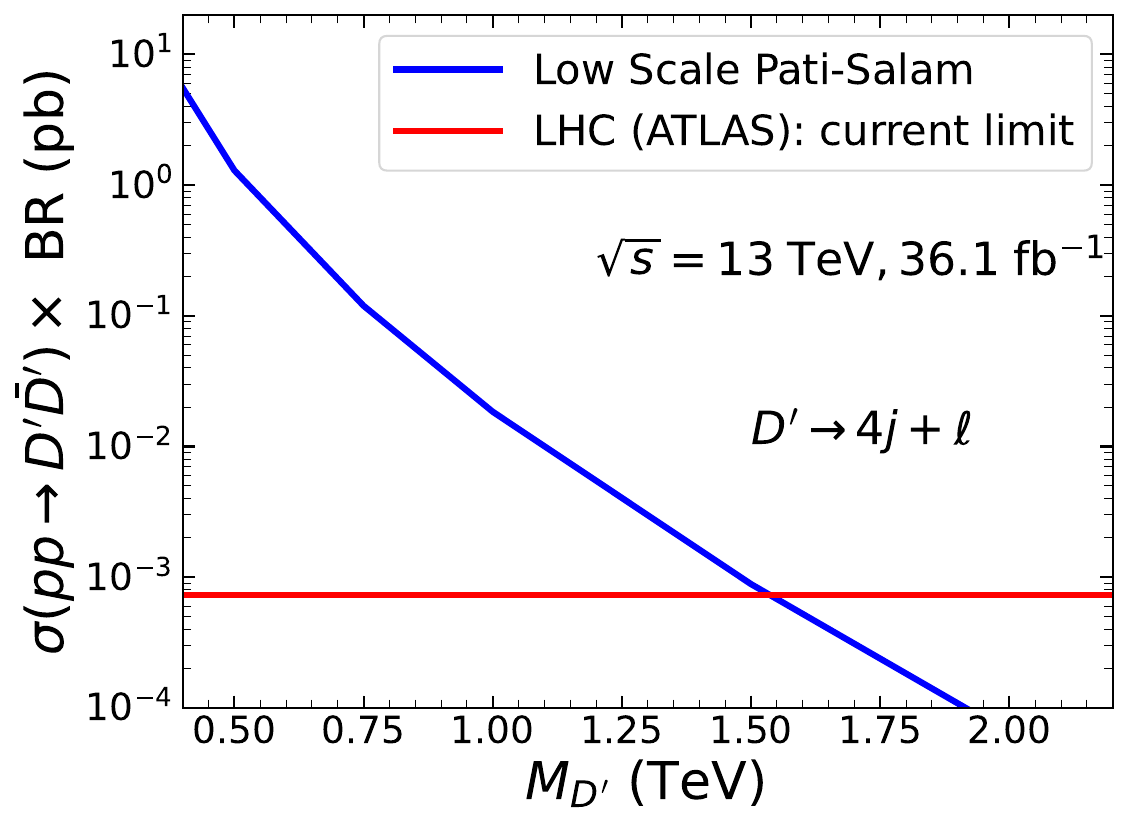}
\includegraphics[scale=0.4]{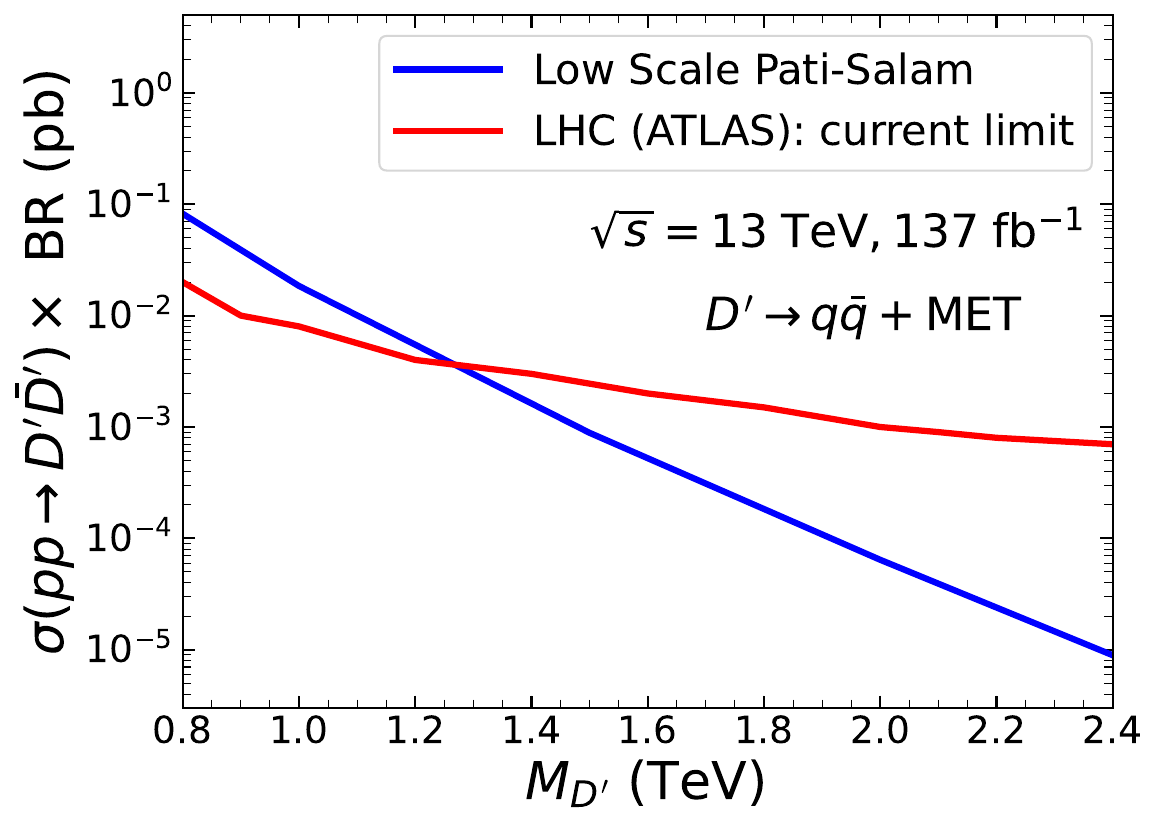}
\caption{Left panel: LHC (model independent) limit on pair-produced VLQ, $pp\to  \textcolor{red}{D'} \overline {\textcolor{red}{D'}}$, decaying into 8 jets and leptons. Right panel: instead, decay of the type $ \textcolor{red}{D'}\to q\overline q+$MET (leading to $4j+$MET in the final state) is assumed.}
\label{fig:LHC:Dprime}
\end{figure}

The process on the left panel in Fig.~\ref{fig:LHC:Dprime:production} somewhat mimics the RPV SUSY model where cascade decays from pair-produced gluinos eventually lead to high jet-multiplicity~\cite{ATLAS:2017oes}. 
For this case of $pp\to \textcolor{red}{D'} \overline{\textcolor{red}{D'}}\to 8j+\ell\overline\ell$, an exact bound depends also on the mass of $\nu^c$. For a dedicated study, see Ref.~\cite{ATLAS:2017oes}. However, taking the model independent limit quoted in  Ref.~\cite{ATLAS:2017oes}, which is applicable for $M_X>M_{\nu^c}$, our study (for simplicity, we set branching ratio to unity) shows that current LHC searches provide $m_{\textcolor{red}{D'}}\gtrsim 1.5$ TeV, as shown in Fig.~\ref{fig:LHC:Dprime}. This estimated bound can be somewhat relaxed in our scenario once all branching ratios are incorporated, as well as by engaging the mass dependence of $M_{\nu^c}$ (which is, however, beyond the scope of this work).

Finally, we also obtain a similar bound (namely, $M_{\textcolor{red}{D'}}\gtrsim 1.3$ TeV) from the process on the right panel of Fig.~\ref{fig:LHC:Dprime:production}, where we consider  $\textcolor{red}{D'}\to q\overline q+$MET (missing transverse energy). This scenario mimics the SUSY model where, after direct gluino pair production, each gluino undergoes a three-body decay to light-flavor quarks and a neutralino, where the latter gives rise to MET~\cite{CMS:2019ybf}. For the estimation of this limit, we again assumed the branching ratio to be unity. The right panel in Fig.~\ref{fig:LHC:Dprime} depicts the production cross section and the current LHC limit, which is taken from Ref.~\cite{CMS:2019ybf} for a vanishing neutralino mass.

\section{ Flavor violating processes }\label{sec:4}

In this section, we analyze various flavor-changing processes that are mediated by the new gauge bosons of the model.  These include meson decays such as $K_L \rightarrow \mu e$ which arise by the tree-level exchanges of the leptoquark gauge boson $X_\mu$, as well as loop-induced processes such as $K^0-\overline{K^0}$ mixing, and leptonic decays of mesons.  We also address charged lepton flavor violation that arises in the model, both at tree-level ($\mu-e$ conversion in nuclei) as well as via loops ($\mu \rightarrow e \gamma, \mu \rightarrow 3 e$, etc).

Before going into the details, let's briefly discuss the main point. In low-scale Pati-Salam models, the strongest flavor constraint usually comes from the $K_L\to \mu e$ channel. In our setup, this bound, along with the other meson decay bounds, is already much weaker at tree level. The suppression mechanism works in three stages: the chiral structure of the couplings, the small mixing induced by the soft breaking of $Z_2$, and then an additional flavor suppression. The first two steps already relax the usual bound substantially, when the remaining step down to the TeV range is achieved through the flavor structure in our benchmark scenarios. In the usual low scale Pati-Salam case, the $K_L\to \mu e$ decay gets a helicity-unsuppressed~(HU) tree-level contribution from the pseudoscalar-current piece, which is proportional to $m_K^2/(m_s+m_d)$, and this is the reason that makes the usual bound on the Pati-Salam scale strong. In contrast, in our model, the tree-level contribution is fully helicity suppressed~(HS), so the amplitude comes only from the axial-current piece and is therefore proportional to the charged-lepton mass. This gives the rough estimate
\begin{equation}
\frac{\Gamma_{\rm HU}}{\Gamma_{\rm HS}}
\sim
4\left[\frac{m_K^2}{(m_s+m_d)m_\mu}\right]^2
\sim 10^3
\end{equation}
for $K_L\to \mu e$. Since the branching ratio scales as $M_X^{-4}$, this structure alone lowers the usual bound on $M_X$ from about $2500$ TeV to about $450$ TeV. Without the soft $Z_2$ breaking, the tree-level decay would be absent. In our model, the soft breaking induces the small $d^c$--$D^c$ mixing, and the decay is generated only through this mixing. Thus, the branching ratio gets an additional suppression factor $(0.2)^4$ (see subsection~\ref{subsec:softly_broken_Z2} for the suppression factor of 0.2), which lowers the $M_X$ bound by another factor of $0.2$, down to about $90$ TeV.  Lowering the scale further down to $M_X\sim 1.1$ TeV asks for an additional flavor suppression beyond the helicity and soft-$Z_2$ effects, as realized in our benchmark scenarios. In kaon decay, the required suppression is roughly of the order $K_{s\mu}K_{de}\lesssim 10^{-4}$. At present, we do not have a dynamical reason for this additional suppression. These benchmark scenarios are not sharply unstable under moderate variations of the relevant entries for the dominant decay bounds. If we vary these entries by 10\%, it changes the lower limit on $M_X$ mildly, from about $1.10$ TeV to at most about $1.16$ TeV. It is also important to emphasize an important feature of our setup. In conventional low-scale Pati-Salam realizations, suppressing the dangerous $K$-meson constraints generally requires imposing special structures directly on both the left- and right-handed mixing matrices. In our case, because of the chiral structure of the setup, the left-handed CKM-like matrix does not need any special structure. Moreover, as we will discuss in benchmark scenario 2, it is also not necessary to impose a special structure on the right-handed CKM-like matrix itself. The required suppression may instead arise from the soft-breaking matrix $x$ (see Eq.~\eqref{eq:rho} for the definition of $x$), while the right-handed CKM-like matrix can have an arbitrary form. For a generic choice of parameters, the tree-level contribution still gives the dominant bound, while the $Z_2$-unsuppressed one-loop contribution depends on a different set of couplings and on exotic fermions running in the loop, and becomes relevant only in some parts of parameter space. As we will see below, this setup makes it possible to satisfy all flavor constraints with the Pati--Salam breaking scale as low as $1.1$ TeV in our benchmark scenarios.

\subsection{Meson decay at tree level}
\label{subsec:Meson decay at tree level}
Due to the $Z_2$ odd nature, $X_\mu$ gauge boson couples SM fermions to new vector-like fermions. In the limit of exact $Z_2$ symmetry, tree-level meson decays are forbidden in our model. However, once the $Z_2$ symmetry is softly broken, a mixing between $d^c$ and $D^c$ is induced, as discussed in Sec.~\ref{subsec:softly_broken_Z2}. These processes are helicity suppressed, since the leptoquark gauge boson $ X_\mu$ does not have a coupling of the type $\overline{d}_L \gamma_\mu e_L X^\mu$ in our setup.
\begin{figure}[h!]
    \centering
    \includegraphics[width=.3\textwidth]{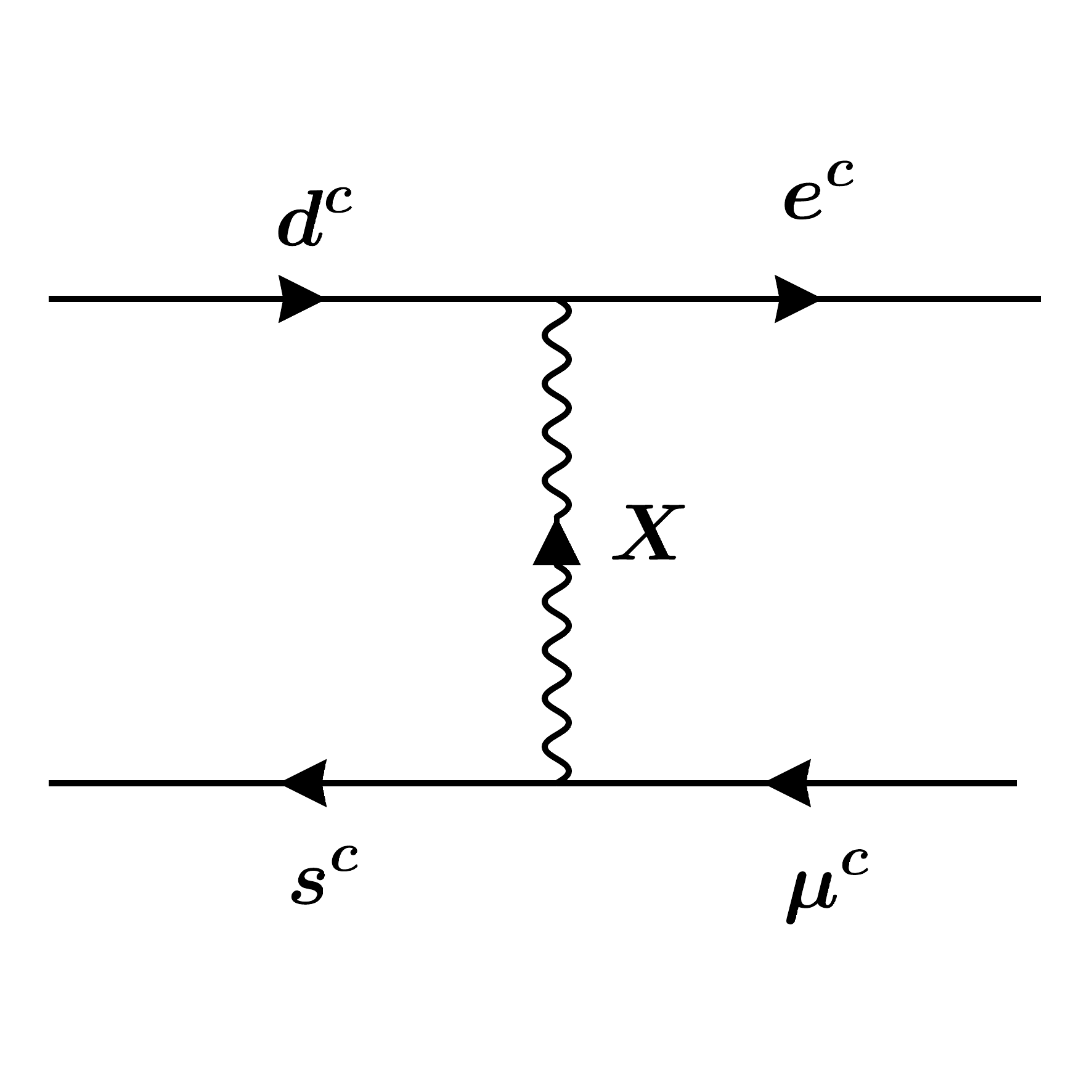}\hfill
     \includegraphics[width=.3\textwidth]{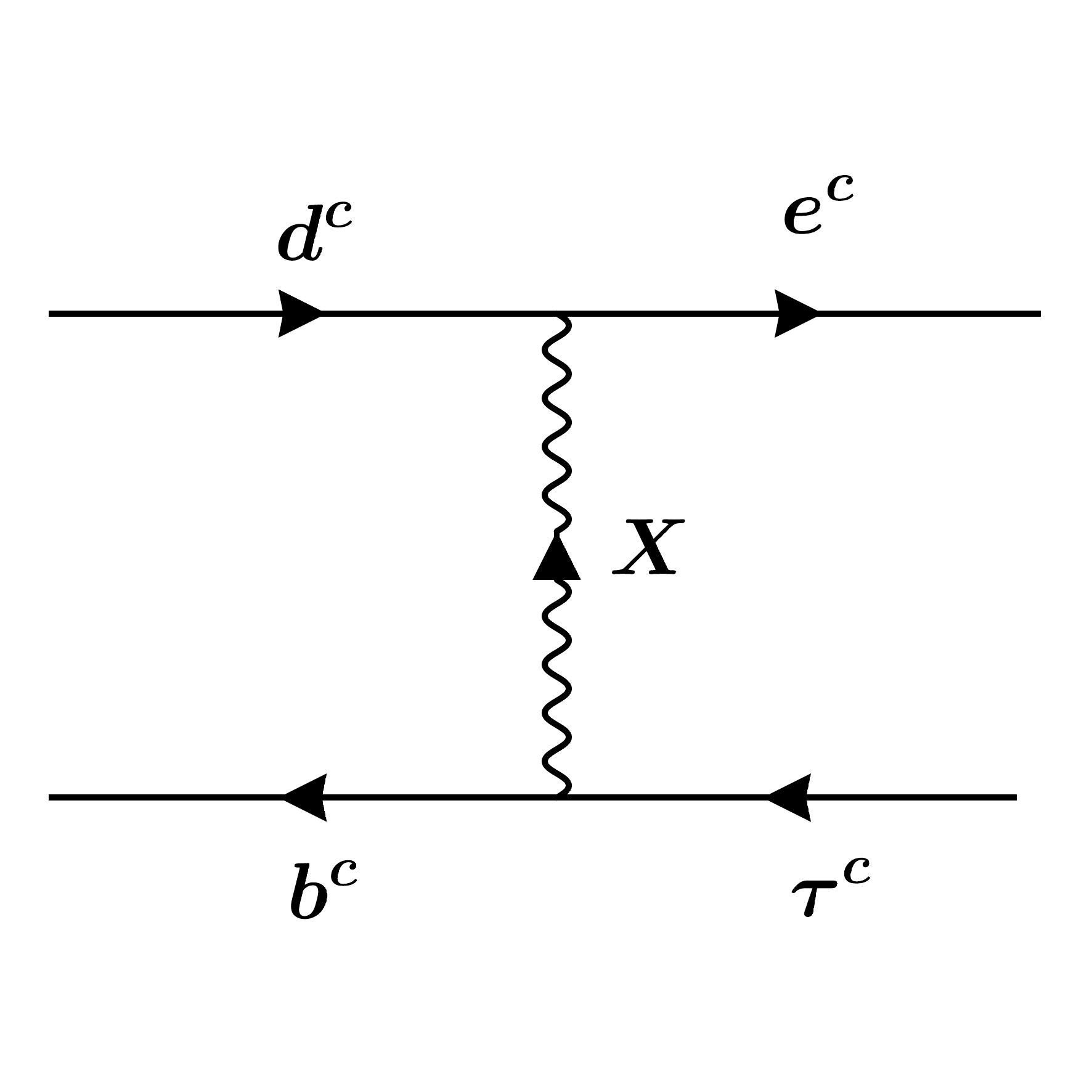}\hfill
      \includegraphics[width=.3\textwidth]{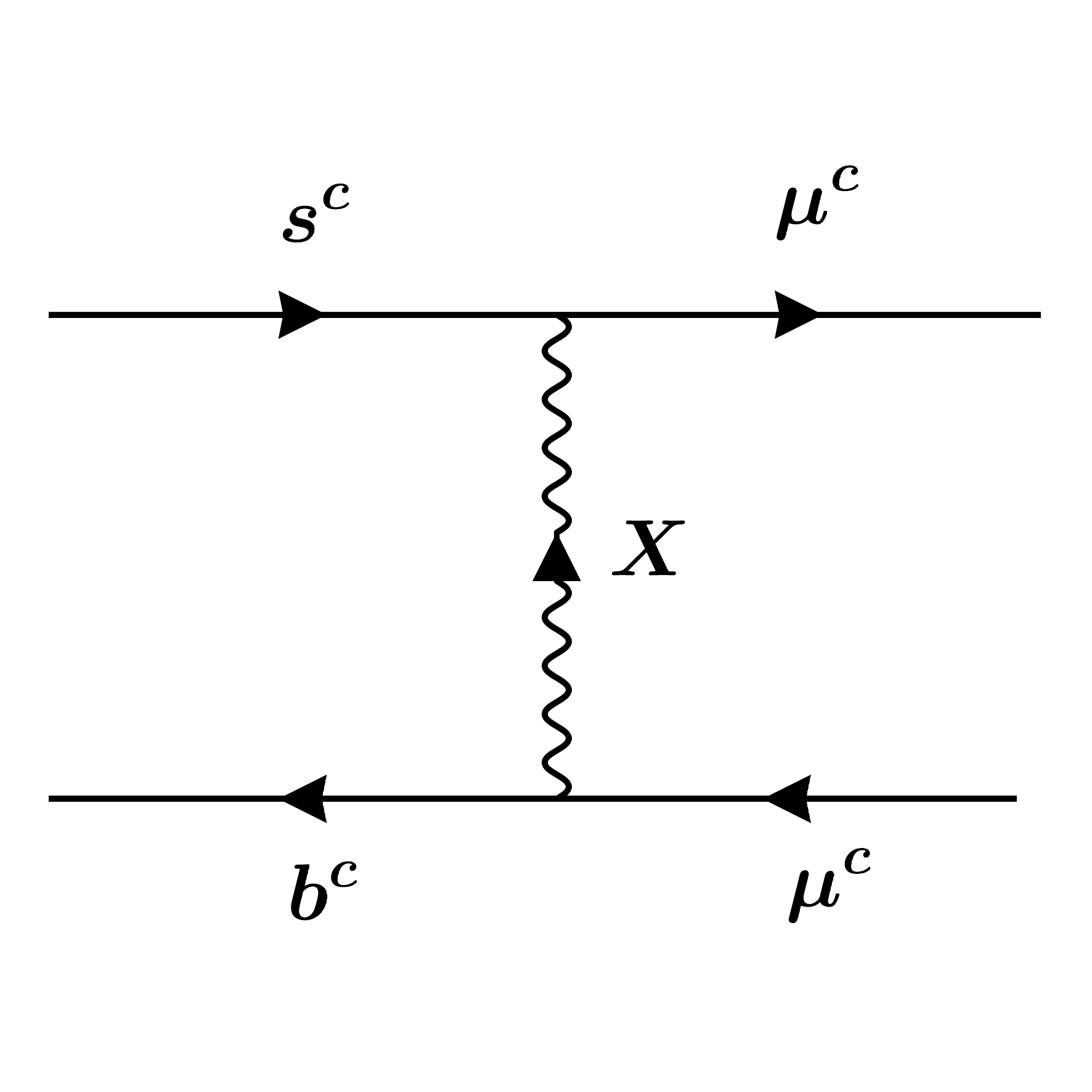}
    \caption{Representative diagrams for meson decay~($K_L \to \mu\, e,~B_d \to \tau\, e,~B_s \to \mu\, \mu$)}
    \label{fig:meson_tree}
\end{figure}
The most stringent constraint on the mass of $X_{\mu}$ arises from neutral meson decay channels with charged-lepton final states of opposite sign and without neutrinos in the final states, as summarized in Table~\ref{tab:FLAVOR}. In our framework, such decays occur only for neutral mesons composed of down-type quarks. The relevant coupling for this process can be found in Eq.~\eqref{eqn:L-L-X}. 
The $3 \times 3$ matrix  $K_{d^ce^c}$ appearing in this equation is not a unitary matrix due to the appearance of the $x$ matrix, unlike in the standard Pati--Salam setup. While the matrix elements of $x$ are taken to be small, $|x_{ij}| \sim 0.2$ (see Eq. (\ref{eq:xij}), the matrix $x$ must have rank-three, since the eigenvalues of $x$ times $\kappa_R$ are the masses of the $\textcolor{red}{D'}$ quarks.  This is because we wish to satisfy the collider limits on the masses of all three families of $\textcolor{red}{D'}$ 
(see Fig.~\ref{fig:LHC:Dprime}).

To get the mass limit on the $X_{\mu}$, let's consider the $X_{\mu}$-induced leptonic decays, \(\Phi_{pq} \rightarrow \ell_i^+ \ell_j^-\)(for examples, see Fig.~\ref{fig:meson_tree}) where the pseudoscalar mesons \( \Phi_{pq}  = (\overline{d_p} d_q)\)  are bound states of quark–antiquark pairs.
The decay width for helicity-suppressed meson decay is given by~\cite{Smirnov:2007hv,Dolan:2020doe}
\begin{align}
\Gamma^{\text{HS}}_{\Phi_{pq} \to \ell^+_i \ell^-_j}& = 
\frac{m_{\Phi_{pq}} \left[ g_4(m_X) \right]^4 f_{\Phi_{pq}}^2}{256 \pi\, m_X^4}
\sqrt{
\left( 1 - \left( \mu_{\ell_i} + \mu_{\ell_j} \right)^2 \right)
\left( 1 - \left( \mu_{\ell_i} - \mu_{\ell_j} \right)^2 \right)
}\notag\\
&\times 
\left(
\left(1 - \mu_{\ell_i}^2 - \mu_{\ell_j}^2 \right)
\left( m_{\ell_i}^2 + m_{\ell_j}^2 \right)
- 4 \mu_{\ell_i} \mu_{\ell_j} m_{\ell_i} m_{\ell_j}
\right)\left(
\left| K_{d^ce^c}^{pi} \right|^2 \left| K_{d^ce^c}^{qj} \right|^2
\right).
\label{eqn:decay_helicity_suppressed}
\end{align}
where $\mu_{\ell_i}=m_{\ell_i}/m_{\Phi_{pq}}$ and $f_{\Phi_{pq}}$ is the decay constant of the pseudoscalar meson.
\begin{table}[t!]
\begin{center}
{\renewcommand{\arraystretch}{1}
\scalebox{1.0}{
\begin{tabular}{|c|c|}
\hline
Process & \,\,Measured value/upper limit  \\ \hline\hline

BR$(K_L^0 \rightarrow e e)$ & $9.0^{+6.0}_{-4.0} \times 10^{-12}\,\,$ \cite{Valencia:1997xe,GomezDumm:1998gw}  \\\hline
BR$(K_L^0 \rightarrow \mu \mu )$ & $6.84^{+0.11}_{-0.11} \times 10^{-12}\,\,\,\,\,\,\,$ \cite{E871:2000wvm} \\\hline
BR$(K_L^0 \rightarrow \mu e )$ & $\,\,< 4.7 \times 10^{-12}$ \cite{BNL:1998apv} \\
\hline\hline
BR$(B_d^0 \rightarrow ee)$ &  $< 2.5\times 10^{-9}$  \cite{LHCb:2020pcv}   \\\hline
BR$(B_d^0 \rightarrow \mu\mu)$ & $\,\,< 3.6 \times 10^{-10}$ \cite{CMS:2019bbr} \\\hline
BR$(B_d^0 \rightarrow \tau\tau)$ & $<2.1\times 10^{-3}$ \cite{LHCb:2017myy} \\\hline
BR$(B_d^0 \rightarrow \mu e)$ & $<1.0\times 10^{-9}$ \cite{LHCb:2017hag} \\\hline
BR$(B_d^0 \rightarrow \tau e)$ &   $< 1.6\times 10^{-5}$  \cite{Belle:2021rod}  \\\hline
BR$(B_d^0 \rightarrow \tau \mu)$ &  $< 1.5\times 10^{-5}$  \cite{Belle:2021rod}  \\
\hline\hline
BR$(B_s^0 \rightarrow ee)$ &  $< 9.4\times 10^{-9}$  \cite{LHCb:2020pcv} \\\hline
BR$(B_s^0 \rightarrow \mu\mu)$ & $3.0^{+0.4}_{-0.4} \times 10^{-9}\,\,\,\,$ \cite{ATLAS:2018cur,CMS:2019bbr,LHCb:2017rmj}\\\hline
BR$(B_s^0 \rightarrow \tau\tau)$ & $<6.8\times 10^{-3}$ \cite{LHCb:2017myy} \\\hline
BR$(B_s^0 \rightarrow \mu e)$ & $<5.4\times 10^{-9}$ \cite{LHCb:2017hag} \\\hline \hline 
$K^0$--$\overline{K^0}: \Delta M_K$&
$(3.484 \pm 0.006)\times 10^{-15}\;\mathrm{GeV}$ \cite{ParticleDataGroup:2024cfk} \\\hline
$D^0$--$\overline D^0: \Delta M_D$&
$(6.25^{+2.70}_{-2.90}) \times 10^{-15}\;\mathrm{GeV}$ \cite{ParticleDataGroup:2024cfk} \\\hline
$B^0_d$--$\overline{B^0_d}: \Delta M_d$&
$(3.334 \pm 0.013) \times 10^{-13}\;\mathrm{GeV}$ \cite{ParticleDataGroup:2024cfk} \\\hline
$B^0_s$--$\overline{B^0_s}: \Delta M_s$&
$(1.1683 \pm 0.0013) \times 10^{-11}\;\mathrm{GeV}$ \cite{ParticleDataGroup:2024cfk} \\\hline\hline

BR($\mu\to e\gamma$) & $<4.2\times 10^{-13}$\ \cite{MEG:2016leq} 
\\
BR($\tau\to e\gamma$) & $<3.3\times 10^{-8}$\ \cite{BaBar:2009hkt} 
\\
BR($\tau\to\mu\gamma)$ & $<4.4\times 10^{-8}$\ \cite{BaBar:2009hkt} 
\\\hline\hline
BR($\mu\to e e e)$ & $<1.0\times 10^{-12}$\ \cite{SINDRUM:1987nra} 
\\
BR($\tau\to e e e)$ & $<2.7\times 10^{-8}$\ \cite{Hayasaka:2010np} 
\\
BR($\tau\to \mu\mu\mu)$ & $<2.1\times 10^{-8}$\ \cite{Hayasaka:2010np} 
\\
BR($\tau^-\to e^-\mu\mu$) & $<2.7\times 10^{-8}$\ \cite{Hayasaka:2010np} 
\\
BR($\tau^-\to \mu^-ee$) & $<1.8\times 10^{-8}$\ \cite{Hayasaka:2010np} 
\\
BR($\tau^-\to e^+\mu^-\mu^-$) & $<1.7\times 10^{-8}$\ \cite{Hayasaka:2010np} 
\\
BR($\tau^-\to \mu^+e^-e^-$) & $<1.5\times 10^{-8}$\ \cite{Hayasaka:2010np} 
\\
\hline
\end{tabular}
}}
\end{center}
\caption{Experimental constraints on rare meson decays,  meson-antimeson oscillations, and other cLFV processes.  }
\label{tab:FLAVOR}
\end{table}
\paragraph{Benchmark scenario-1:} For a simple numerical illustration, we first consider a case where the strongest kaon bound is absent for a specific flavor choice. This allows us to see how the remaining meson decay constraints bound the model when the kaon decay is absent. At the same time, the helicity suppression together with the softly broken $Z_2$ structure remains essential for realizing the low-scale solution. Let's first take the matrix \(x\) to be proportional to \(\mathbb{I}\), which means \((\mathbb{I}+x^\dagger x)^{-1/2} x^\dagger\approx 0.2\,\mathbb{I}\) (see subsection~\ref{subsec:softly_broken_Z2} for an explanation for the suppression of 0.2). In that case, \(K_{d^ce^c}\) which follows from Eq.~\eqref{eqn:Kdcec} is
\begin{equation}
   K_{d^ce^c}\approx -0.2\,V_{d^ce^c},
\end{equation}
with \(V_{d^ce^c}=V_{d^c}^\dagger\, V_{e^c}\) for which we assume a CKM matrix-like form, with the angle $\theta_{13}$ set to  $\pi/2$, given by 
\begin{equation}
V_{d^ce^c} =
\left(
\begin{array}{ccc}
0 & 0 & 1 \\
- s_{12}c_{23}- c_{12} s_{23}e^{i \delta}  & 
c_{12} c_{23} - e^{i \delta} s_{12} s_{23} & 0 \\
- c_{12} c_{23} + e^{-i \delta} s_{12} s_{23} & 
- c_{23} s_{12} - e^{-i \delta} c_{12} s_{23} & 0
\end{array}
\right).
\end{equation}
\begin{figure}[t!]
    \centering
    \includegraphics[width=0.65\textwidth]{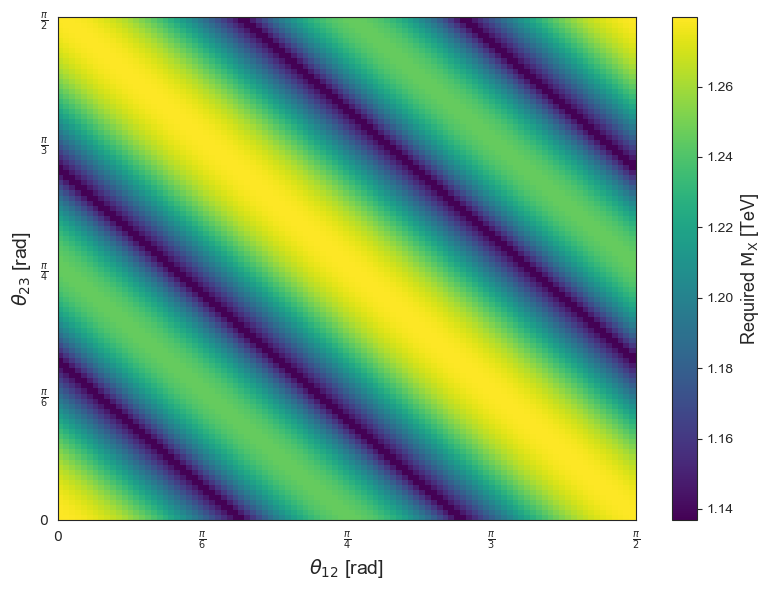}
    \caption{Minimum leptoquark gauge boson mass $M_X$ that is required to satisfy all meson decay constraints, as a function of the mixing angles $\theta_{12}$ and $\theta_{23}$, with fixed $\theta_{13}=\pi/2,\,\delta = 0$.}
    \label{fig:mx_theta_scan}
\end{figure}
Here $s_{ij} = \sin\theta_{ij}$ and $c_{ij} = \cos\theta_{ij}$. 
The motivation for this choice, with a third angle $\theta_{13}$ in the unitary matrix set to $\pi/2$, is clear from   Table~\ref{tab:FLAVOR} which shows that this choice eliminates the most stringent limit on the mass of $X_\mu$ arising from the decay $K_L \rightarrow \mu e$. We can perform a numerical scan of the angle space $(\theta_{12}, \theta_{23})$ for fixed $\theta_{13}=\pi/2$, and fixed CP phase $\delta = 0$, that will compute the minimum $M_X$ required to satisfy all the experimental bounds simultaneously~(see Table~\ref{tab:FLAVOR}). The resulting pattern is shown in Fig.~\ref{fig:mx_theta_scan}, where the color-bar corresponds to \(M_X\) coming from the strongest constraint among the relevant flavor-violating meson decays.

Now, we can optimize all the processes in order to get the lowest allowed limit on \(M_X\) for the specific configuration of \(\theta_{12},\theta_{23},\delta \). We find that the anglular configuration $\theta_{12} = 0.16\,\text{rad}$, $\theta_{23} = 0.36\,\text{rad}$, and $\delta = 0.81\,\text{rad}$ provides the lowest value of the leptoquark gauge boson mass $M_X$ consistent with the five decay constraints, which is $M_X \geq 1.1\,\text{TeV}$. As seen in Table~\ref{table:meson_decay}, this bound is primarily driven by the rare decay process $B_s \to \mu\mu$, with $B_s \to \mu e$ providing a nearly identical lower limit. The constraints from $B_d \to \tau e$, $B_d \to \tau\mu$, and $B_s \to ee$ are significantly weaker and are automatically satisfied.
\begin{table}[h!]
\centering
\renewcommand{\arraystretch}{1.2}
\begin{tabular}{|c|c|c|}
\hline
Decay process & Limit on $M_X$ (GeV)\\
\hline
1. $B_d \to \tau e$      & $\geq 5.7\times 10^{2}$ \\
2. $B_d \to \tau \mu$    & $\geq 4.1\times 10^{2}$ \\
3. $B_s \to e e$         & $\geq 58.4$ \\
4. $B_s \to \mu \mu$     & $\geq 1.1 \times 10^{3}$ \\
5. $B_s \to \mu e$       & $\geq 1.1 \times 10^{3}$ \\
\hline
\end{tabular}
\caption{Lower bounds on \( M_X \) from various meson decay processes (Benchmark-1)}
\label{table:meson_decay}
\end{table}
\paragraph{Benchmark scenario-2:} This benchmark is the more general representative realization of the low-scale solution in our setup, and it will also be the one used in the downstream cLFV analysis. An interesting feature of our model is that, to suppress $K$-meson decays, imposing a specific structure on the CKM-like matrix is unnecessary. The required flavor suppression can arise from the matrix $x$ instead, which is proportional to the ${\color{red} D'}$ mass matrix. To illustrate this, let's take $V_{d^c}$ to be similar to the CKM matrix:
\begin{equation}
  V_{d^c}=
\begin{pmatrix}
0.9745 & 0.2243 & 0.00143 - 0.00367\,i\\
-0.2241 - 0.00016\,i & 0.9736 - 0.00004\,i & 0.04215\\
0.00839 - 0.00350\,i & -0.0410 - 0.00072\,i & 0.9991
\end{pmatrix},
\label{eqn:benchmark2}
\end{equation}
and we take $V_{e^c} \simeq \mathbb{I}_3$ for definiteness. We now take $x$ to be a general $3 \times 3$ matrix, subject to the condition that it has full rank. This would guarantee that all three of the ${\color{red} D'}$ quarks have similar masses of order 1.5 TeV. Using the helicity-suppressed decay formula given in Eq.~\eqref{eqn:decay_helicity_suppressed}, we compute the branching ratios for all 13 flavor-violating meson decay channels listed in Table~\ref {tab:FLAVOR}, and minimize the $X_\mu$ gauge boson mass consistent with all bounds. We summarize the resulting lower bounds on $M_X$ for this benchmark scenario in Table~\ref {tab:benchmark-2}.
\begin{table}[h!]
\centering
\setlength{\tabcolsep}{8pt}
\renewcommand{\arraystretch}{1.15}
\begin{tabular}{|l|r@{\,$\ge$\,}l|}
\hline
\multicolumn{1}{|c|}{Decay process} &
\multicolumn{2}{c|}{Limit on $M_X$ (GeV)} \\
\hline
$K_L \to ee$         & & $47$ \\
$K_L \to \mu\mu$     & & $8.1\times 10^2$ \\
$K_L \to \mu e$      & & $9.1\times 10^{2}$ \\
\hline
$B_d \to ee$         & & $2.3$ \\
$B_d \to \mu\mu$     & & $48 $ \\
$B_d \to \tau\tau$   & & $0.10$ \\
$B_d \to \mu e$      & & $31$ \\
$B_d \to \tau e$     & & $5.8\times 10^{2}$ \\
$B_d \to \tau\mu$    & & $4.3\times 10^{2}$ \\
\hline
$B_s \to ee$         & & $57.4$ \\
$B_s \to \mu\mu$     & & $1.1\times 10^{3}$ \\
$B_s \to \tau\tau$   & & $0.04$ \\
$B_s \to \mu e$      & & $1.1\times 10^{3}$ \\
\hline
\end{tabular}
\caption{Lower bounds on $M_X$ from meson decays (Benchmark-2).}
\label{tab:benchmark-2}
\end{table}
For the best-fit point, we obtain a representative numerical solution for the matrix \(x\):
\begin{equation}
x \;\approx\;
\begin{pmatrix}
 0.0173 + 0.0108\,\mathrm{i}  &\quad 0.0690 + 0.0495\,\mathrm{i}  &\quad -0.180 + 0.0062\,\mathrm{i} \\
 0.0401 + 0.0020\,\mathrm{i}  &\quad 0.1750 + 0.0059\,\mathrm{i}  &\quad  0.0711 - 0.0506\,\mathrm{i} \\
 -0.2110                      &\quad 0.0487                      &\quad -0.00190 - 0.000687\,\mathrm{i}
\end{pmatrix},
\label{eqn:x_bench2}
\end{equation}
corresponding to physical eigenvalues \(\sigma(x) \approx \{0.22,\,0.20,\,0.20\}\). These eigenvalues, multiplied by $\kappa_R \sim 6\, \rm TeV$ are the masses of the ${\color{red} D'}$ quarks. The corresponding CKM-like matrix $K_{d^ce^c}$ takes the form
\begin{equation}
\resizebox{\textwidth}{!}{$
K_{d^ce^c}\approx -
\begin{pmatrix}
 9.48\times10^{-5} & -7.08\times10^{-5} & 2.12\times10^{-1} \\
 -7.70\times10^{-2} + 4.95\times10^{-2}\,\mathrm{i} &
 -1.73\times10^{-1} + 8.05\times10^{-3}\,\mathrm{i} &
 2.54\times10^{-5} + 2.10\times10^{-5}\,\mathrm{i} \\
 1.73\times10^{-1} + 8.05\times10^{-3}\,\mathrm{i} &
 -7.70\times10^{-2} - 4.95\times10^{-2}\,\mathrm{i} &
 1.12\times10^{-4} + 2.18\times10^{-5}\,\mathrm{i}
\end{pmatrix}.
$}
\label{eqn:optimized_K_mu_e}
\end{equation}
The most stringent constraints in this benchmark also come from the
$B_s\to\mu\mu$ and $B_s\to\mu e$ channels for this benchmark point, which require
\begin{equation}
  M_X \;\gtrsim\; 1.1\times 10^3\,\text{GeV}.
\end{equation}
We wish to make an important comment here: In this second benchmark, when the flavor structure is pushed into $x$, $K_{d^ce^c}=-V_{d^c}^\dagger \, (\mathbb{I}+x^\dagger x)^{-1/2} x^\dagger \, V_{e^c}$ is no longer unitary. A priori, dipole amplitudes in charged lepton flavor-violating (cLFV) processes can be in danger because of the absence of a GIM suppression owing to the non-unitarity. For example, the dipole amplitude for $\mu \to e \gamma$ will depend on the off–diagonal combination \(S \equiv \sum_{q^c} (K)_{q^ce^c}\,(K)^{*}_{q^c\mu^c}
\) when a light state propagates in the loop. (Loops with an internal exotic fermion carry free $X_{\mu}$–$F$ couplings and
do not involve $K_{d^ce^c}$, so they are unaffected by its non–unitarity.) Accordingly, in optimizing the matrices in Eq.~\eqref{eqn:optimized_K_mu_e}, we required $S$ to be numerically small, so that the predicted cLFV processes comfortably satisfy existing bounds. Therefore, we will adopt the same $K_{d^ce^c}$ quoted in Eq.~\eqref{eqn:optimized_K_mu_e} in all downstream cLFV analyses; the detailed \(\mu\!\to e\gamma\), \(\mu\!\to eee\), and \(\mu\!\text{–}e\) conversion formulas and limits are presented in Sec.~\ref{subsec:Limit on m_X from CLFV processes}.

\subsection{$K_L \to \mu e$ at one loop level}

\begin{figure}[H]
    \centering
    \begin{subfigure}[b]{0.32\textwidth}
        \centering
        \includegraphics[width=\textwidth]{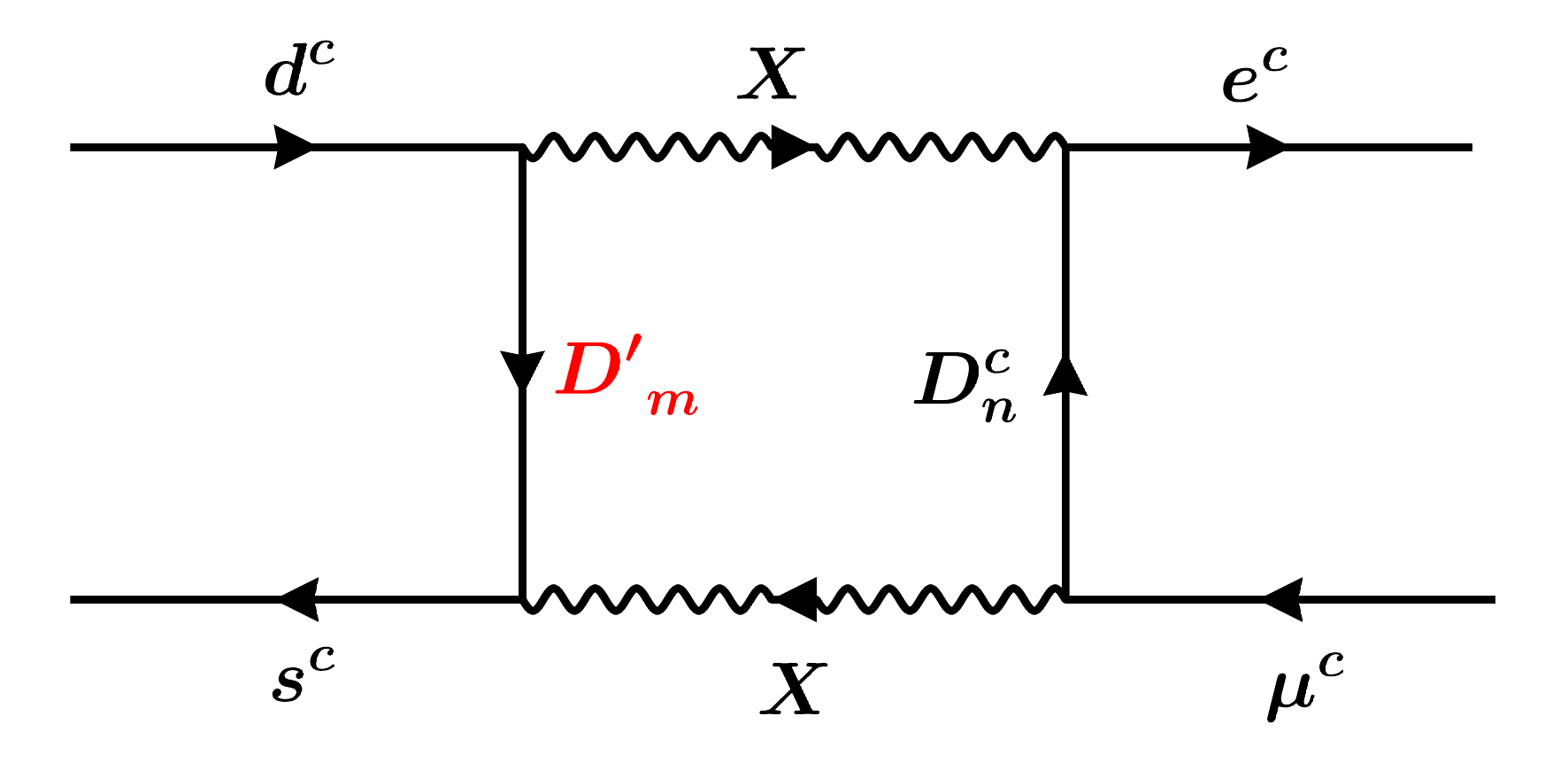}
        \caption{}
    \end{subfigure}
    \hfill
    \begin{subfigure}[b]{0.32\textwidth}
        \centering
        \includegraphics[width=\textwidth]{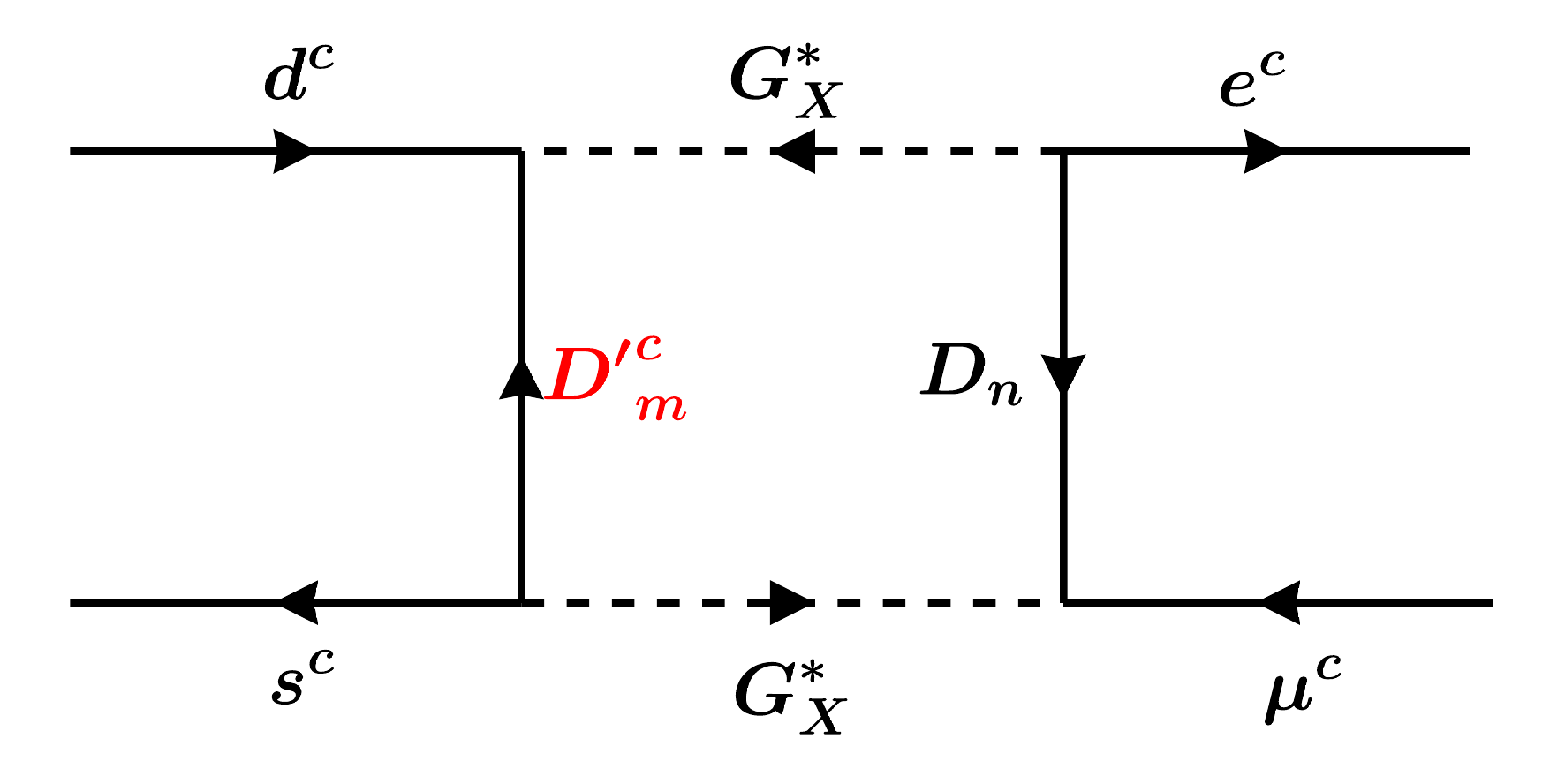}
        \caption{}
    \end{subfigure}
        \hfill
    \begin{subfigure}[b]{0.32\textwidth}
        \centering
        \includegraphics[width=\textwidth]{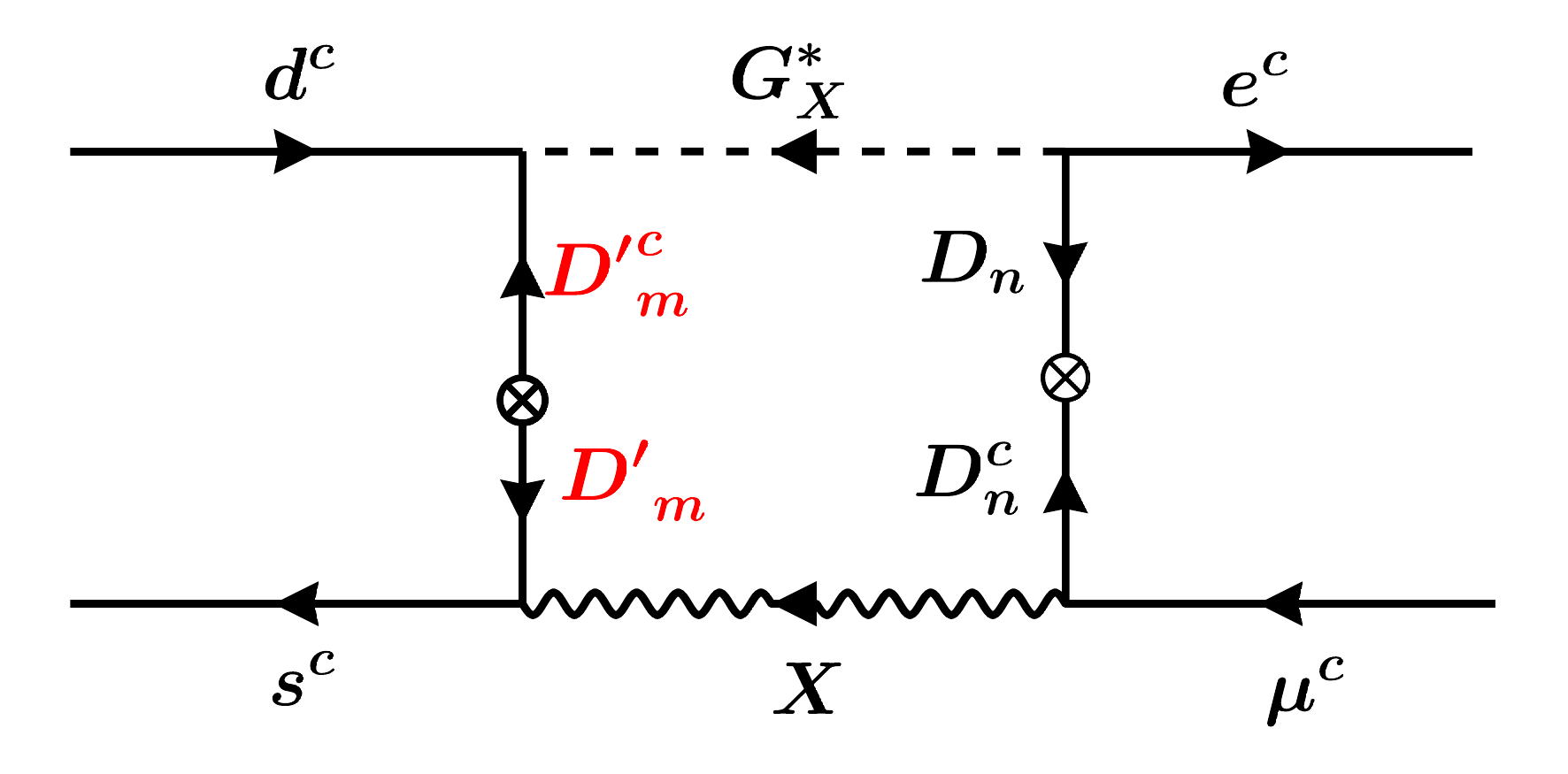}
        \caption{}
    \end{subfigure}
    \begin{subfigure}[b]{0.32\textwidth}
        \centering
        \includegraphics[width=\textwidth]{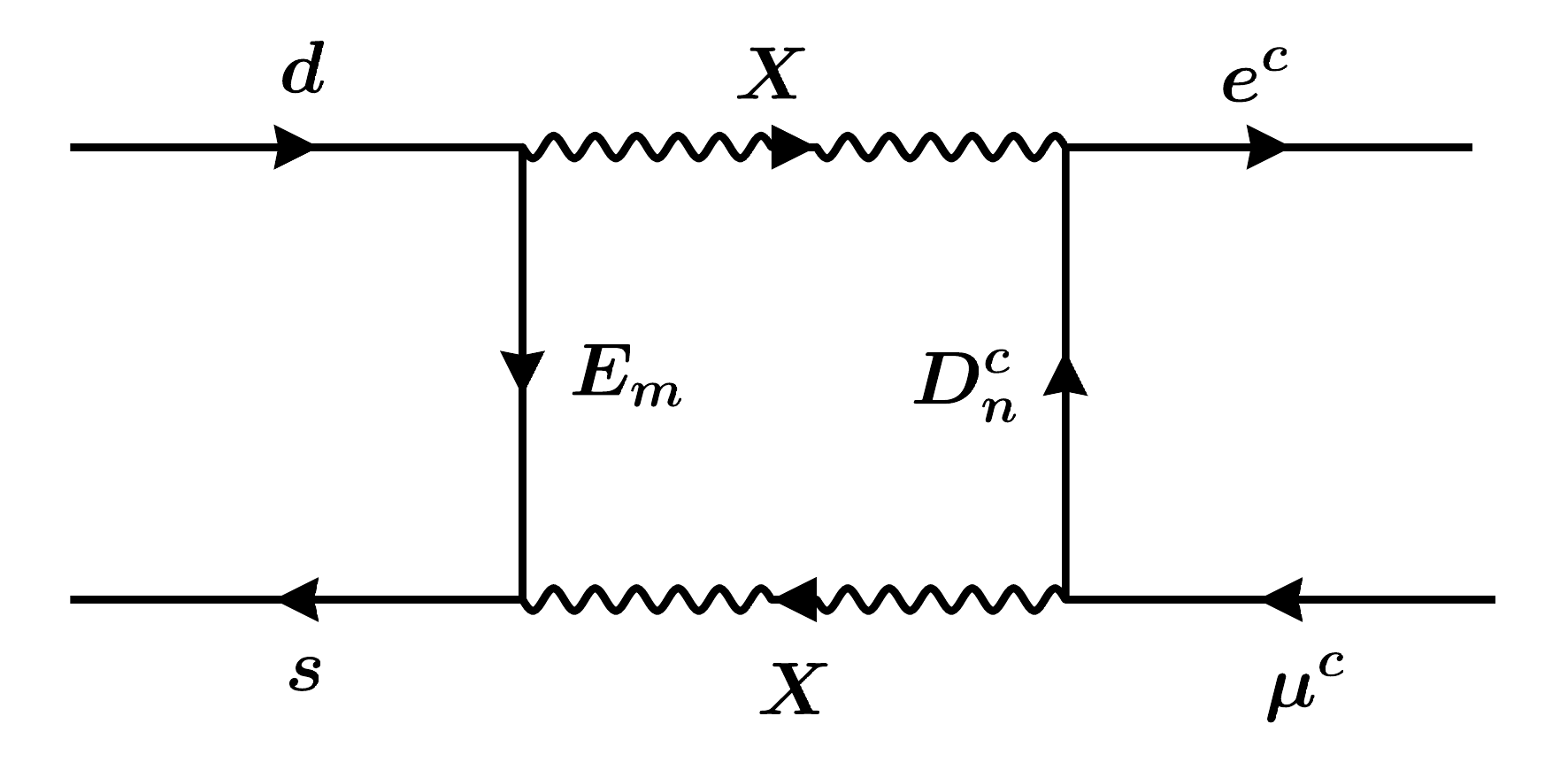}
        \caption{}
    \end{subfigure}
    \hfill
    \begin{subfigure}[b]{0.32\textwidth}
        \centering
        \includegraphics[width=\textwidth]{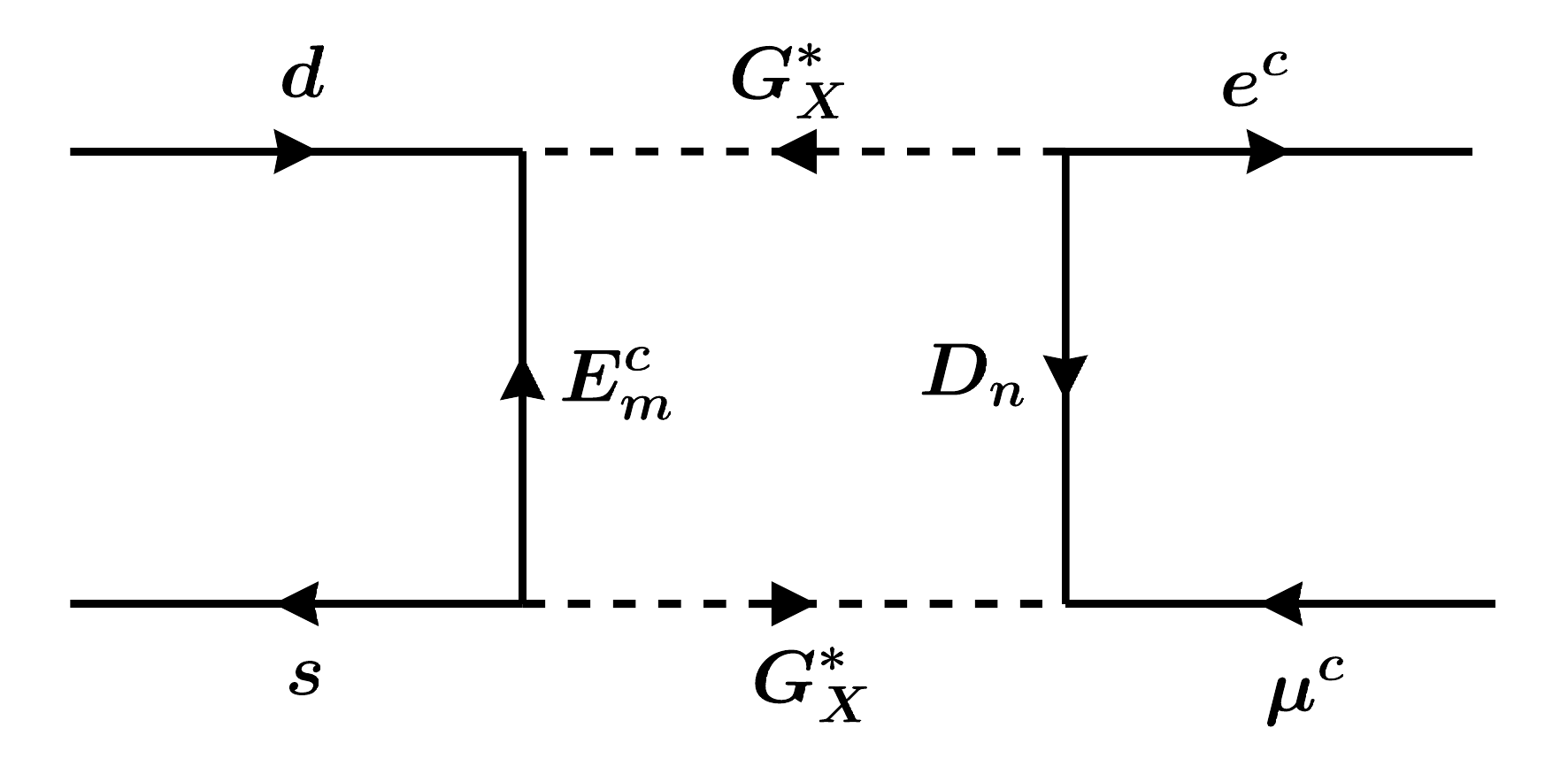}
        \caption{}
    \end{subfigure}
        \hfill
    \begin{subfigure}[b]{0.32\textwidth}
        \centering
        \includegraphics[width=\textwidth]{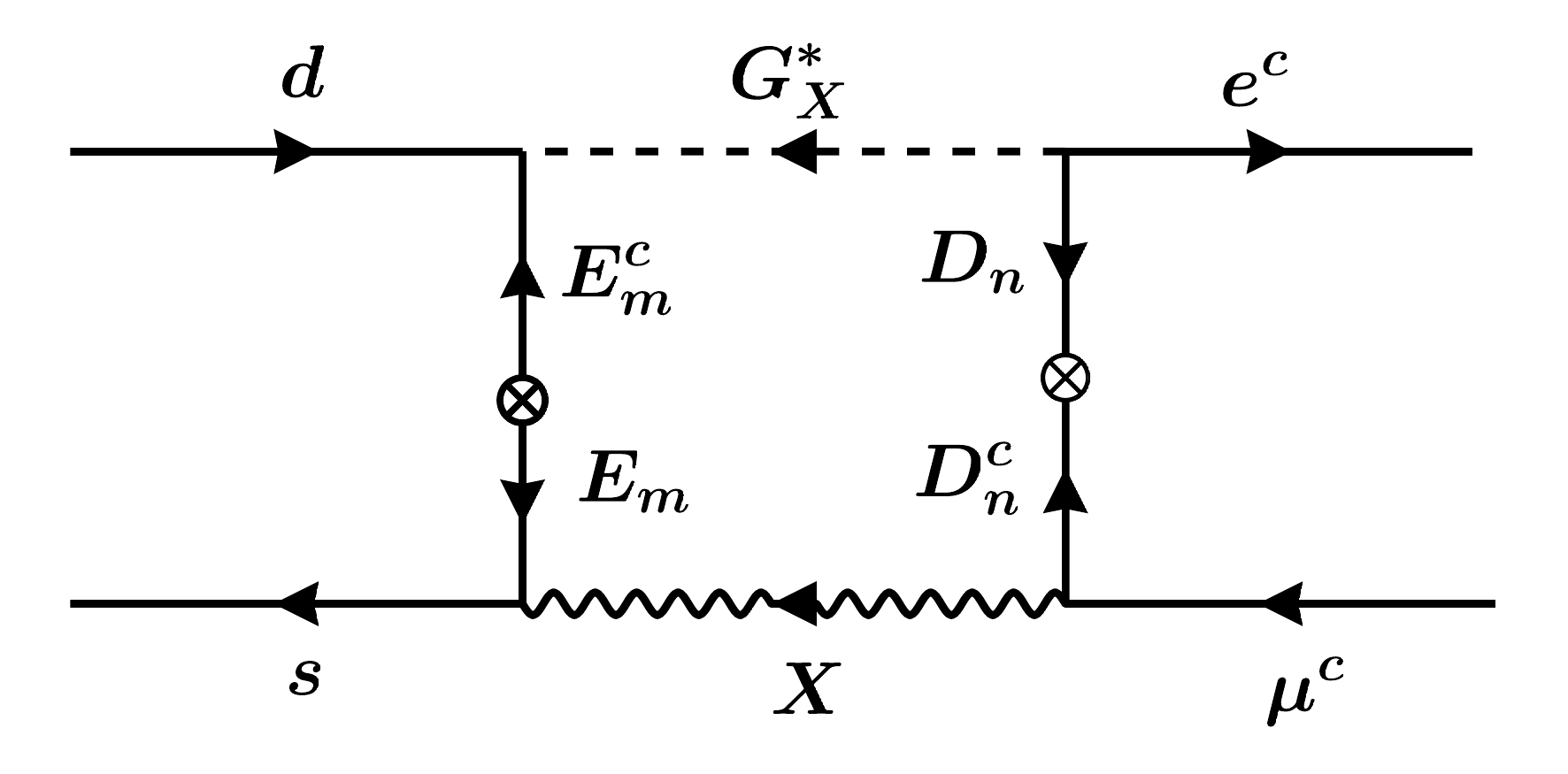}
        \caption{}
    \end{subfigure}
   \caption{Box diagrams with leptoquark gauge boson and its Goldstone mode contributing to the $K_L \to \mu e$ decay process. Here, $m,n$ are flavor indices. For topology (c) and (f), there is an additional diagram obtained by interchanging $X_\mu$ and $G_X$ on the internal lines. These will give identical contributions to the Wilson coefficients, which has been included in our results.}
    \label{fig:box-diagrams}
\end{figure}

We should also consider meson decays into lepton pairs at the loop level, because we can have box diagrams containing leptoquark gauge boson and new vector-like quarks, where the suppression coming from the $Z_2$ soft breaking will be absent. It is not obvious that the one-loop contribution will be subdominant for all of the parameter space of the model. We focus on the $K_L \to \mu e$ process because it provides the most stringent limit with the CKM-like matrices taken to be identical to the generic CKM matrix (the same structure used in Eq.~\eqref{eqn:benchmark2} for \(V_{d^c}\)). We will perform the calculations in ’t
Hooft-Feynman gauge, where one should also include contributions from the unphysical Goldstone bosons, which is reflected in Fig.~\ref{fig:box-diagrams}. We consider the effective operators under the approximation of vanishing external momenta and negligible light quark masses:

\begin{equation}
\mathcal L_{\rm eff} \;\supset\;
C_{RR}\,(\overline{s^c} \gamma_\mu d^c)(\overline{\mu^c} \gamma^\mu e^c)
+ C_{LR}\,(\bar s \gamma_\mu d)(\overline{\mu^c} \gamma^\mu e^c)+\mathrm{h.c.},
\end{equation}
\noindent with the leading-order expressions for the Wilson coefficients:
\begin{align}
    C_{RR}^{(a+b+c)}&= -2\left(\frac{g_4}{\sqrt{2}} \right)^4 \frac{1}{4\pi^2 M_X^2}\left( \sum_{m,n} \xi_{\textcolor{red}{D'_m}} \eta_{D^c_n}A(r_{\textcolor{red}{D'_m}}, r_{D_n})+\frac{1}{32}\sum_{m,n} \xi_{\textcolor{red}{D'_m}} \eta_{D^c_n}A(r_{\textcolor{red}{D'_m}}, r_{D_{n}})r_{\textcolor{red}{D'_m}} r_{D_{n}}\right.\notag\\
    &\left.~~~~~~~~~~~~~~~~~~~~~~~~~~~~~~~~~~~~~~~-\sum_{m,n} \xi_{\textcolor{red}{D'_m}} \eta_{D^c_n}B(r_{\textcolor{red}{D'_m}}, r_{D_{n}})r_{\textcolor{red}{D'_m}} r_{D_{n}}\right),\nonumber\\
     C_{LR}^{(d+e+f)}&= -2\left(\frac{g_4}{\sqrt{2}} \right)^4 \frac{1}{4\pi^2 M_X^2} \left(\sum_{m,n} \xi_{E_{m}} \eta_{D^c_{n}}A(r_{E_{m}}, r_{D_{n}})+\frac{1}{32}\sum_{m,n} \xi_{E_{m}} \eta_{D^c_{n}}A(r_{E_{m}}, r_{D_{n}})r_{E_{m}} r_{D_{n}}\right.\nonumber\\
 &\left.~~~~~~~~~~~~~~~~~~~~~~~~~~~~~~~~~~~~~~~-\sum_{m,n} \xi_{E{m}} \eta_{D^c_{n}}B(r_{E_{m}}, r_{D_{n}})r_{E_{m}} r_{D_{n}}\right).
\end{align}
Here we have used the definitions \(\xi_{\textcolor{red}{D'_m}}=K_{s^c\textcolor{red}{D'_m}}K_{d^c\textcolor{red}{D'_m}}^*\), $\xi_{E_m}=V_{sE_m}V_{dE_n}^*$, $\eta_{D^c_n}=K_{D_n^c\mu^c}K_{D_n^ce^c}^*$, $r_c=M_c^2/M_X^2$, and
\begin{align}
    A(x_i,x_j)&=\frac{1}{(1-x_i)(1-x_j)}+\frac{1}{(x_i-x_j)}\left[\frac{x_i^2}{(1-x_i)^2}\mathrm{ln}x_i-\frac{x_j^2}{(1-x_j)^2}\mathrm{ln}x_j\right],\nonumber\\
   B(x_i,x_j)&=-\frac{1}{(1-x_i)(1-x_j)}+\frac{1}{(x_i-x_j)}\left[\frac{x_j}{(1-x_j)^2}\mathrm{ln}x_j-\frac{x_i}{(1-x_i)^2}\mathrm{ln}x_i\right]. 
   \label{eqn:Inami-Lim}
\end{align}
The notation \( C_{RR}^{(a+b+c)} \) indicates that diagrams (a), (b), and (c) in Fig.~\ref{fig:box-diagrams} contribute to $C_{RR}$. We use the standard definition of the axial current matrix element~\cite{Valencia:1994cj},
\begin{equation}
\langle 0 |\, \bar d \gamma^\mu \gamma_5 s \,|K(p_K)\rangle 
= i f_K p_K^\mu ,
\end{equation}
where $f_K$ is the kaon decay constant. 
\begin{figure}[h!]
    \centering
    \includegraphics[width=0.65\textwidth]{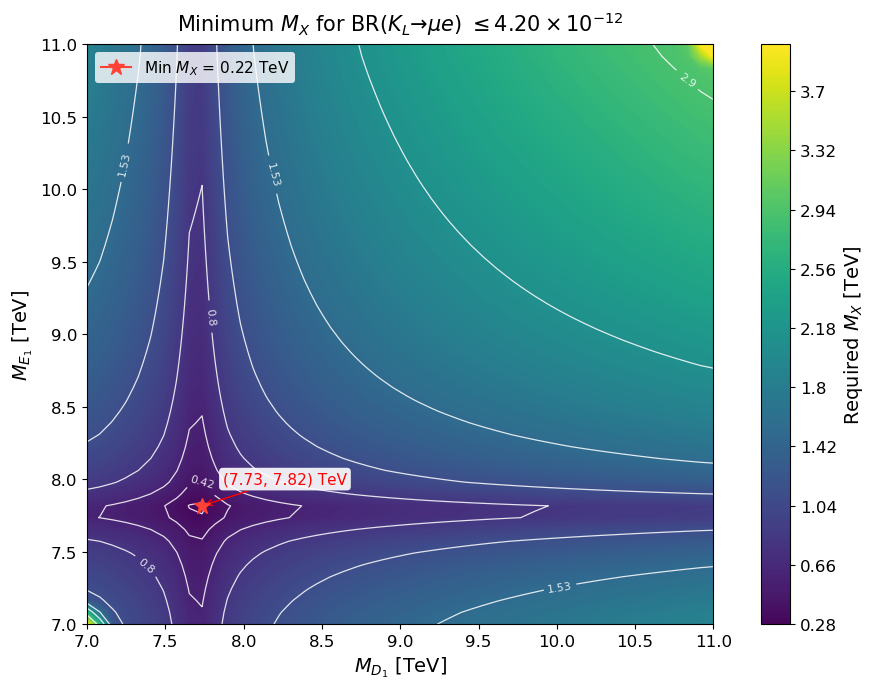}
    \caption{Heatmap showing the contour of $M_X$ (TeV) required to suppress the branching ratio $\mathrm{Br}(K_L \to \mu^\pm e^\mp) < 4.7 \times 10^{-12}$, as a function of vector-like fermion mass $M_{D_1}$. For this benchmark, we assume all the heavy fermions have similar masses, and the CKM-like matrices have structures aligned with the standard CKM matrix.  }
    \label{fig:kL_one_loop_decay}
\end{figure}
The decay width at one-loop in our setup can be written as 
\begin{equation}
\Gamma^{\mathrm{one-loop}}_{K\to \mu e}
=\frac{f_K^{\,2}\,m_K}{64\pi}\;
\lambda^{1/2}\!\big(1,r_\mu,r_e\big)\;
\big(1-r_\mu-r_e\big)\;
\big(m_\mu - m_e\big)^2\;
\big|C_{LR}-C_{RR}\big|^2
\end{equation}
with the definitions of dimensionless ratios \(r_\ell \equiv \frac{m_\ell^2}{m_K^2}~(\ell=\mu,e),\) and the dimensionless Källén function \(\lambda(1,x,y)\equiv 1 + x^2 + y^2 - 2\,(x + y + xy).\) Our four–fermion operators contain only vector/axial quark currents, and because of their non-singlet Ward identities, which enforce $Z_V=Z_A=1$ (zero anomalous dimensions) for $V^\mu=\bar q\gamma^\mu T_a q$ and $A^\mu=\bar q\gamma^\mu\gamma_5 T_a q$, all long-distance QCD can be absorbed into $f_K$. By contrast, scalar/pseudo-scalar densities would require a running factor~\cite{Buras:1998raa,Peskin:1995ev}.

Now that we have the setup in place, we can proceed with a benchmark parameter scan to gain an understanding of the variation in the PS scale with vector-like fermion masses involved in the loop. As mentioned before, we will take the benchmark CKM-like matrices (\(\xi_{E_m}, \eta_{D_n}\) etc) to be identical to the CKM matrix. We want to emphasize the point that the entries of the CKM-like matrices entering the amplitude are free input parameters here. In fact, for many choices (such as a diagonal structure), the new physics contributions to meson decay can be minimal. But we are interested in asking the following question: assuming that the mixing is not fine-tuned to minimize the contribution to \(K_L \to \mu e\), how low can \( M_X \) be while satisfying the experimental flavor constraints? Fig.~\ref{fig:kL_one_loop_decay} presents one benchmark scenario, where we vary the exotic fermion masses for a single generation, under the assumption that \(M_{\textcolor{red}{D'}}=(1.5,1.7,2.1)~\mathrm{TeV}\) and \(M_D,~M_E\) exotics lie around a common scale. Specifically, we scan over \( M_{D}= (M_{D_1},~7.7,~8.5)~\mathrm{TeV} \) and 
\( M_E = (M_{E_1},~7.9,~8.6)~\mathrm{TeV} \). We have shown the contours of \(M_X\) up-to 4 TeV satisfying the constraints with different choices of \((M_{D_1},M_{E_1})\).

\subsection{Limit on $M_X$ from mass difference of neutral mesons}
Neutral meson systems, $\Phi^0-\bar{\Phi}^0$ with $\Phi^0\equiv \left( K^0, D^0, B_s^0,~B_d^0\right)$ are known to exhibit oscillations due to neutral current interactions that change flavor by two units~($\Delta F=2$). The propagating physical neutral mesons are not CP eigenstates but rather mass eigenstates, $M_H$ and $M_L$, which are combinations of $\Phi^0$ and $\bar{\Phi}^0$. 
The small mass difference $\Delta M\equiv M_H - M_L=2~\mathrm{Re}\langle\bar{\Phi}^0|\mathcal{H}^{\Delta F = 2}_\text{eff}|\Phi^0\rangle$ arises from off-diagonal contributions in the meson mixing matrix. In the SM, this occurs at the one-loop level through the familiar box diagrams involving $W$ and up-type quarks and is loop and CKM suppressed. So, these mass differences can be highly sensitive to new-physics contributions, especially if the scenario introduces new sources of flavor violation or right-handed currents.

In our model, additional contributions to the $\Delta F=2$ processes are induced through box diagrams involving $X_{\mu}, W'_{\mu}$, and vector-like fermions. Thus, it is necessary to  $\Delta M$ for some benchmark scenarios to check whether it sets a more stringent limit on the Pati-Salam scale compared to tree-level meson decay and direct collider bounds. 

The effective Hamiltonian for these processes can be written as:
\begin{equation}
\mathcal{H}^{\Delta F = 2}_\text{eff} = \sum_{i=1}^{5} C_i\, Q_i^{q_i q_j} + \sum_{i=1}^{3} \tilde{C}_i\, \tilde{Q}_i^{q_i q_j},
\label{eqn:H_eff_gen}
\end{equation}
where $q_i q_j = ds, cu, bd, bs$ label the quark flavors involved in the meson mixing. The operator basis is given by~\cite{DiLuzio:2019jyq,UTfit:2007eik,Bauer:2009cf,Dcruz:2023mvf}:
\begin{align}
Q_1^{q_i q_j} &= (\bar{q}_{L i}^\alpha \gamma^\mu q_{L j}^\alpha)(\bar{q}_{L i}^\beta \gamma_\mu q_{L j}^\beta), &
\widetilde{Q}_1^{q_i q_j} &= (\bar{q}_{R i}^\alpha \gamma^\mu q_{R j}^\alpha)(\bar{q}_{R i}^\beta \gamma_\mu q_{R j}^\beta), \nonumber \\
Q_2^{q_i q_j} &= (\bar{q}_{R i}^\alpha q_{L j}^\alpha)(\bar{q}_{R i}^\beta q_{L j}^\beta), &
\widetilde{Q}_2^{q_i q_j} &= (\bar{q}_{L i}^\alpha q_{R j}^\alpha)(\bar{q}_{L i}^\beta q_{R j}^\beta), \nonumber \\
Q_3^{q_i q_j} &= (\bar{q}_{R i}^\alpha q_{L j}^\beta)(\bar{q}_{R i}^\beta q_{L j}^\alpha), &
\widetilde{Q}_3^{q_i q_j} &= (\bar{q}_{L i}^\alpha q_{R j}^\beta)(\bar{q}_{L i}^\beta q_{R j}^\alpha), \nonumber \\
Q_4^{q_i q_j} &= (\bar{q}_{R i}^\alpha q_{L j}^\alpha)(\bar{q}_{L i}^\beta q_{R j}^\beta), & & \nonumber \\
Q_5^{q_i q_j} &= (\bar{q}_{R i}^\alpha q_{L j}^\beta)(\bar{q}_{L i}^\beta q_{R j}^\alpha). & &
\label{eqn:eff-operators}
\end{align}
where \(q_{L,R} = P_{L,R}~q\), with \(P_{L,R} = (1 \mp \gamma^5)/2\), and \(\alpha, \beta\) are color indices. As we will see in the next section, the operators $Q_1$, $\widetilde Q_1$, $Q_4$, and $Q_5$ are generated at the
Pati--Salam scale in our setup. Their hadronic matrix elements
can be conveniently expressed in terms of bag parameters $B_i$.
For a neutral meson system $\Phi^0\equiv \Phi^0_{q_i q_j}$ one has~\cite{Bauer:2009cf}

\begin{align}
\langle \bar{\Phi}^0 | Q_1(\mu) | \Phi^0 \rangle &= \langle \bar{\Phi}^0 | \widetilde{Q_1}(\mu) | \Phi^0 \rangle=\frac{1}{3} M_\Phi f_\Phi^2 B_1(\mu),
\label{eqn:meson_hadronic_Q_1}\\
\langle \bar{\Phi}^0 | Q_4(\mu) |\Phi^0 \rangle &= \frac{1}{4} \left( \frac{M_\Phi}{m_{q_i}(\mu) + m_{q_j}(\mu)} \right)^2 M_\Phi f_\Phi^2 B_4(\mu),
\label{eqn:meson_hadronic_Q_4}\\
\langle \bar{\Phi}^0 | Q_5(\mu) | \Phi^0 \rangle &= \frac{1}{12} \left( \frac{M_\Phi}{m_{q_i}(\mu) + m_{q_j}(\mu)} \right)^2 M_\Phi f_\Phi^2 B_5(\mu).
\label{eqn:meson_hadronic_Q_5}
\end{align}
where \( M_\Phi \) is the mass of the \( \Phi^0 \) meson, \( f_\Phi \) is its decay constant. Here, the meson states are normalized to \(\langle \Phi^0|\Phi^0 \rangle=\langle \bar{\Phi}^0|\bar{\Phi}^0 \rangle=1\). The definitions in Eq.~\eqref{eqn:meson_hadronic_Q_1}--\eqref{eqn:meson_hadronic_Q_5} are such that in the vacuum-insertion approximation~(VIA) one can write,
\begin{align}
  \bigl[B_1(\mu)\bigr]_{\text{VIA}} &= 1, \\[4pt]
  \bigl[B_4(\mu)\bigr]_{\text{VIA}} &= 1
    + \frac{1}{6}
      \left(\frac{m_{q_i}(\mu) + m_{q_j}(\mu)}{m_\Phi} \right)^2 ,\\
  \bigl[B_5(\mu)\bigr]_{\text{VIA}} &= 1
    + \frac{3}{2}
      \left(\frac{m_{q_i}(\mu) + m_{q_j}(\mu)}{m_\Phi} \right)^2 .
\end{align}
The central values for the \( B_i(\mu) \) and the other relevant parameters to calculate \(\Delta M\) for $\Phi^0\equiv \left( K^0, D^0, B_s^0,~B_d^0\right)$ have been provided in Appendix~\ref{subsection: Input parameters for K-K oscillation}. 

Now, to connect new contributions at high scales to low-energy observables such as neutral meson mass differences, it is necessary to evolve the Wilson coefficients $C_i$ from Eq.~\eqref{eqn:H_eff_gen}, which is defined at the Pati-Salam scale $(\Lambda \sim   M_X)$, down to the appropriate hadronic scale $\mu$ where the matrix elements are evaluated. This renormalization group running will account for QCD effects, which can mix operators and change the values of the coefficients. The hadronic scale depends on the meson that we are considering: for examples, for kaons we should use $\mu_K = 2\,\text{GeV}$, for charm mesons $\mu_D = 2.8\,\text{GeV}$, and for bottom mesons $\mu_B = 4.6\,\text{GeV}$, which align with the lattice QCD computations~\cite{UTfit:2007eik}. The analytical formula for the running of the coefficients can be represented as~\cite{Ciuchini:1998ix}:
\begin{equation}
C_r(\mu) = \sum_i \sum_s \left( b_i^{(r,s)} + \eta\, c_i^{(r,s)} \right) \eta^{a_i} C_s(\Lambda),
\label{eqn:Wilson_running}
\end{equation}
where $\eta = \alpha_s(\Lambda)/\alpha_s(m_t)$, and the parameters $a_j, b_j^{(r,i)}, c_j^{(r,i)}$ are called the magic numbers. With \(\Lambda= 4.3\) TeV, \(\alpha_s(m_Z)=0.118\), and \(m_t^{\mathrm{\overline{MS}}}(m_t)=162.7\) GeV, we get \(\eta=\alpha_s(\Lambda)/\alpha_s(m_t)=0.078/0.108=0.716\). The magic numbers for \(\Phi^0\) meson systems have been provided in Appendix~\ref{subsection: Input parameters for K-K oscillation}. The magic numbers that are used to evolve the coefficients \( \widetilde{C}_{1-3} \) are identical to those for \( C_{1-3} \), since QCD running is independent of chirality structure in these cases. Neglecting this running may introduce larger theoretical uncertainties, and there will be possibilities that we may end up with incorrect estimations of $\Delta M$ due to the strong operator mixing under QCD. So, we will consider all possible induced four-fermion operators \( Q_i \) at the Pati-Salam scale, compute the corresponding Wilson coefficients \( C_i(\Lambda) \) at that scale, and evolve them to the hadronic scale using Eq.~\eqref{eqn:Wilson_running}. This will enable us to consistently evaluate their contribution to the meson mass splittings at low energies. At the hadronic scale, for \(K^0\) system, we find, 
\begin{align}
  C_2(\mu_K) &= 2.82\,C_2(\Lambda), &
  C_3(\mu_K) &= -5.01\times 10^{-3}\,C_2(\Lambda),\notag\\
  C_4(\mu_K) &= 5.66\,C_4(\Lambda) + 1.64\,C_5(\Lambda), &
  C_5(\mu_K) &= 0.22\,C_4(\Lambda) + 0.87\,C_5(\Lambda).
  \label{wilson_running_K}
\end{align}
For the \(D^0\) meson, we find the coefficients to be
\begin{align}
  C_2(\mu_D) &= 2.54\,C_2(\Lambda), &
  C_3(\mu_D) &= -2.5\times 10^{-2}\,C_2(\Lambda),\notag\\
  C_4(\mu_D) &= 4.70\,C_4(\Lambda) + 1.27\,C_5(\Lambda), &
  C_5(\mu_D) &= 0.17\,C_4(\Lambda) + 0.87\,C_5(\Lambda).
\end{align}
And for the \(B^0_{d,s}\) systems, we get
\begin{align}
  C_2(\mu_B) &= 2.20\,C_2(\Lambda), &
  C_3(\mu_B) &= -4\times 10^{-2}\,C_2(\Lambda),\notag\\
  C_4(\mu_B) &= 3.67\,C_4(\Lambda) + 0.92\,C_5(\Lambda), &
  C_5(\mu_B) &= 0.11\,C_4(\Lambda) + 0.87\,C_5(\Lambda).
\end{align}

\subsubsection{$K^0-\bar{K}^0$ oscillation}
We first compute the leptoquark gauge-boson (\(X_{\mu}\)) contributions to the $\Delta S=2$ effective Hamiltonian. We can read the Feynman rules for the relevant box diagrams from Eqs.~\eqref{eqn:L-L-X}--\eqref{eqn:L-H-GX}.

\begin{figure}[!h]
  \centering
  \begin{subfigure}[b]{0.32\textwidth}\includegraphics[width=\linewidth]{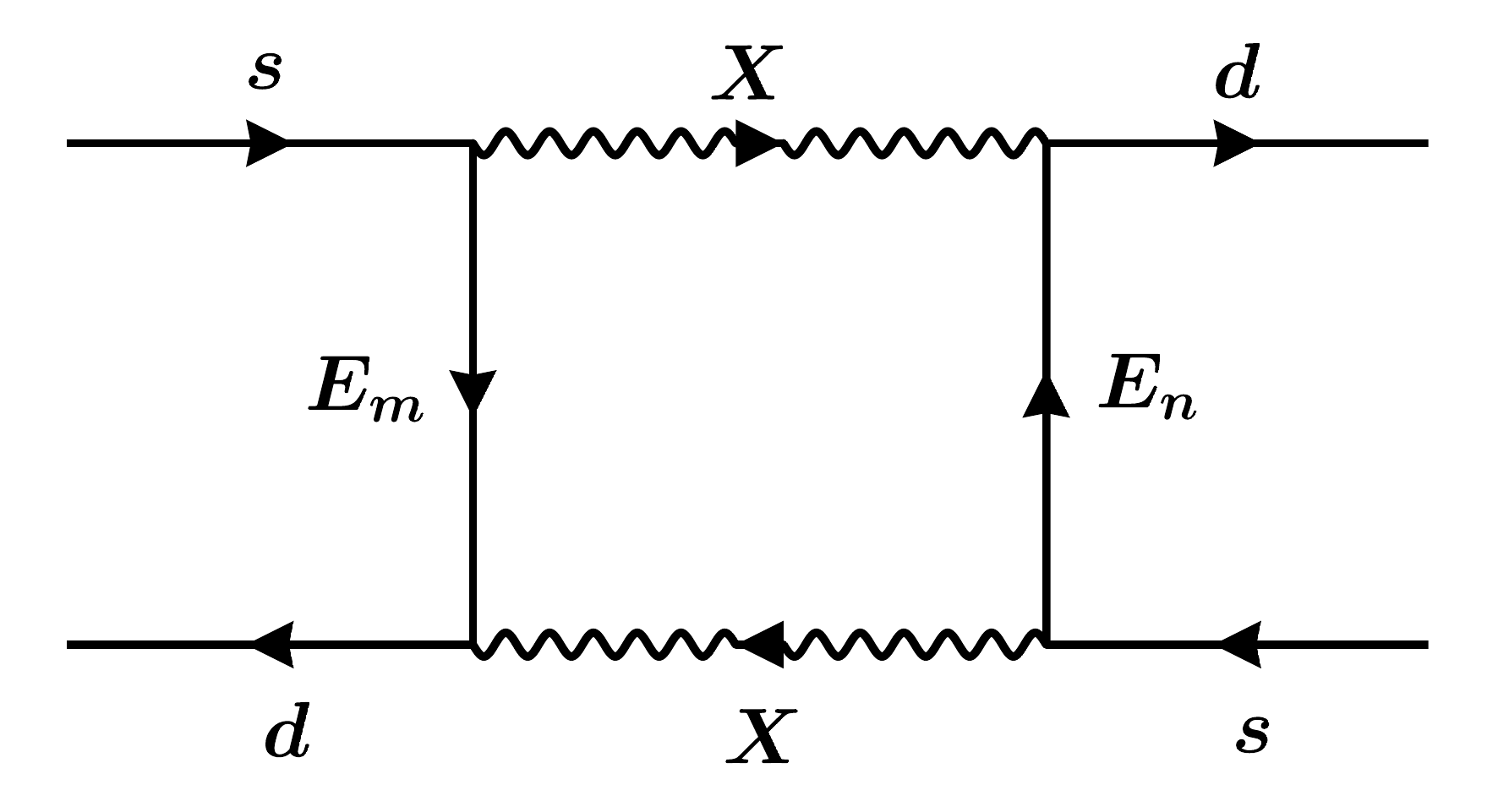}
 \caption{}\end{subfigure}\hfill
  \begin{subfigure}[b]{0.32\textwidth}\includegraphics[width=\linewidth]{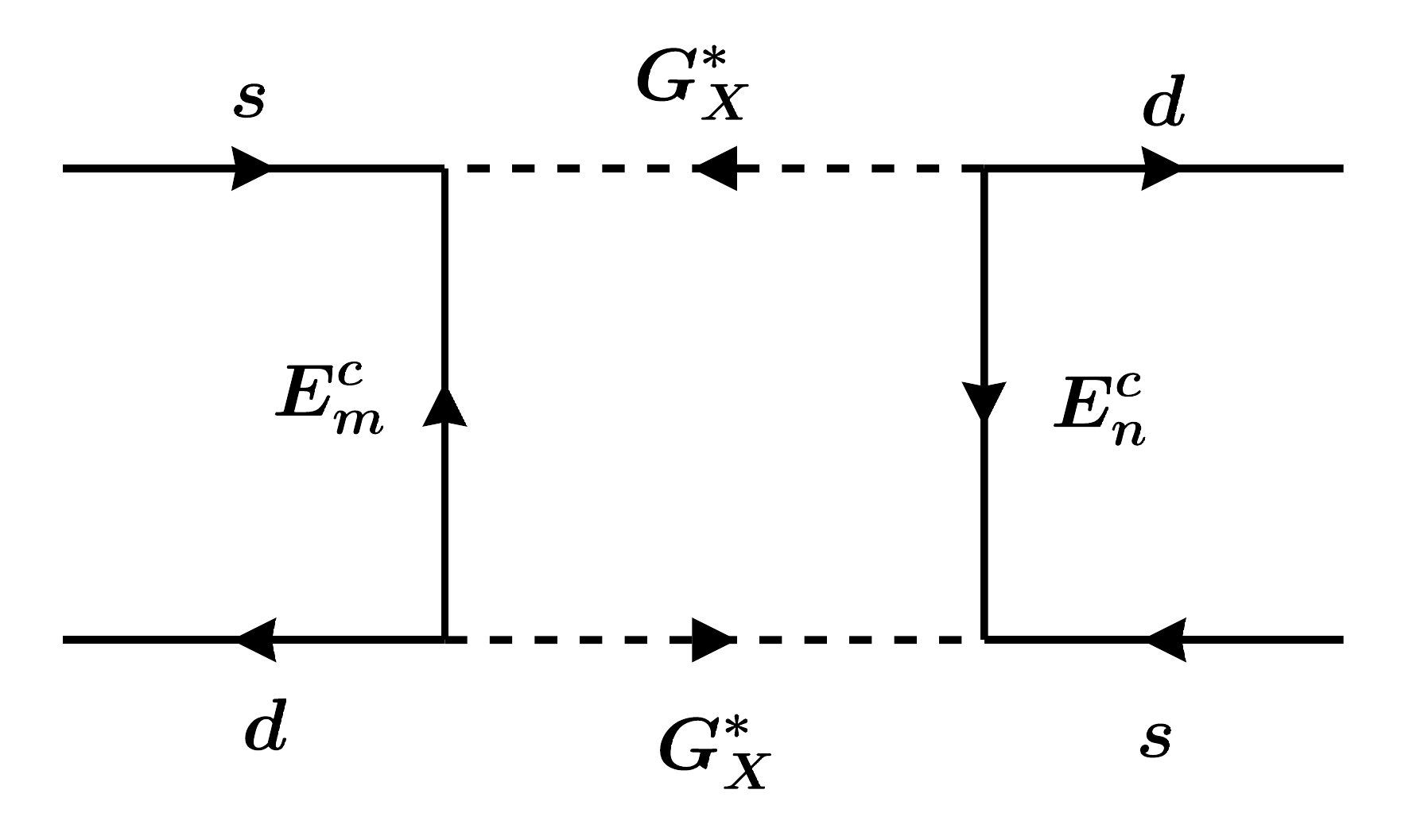}\caption{}\end{subfigure}\hfill
  \begin{subfigure}[b]{0.32\textwidth}\includegraphics[width=\linewidth]{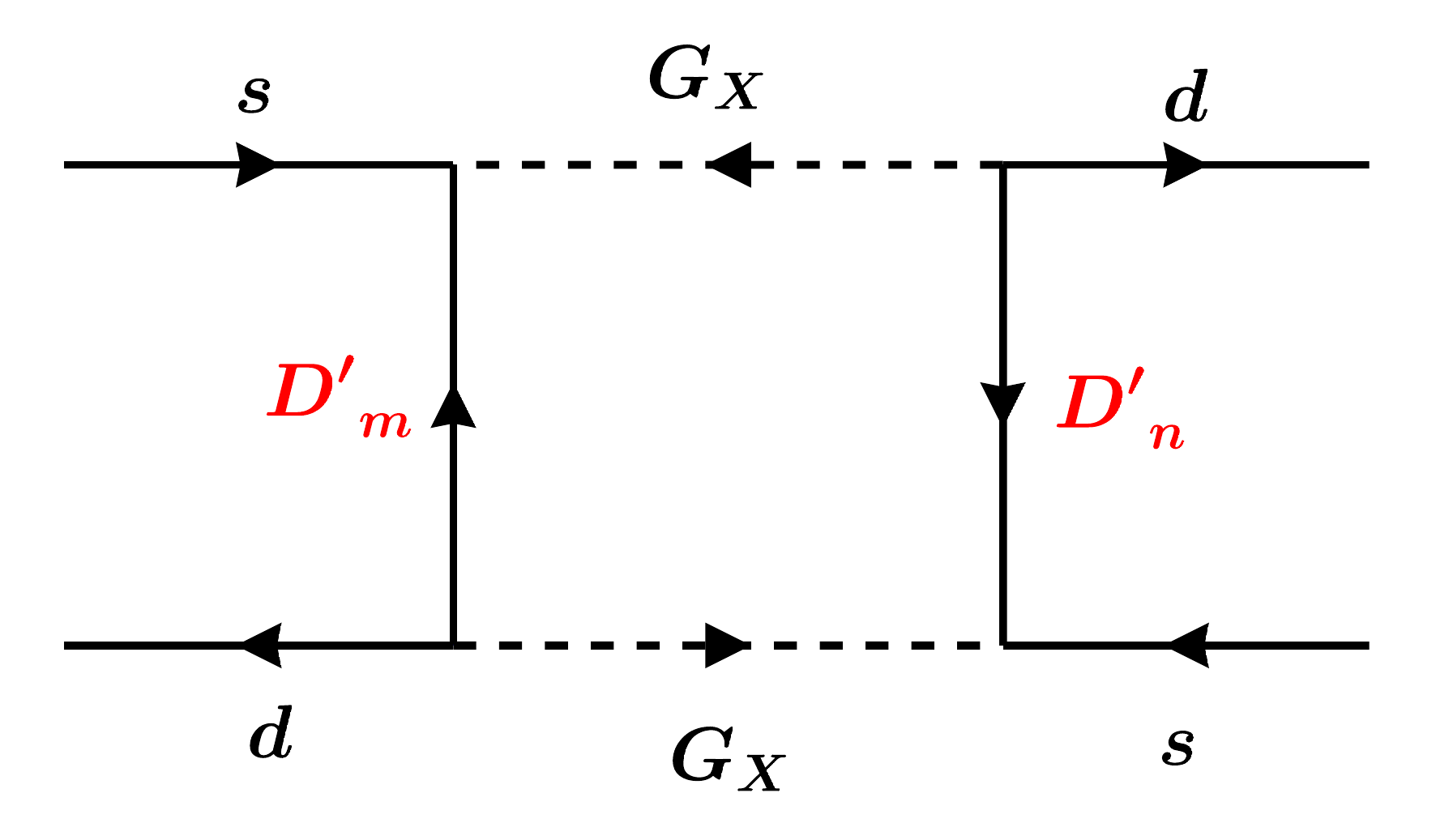}\caption{}\end{subfigure}\\[3pt]
  \begin{subfigure}[b]{0.32\textwidth}\includegraphics[width=\linewidth]{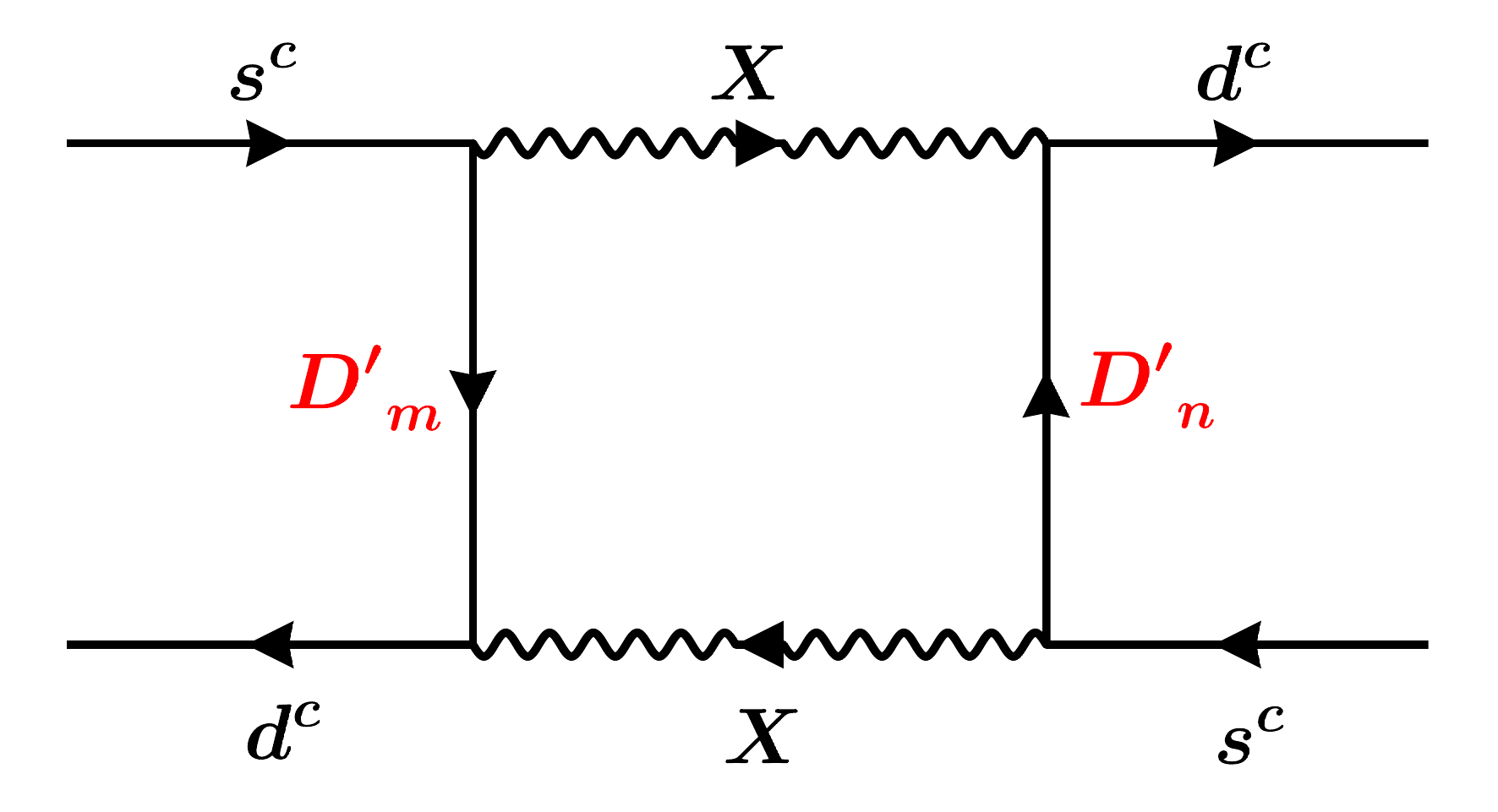}\caption{}\end{subfigure}\hfill
  \begin{subfigure}[b]{0.32\textwidth}\includegraphics[width=\linewidth]{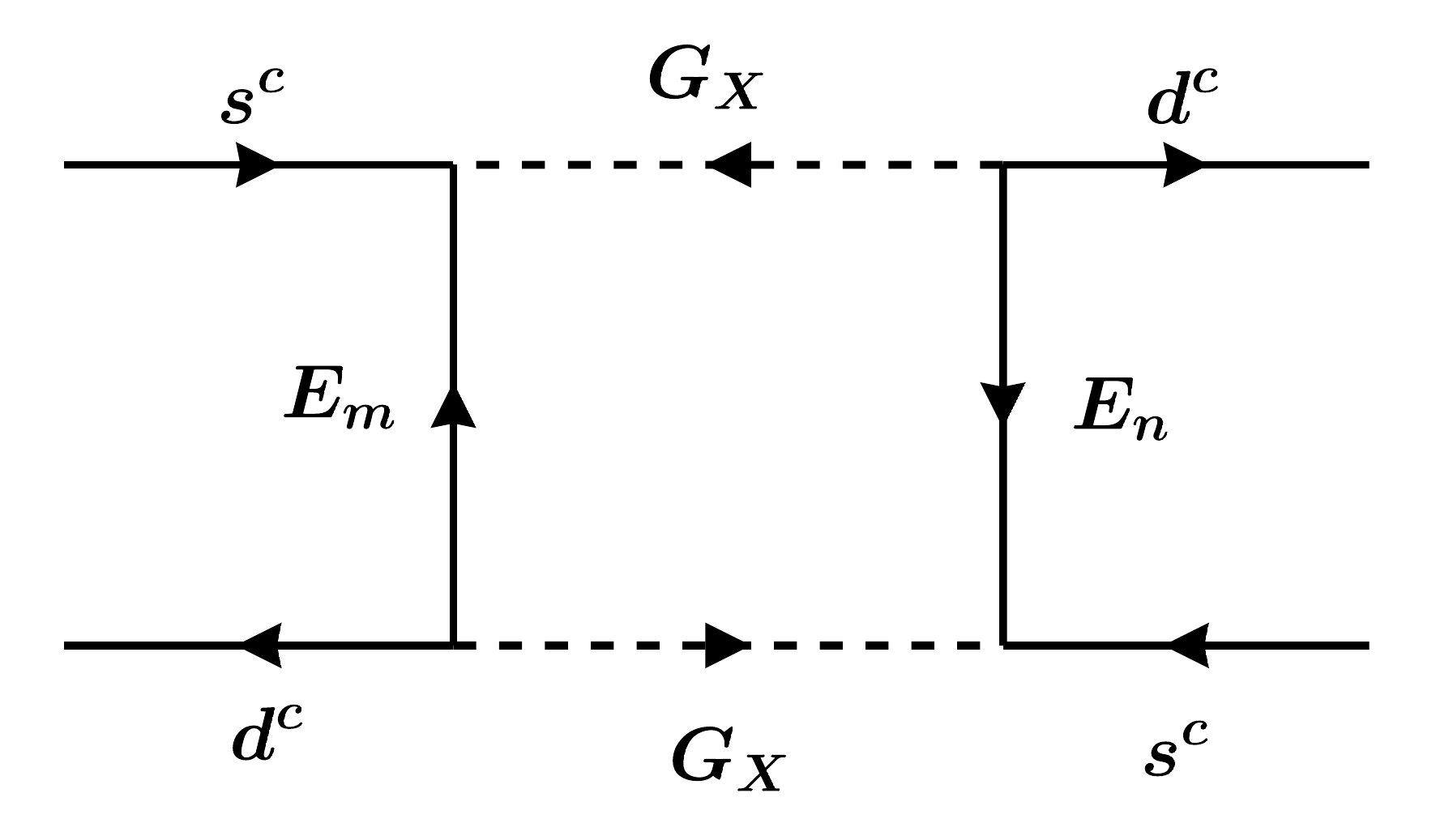}\caption{}\end{subfigure}\hfill
  \begin{subfigure}[b]{0.32\textwidth}\includegraphics[width=\linewidth]{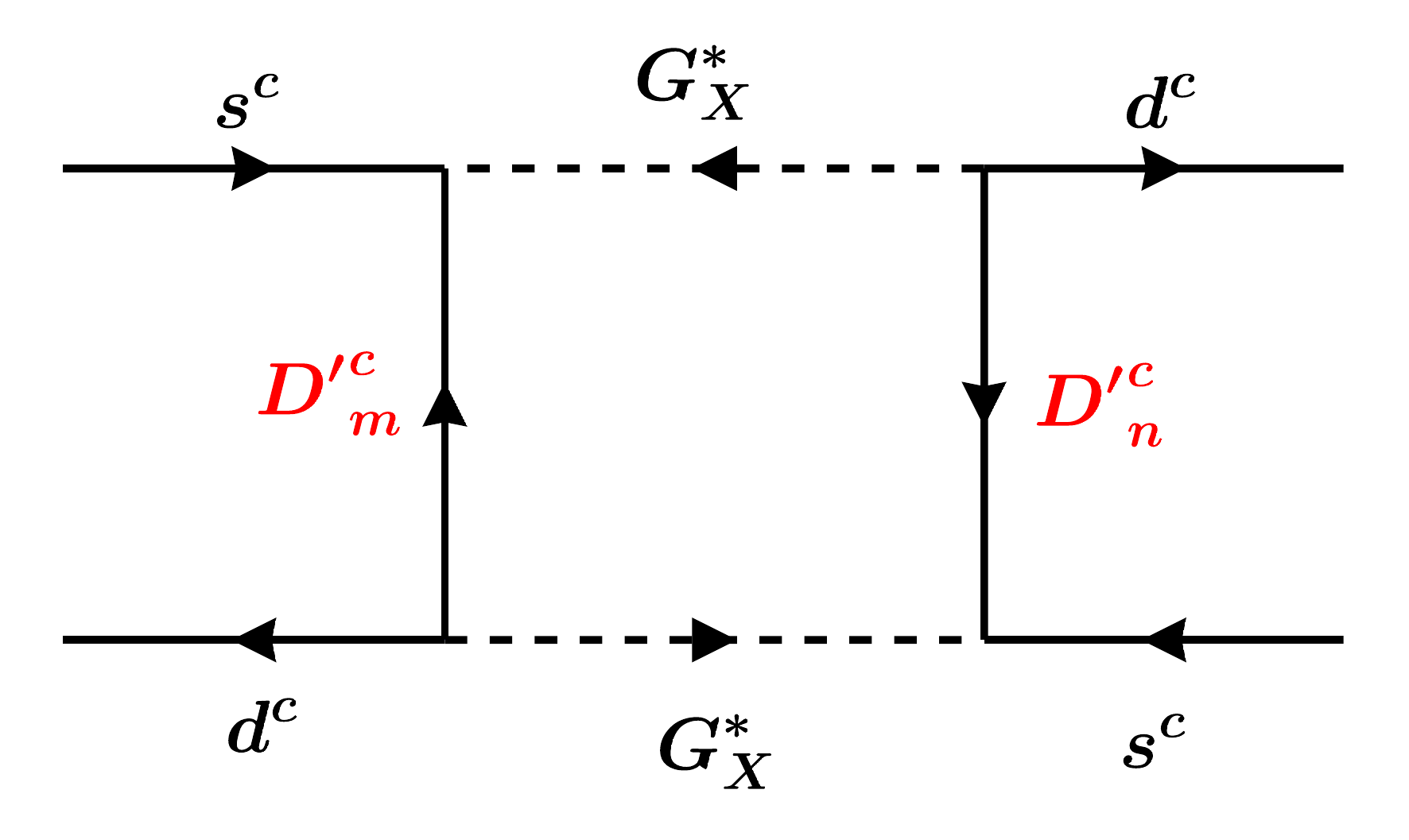}\caption{}\end{subfigure}\\[3pt]
  \begin{subfigure}[b]{0.32\textwidth}\includegraphics[width=\linewidth]{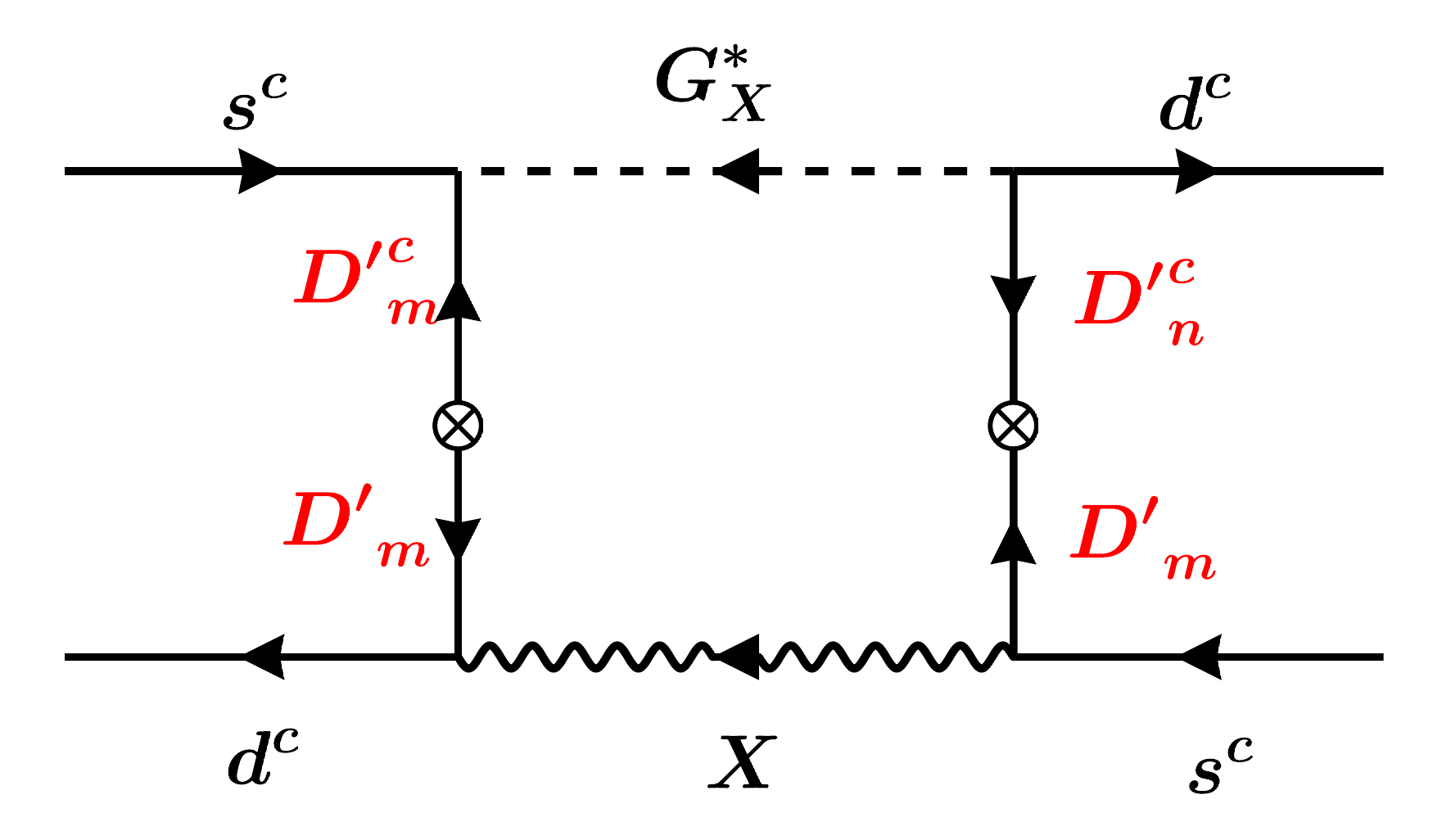}\caption{}\end{subfigure}\hfill
  \begin{subfigure}[b]{0.32\textwidth}\includegraphics[width=\linewidth]{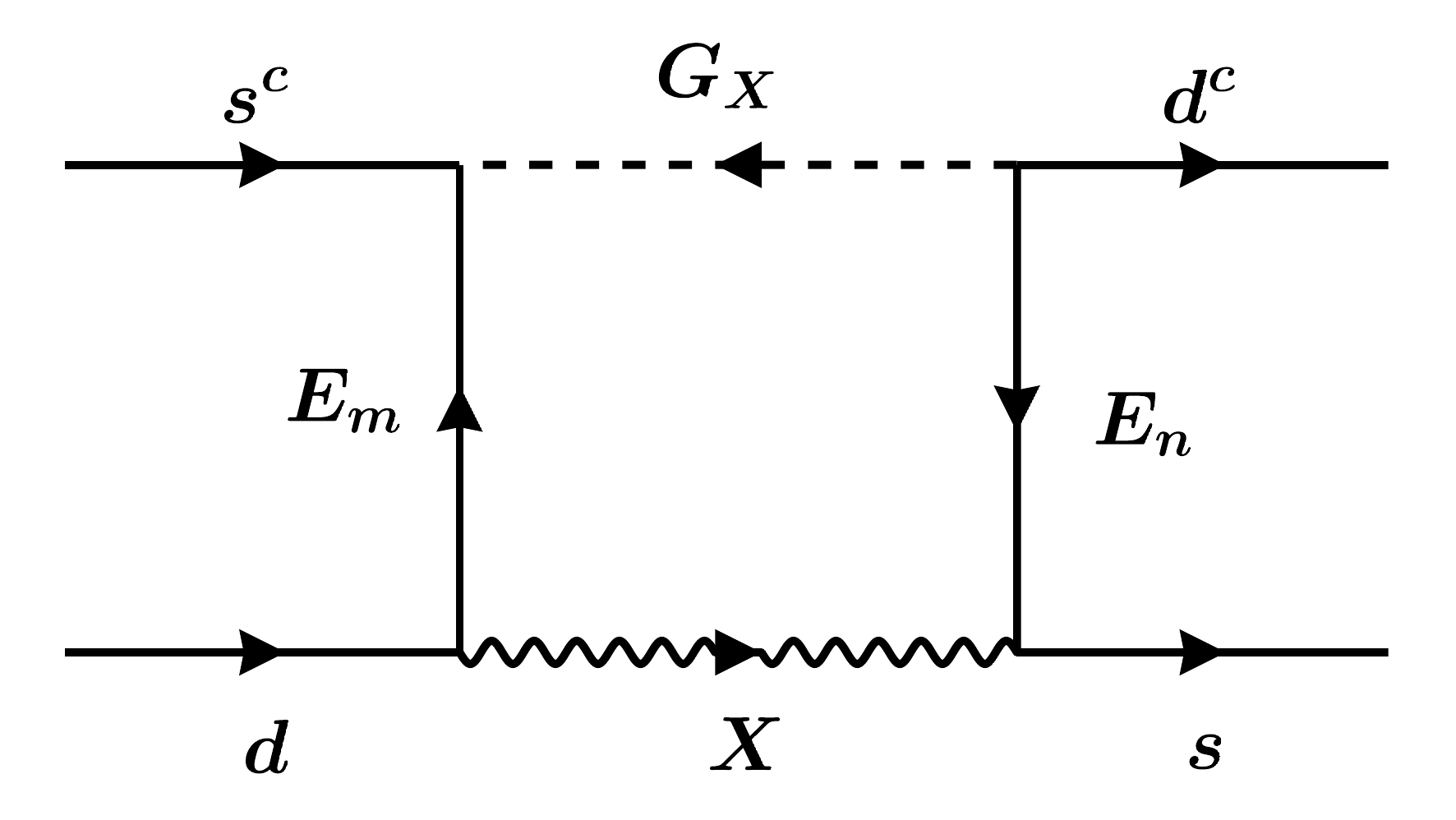}\caption{}\end{subfigure}\hfill
  \begin{subfigure}[b]{0.32\textwidth}\includegraphics[width=\linewidth]{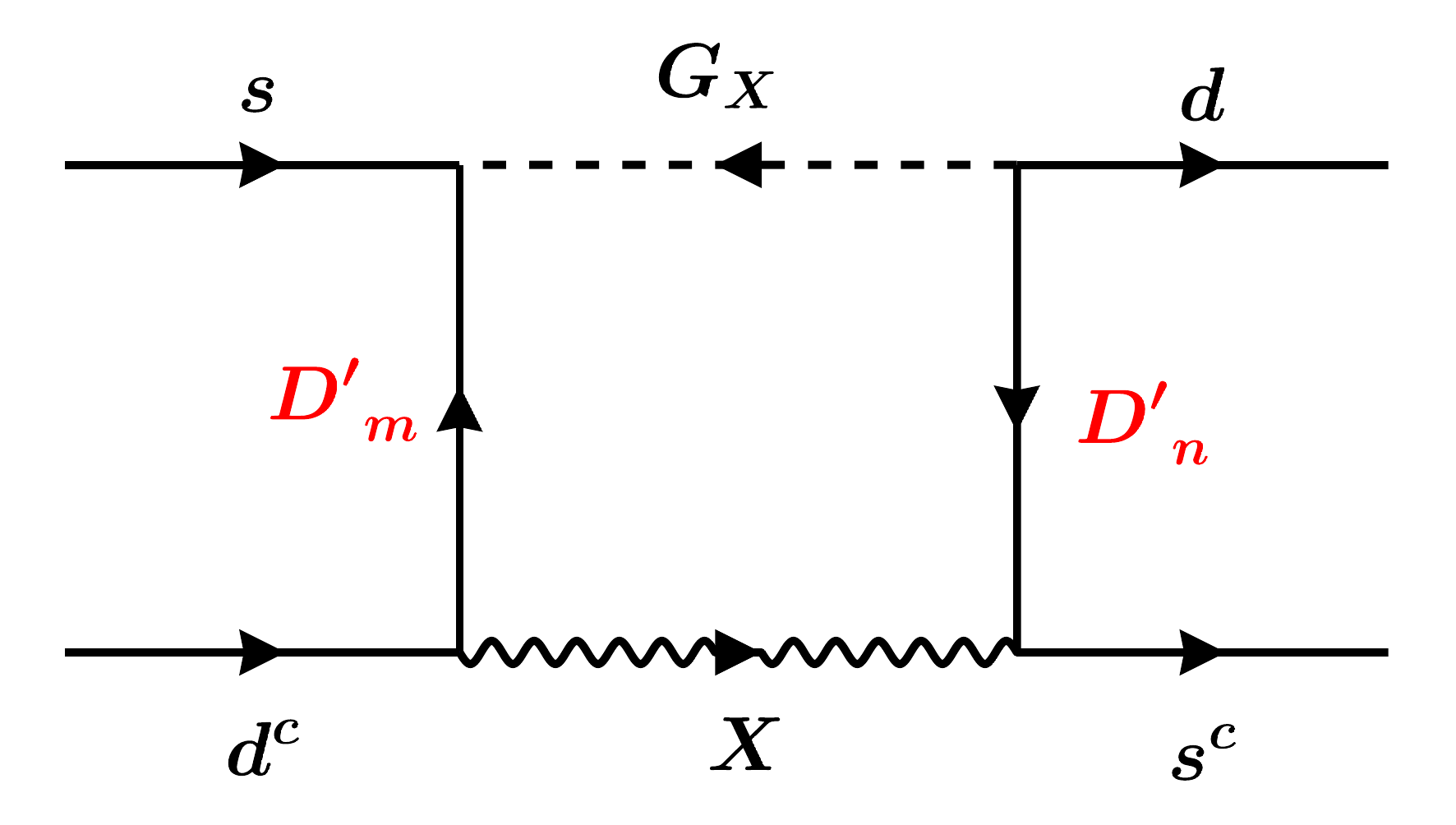}\caption{}\end{subfigure}
  \begin{subfigure}[b]{0.32\textwidth}\includegraphics[width=\linewidth]{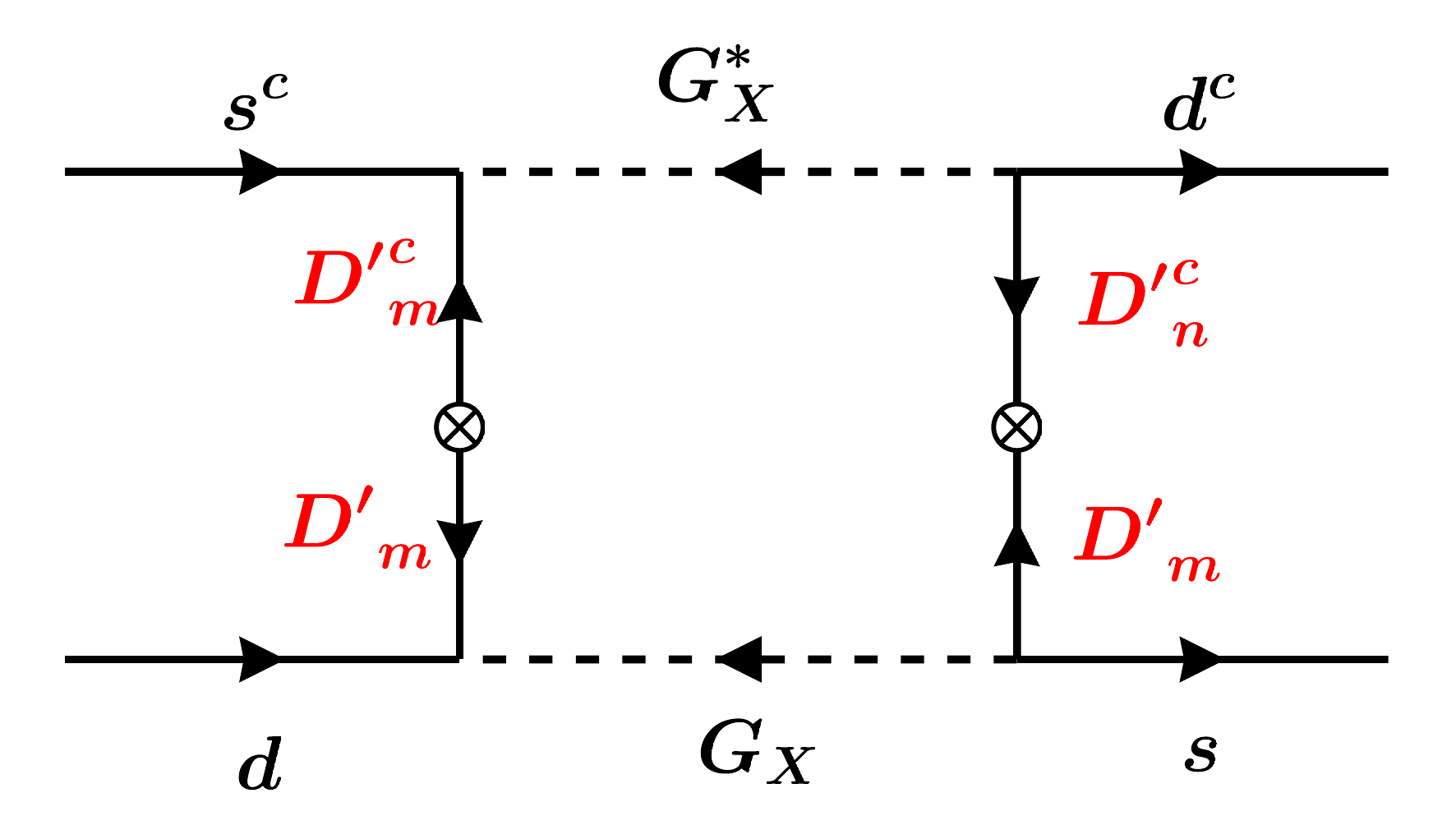}\caption{}\end{subfigure}\hfill
  \begin{subfigure}[b]{0.32\textwidth}\includegraphics[width=\linewidth]{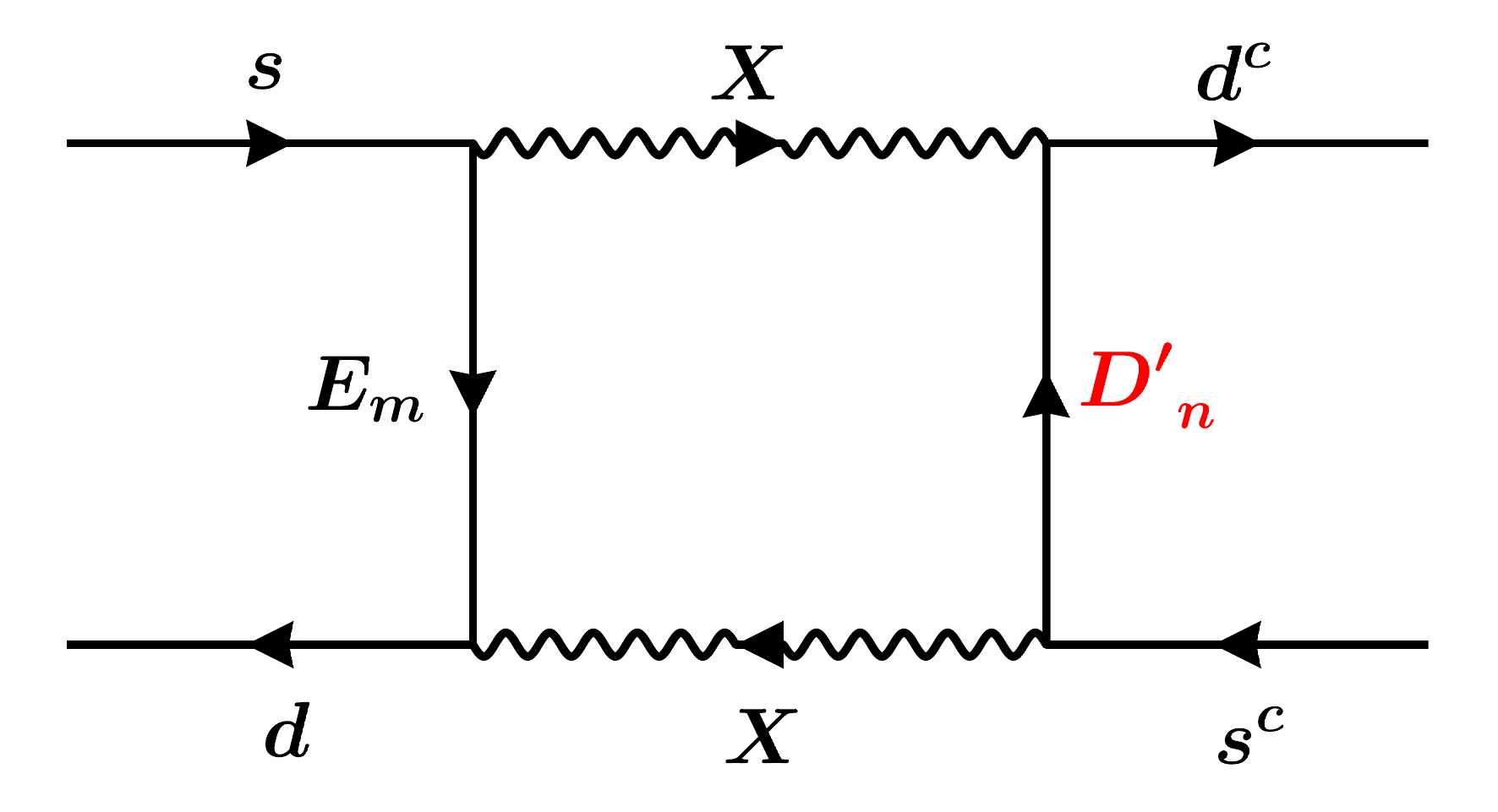}\caption{}\end{subfigure}\hfill
  \begin{subfigure}[b]{0.32\textwidth}\includegraphics[width=\linewidth]{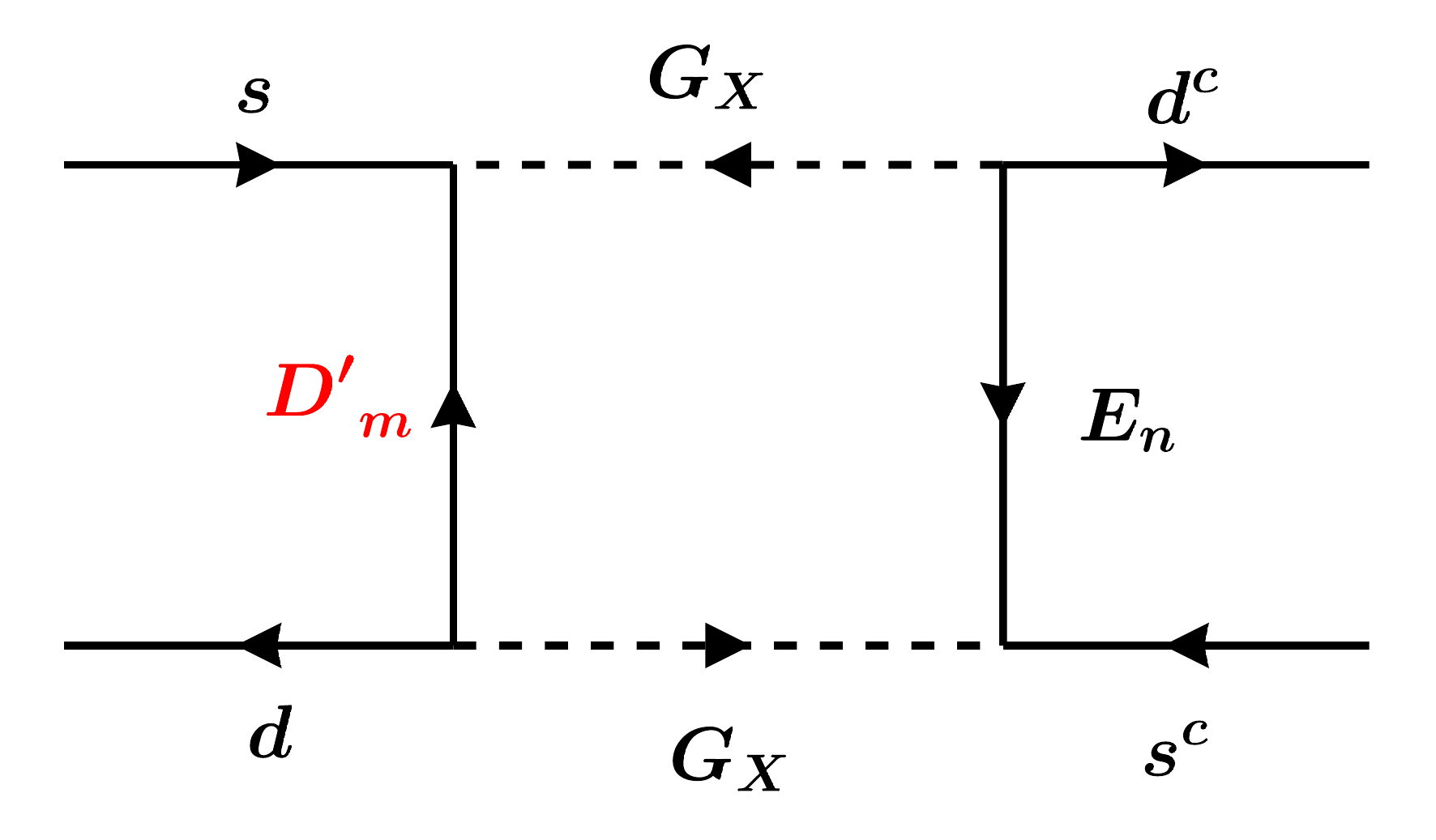}\caption{}\end{subfigure}\\[3pt]
  \begin{subfigure}[b]{0.32\textwidth}\includegraphics[width=\linewidth]{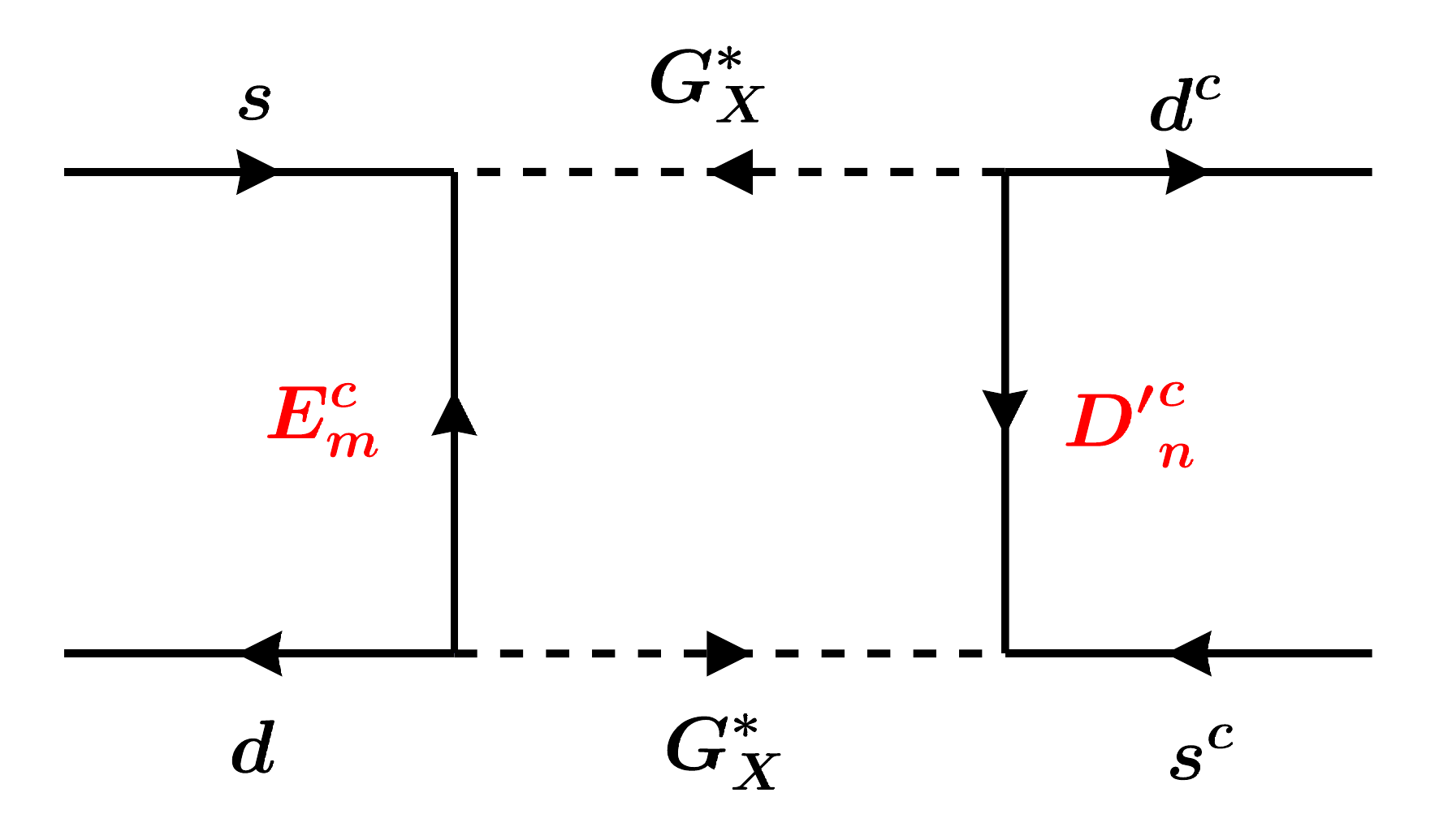}\caption{}\end{subfigure}
  \caption{Representative Feynman diagrams contributing to the $K$-meson oscillation. Here, $ m,~n$ are the flavor indices for the vector-like fermions, and the color indices are implicit.}
  \label{fig:K-K_box_diagrams}
\end{figure}

\begin{figure}[t!]
  \centering
  \begin{subfigure}[b]{0.32\textwidth}
    \centering
    \includegraphics[width=\linewidth]{FIG/K-1.pdf}
    \caption{(a-1)}
    \label{fig:K-decomp-a1}
  \end{subfigure}
  \hspace{3cm}
  \begin{subfigure}[b]{0.32\textwidth}
    \centering
    \includegraphics[width=\linewidth]{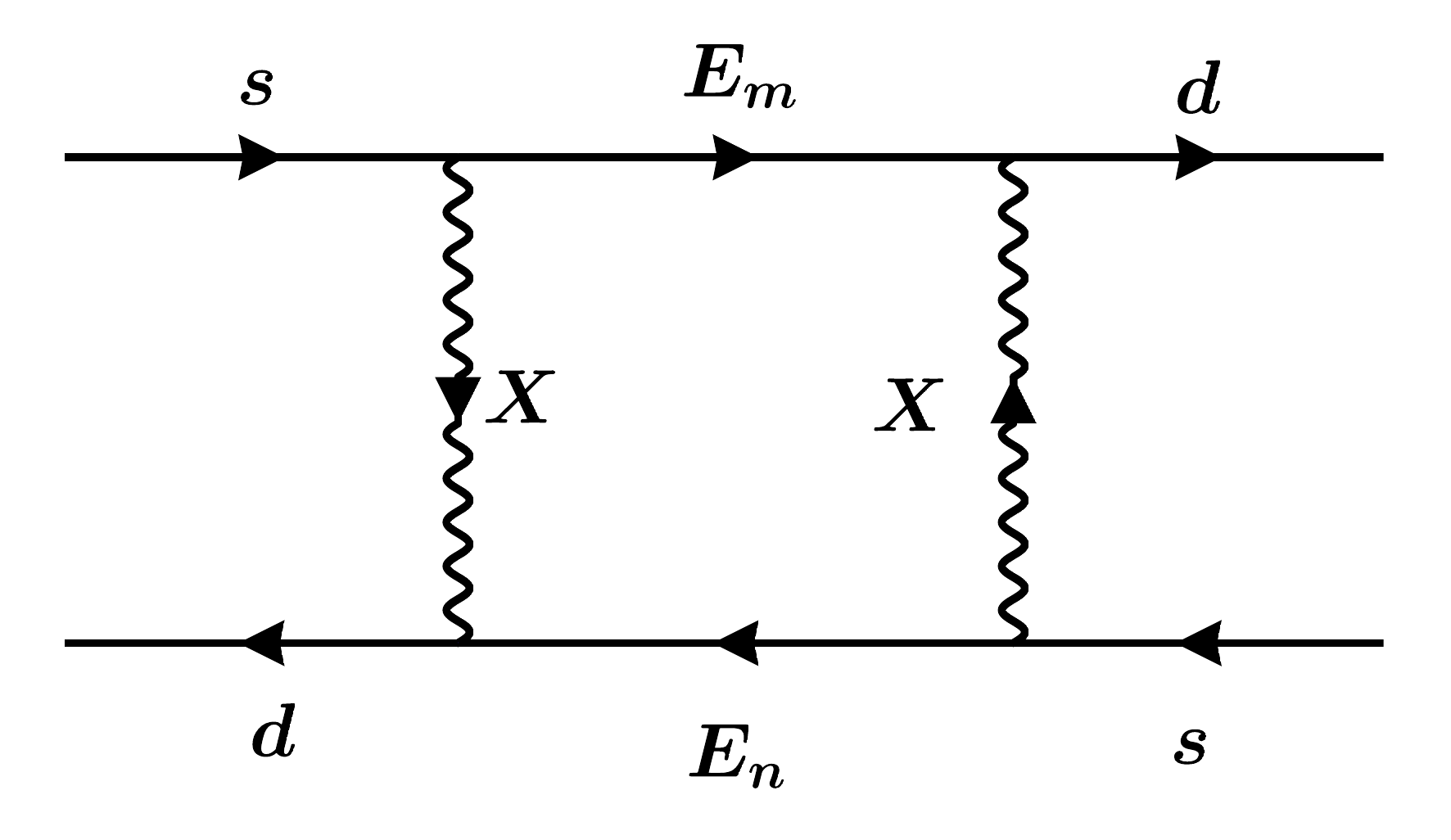}
    \caption{(a-2)}
    \label{fig:K-decomp-a2}
  \end{subfigure}
  \caption{Decomposition of the box topology in
    Fig.~\ref{fig:K-K_box_diagrams}(a). The ``annihilation'' term of diagram (a-1)
    is equal to the ``scattering'' term of diagram (a-2).}
  \label{fig:K-K_box_decomposition}
\end{figure}

Fig.~\ref{fig:K-K_box_diagrams} shows the box diagrams involving the leptoquark gauge boson and its Goldstone mode, as well as the mixed gauge-Goldstone diagrams that will induce the relevant Wilson coefficients at the Pati-Salam scale. Each topology in Fig.~\ref{fig:K-K_box_diagrams} should be understood as the sum of two diagrams with the crossing of the internal lines. For example, the box in Fig.~\ref{fig:K-K_box_diagrams} (a) corresponds to the sum of the ``annihilation'' and ``scattering'' diagrams shown in Fig.~\ref{fig:K-K_box_decomposition} (a-1) and Fig.~\ref{fig:K-K_box_decomposition} (a-2). For the mixed gauge–Goldstone diagrams in Fig.~\ref{fig:K-K_box_diagrams} (g)–(i), there is, in addition, a second diagram that we obtain by interchanging $X_\mu$ and $G_X$ on the internal lines. This diagram makes the same contribution to the Wilson coefficients, and we account for it through an overall multiplicity factor, even though we show only one representative diagram. The same comments hold for \(D^0-\bar{D^0}\) mixing topologies as well.

As mentioned earlier, we see that the operators \( Q_1, \widetilde{Q}_1, Q_4 \), and \( Q_5 \) contribute to \(K\)-meson mixing in our setup. The leading-order expressions for the corresponding Wilson coefficients are computed under the approximation of vanishing external momenta and negligible light quark masses:
\begingroup
\allowdisplaybreaks
\begin{align}
    C_1^{(a+b+c)}&= -\left( \frac{g_4}{\sqrt{2}} \right)^4 
    \frac{1}{4\pi^2 M_X^2} \Bigg[
      \sum_{m,n} \xi_{E_m} \xi_{E_n} A(t_{E_m}, t_{E_n}) 
      + \frac{1}{16} \sum_{m,n} \xi_{E_m} \xi_{E_n}
        A(t_{E_m}, t_{E_n}) t_{E_m} t_{E_n}
    \Bigg],\\[4pt]
    \tilde{C}_1^{(d+e+f+g)}&= -2\left(\frac{g_4}{\sqrt{2}} \right)^4 
    \frac{1}{4\pi^2 M_X^2} \Bigg[
      \sum_{m,n} \xi_{\textcolor{red}{D'_m}} \xi_{\textcolor{red}{D'_n}}
        A(t_{\textcolor{red}{D'_m}}, t_{\textcolor{red}{D'_n}})
      + \frac{1}{16} \sum_{m,n} \xi_{\textcolor{red}{D'_m}} \xi_{\textcolor{red}{D'_n}}
        A(t_{\textcolor{red}{D'_m}}, t_{\textcolor{red}{D'_n}})
        t_{\textcolor{red}{D'_m}} t_{\textcolor{red}{D'_n}} \notag\\
    &\hspace{4.2cm}
      + \frac{1}{4} \sum_{m,n} \xi_{\textcolor{red}{D'_m}} \xi_{\textcolor{red}{D'_n}}
        B(t_{\textcolor{red}{D'_m}}, t_{\textcolor{red}{D'_n}})
        t_{\textcolor{red}{D'_m}} t_{\textcolor{red}{D'_n}}
    \Bigg],\\[4pt]
    C_4^{(h+i+j)}&=+\left(\frac{g_4}{\sqrt{2}} \right)^4 
    \frac{1}{8\pi^2 M_X^2} \frac{m_d m_s}{M_X^2}
    \Bigg[
      \sum_{m,n} \xi_{\textcolor{red}{D'_m}} \xi_{\textcolor{red}{D'_n}}
        A(t_{\textcolor{red}{D'_m}}, t_{\textcolor{red}{D'_n}})
      + \sum_{m,n} \xi_{E_m} \xi_{E_n}
        A(t_{E_m}, t_{E_n})\notag \\
    &\hspace{4.2cm}
      + \sum_{m,n} \xi_{\textcolor{red}{D'_m}} \xi_{\textcolor{red}{D'_n}}
        B(t_{\textcolor{red}{D'_m}}, t_{\textcolor{red}{D'_n}})
        t_{\textcolor{red}{D'_m}} t_{\textcolor{red}{D'_n}}
    \Bigg],\\[4pt]
    C_5^{(k+l+m)}&=-2 \left(\frac{g_4}{\sqrt{2}} \right)^4 
    \frac{1}{16\pi^2 M_X^2} \Bigg[
      \sum_{m,n} \xi_{E_m} \xi_{\textcolor{red}{D'_n}}
        A(t_{E_m}, t_{\textcolor{red}{D'_n}})
      + \frac{1}{4} \sum_{m,n} \xi_{E_m} \xi_{\textcolor{red}{D'_n}}
        A(t_{E_m}, t_{\textcolor{red}{D'_n}})
        t_{E_m} t_{\textcolor{red}{D'_n}}
    \Bigg].
\label{eqn:Wilson_K_oscillation}
\end{align}
\endgroup

\captionsetup[subfigure]{labelformat=empty}
\captionsetup[subfigure]{labelformat=parens}

\begin{figure}[h!]
    \centering
    \includegraphics[width=0.65\textwidth]{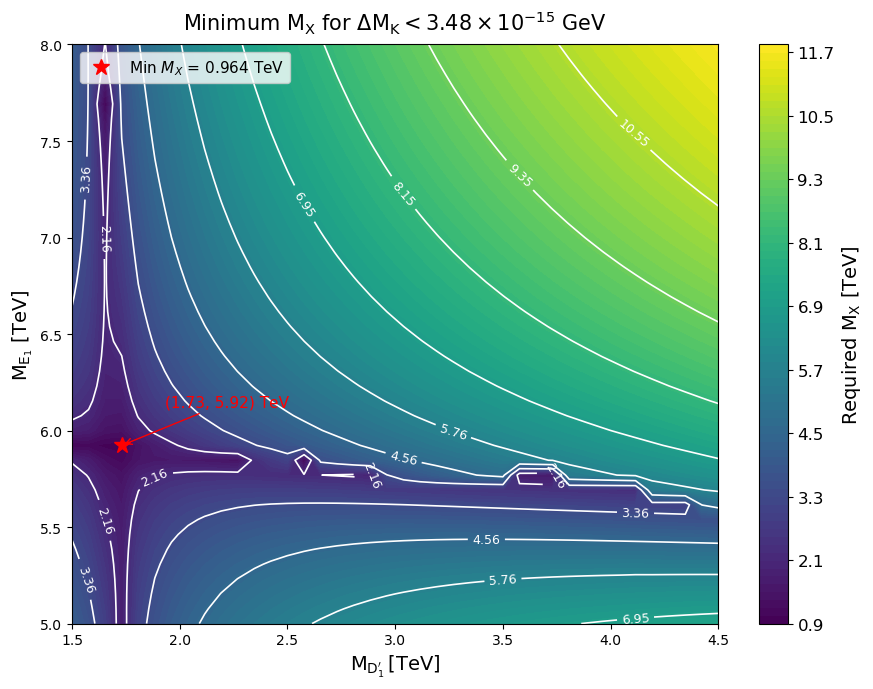}
\caption{Minimum leptoquark mass $M_X$ (in TeV) required by the
$K^0$–$\bar K^0$ mixing bound $\Delta M_K < 3.48\times10^{-15}\,\mathrm{GeV}$, shown as a function of the exotic–fermion masses $M_{\textcolor{red}{D'_1}}$ and $M_{E_1}$. The CKM-like mixing matrices are fixed according to Benchmark~2, Eq.~\eqref{eqn:benchmark2}, and the star denotes the global minimum in the scan.}

    \label{fig:K-K_oscillation}
\end{figure}
\noindent Here we have defined \(\xi_{\textcolor{red}{D'_m}}=K_{d^c\textcolor{red}{D'_m}}K_{s^c\textcolor{red}{D'_m}}^*\), $\xi_{E_m}=V_{dE_m}V_{sE_n}^*$, $t_{\alpha}=M_{\alpha}^2/M_X^2$, and, \( A(x_i,x_j)\), \( B(x_i,x_j)\) is already defined in Eq.~\eqref{eqn:Inami-Lim}. As before, the notation \( C_1^{(a+b+c)} \) indicates that diagrams (a), (b), and (c) in Fig.~\ref{fig:K-K_box_diagrams} generate \( Q_1 = (\bar{q}_{L i}^\alpha \gamma^\mu q_{L j}^\alpha)(\bar{q}_{L i}^\beta \gamma_\mu q_{L j}^\beta) \) operator. We should make several comments here: 

(i) The extra factor of 2 in the contribution from \( \widetilde{C}_1 \) compared to \(C_1\) comes after considering the color flow. Specifically, after decomposing diagrams d, f, and g, we see that 8 diagrams contribute to the right-chirality structure, compared to only  4 diagrams for the left-chirality diagram~(a).

(ii) Even though the Wilson coefficient \(C_4^{(h+i+j)}\) is proportional to light quark masses as \( m_d m_s / M_X^2 \), Eq.~\eqref{wilson_running_K} shows that it can induce \( C_5 \) at the hadronic scale through QCD running, which will contribute to the mixing.

(iii) The factor of 2 in \( C_5^{(k+l+m)} \) appears from matching the induced operator \( (\bar{q}_{L i}^\alpha \gamma^\mu q_{L j}^\alpha)(\bar{q}_{R i}^\beta \gamma_\mu q_{R j}^\beta) \) to the operator \( Q_5 \) using a Fierz transformation:
\[
(\bar{q}_{L i}^\alpha \gamma^\mu q_{L j}^\alpha)(\bar{q}_{R i}^\beta \gamma_\mu q_{R j}^\beta)
= 2(\bar{q}_{L i}^\alpha q_{R j}^\beta)(\bar{q}_{R i}^\beta q_{L j}^\alpha).
\]
With all the ingredients in our hands, we can go ahead and check a numerical example by considering the CKM-like matrices~(\(\xi_{E_m},\xi_{\textcolor{red}{D'_m}}\)) to have the structure like in Eq.~\eqref{eqn:benchmark2}~(Benchmark-2). Fig.~\ref{fig:K-K_oscillation} presents this benchmark, where we vary the vector-like fermion masses for a single generation, assuming that the exotics have a similar scale. Specifically, we scan over \( M_{\textcolor{red}{D'}} = (M_{\textcolor{red}{D'_1}},~1.7,~1.5)~\mathrm{TeV} \) and 
\( M_E = (M_{E_1},~5.9,~6.5)~\mathrm{TeV} \). We find that in this setup, \( M_X \) can be as low as \(\sim\) 964~GeV, for the choice \( (M_{\textcolor{red}{D'_1}}, M_{E_1}) = (1.73,5.92)~\mathrm{TeV} \).

\subsubsection{$D^0-\bar{D}^0$ oscillation}
\begin{figure}[t!]
    \centering
    \begin{subfigure}[b]{0.32\textwidth}
        \centering
        \includegraphics[width=\textwidth]{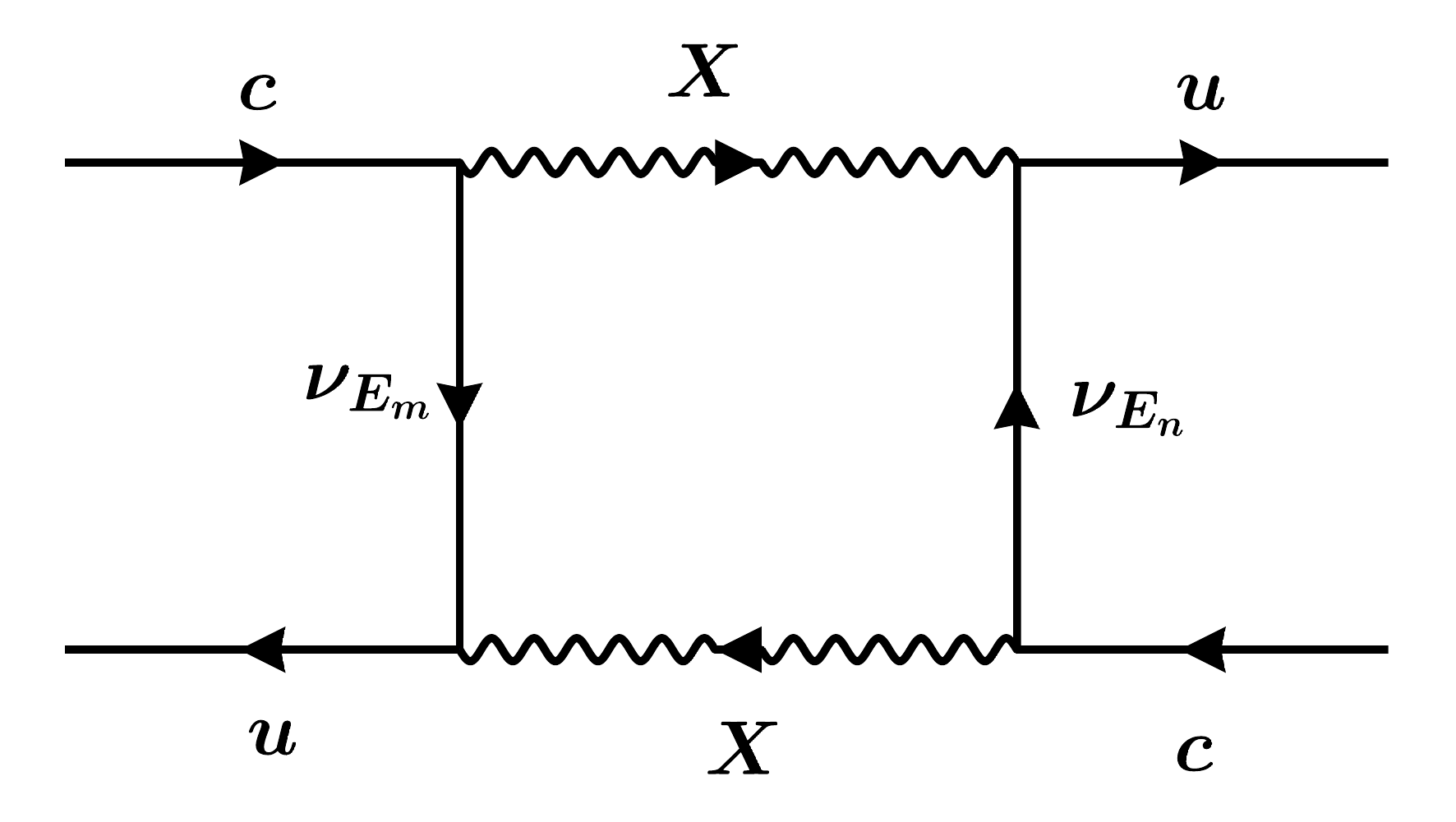}
        \caption{}
    \end{subfigure}
    \hfill
    \begin{subfigure}[b]{0.32\textwidth}
        \centering
        \includegraphics[width=\textwidth]{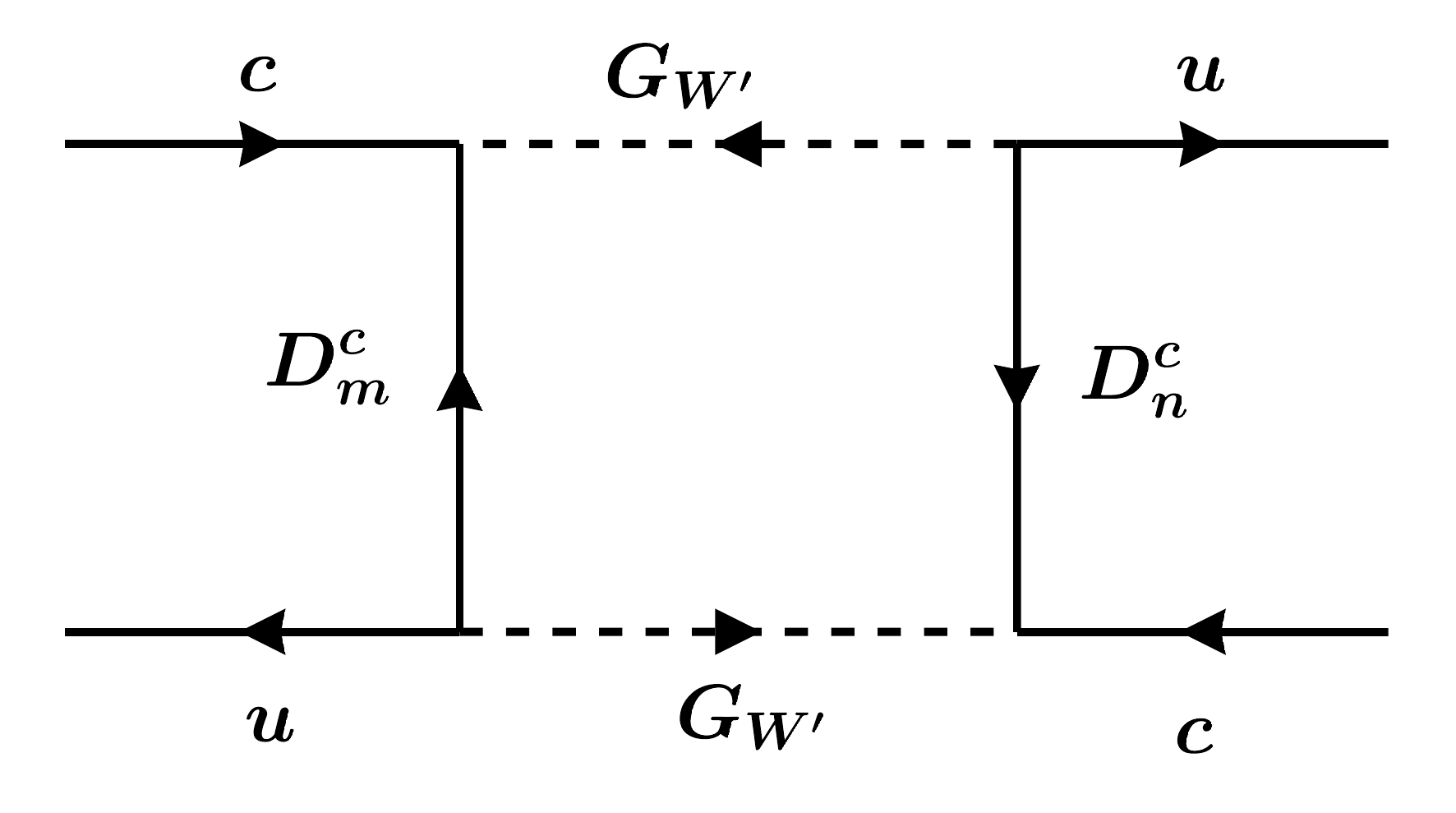}
        \caption{}
    \end{subfigure}
    \begin{subfigure}[b]{0.32\textwidth}
        \centering
        \includegraphics[width=\textwidth]{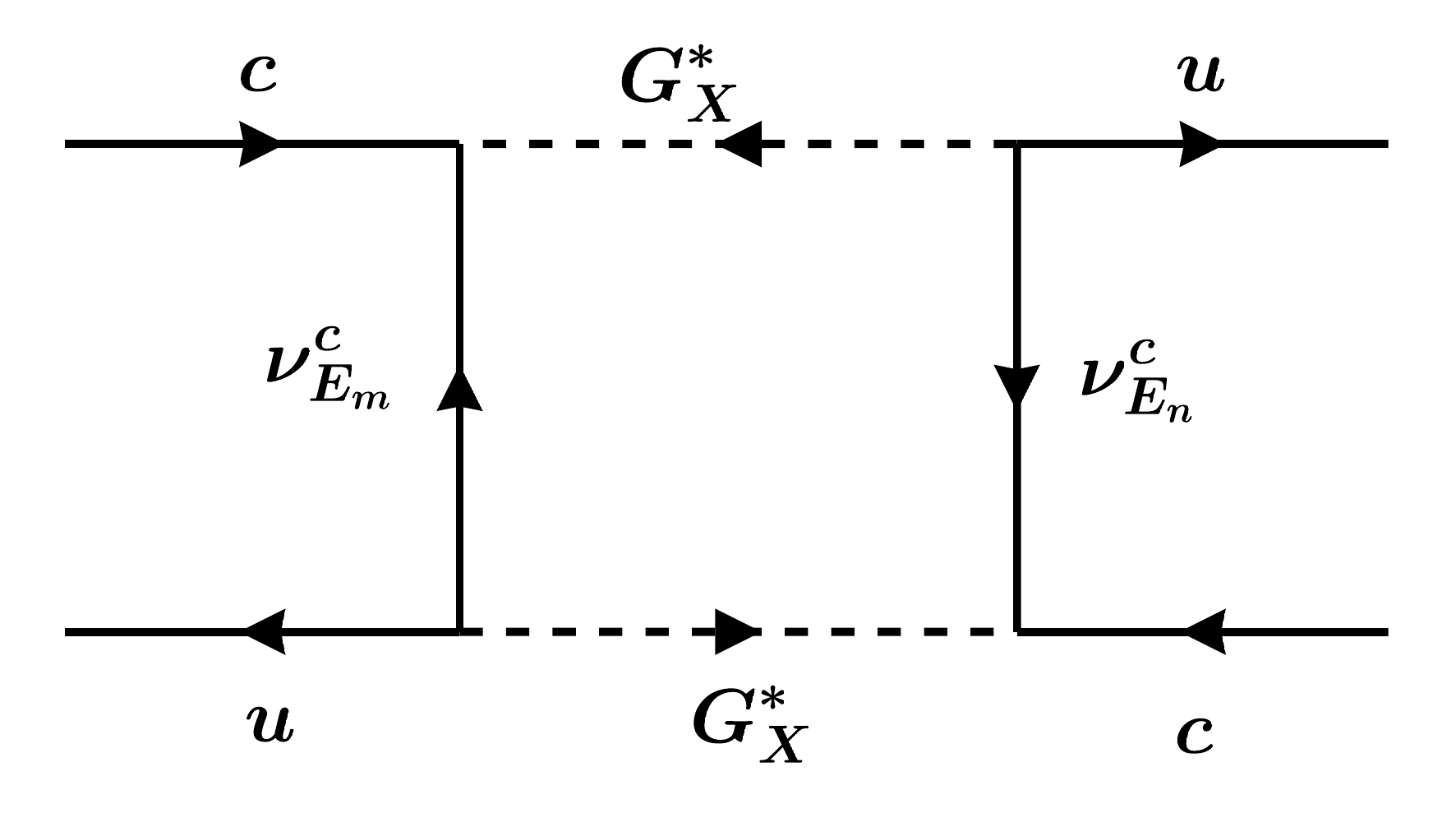}
        \caption{}
    \end{subfigure}

    \vspace{0.4cm} 

    \begin{subfigure}[b]{0.32\textwidth}
        \centering
        \includegraphics[width=\textwidth]{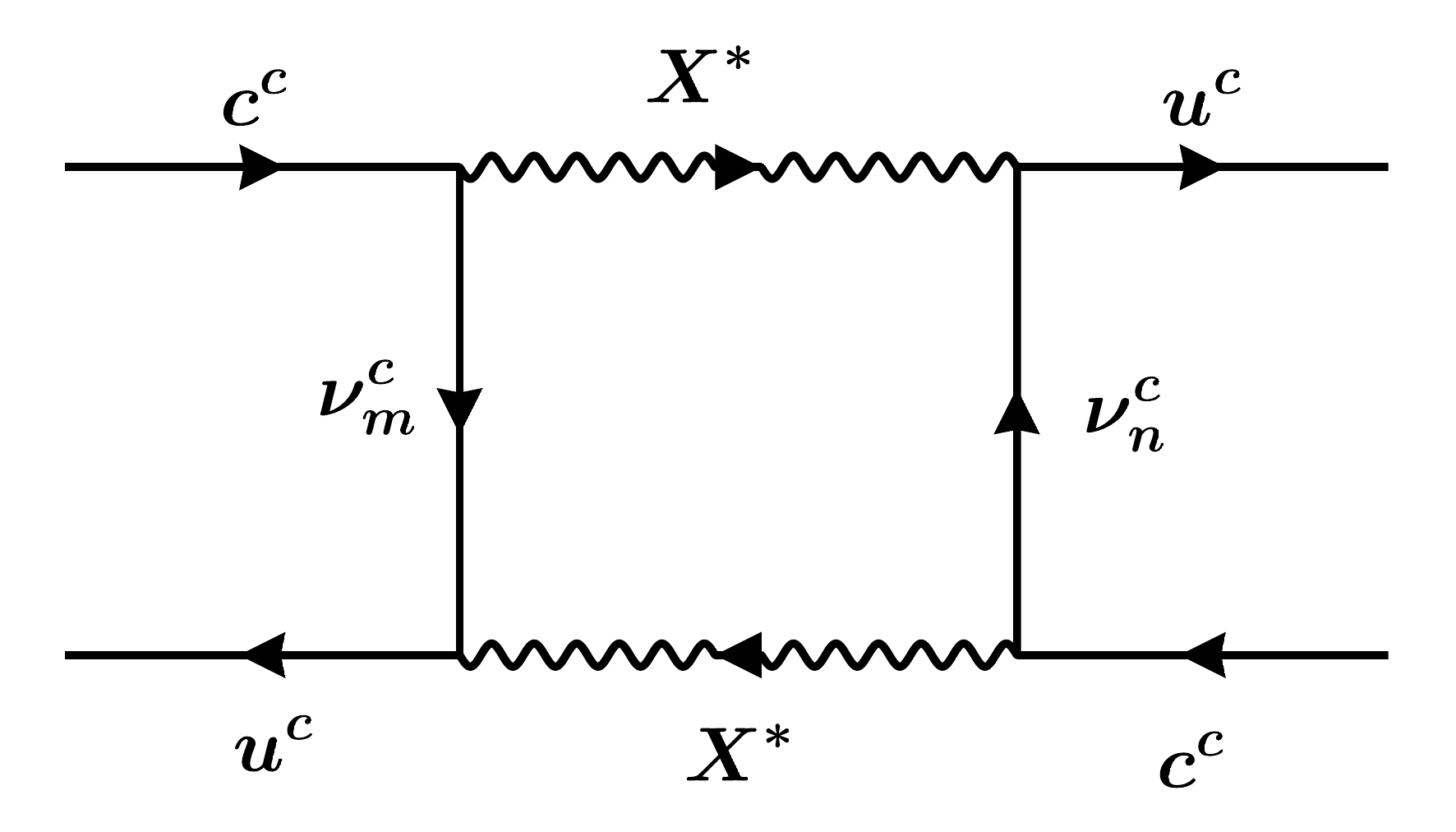}
        \caption{}
    \end{subfigure}
    \hfill
    \begin{subfigure}[b]{0.32\textwidth}
        \centering
        \includegraphics[width=\textwidth]{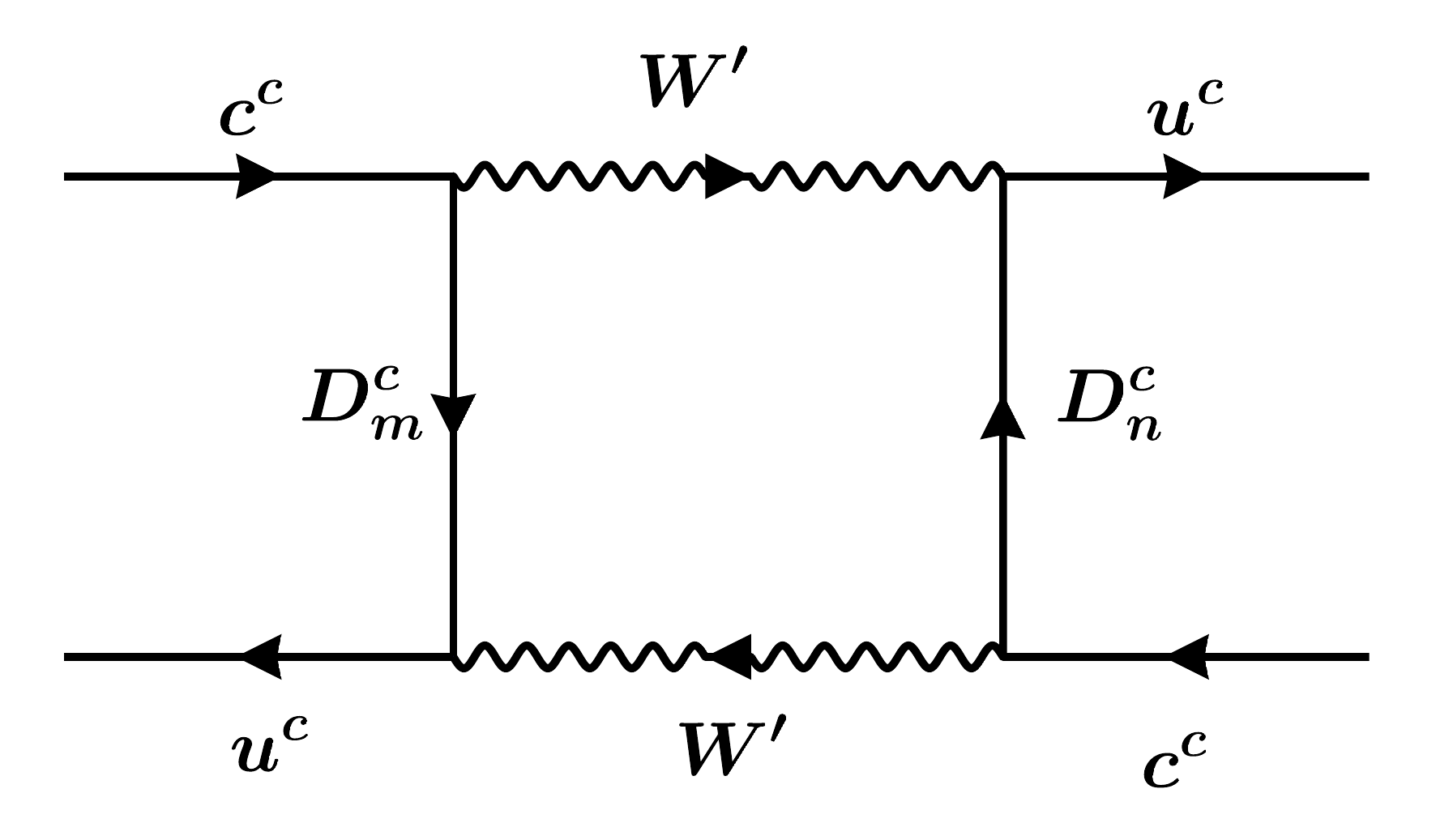}
        \caption{}
    \end{subfigure}
    \begin{subfigure}[b]{0.32\textwidth}
        \centering
        \includegraphics[width=\textwidth]{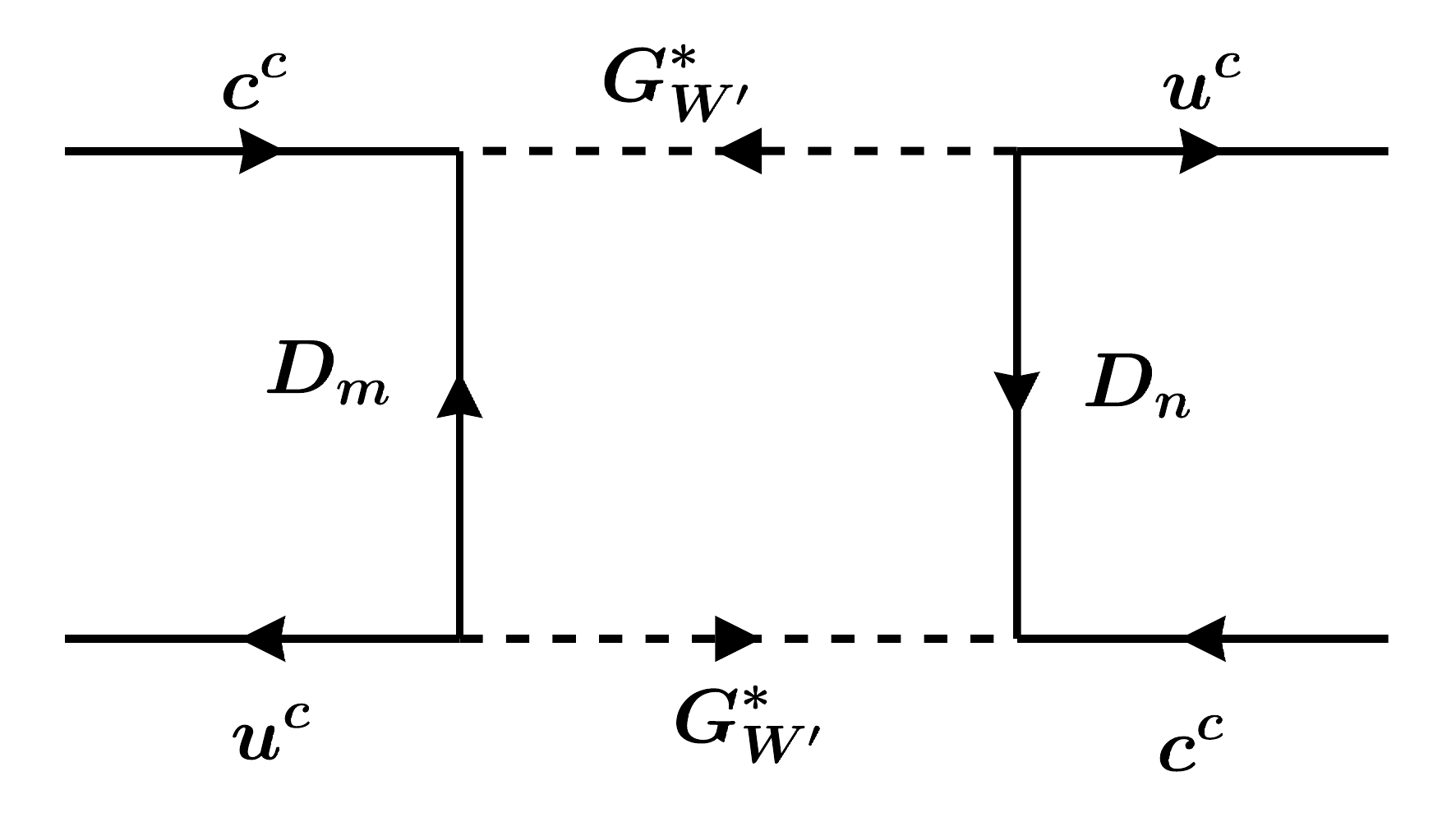}
        \caption{}
    \end{subfigure}

    \vspace{0.4cm}

    \begin{subfigure}[b]{0.32\textwidth}
        \centering
        \includegraphics[width=\textwidth]{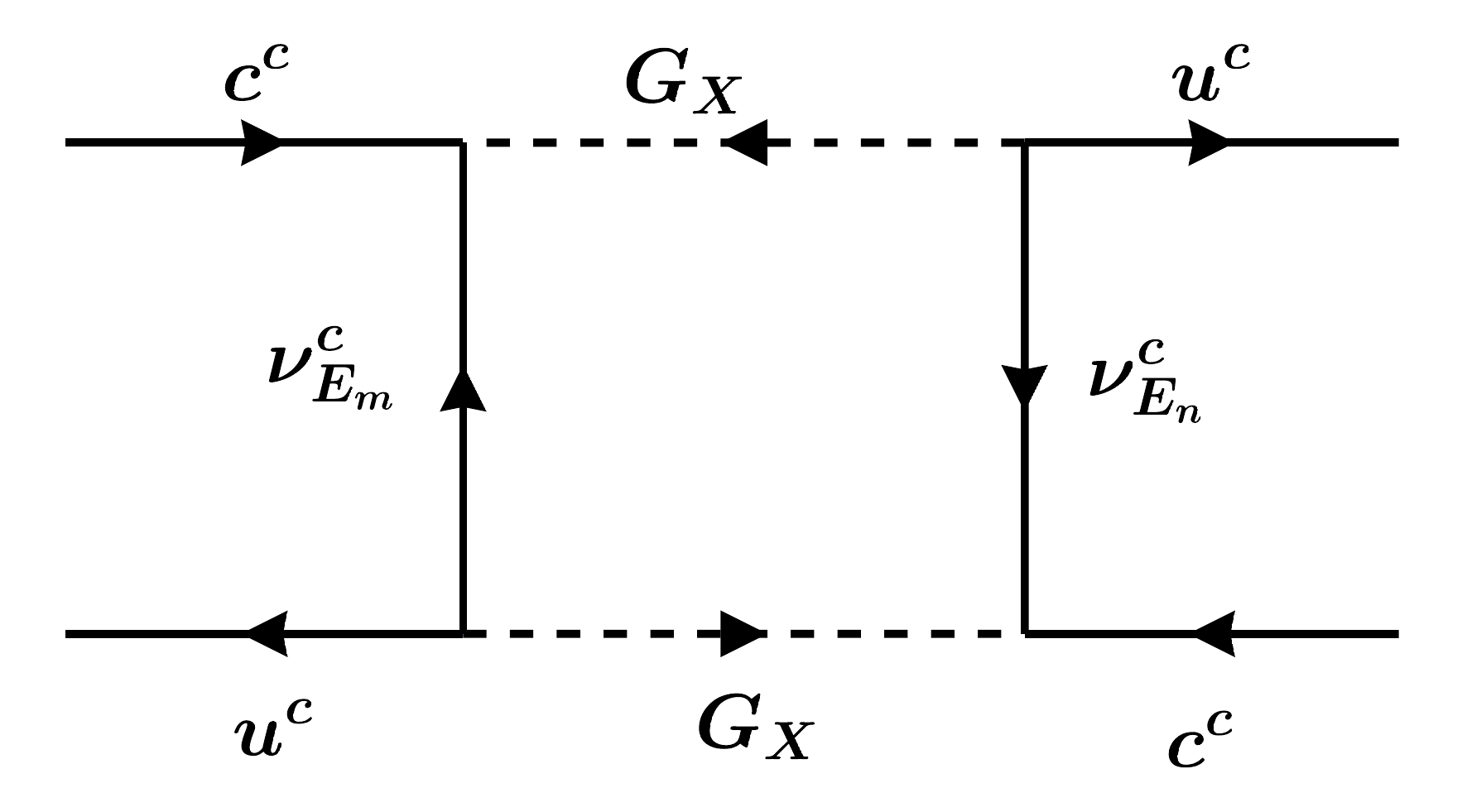}
        \caption{}
    \end{subfigure}
    \hfill
    \begin{subfigure}[b]{0.32\textwidth}
        \centering
        \includegraphics[width=\textwidth]{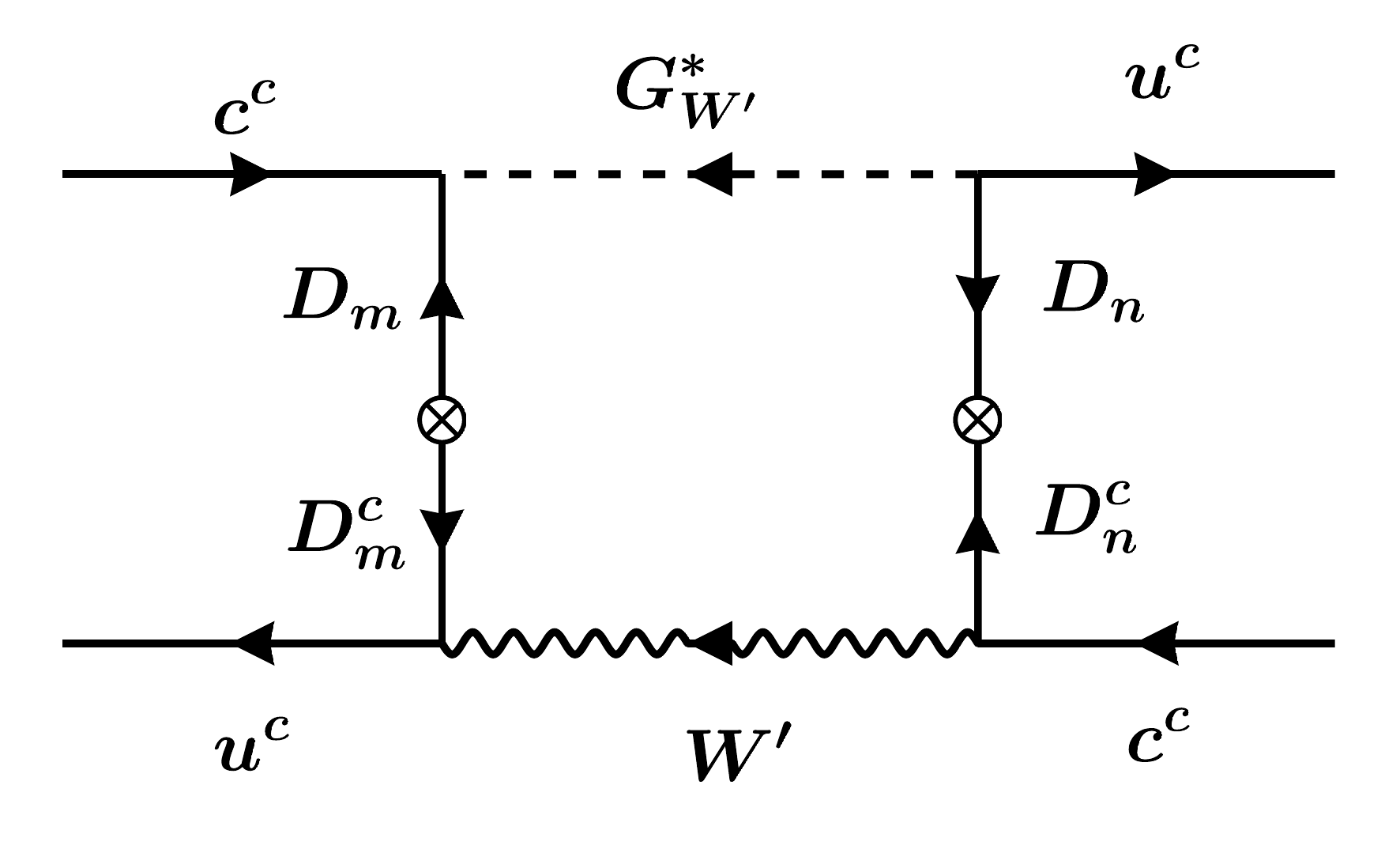}
        \caption{}
    \end{subfigure}
    \begin{subfigure}[b]{0.31\textwidth}
        \centering
        \includegraphics[width=\textwidth]{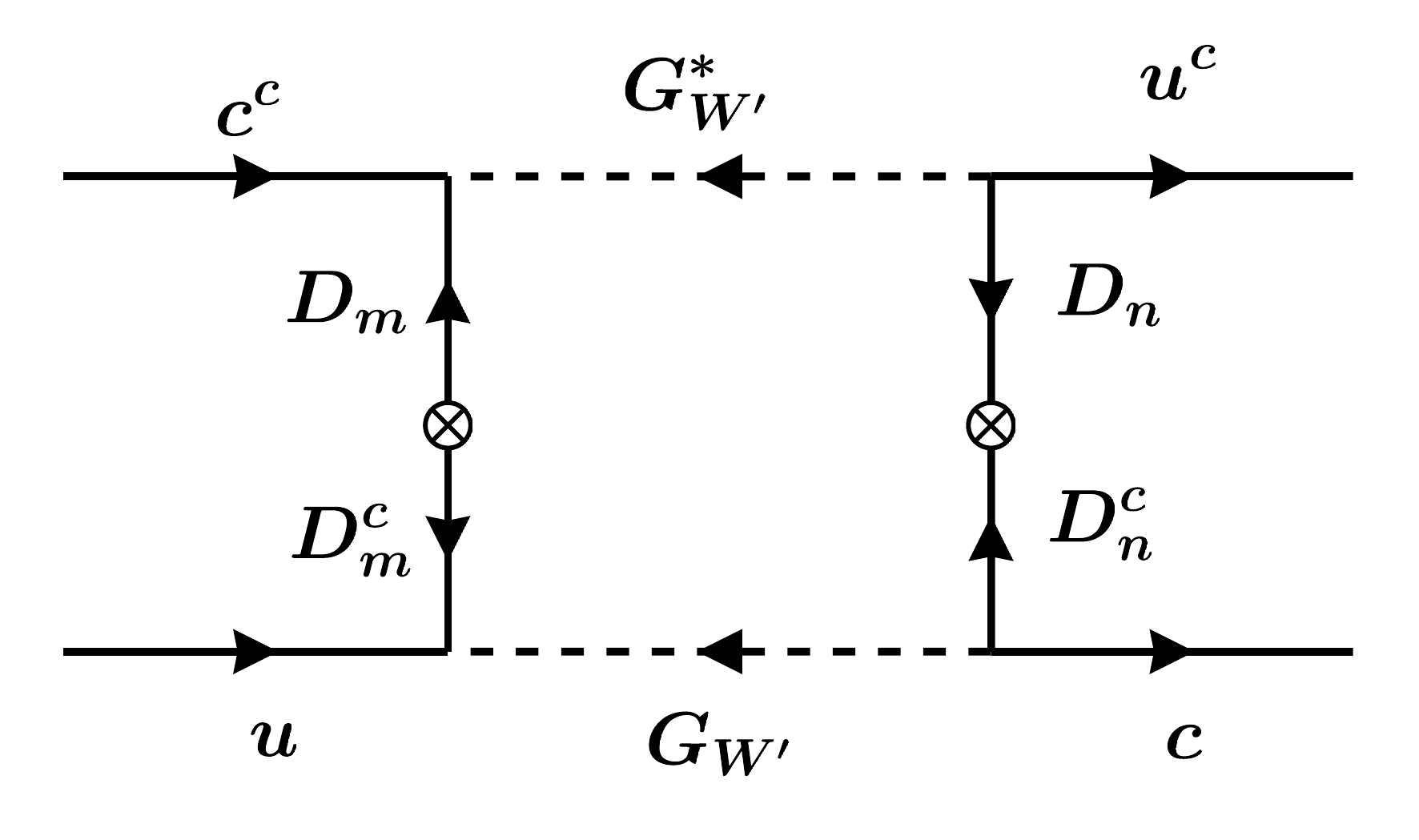}
        \caption{}
    \end{subfigure}

       \vspace{0.4cm}

    \begin{subfigure}[b]{0.32\textwidth}
        \centering
        \includegraphics[width=\textwidth]{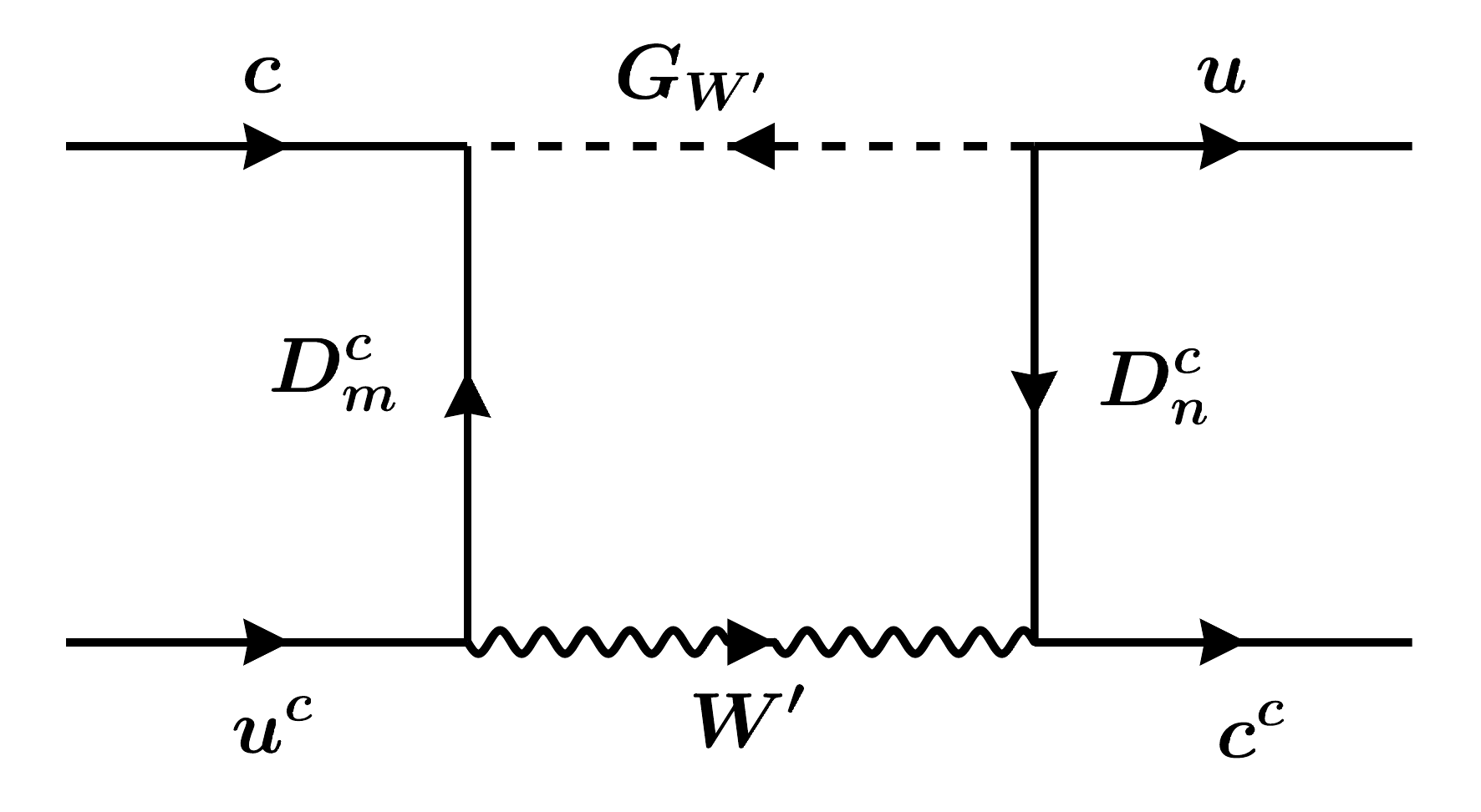}
        \caption{}
    \end{subfigure}
    \begin{subfigure}[b]{0.32\textwidth}
        \centering
        \includegraphics[width=\textwidth]{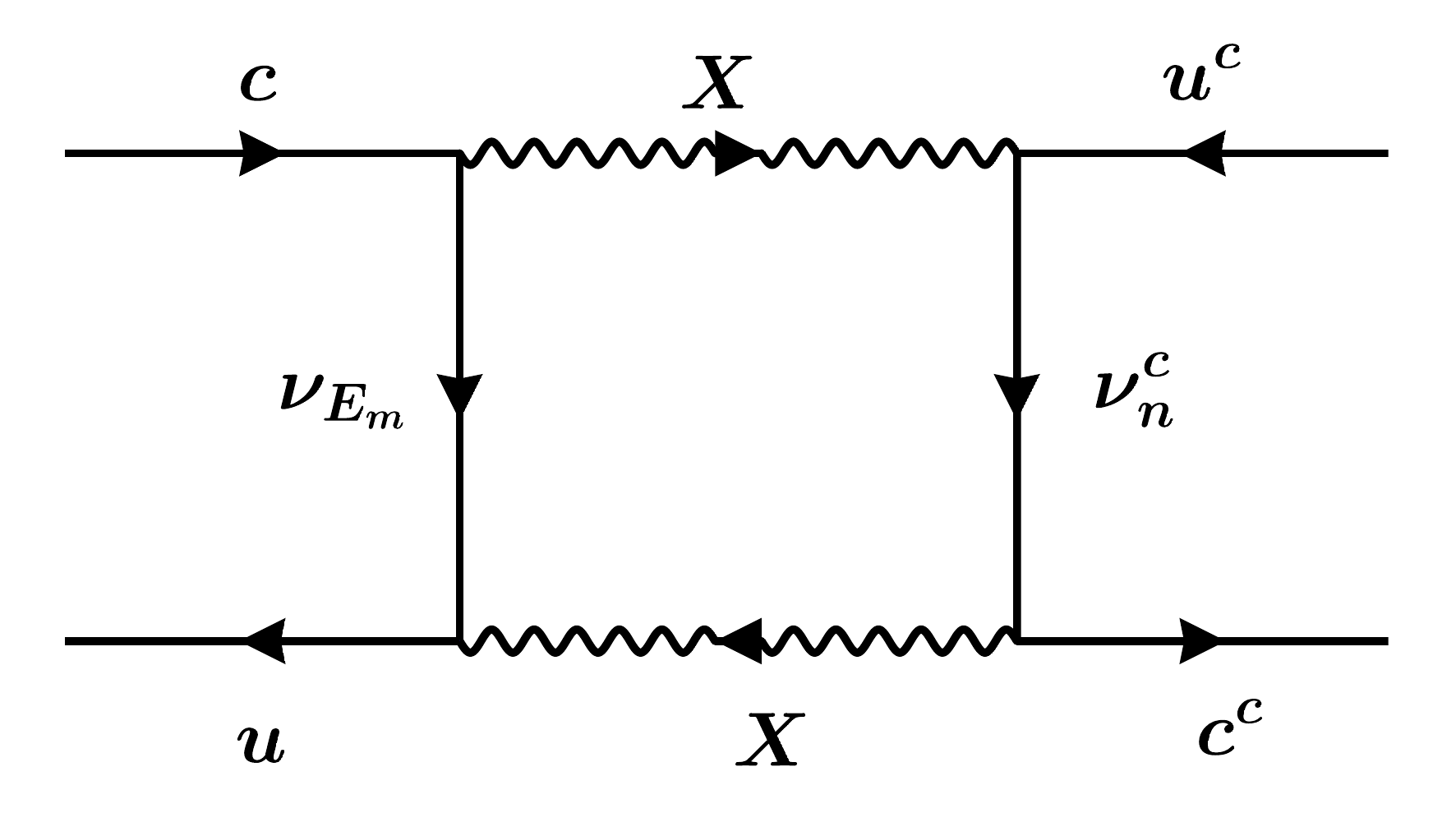}
        \caption{}
    \end{subfigure}

    \caption{Representative Feynman diagrams contributing to the $D$-meson oscillation with heavy fermions running in the loop. Here, $m,~n$ are the flavor indices for the exotic fermions, and color indices are implicit.}
    \label{fig:D-D_box_diagrams}
\end{figure}

The recipe to obtain a lower limit on leptoquark gauge boson mass, $M_X$, from $D^0-\bar{D}^0$ oscillation is the same as $K^0-\bar{K}^0$ oscillation. First, we read the Feynman rules from Eq.~\eqref{eqn:L-H-GX}--\eqref{eq:Wprime} that identify the diagrams that contribute to \(D^0-\bar{D^0}\) mixing in Fig.~\ref{fig:D-D_box_diagrams}. Now, the \(W'\) gauge boson will have new contributions to \( D^0\text{--}\bar{D}^0 \) mixing, which were absent in the \( K^0\text{--}\bar{K}^0 \) case. The contribution from the \( W\text{–}W' \) box diagrams is negligible in our setup, as it will be suppressed by the small soft $Z_2$ breaking VEV ${\color{magenta} w}$.

We now present the individual new physics contributions from the gauge boson, the Goldstone boson, and the mixed gauge-Goldstone boson. As shown in Fig.~\ref{fig:D-D_box_diagrams}, the operators $Q_1$, $\widetilde{Q}_1$, $Q_4$, and $Q_5$ are induced similar to the $K^0$--$\bar{K}^0$ oscillation. We can write down the leading-order expressions for Wilson coefficients following the conventions mentioned in Sec.~\ref{subsec:Interactions of the gauge bosons 
with light and heavy fermions}:
\begin{align}
      C_1^{(a+b+c)}
  &= -\left( \frac{g_4}{\sqrt{2}} \right)^4
     \frac{1}{4\pi^2 M_X^2}
     \Bigg[
       \sum_{m,n} \zeta_{E_m} \zeta_{E_n}
         A(t_{E_m}, t_{E_n}) 
       + \frac{1}{16} \sum_{m,n}
         \zeta_{E_m} \zeta_{E_n}
         A(t_{E_m}, t_{E_n})\,
         t_{E_m} t_{E_n}
     \Bigg],
\end{align}

\begingroup
\allowdisplaybreaks
\begin{align}
  \widetilde C_1^{(d+e+f+g+h)}
  &= -\left( \frac{g_4}{\sqrt{2}} \right)^4
     \frac{1}{4\pi^2 M_X^2}
     \sum_{m,n} \zeta_{\nu^c_m} \zeta_{\nu^c_n}
       A(t_{\nu^c_m}, t_{\nu^c_n})\notag\\
  &\quad
     -\left( \frac{g_R}{\sqrt{2}} \right)^4
      \frac{1}{4\pi^2 M_{W'}^2}
     \Bigg[
       \sum_{m,n} \zeta_{D^c_m} \zeta_{D^c_n}
         A(r_{D^c_m}, r_{D^c_n})\notag\\
  &\qquad\qquad\qquad
       + \frac{1}{16} \sum_{m,n} \zeta_{D^c_m}\zeta_{D^c_n}
         A(r_{D_m}, r_{D_n})\, r_{D_m} r_{D_n}\notag \\
         &\qquad\qquad\qquad
       + \frac{1}{2} \sum_{m,n} \zeta_{D^c_m} \zeta_{D^c_n}
         B(r_{D_m},r_{D_n})\, r_{D_m} r_{D_n}
     \Bigg],\\
  C_4^{(i+j)}
  &= \left( \frac{g_R}{\sqrt{2}} \right)^4
     \frac{1}{16\pi^2 M_{W'}^2}
     \frac{m_u m_c}{M_{W'}^2}
     \Bigg[
       \sum_{m,n} \zeta_{D^c_m} \zeta_{D^c_n}
         B(r_{D_m}, r_{D_n})\, r_{D_m} r_{D_n}\notag\\
  &\qquad\qquad\qquad
       + 2\sum_{m,n} \zeta_{D^c_m} \zeta_{D^c_n}
         A(r_{D_m}, r_{D_n})
     \Bigg],\\
  C_5^{(k)}
  &= -2 \left( \frac{g_4}{\sqrt{2}} \right)^4
      \frac{1}{16\pi^2 M_X^2}
      \sum_{m,n} \zeta_{\nu_{E_m}} \zeta_{\nu_{E_n}}
        A(t_{\nu_{E_m}}, t_{\nu_{E_n}}).
\label{eqn:Wilson_D_oscillation}
\end{align}
\endgroup


\begin{figure}[th]
    \centering
    \includegraphics[width=0.65\textwidth]{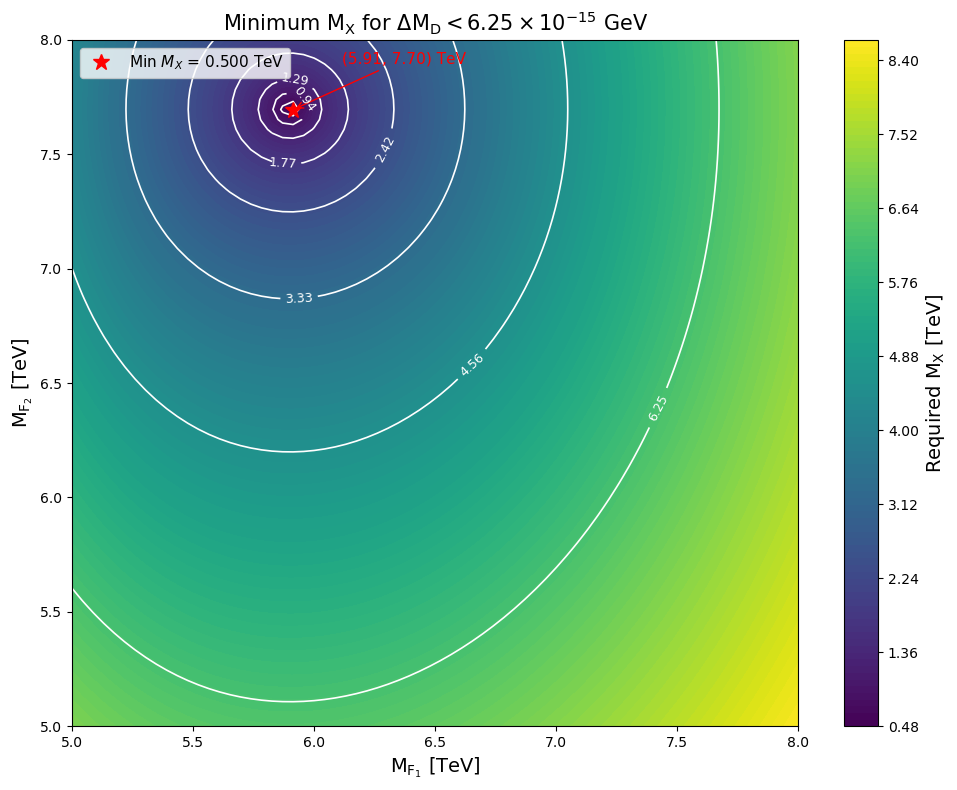}
    \caption{Heatmap showing the minimum value of \( M_X \) (in TeV) required to satisfy the constraint \( \Delta M_D < 6.25 \times 10^{-15}~\mathrm{GeV}\), as a function of the exotic fermion masses. For benchmark scenarios where the heavy fermions \( \nu_E,\nu^c, D\) are of similar mass scale and the CKM-like mixing matrices are followed by benchmark-2, the required value of \( M_X \) can be as low as \( \sim 500~\mathrm{GeV} \).
 }
    \label{fig:D-D_oscillation}
\end{figure}

Here we use the definition $\zeta_{E_m}=V_{uE_m}V_{cE_m}^*$, $\zeta_{\nu^c_{m}}=V_{u\nu^c_{m}}V_{c\nu^c_{m}}^*$, 
$\zeta_{D^c_m}=K_{D^c_mu}^*K_{D^c_mc}$, $t_{\alpha}=M_{\alpha}^2/M_X^2,~r_{\alpha}=M_{\alpha}^2/M_{W'}^2$, and, \(A(x_i,x_j)\), \( B(x_i,x_j)\) are already defined in Eq.~\eqref{eqn:Inami-Lim}.
Now, to get a rough numerical estimation, we take the same CKM-like structure as in Eq.~\eqref{eqn:benchmark2}, and perform a parameter scan over \( M_{\nu_E,\nu^c}\sim M_1 = (M_{F_1}, 5.9, 6.2)~\mathrm{TeV} \) and \( M_D\sim M_2 = (M_{F_2}, 7.7, 6.5)~\mathrm{TeV} \). We find that \( M_X \) can be as low as 500~GeV for the choice \((M_{F_1}, M_{F_2}) = (5.91, 7.70)~\mathrm{TeV},\) while satisfying the experimental constraint \(\Delta M_D < 6.25 \times 10^{-15}~\mathrm{GeV},\) as illustrated in Fig.~\ref{fig:D-D_oscillation}.

\subsubsection{$B_q^0-\bar{B_q}^0$ oscillation}
The calculation of new contributions to \( B_q^0 - \bar{B}_q^0 \) oscillation ($q=d,s$) closely follows that of the \( K^0 - \bar{K}^0 \) system. The only difference is that it involves different generations of the down-type quarks. As a result, the hadronic matrix elements and the magic numbers would differ, as given in Appendix~\ref{subsection: Input parameters for B-B oscillation}.
\begin{figure}[!h]
\centering
\includegraphics[scale=0.35]{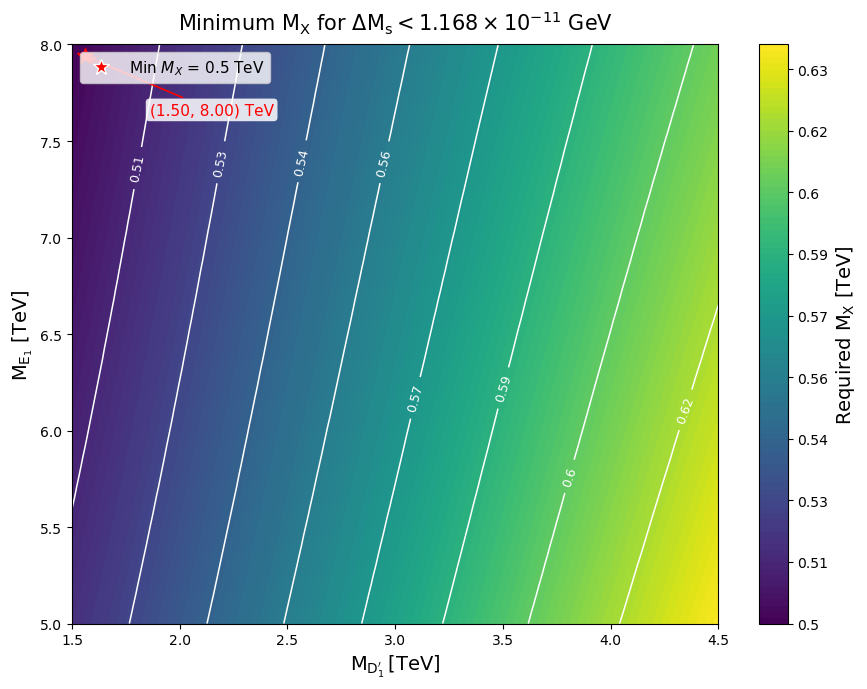}
\includegraphics[scale=0.35]{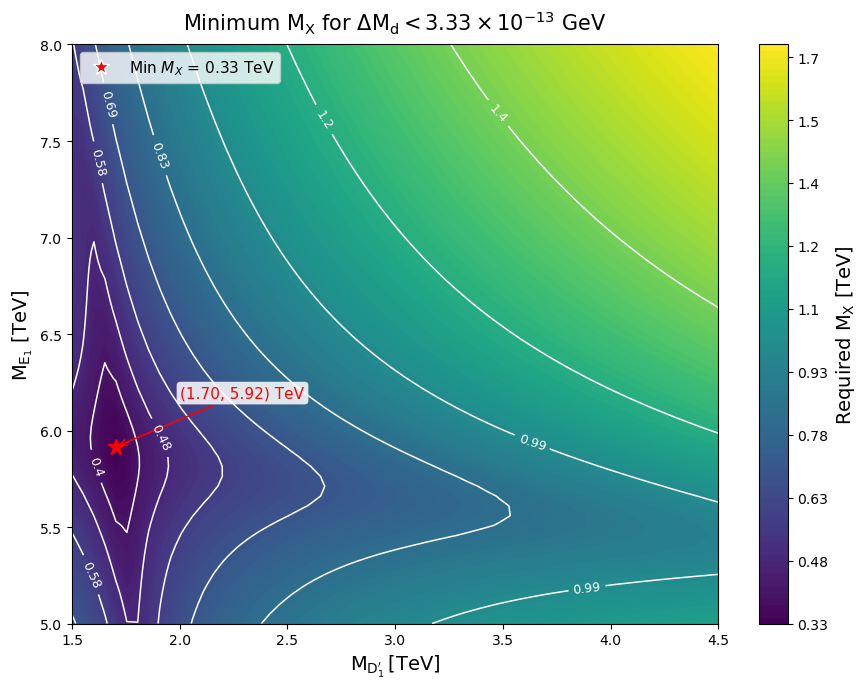}
\caption{With a comparable benchmark scenario as $K^0$--$\bar{K}^0$ oscillation, we show the required value of $M_X$ with the variation of the vector-like fermion masses for $B_s^0-\bar{B}_s^0$ oscillation~(left), and for $B_d^0-\bar{B}_d^0$ oscillation~(right).}
\label{fig:Bq_oscillation}
\end{figure}

For the benchmark scenario, we adopt the CKM-like matrices structured as defined in Eq.~\eqref{eqn:benchmark2} and perform a scan over exotic fermion masses: \( M_{\textcolor{red}{D'}} = (M_{\textcolor{red}{D'_1}},~1.7,~1.5)~\mathrm{TeV} \) and 
\( M_E = (M_{E_1},~5.9,~6.5)~\mathrm{TeV} \). We observe from Fig.~\ref{fig:Bq_oscillation} that the required value of \( M_X \) can be as low as \( \sim 500~\mathrm{GeV} \) from \( B_s^0-\bar{B}_s^0\) oscillation and as \( \sim 330~\mathrm{GeV} \) from \(B_d^0- \bar{B}_d^0\) oscillation.

\subsubsection{Tree-level meson mixing processes}
Here we present the tree-level meson mixing processes mediated by heavy neutral gauge boson and neutral Higgs boson interactions that change flavor. They will contribute to \(\bar{K}^0-K^0\) and \(\bar{B}^0-B^0\) mass differences.

\noindent \textbf{Neutral Higgs boson effects:} Since we are interested in the stringent limits that can be placed on \(\kappa_R\) through meson oscillation, let us consider \(\widetilde{\chi}(2,2,1)\), and ignore the possible mixings of \(\widetilde{\chi}\) with other fields which has been discussed in Sec.~\ref{sec:Higgs} extensively.
\begin{figure}[!h]
\centering
\includegraphics[scale=0.15]{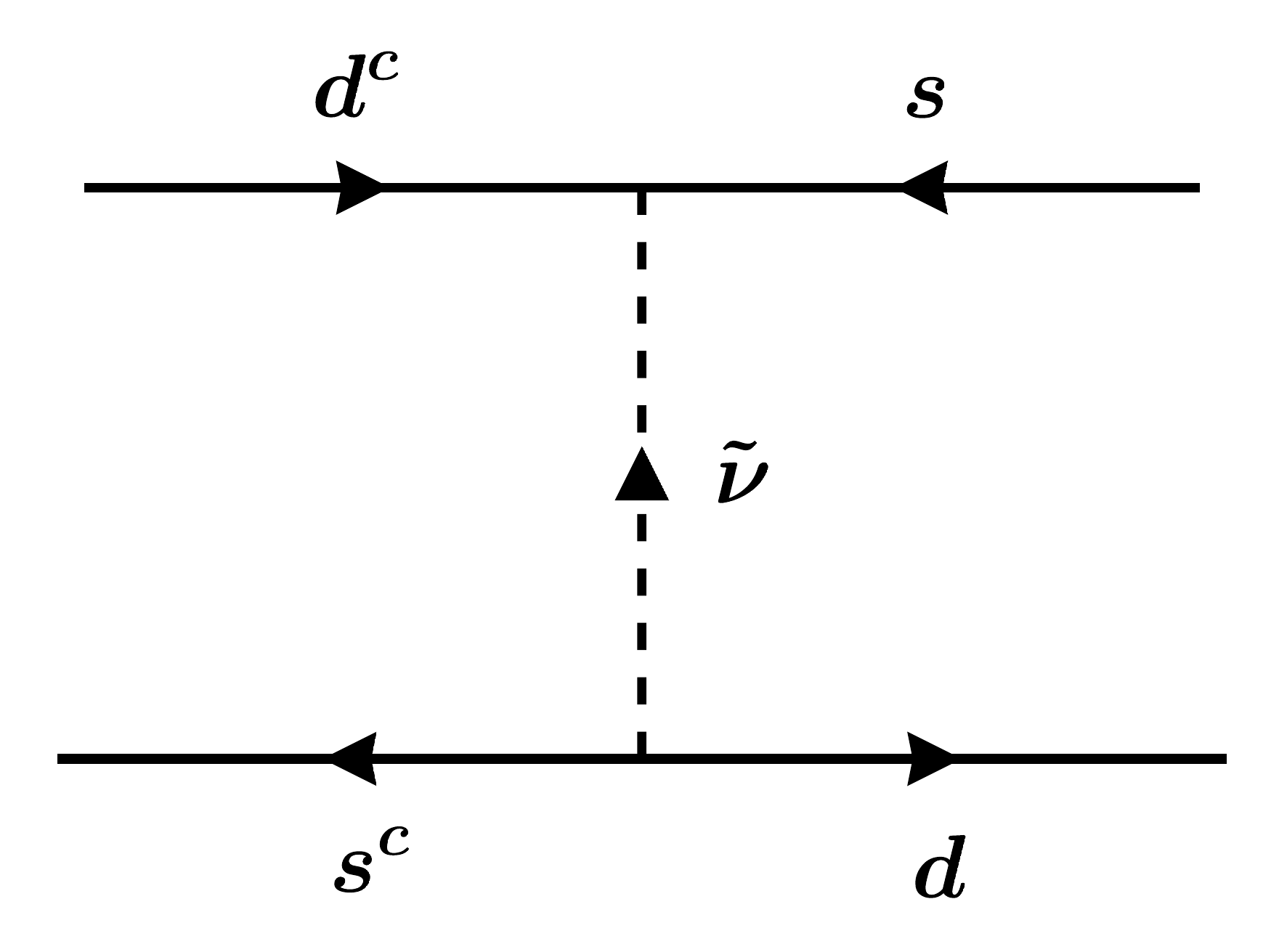}
\hspace{1 cm}
\includegraphics[scale=0.20]{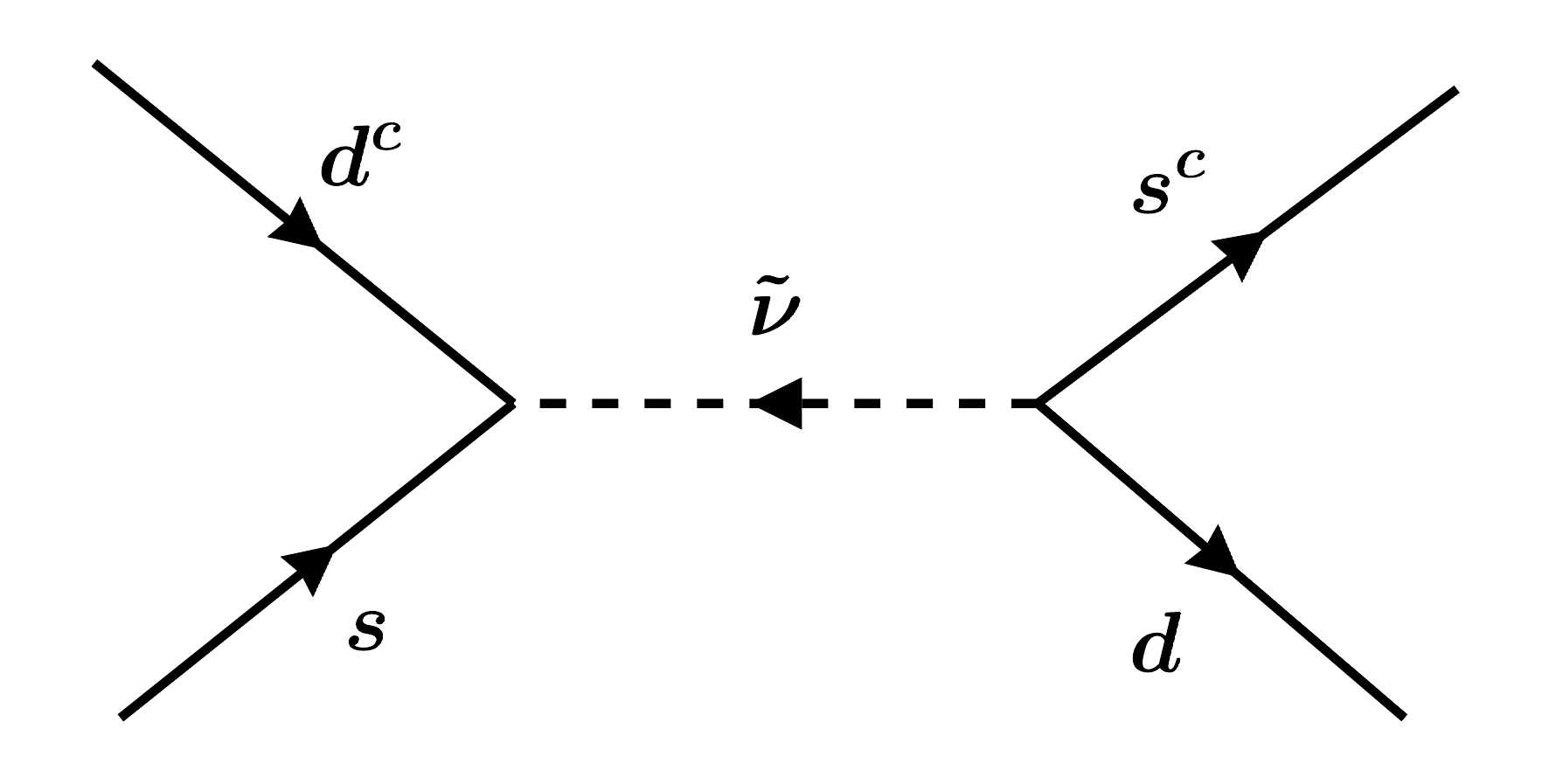}
\caption{For example, \(\bar{K}^0-K^0\) mixing diagrams induced by neutral Higgs bosons.}
\label{fig:neutral_Higgs_contribution}
\end{figure}
It is obvious from Eq.~\eqref{eqn:L-Yukawa} that \(\widetilde{\nu}_{E}^{c}\) interactions are flavor diagonal and the absence of scalar-pseudo-scalar mixing allows us to write $\widetilde{\nu} =\mathrm{Re}[\widetilde{\nu}] + i~ \mathrm{Im}[\widetilde{\nu}]$. In the limit of soft $Z_2$ breaking VEV \(\textcolor{magenta}{w}=0\), the physical states become \(H_I= \mathrm{Im}[\widetilde{\nu}]\) and \(H_R= \mathrm{Re}[\widetilde{\nu}]\) with mass \(m_{H_I}^2\approx m_{H_R}^2\approx \alpha\kappa_R^2\) where \(\alpha\) parameter can be determined from the Higgs-boson self interactions~\cite{Senjanovic:1979cta}. Therefore, the relevant interaction becomes (for example, for \(\Delta S=1\) processes)
\begin{equation}
    \mathcal{L}_Y^0= \dfrac{-ig_L}{\sqrt{2}M_{W_L}}(\sum_{i=u,c,t}\lambda_im_i)\left(H_I\bar{s}\gamma_5d-H_R\bar{s}d\right),
\end{equation}
with \(\lambda_i=V_{si^c}^{*}K_{i^cd^c}\) where \(V_{si}=(V_{d}^{\dagger}V_{u})_{si}\) and \(K_{i^cd^c}=K_{d^ci^c}^{\dagger}=(-V_{u^c}^{\dagger}\,x(\mathbb{I}+x^{\dagger}x)^{-1/2}V_{d^c})_{i^cd^c}\) as defined in Eq.~\eqref{eqn:Kdcuc}. Now we can write the effective Hamiltonian leading to meson mixing using the conventional effective operators defined in Eq.~\eqref{eqn:eff-operators} as :
\begin{align}
    \mathcal{H}_{\mathrm{eff}}^{\tilde{\nu}}&=\left(\frac{C^2(\Lambda)}{2m_{H_I}^2}+\frac{C^2(\Lambda)}{2m_{H_R}^2}\right)\left(Q_2^{ds}+\widetilde{Q}_2^{ds}\right)+\left(\frac{C^2(\Lambda)}{m_{H_R}^2}-\frac{C^2(\Lambda)}{m_{H_L}^2}\right)Q_4^{ds},
\end{align}
where \(\Lambda\sim \kappa_R\) and \( C = \dfrac{g_L}{M_{W_L}} ( \sum\limits_{i=u,c,t} \lambda_i m_i) \)
 and we evolve the Wilson coefficients from the high scale \( \kappa_R \) down to the hadronic scale using renormalization group running. To estimate this effect, let us take \(m_{H_I}\simeq m_{H_R}\simeq 2~\text{TeV}\), \(V_{si}\) given by the CKM matrix, 
\(V_{u^c},\,V_{d^c}\) as identity, and the benchmark matrix \(x\) from Eq.~\eqref{eqn:x_bench2}. For this choice, we obtain 
\begin{align}
    \Delta M_K^{(\tilde{\nu})} \simeq 2.8\times 10^{-15}~\text{GeV},
\end{align} 
which is satisfied by the \(K\)-meson mixing bound (see Table~\ref{tab:FLAVOR}). An interesting point to note here is that the flavor-changing matrix \(x\) allows us to realize neutral Higgs bosons at a lower scale than the conventional limits from meson mixing in left--right symmetric models, where he limit is about 20 TeV~\cite{Mohapatra:1983ae,Zhang:2007da}.
\begin{figure}[!h]
\centering
\includegraphics[scale=0.15]{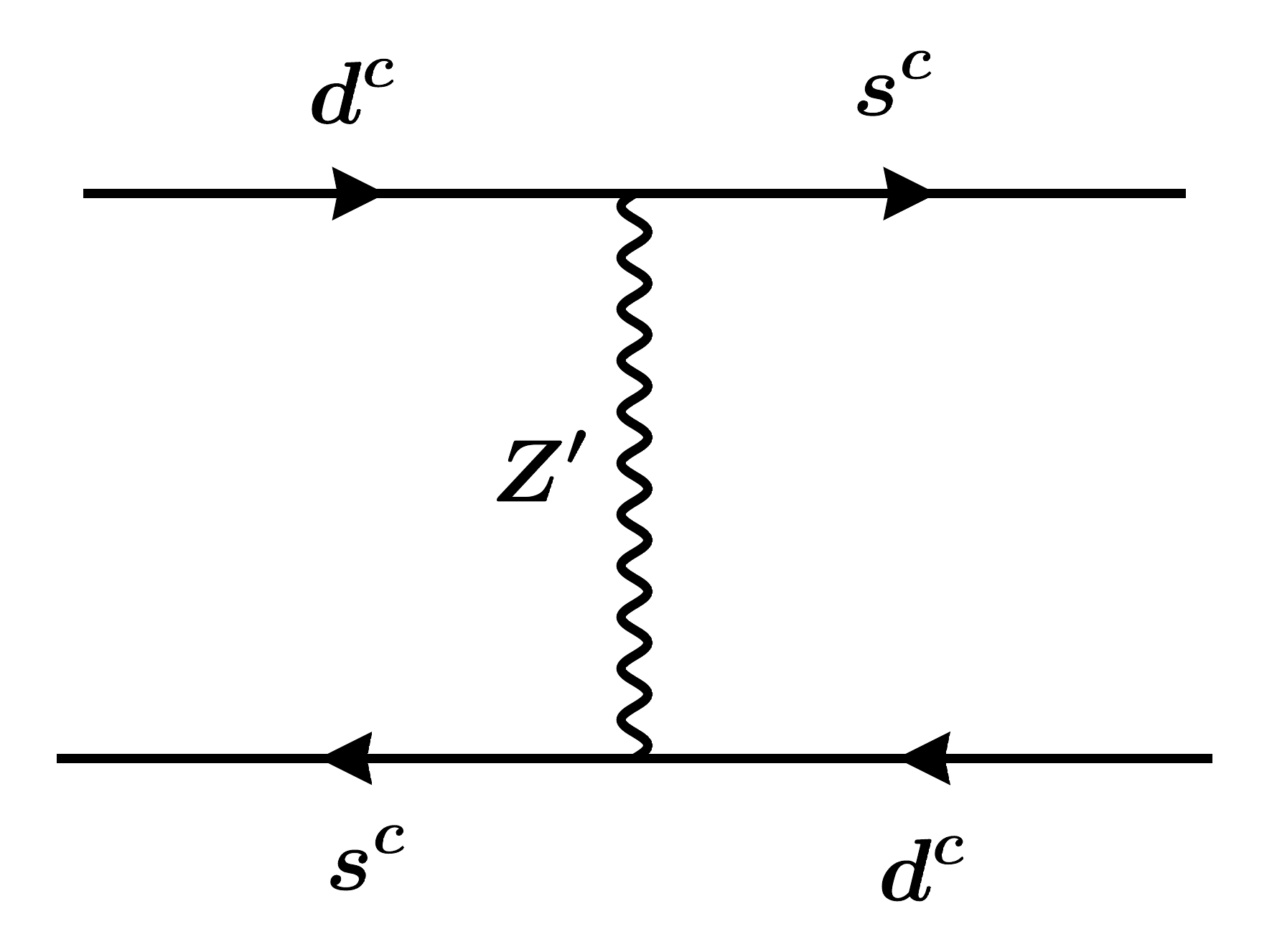}
\includegraphics[scale=0.15]{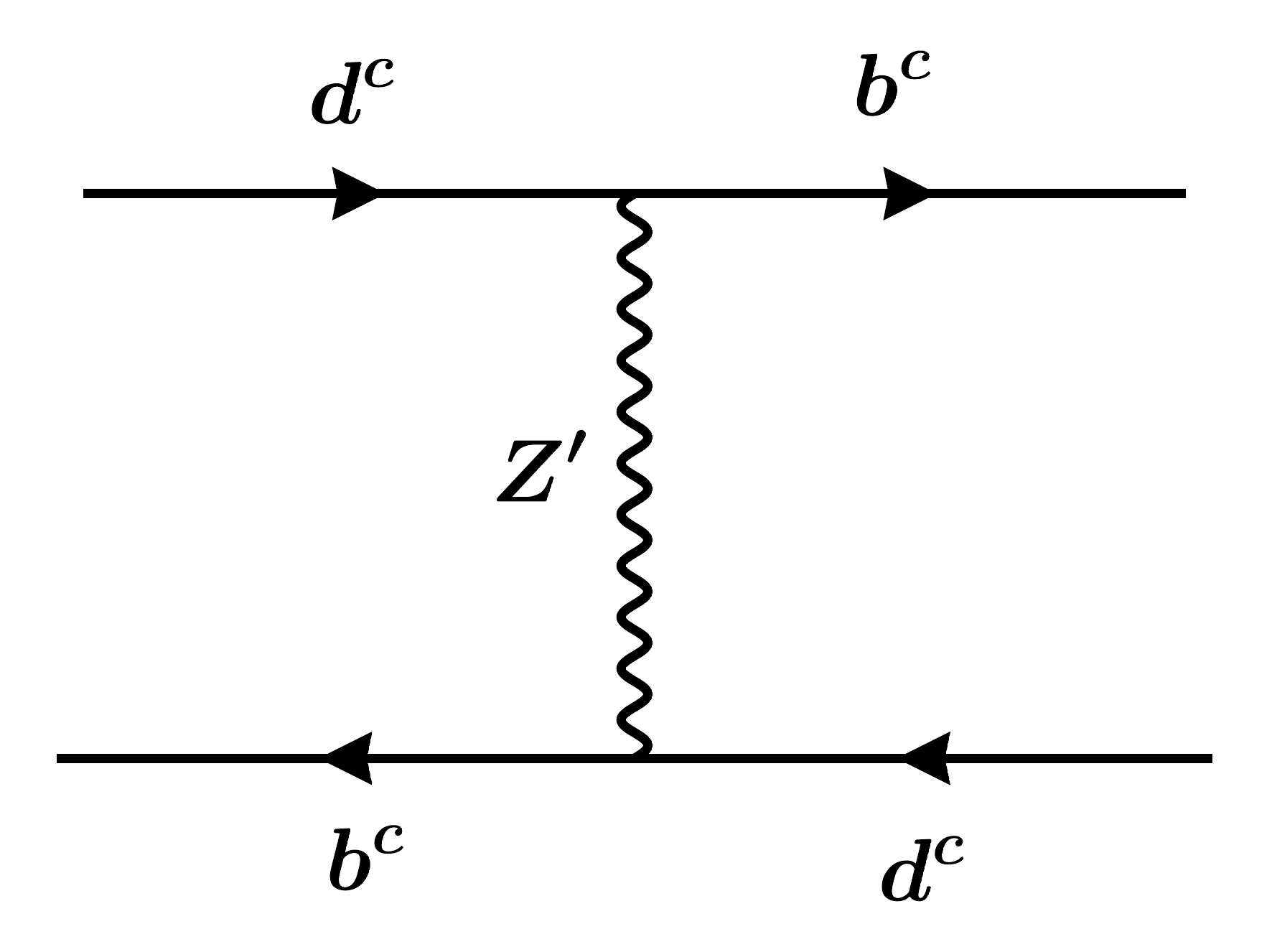}
\includegraphics[scale=0.15]{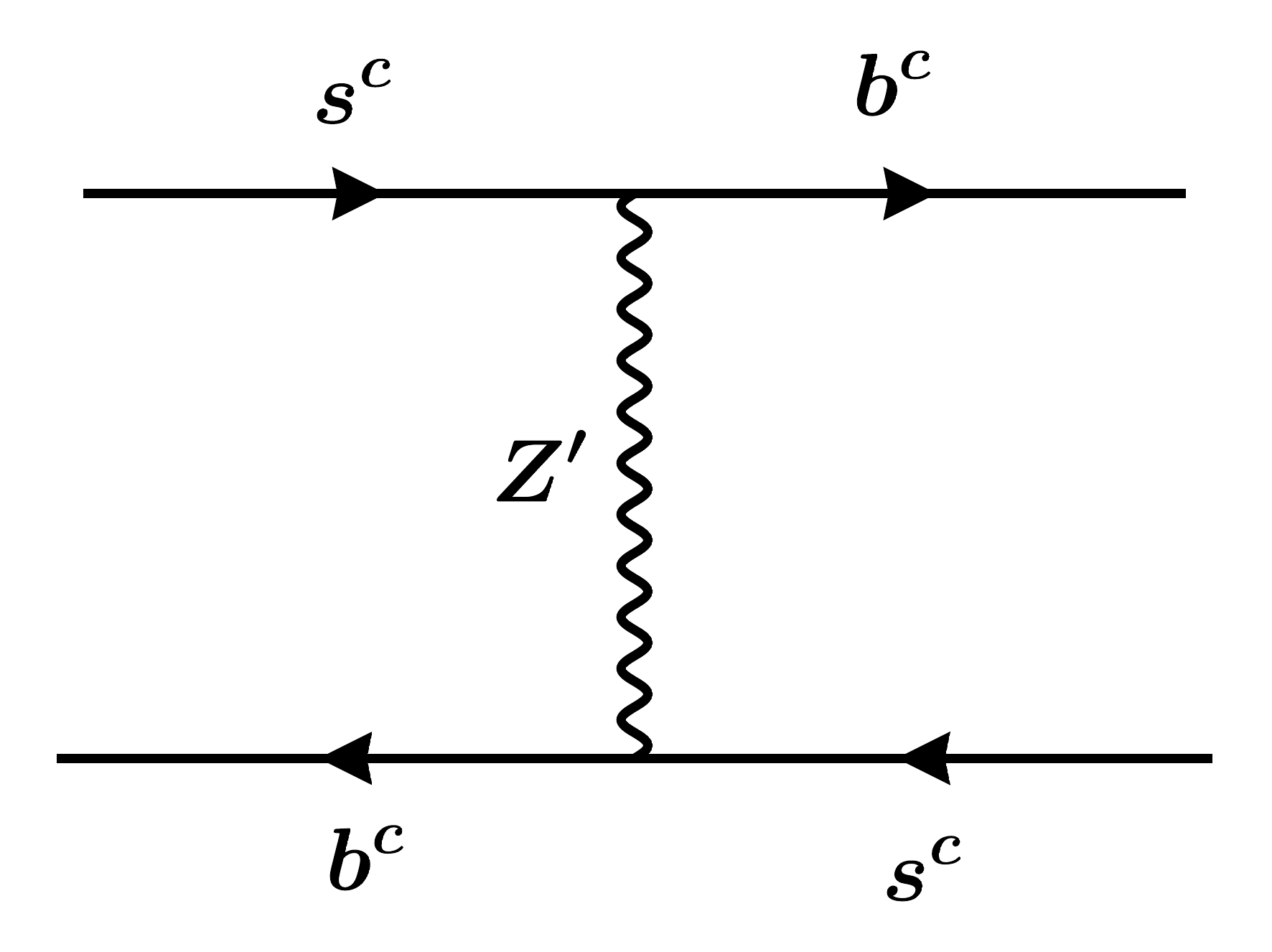}
\caption{\(K\)- and \(B\)-meson mixing diagrams induced by \(Z'\).}
\label{fig:neutral_Higgs_contribution}
\end{figure}

\noindent\textbf{Neutral heavy gauge boson \(Z'\) effects:}
As we have seen already, because of the soft breaking of $Z_2$ symmetry, the fields \(d^c\) and \(D^c\) mix, and as a consequence, the neutral gauge boson \(Z'\) has flavor changing effect (\(d^c\) and \(D^c\) do not have the same \(T_{3R}\) charge) and contributes to \(K\)- and \(B\)-meson mixing at tree level through its coupling with \(d^c\) shown in Eq.~\eqref{eqn:Pdc}. However, this contribution is highly suppressed in our setup. The tree-level exchange will induce the operator \(\widetilde Q_1\), so that
\begin{equation}
  \mathcal{H}_{\mathrm{eff}}^{Z'}
  = \widetilde C_1\,\widetilde Q_1, 
  \qquad
  \widetilde C_1 \sim 
  \frac{g_R^2}{2 M_{Z'}^2}
  (P^{d^c})^{\dagger} P^{d^c},
\end{equation}
where \(P^{d^c}\) is defined in Eq.~\eqref{eqn:Pdc}. For a numerical estimation we take \(M_{Z'} = 5.6~\text{TeV}\), \(V_{d^c}\) to be the same structure as in Eq.~(\ref{eqn:benchmark2}), and let's use the benchmark
matrix \(x\) shown in Eq.~\eqref{eqn:x_bench2}. Including the appropriate hadronic matrix elements, we find,
\begin{equation}
  \Delta M_{K}^{(Z')} \sim 1.3\times 10^{-21}~\text{GeV}, \qquad
  \Delta M_{B_d}^{(Z')} \sim 1.2\times 10^{-17}~\text{GeV}, \qquad
  \Delta M_{B_s}^{(Z')} \sim 5.8\times 10^{-18}~\text{GeV},
\end{equation} 
and all of them are well within the experimental limits.

There are also flavor-changing processes induced by the $Z$-boson at tree-level in the model.  The relevant couplings are given in Eq. (\ref{eq:ZFCNC}). However, these processes are highly suppressed, as the off-diagonal couplings of the $Z$, such as $\overline{d}_L \gamma_\mu s Z^\mu$ have coefficients of order $Y_d^2 v^2/\kappa_R^2 \lesssim 10^{-8}$ (with the largest $Y_d$ identified as $Y_b \simeq 1.6 \times 10^{-2}$).  The resulting contributions to $K^0-\overline{K^0}$ mixing, for example, are several orders of magnitude below the experimental limit.

\subsection{Limit on $M_X$ from cLFV through leptoquark gauge boson \(X_{\mu}\)}
\label{subsec:Limit on m_X from CLFV processes}
Charged lepton-flavor violation~(cLFV) processes such as $\mu \rightarrow e \gamma$ decay are highly suppressed in the SM, with a branching ratio in the unobservable range of order $\sim 10^{-54}$, induced by non-zero neutrino masses and mixing angles~\cite{Petcov:1976ff,MEGII:2025}. Thus, any positive signal in the experimental searches for such processes will be clear evidence for a new dynamics. In this section, we present the detailed calculations for lepton-flavor violation~(LFV) processes through leptoquark gauge boson \(X_{\mu}\) exchange in the muonic sector, focusing on $\mu \to e \gamma$ radiative decay, $\mu\to e e e$ decay, and $\mu$-$e$ conversion in a muonic atom in our setup. 
\subsubsection{$\mu \to e \gamma$ decay}
\label{subsubsec:mueg}
One of the most studied processes of LFV is $\mu\to e \gamma$ decay. This decay probes the dipole operator with a chirality flip on the external muon and can provide dominant constraints on flavor structure in scenarios with quark–lepton correlation at the TeV scale.
\begin{figure}[h!]
    \centering
    \begin{subfigure}[b]{0.32\textwidth}
        \centering
        \includegraphics[width=\textwidth]{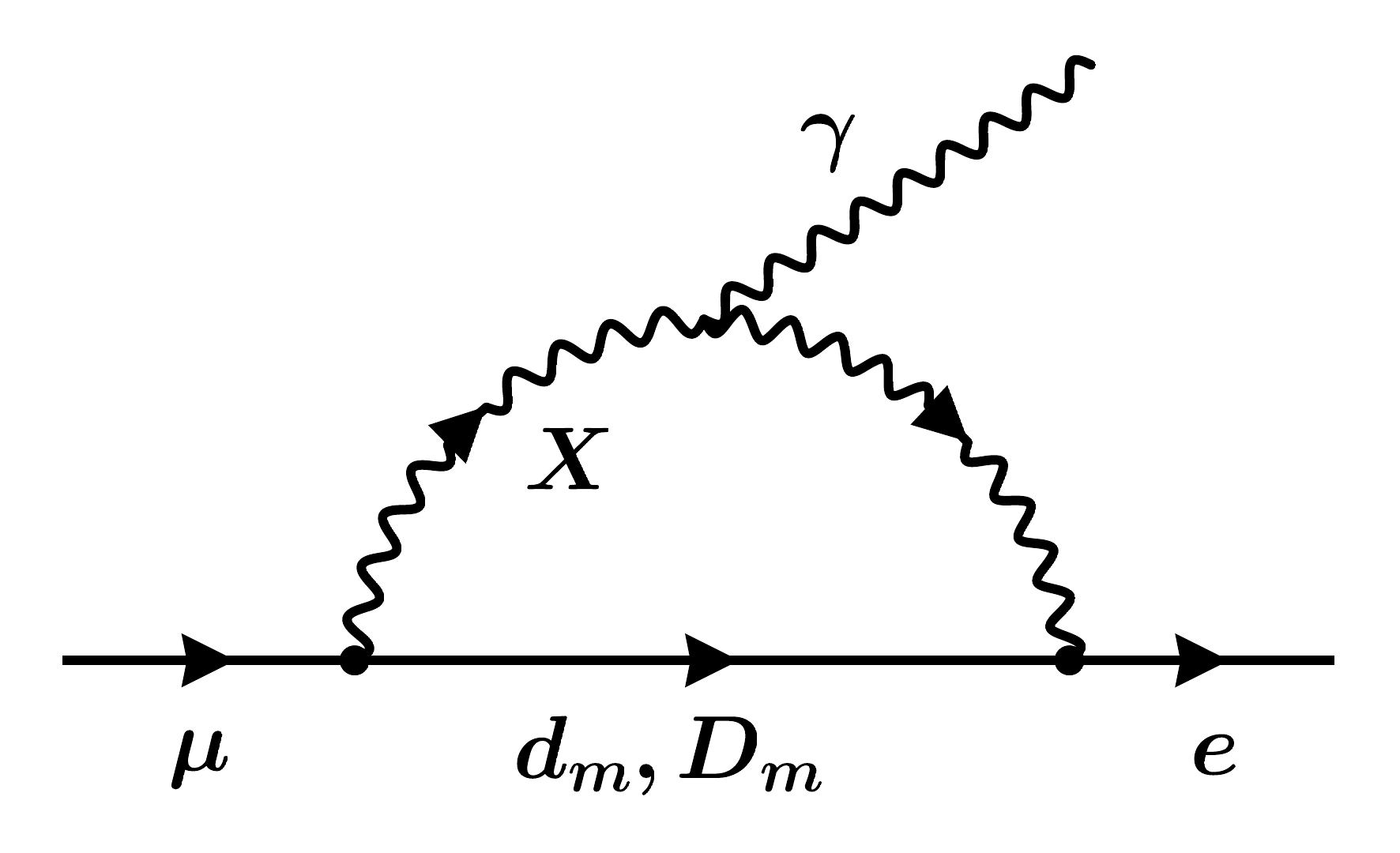}
        \caption{}
    \end{subfigure}
    \hfill
    \begin{subfigure}[b]{0.32\textwidth}
        \centering
        \includegraphics[width=\textwidth]{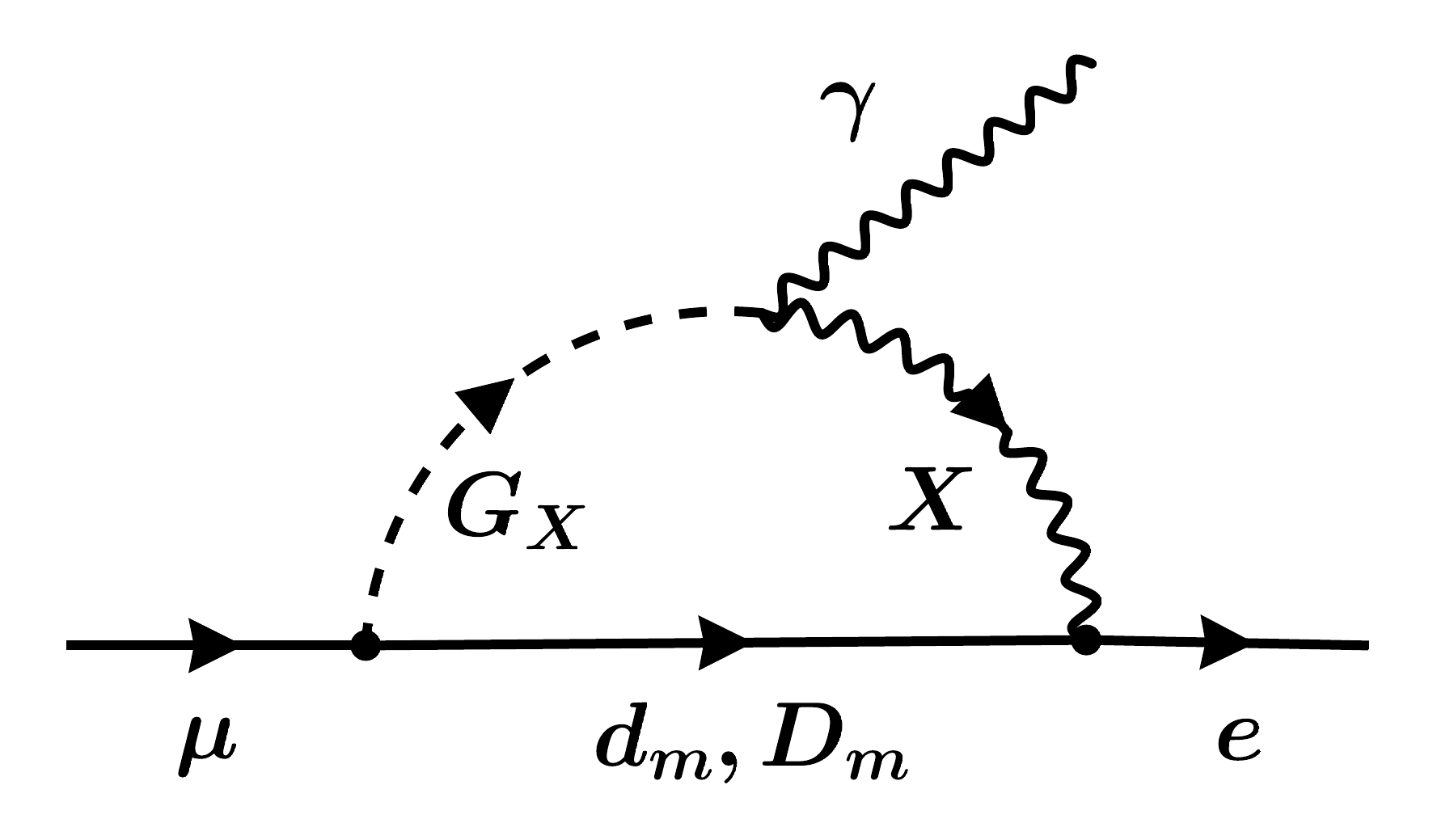}
        \caption{}
    \end{subfigure}
    \begin{subfigure}[b]{0.32\textwidth}
        \centering
        \includegraphics[width=\textwidth]{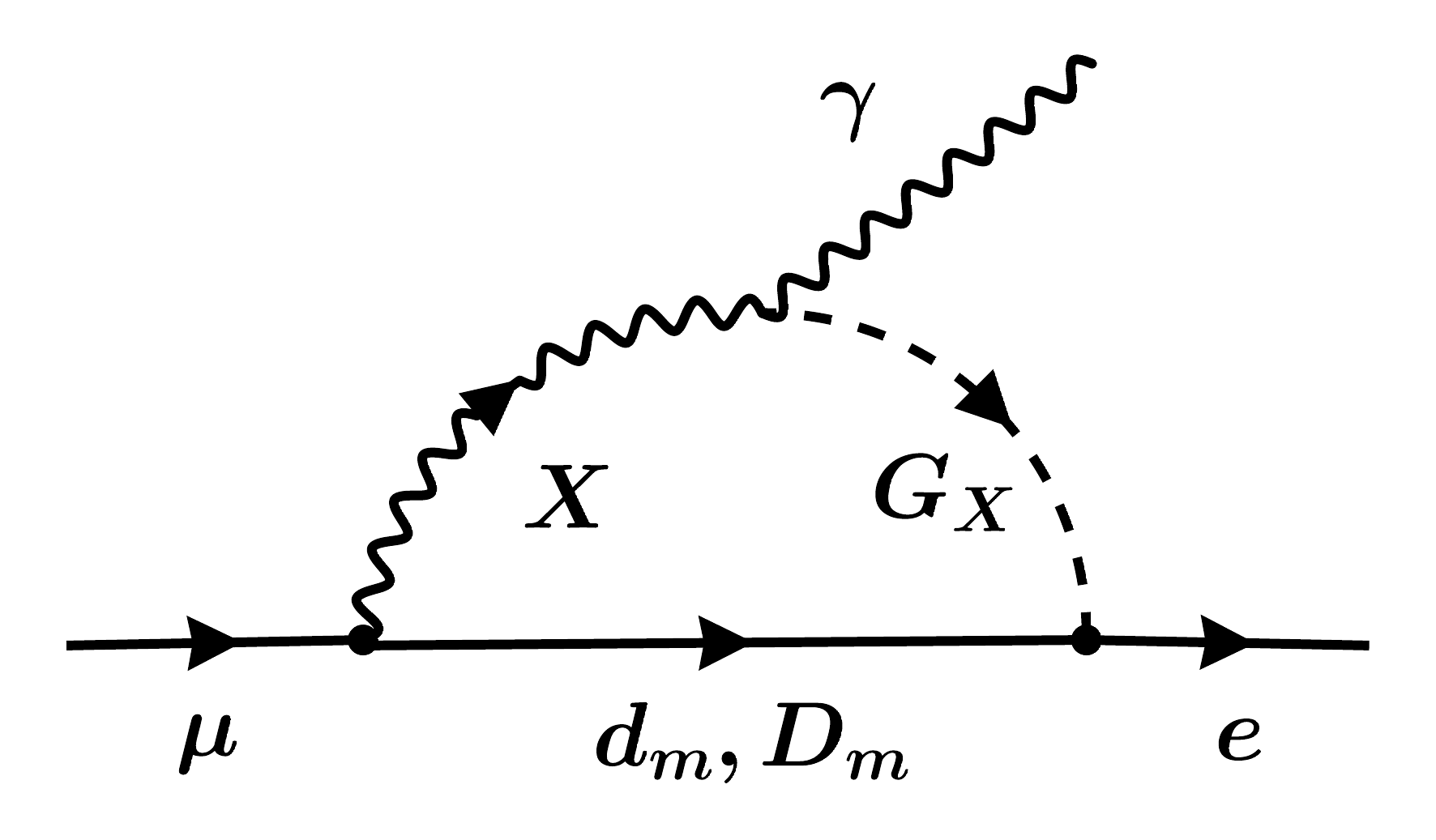}
        \caption{}
    \end{subfigure}

    \vspace{0.4cm} 

    \begin{subfigure}[b]{0.32\textwidth}
        \centering
        \includegraphics[width=\textwidth]{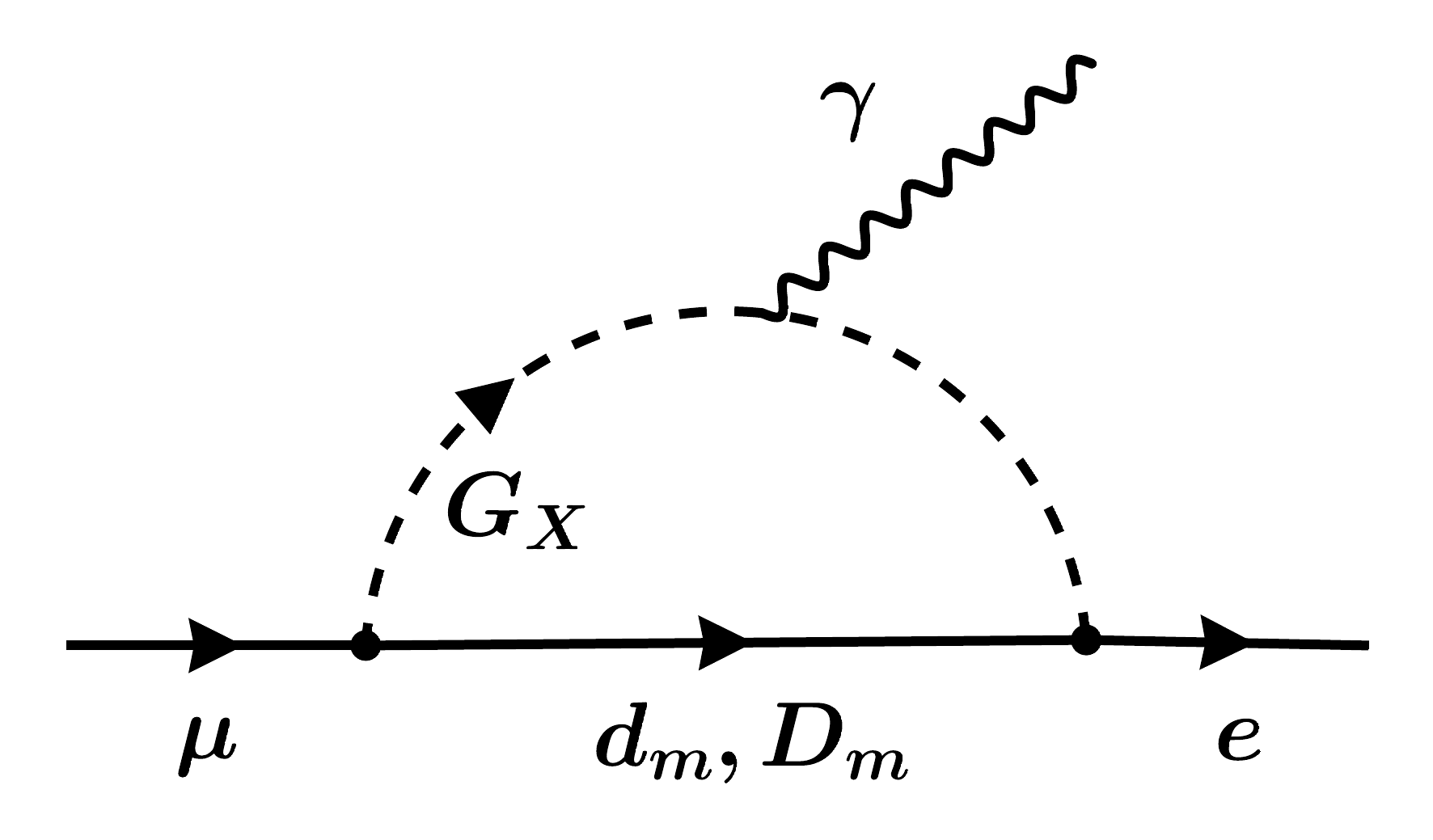}
        \caption{}
    \end{subfigure}
    \hfill
    \begin{subfigure}[b]{0.32\textwidth}
        \centering
        \includegraphics[width=\textwidth]{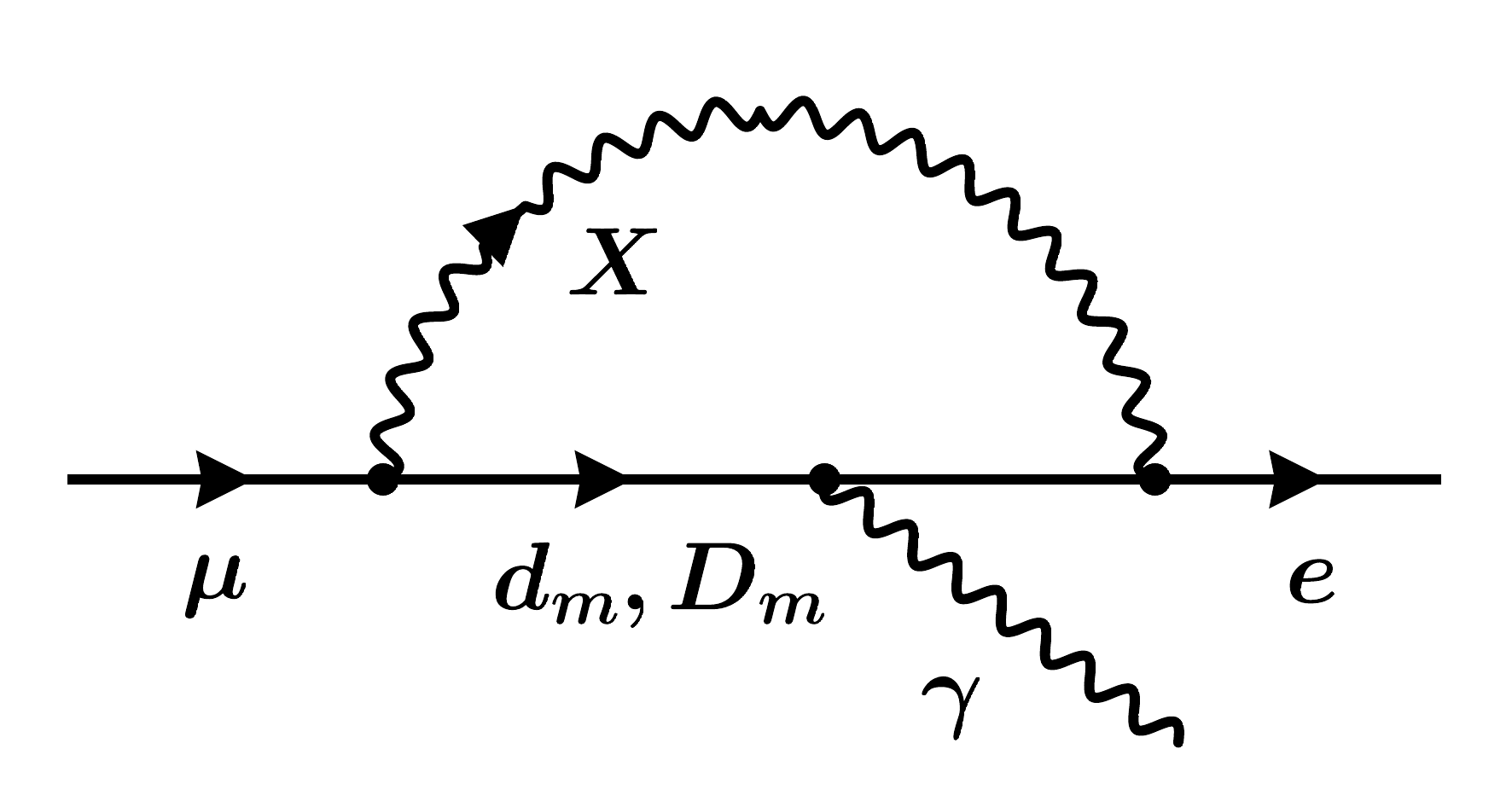}
        \caption{}
    \end{subfigure}
    \begin{subfigure}[b]{0.32\textwidth}
        \centering
        \includegraphics[width=\textwidth]{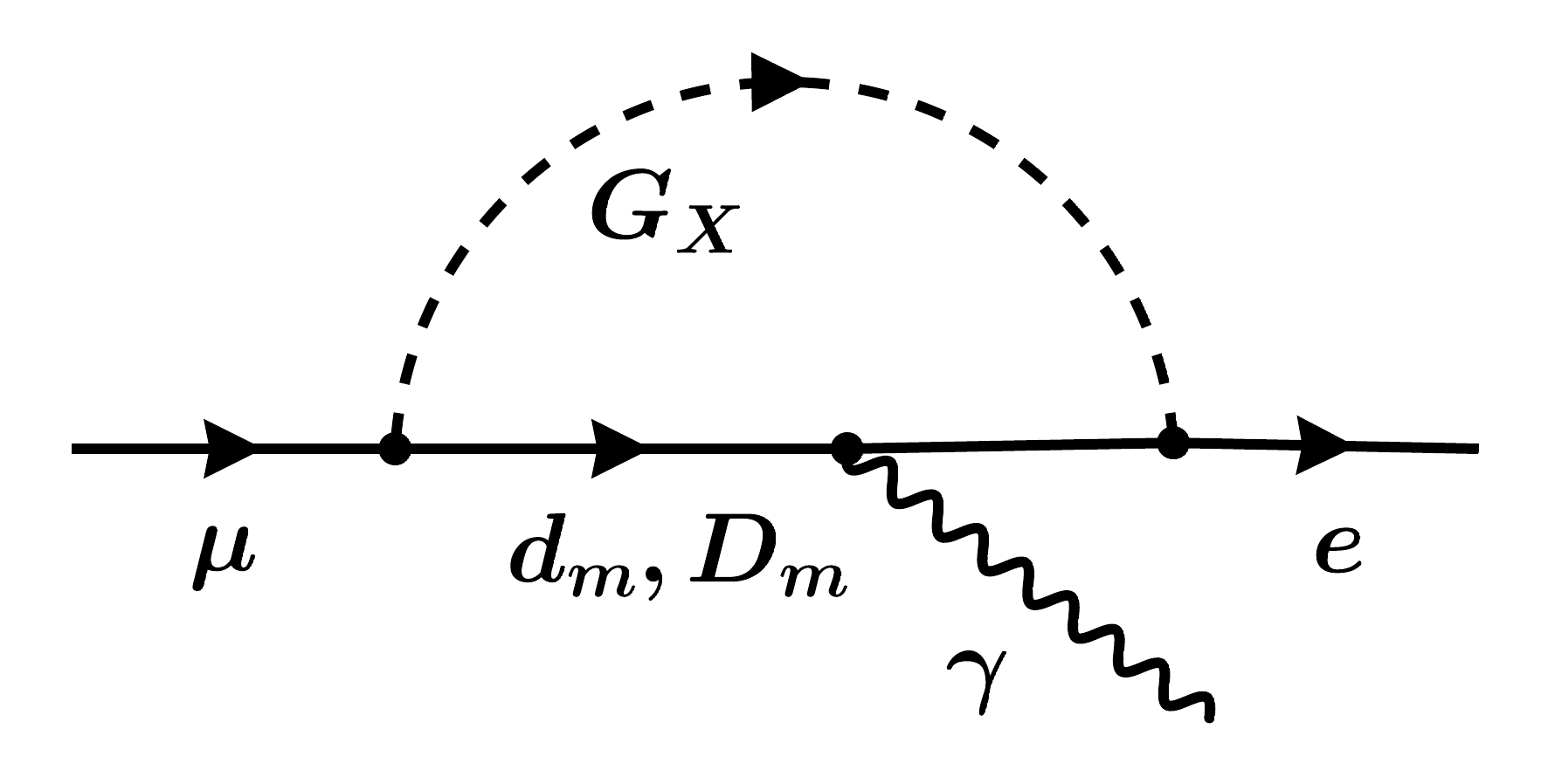}
        \caption{}
    \end{subfigure}
    \caption{One loop contributions from leptoquark gauge boson to \(\mu^c \to e^c \gamma\) amplitude in Feynman–’t
    Hooft gauge. The chirality of the fermions and couplings can be understood from Eqs.~\eqref{eqn:L-L-X} and \eqref{eqn:H-L-X}.}
    \label{fig:mu_e_gamma_diagrams}
\end{figure}
The amplitude for the physical decay process \(\mu(p_1)\to e(p_2)+\gamma(q)\) can be written purely in terms of the gauge–invariant dipole form,
\begin{equation}
T(\mu\to e \gamma)
=\epsilon_\lambda^*(q)\;
\bar u_e(p_2)\!\left[i\sigma^{\lambda\nu}q_\nu\big(\sigma_L P_L+\sigma_R P_R\big)\right]u_\mu(p_1),
\label{eqn:Tmuegamma}
\end{equation}
where \(\epsilon_{\lambda}^*(q)\) is the polarization vector of the outgoing photon, \(\sigma^{\lambda\nu}=(i/2)[\gamma^{\lambda},\gamma^{\nu}]\) is the antisymmetric tensor, and \(
P_{R,L}=(1\pm\gamma_{5})/2\) are the chirality projectors. The form factors \(\sigma_{L,R}\) are obtained by computing the diagrams shown in Fig.~\ref{fig:mu_e_gamma_diagrams}. 

In the limit \(m_e\!\to\!0\), the decay width can be written as~\cite{Cheng:2000ct,Lavoura:2003xp}
\begin{equation}
\Gamma(\mu\to e\gamma)\;=\;\frac{m_\mu^3}{16\pi}\,\big(|\sigma_L|^2+|\sigma_R|^2\big).
\end{equation}
In our setup, \(X_{\mu}\) couples only to right-handed leptons (their left-handed partners are not charged under \(SU(4)_C\)), as can be seen from the relevant Lagrangian terms in Eqs.~\eqref{eqn:L-L-X} and \eqref{eqn:H-L-X}. Thus, the chirality flip must occur on the external lepton legs in these diagrams. From Eq.~\eqref{eqn:Tmuegamma}, this implies \(\sigma_L \propto m_{\mu}\) and \(\sigma_R \propto m_{e}\). Taking \(m_e \simeq 0\), it is justified to neglect \(\sigma_R\) and to work in the limit \(|\sigma_L|^2 \gg |\sigma_R|^2\).

\begin{figure}[!h]
    \centering
    \includegraphics[width=0.65\textwidth]{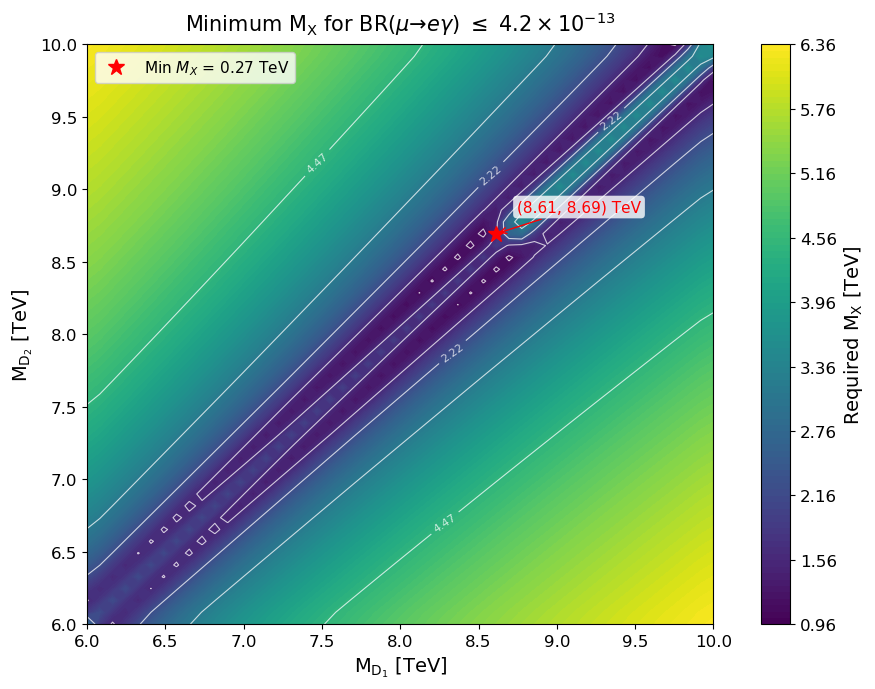}
\caption{Minimum $M_X$ required by the
$\mathrm{BR}(\mu\!\to\!e\gamma)\le 4.2\times10^{-13}$ across the $(M_{D_1},M_{D_2})$ plane (with $M_{D_3}=6.7~\mathrm{TeV}$ fixed).
The colorbar indicates the required $M_X$ in TeV, with the white lines representing iso–$M_X$ contours. The red star marks the global minimum on the scan.    }
 \label{fig:mu_to_e_gamma}
\end{figure}

We perform the calculations in ’t Hooft-Feynman gauge instead of unitary gauge following Ref.~\cite{Lavoura:2003xp}. Summing the vector and would–be Goldstone boson exchanges provides a finite, gauge–invariant result, which can be expressed as follows: 
\begin{equation}
    \sigma_L=\frac{N_c\,e\,g_4^2}{2}\sum_{m}\Big[\rho_m^d\Big(\,Q_{d}\,\mathcal{I}_1(t_m)+Q_X\,\mathcal{I}_2(t_m)\Big)+\rho_m^D\Big(\,Q_{D}\,\mathcal{I}_1(t_m)+Q_X\,\mathcal{I}_2(t_m)\Big)\Big].
    \label{eqn:sigma_L}
\end{equation}
Here \(m\) is the flavor index, \(t_m=\dfrac{M_{m}^2}{M_X^2}\), \(Q_{d}=Q_{D}=-1/3, Q_X=2/3\) are the electric charges of the internal down–type fermion and the gauge boson \(X_{\mu}\) respectively, in the units of \(e\), and \(N_c=3\) is the number of color flow in the loop. The loop kernels can be expressed as: 
\begin{align}
    \mathcal{I}_1(t) \;&=\; \frac{m_\mu}{(-i)\,16\pi^2\,M_X^2}\,\left(\frac{-5t^3+9t^2-30t+8}{12\,(t-1)^3} + \frac{3t^2\ln t}{2\,(t-1)^4}\right),
    \label{eqn:yPsi}\\
    \mathcal{I}_2(t) \;&=\; \frac{m_\mu}{(-i)\,16\pi^2\,M_X^2}\left(\frac{-4t^3+45t^2-33t+10}{12\,(t-1)^3} - \frac{3t^3\ln t}{2\,(t-1)^4}\right).
    \label{eqn:yX}
\end{align}
where the three-gauge-boson vertex~(see Fig.~\ref{fig:mu_e_gamma_diagrams} (a)) of the outgoing photon \(\gamma_{\alpha}\) with momentum $q$, incoming \(X_{\mu}\) and \(X_{\nu}^*\) with momentum \(p\)\ and \(\bar{p}\) has the Feynman rule:~~\(
i e\,Q_X\Big(\, g^{\mu\nu}(p-\bar p)^{\alpha}
 - g^{\alpha\mu}(q+p)^{\nu}
 + g^{\alpha\nu}(q+\bar p)^{\mu} \Big) \,\). And the coupling of the vertex of \(\gamma_{\mu}\) with incoming \(X_{\alpha}^*\) and \(G_X\) has the form \(ie\,Q_XM_Xg^{\mu\alpha}\). 
 
As for the coefficient \(\rho_m\) in Eq.~\eqref{eqn:sigma_L}, we decompose it into two physically distinct pieces, depending on whether a light quark \((\rho_m^d)\) or a heavy quark \((\rho_m^D)\) runs in the loop.
When light quarks, \(d_m^c=(d^c,s^c,b^c)\), are flowing internally, Eq.~\eqref{eqn:L-L-X} gives us the form of \(\rho_m\):
 \begin{equation}
   \rho_m^d = K^{\dagger}_{d^c_me^c}K_{d^c_m\mu^c}
   \label{eqn:rhoL}
 \end{equation}
where the definition of \(K_{d^ce^c}\) can be read from Eq.~\eqref{eqn:Kdcec} which is not necessarily unitary~(see the discussion in section~\ref{subsec:Meson decay at tree level}). A sizable deviation of \(K_{d^ce^c}\) from unitarity will not guarantee GIM cancellation in Eq.~\eqref{eqn:sigma_L} anymore, and it can have a large impact on cLFV, since the loop kernels in Eqs.~\eqref{eqn:yPsi}-\eqref{eqn:yX} will pick up contributions which are not suppressed by the small quark masses. This type of enhanced cLFV rates at observable levels has been studied for low-scale right-handed neutrinos~\cite{Antusch:2006vwa,Abada:2007ux}. As discussed earlier in Sec.~\ref{subsec:Meson decay at tree level}, while optimizing the for meson decays, we already took this situation into account, and made sure that the optimized \(K\)~(see Eq.~\eqref{eqn:optimized_K_mu_e}) simultaneously takes care of the meson decays and the cLFV processes while lowering the scale of the PS gauge boson. And the matrix \(x\) carries the information of the overall suppression coming from the softly broken \(Z_2\) symmetry, which further limits direct couplings of $X$ to purely SM states.

When heavy vector-like quarks \(D_m^c=(D_1^c, D_2^c, D_3^c)\) run in the loops, we get the form of \(\rho_m\) from Eq.~\eqref{eqn:H-L-X}:
  \begin{equation}
   \rho_m^D =  K^{\dagger}_{D^c_me^c}K_{D^c_m\mu^c},
   \label{eqn:rho_H}
 \end{equation}
where \(K^{\dagger}_{D^ce^c}\) can be read from Eq.~\eqref{eqn:kDcec}. When evaluating $\rho_m^D$, we use the benchmark-2 structure as in Eq.~\eqref{eqn:benchmark2} for \(K_{D^ce^c}\) in the scans and we present the results in Fig.~\ref{fig:mu_to_e_gamma}. Now, to have a quantitative estimation,  we scan over $(M_{D_1},M_{D_2})$ while fixing $M_{D_3}=6.7~\text{TeV}$. At each point, we compute the $\mu \rightarrow e \gamma$ branching ratio using the expression
\begin{equation}
\mathrm{BR}(\mu\to e\gamma)\;=\;\frac{m_\mu^3}{16\pi\,\Gamma_\mu}\,\big|\sigma_L\big|^2,
\end{equation}
with \(\Gamma_\mu=G_F^2 m_\mu^5/(192\pi^3)\) and determine the minimum $M_X$ required at that point by solving the constraint $\mathrm{BR}\le 4.2\times 10^{-13}$ (see Table~\ref{tab:FLAVOR}). Fig.~\ref{fig:mu_to_e_gamma} shows the resulting contours of the required $M_X$ throughout the $(M_{D_1},M_{D_2})$ plane. 

One comment about the diagonal valley that we see in this plot: it is not any special requirement of lowering the PS scale; rather, it is an artifact of the choice for CKM-like matrices. In detail, with the benchmark choice for our case, the dominant contribution comes when heavy quarks \(D^c_m\) run in the loop. The CKM–weighted dipole sum from Eq.~\eqref{eqn:sigma_L} in that case reads as,
\[
\sigma_L\approx \sum_{m=1}^{3}\rho_m^D F_m,\qquad  
F_m\equiv Q_{D}\,\mathcal{I}_1(t_m)+Q_X\,\mathcal{I}_2(t_m)~.
\]
Unitarity requires $\sum_m\rho_m^D=0$. Along the degenerate line $M_{D_1}=M_{D_2}$ (so $F_1=F_2\equiv F$), we have
\[
\sigma_L=(\rho_1^D+\rho_2^D)F+\rho_3^D F_3 = \rho_3^D(F_3-F),
\]
leaving only the small weight $\rho_3^D\approx V_{D_3^ce^c}^\ast V_{D_3^c\mu^c}$ times a loop difference, that is the reason we see the valley.  Moving away from the degeneracy with $F_{1,2}=F\pm\delta F$, will lift $|\sigma_L|$,
\[
\sigma_L\simeq \rho_3^D(F_3-F) + (\rho_1^D-\rho_2^D)\,\delta F.
\]
Since we will be using the same benchmark throughout, this pattern will also appear in the subsequent sections.
\subsubsection{$\mu\to e e e$}
\label{subsubsec:mueee}
\begin{figure}[t!]
    \centering
    \begin{subfigure}[b]{0.32\textwidth}
        \centering
        \includegraphics[width=\textwidth]{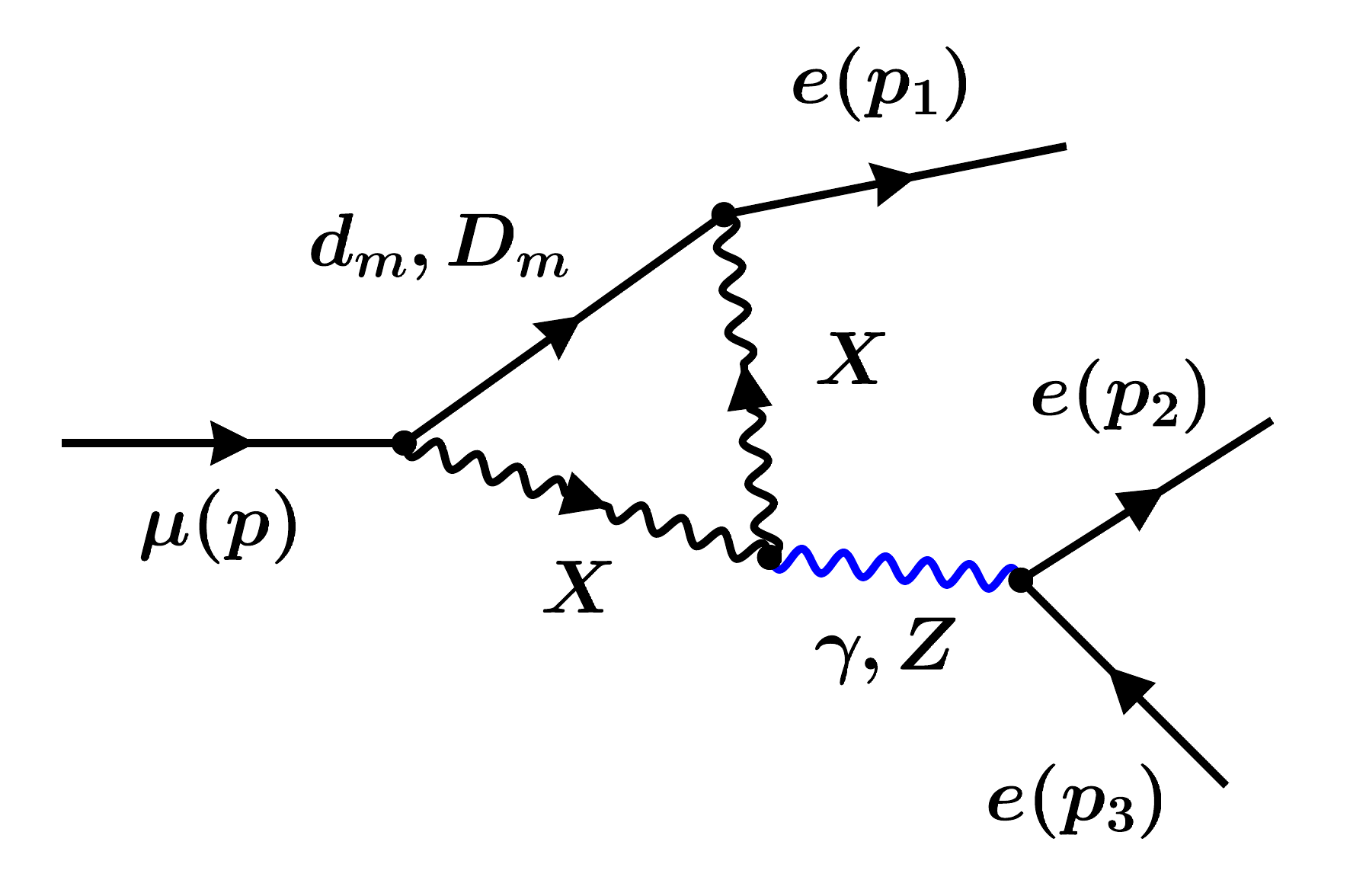}
        \caption{}
    \end{subfigure}
    \hfill
    \begin{subfigure}[b]{0.32\textwidth}
        \centering
        \includegraphics[width=\textwidth]{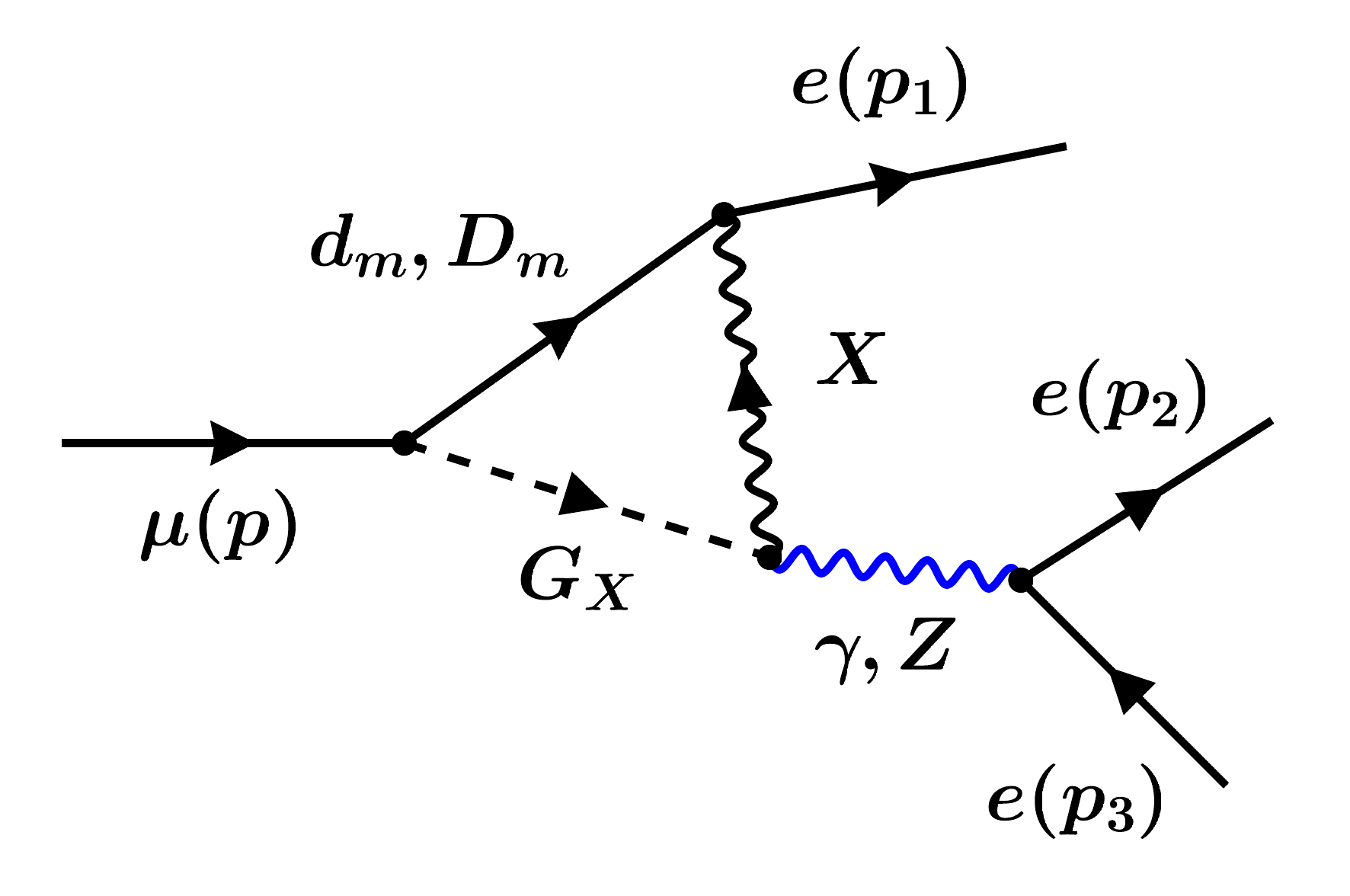}
        \caption{}
    \end{subfigure}
    \begin{subfigure}[b]{0.32\textwidth}
        \centering
        \includegraphics[width=\textwidth]{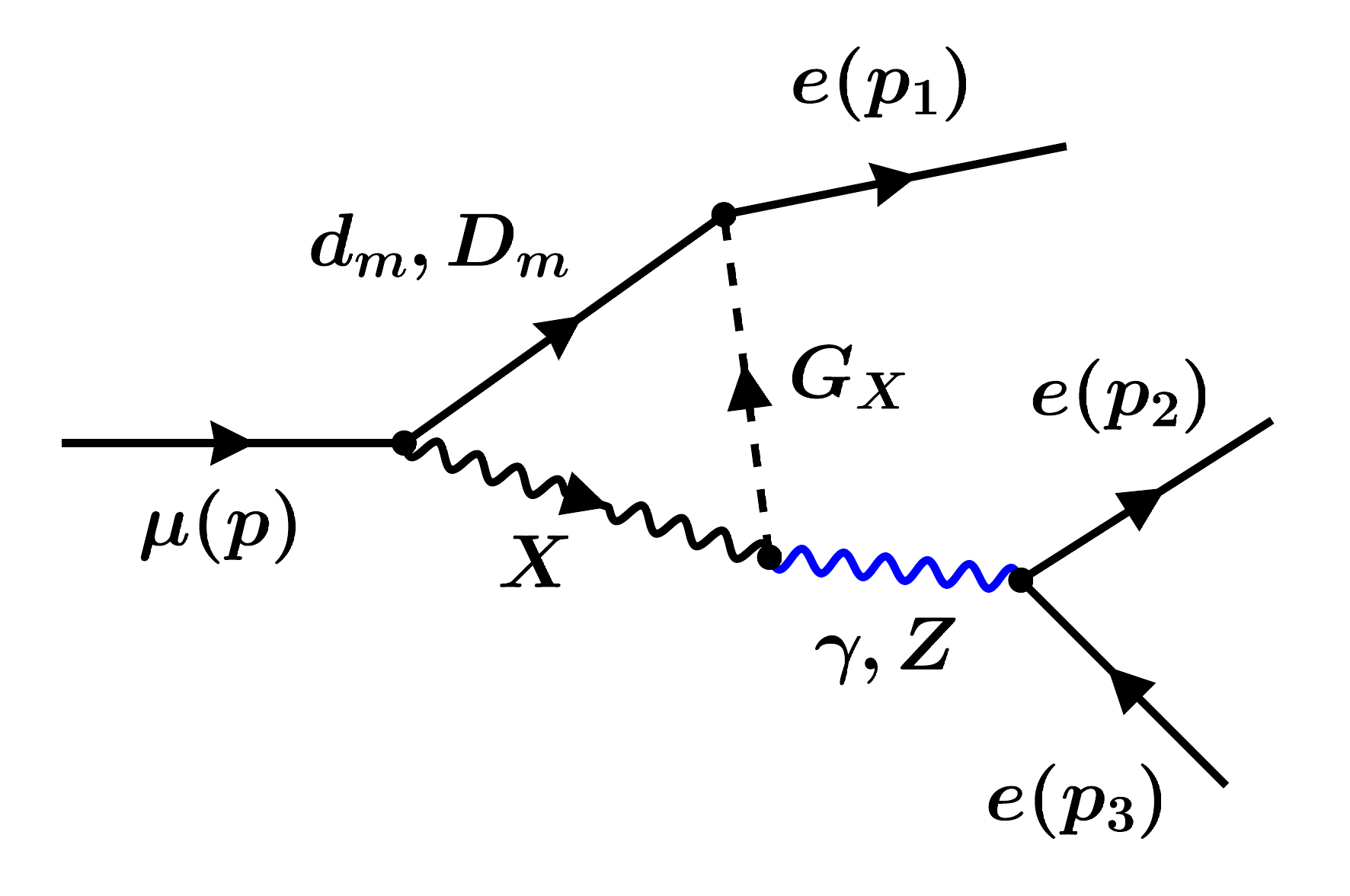}
        \caption{}
    \end{subfigure}

    \vspace{0.4cm} 

    \begin{subfigure}[b]{0.32\textwidth}
        \centering
        \includegraphics[width=\textwidth]{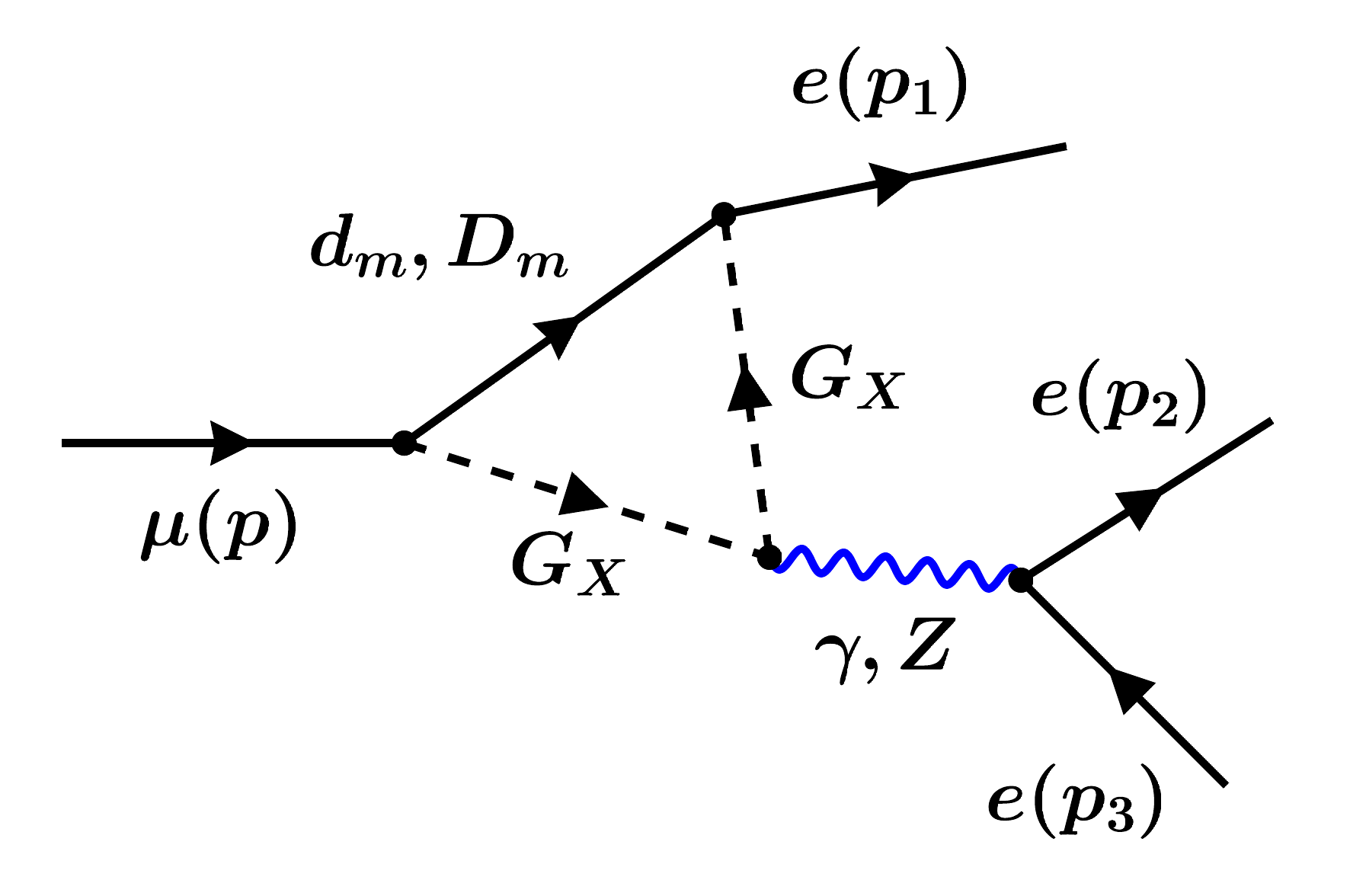}
        \caption{}
    \end{subfigure}
    \hfill
    \begin{subfigure}[b]{0.32\textwidth}
        \centering
        \includegraphics[width=\textwidth]{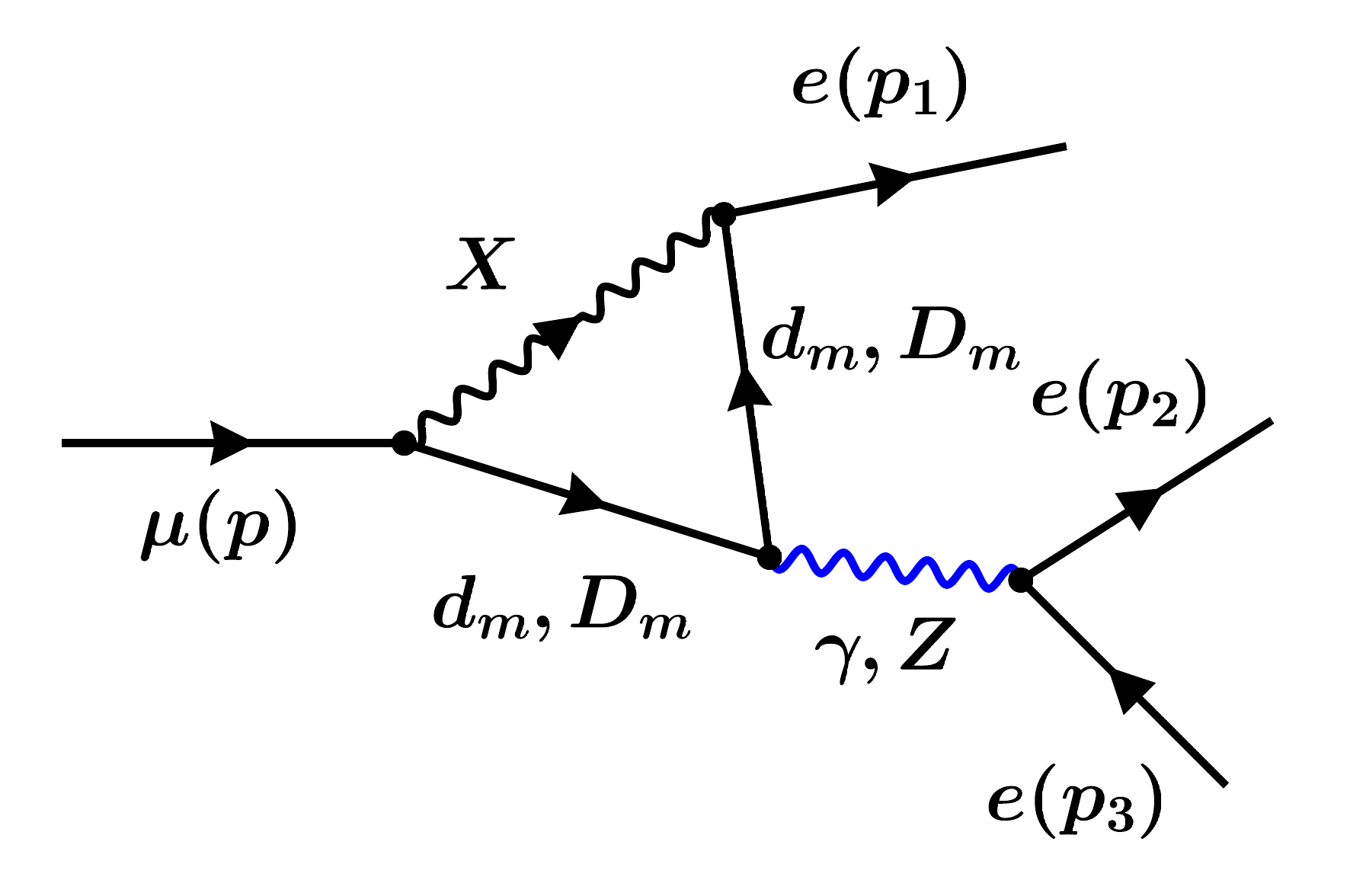}
        \caption{}
    \end{subfigure}
    \begin{subfigure}[b]{0.32\textwidth}
        \centering
        \includegraphics[width=\textwidth]{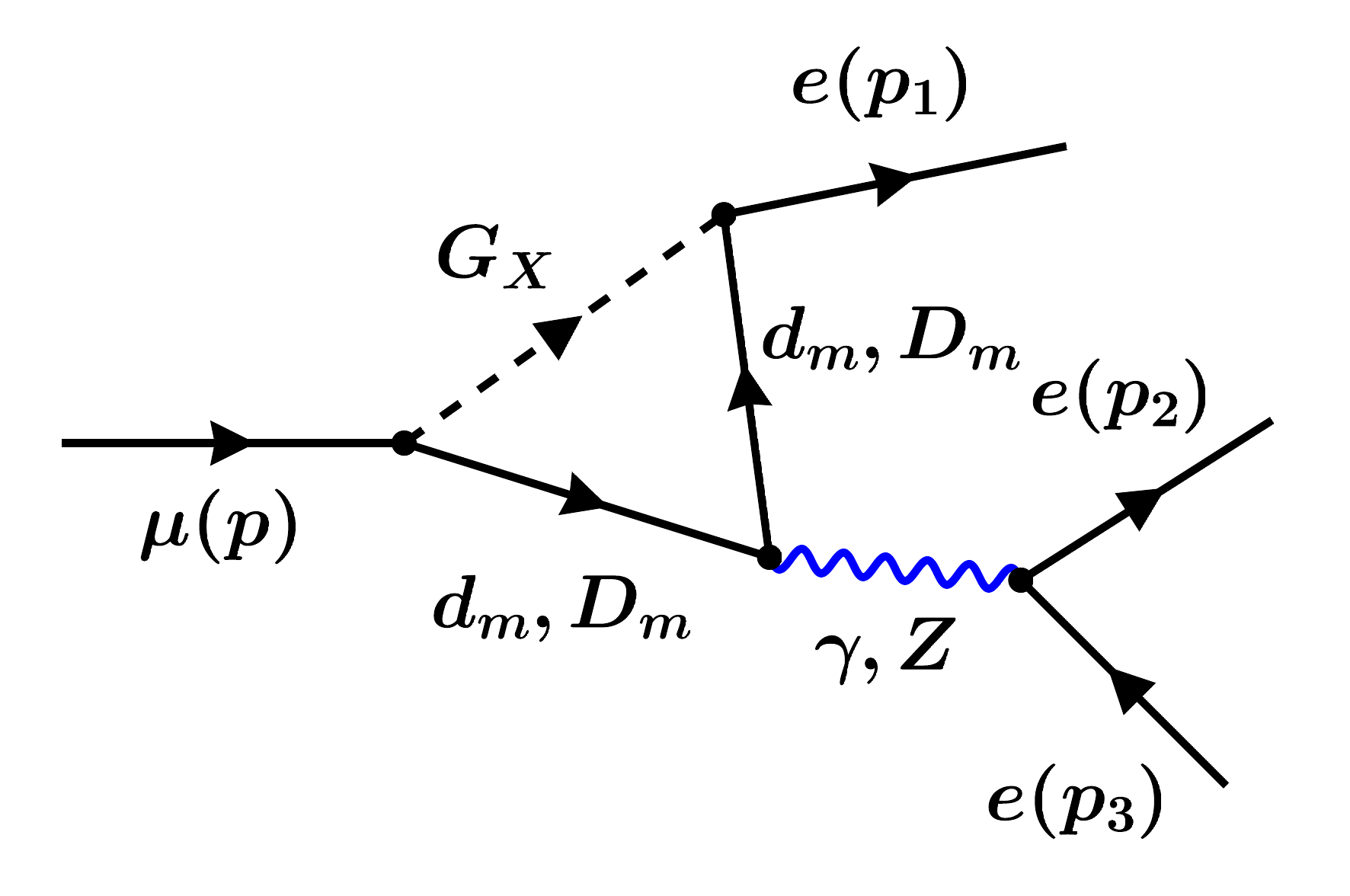}
        \caption{}
    \end{subfigure}

    \vspace{0.4cm}

    \begin{subfigure}[b]{0.2\textwidth}
        \centering
        \includegraphics[width=\textwidth]{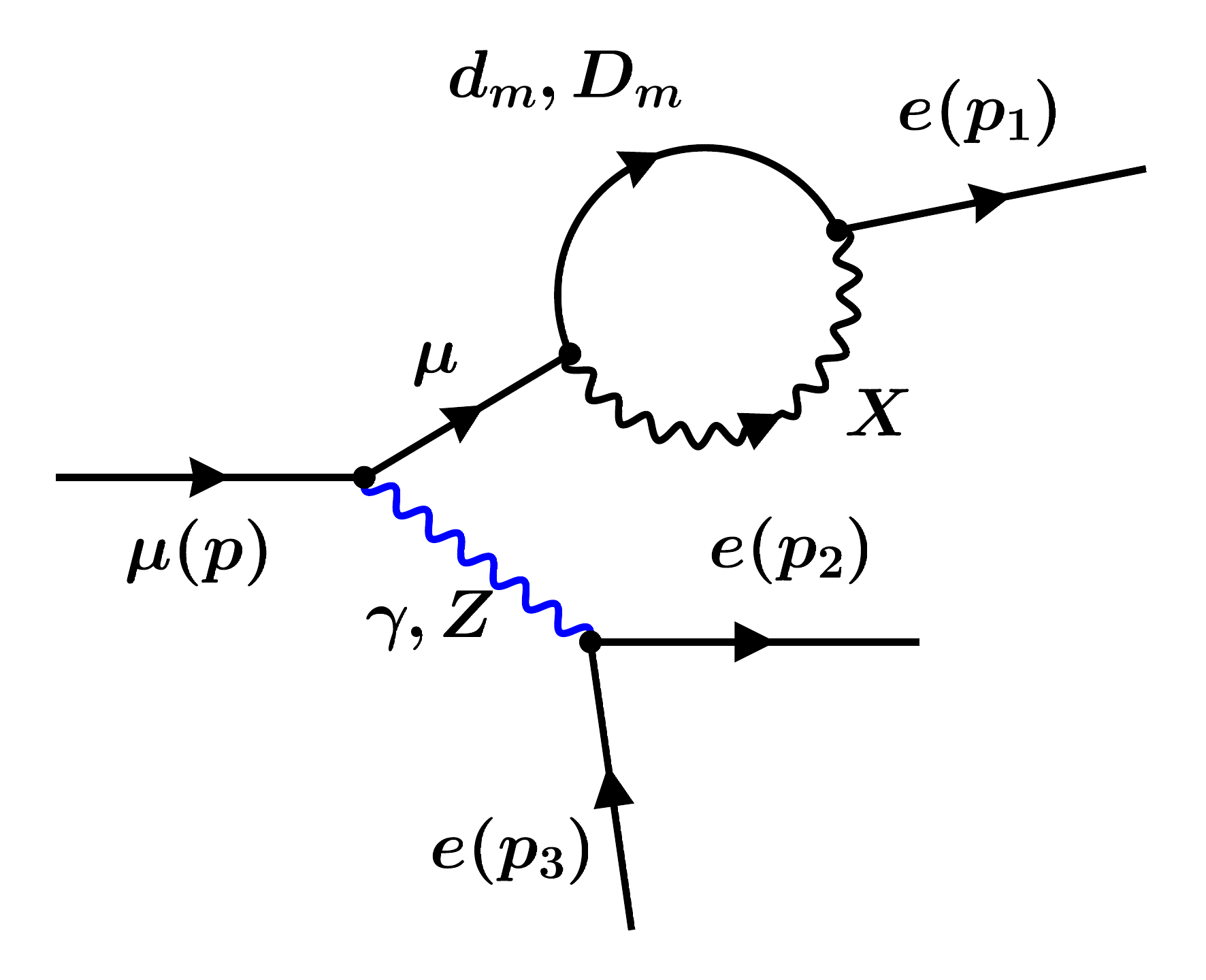}
        \caption{}
    \end{subfigure}
    \hfill
    \begin{subfigure}[b]{0.2\textwidth}
        \centering
        \includegraphics[width=\textwidth]{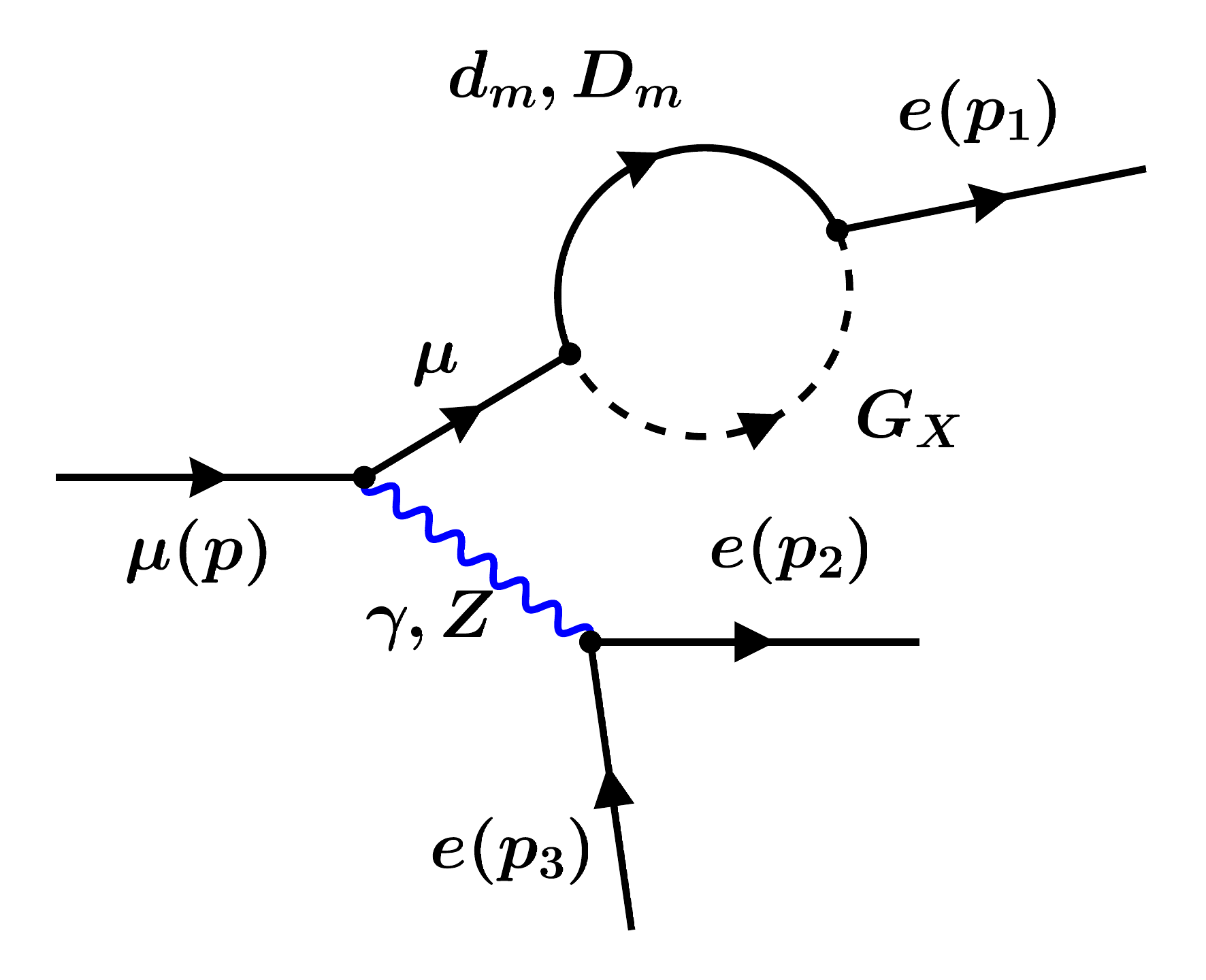}
        \caption{}
    \end{subfigure}
    \hfill
    \begin{subfigure}[b]{0.28\textwidth}
        \centering
        \includegraphics[width=\textwidth]{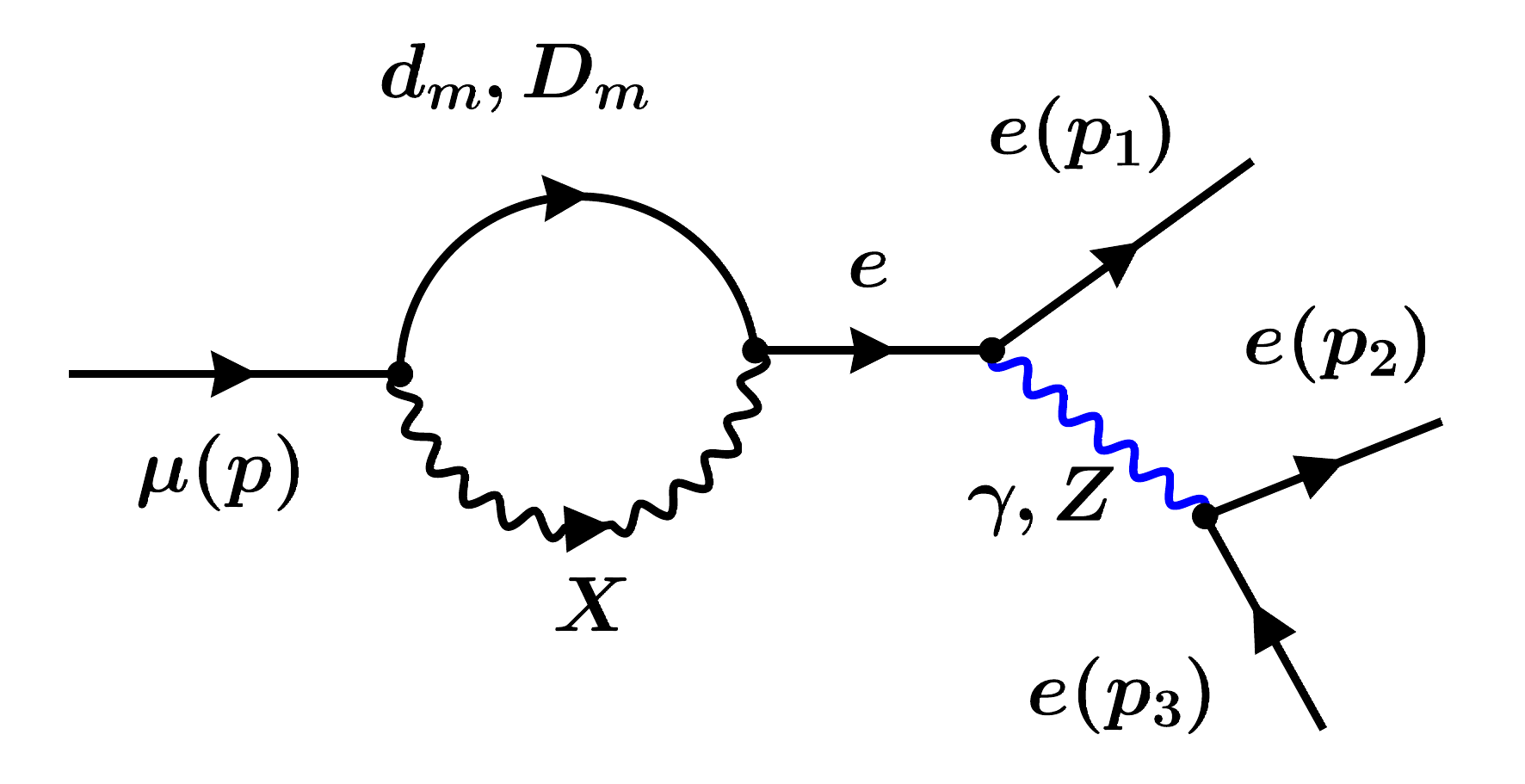}
        \caption{}
    \end{subfigure}
        \hfill
    \begin{subfigure}[b]{0.28\textwidth}
        \centering
        \includegraphics[width=\textwidth]{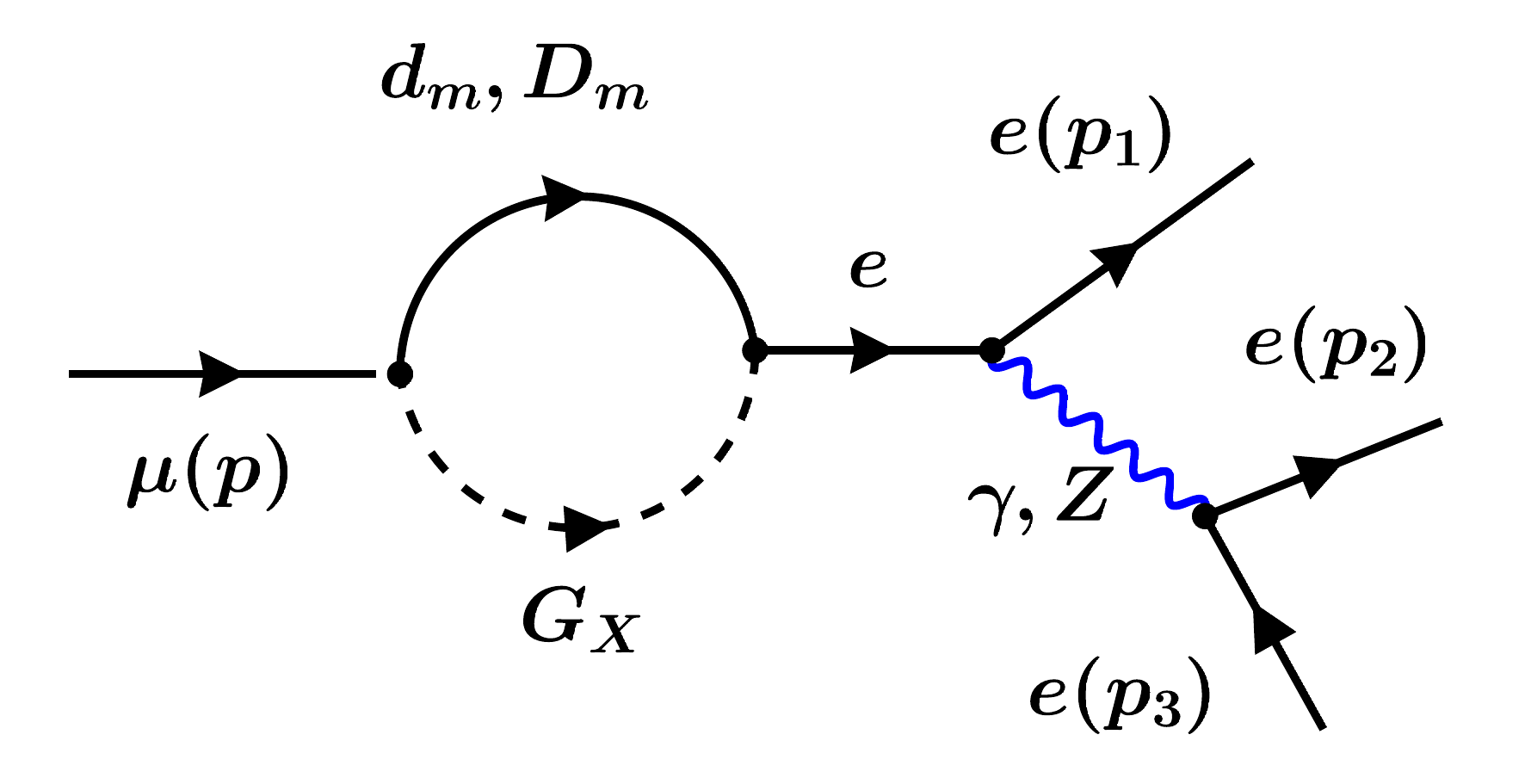}
        \caption{}
    \end{subfigure}

    \caption{Penguin topologies contributing to the \(\mu \to eee\) decay through monopole and dipole transitions. \(m\) indicates the flavor index of the internal fermion field. The chirality and the coupling for the external fermion lines can be understood from Eqs.~\eqref{eqn:L-L-X} and \eqref{eqn:H-L-X}. The set of diagrams related by $p_1\leftrightarrow p_2$ is implied.} 
    \label{fig:mu_eee_penguins}
\end{figure}
This flavor-violating decay arises at one-loop in our setup~(see Fig.~\ref{fig:mu_eee_penguins}--\ref{fig:mu_eee_boxes}). Unlike $\mu\to e\gamma$, this mode receives contributions from the off-shell photon. In addition, $Z$ penguins and box diagrams also contribute to this process, which involve leptoquark gauge bosons in the loop. As a result, even when the dipole form factor $\sigma_L$ is suppressed and $\mu\to e\gamma$ is effectively absent, these contributions can make $\mu\to eee$ competitive and more constraining.

We start with the effective Lagrangian (at the scale \(m_{\mu}\)) which contributes to the \(\mu \to eee\) process~\cite{Kuno:1999jp}:
\begin{align}
\mathcal{L}_{\mu\to eee}
&= -\frac{4G_F}{\sqrt{2}}\Big[
m_\mu A_R\,\overline{\mu_L}\sigma^{\mu\nu}e_R\,F_{\mu\nu}
+ m_\mu A_L\,\overline{\mu_R}\sigma^{\mu\nu}e_L\,F_{\mu\nu}\notag  \\
&\quad 
+\,h_1(\overline{\mu_R}e_L)(\overline{e_R}e_L)
+ h_2(\overline{\mu_L}e_R)(\overline{e_L}e_R)
+ h_3(\overline{\mu_R}\gamma^\mu e_R)(\overline{e_R}\gamma_\mu e_R)
+ h_4(\overline{\mu_L}\gamma^\mu e_L)(\overline{e_L}\gamma_\mu e_L)\notag \\
&\quad
+\,h_5(\overline{\mu_R}\gamma^\mu e_R)(\overline{e_L}\gamma_\mu e_L)
+ h_6(\overline{\mu_L}\gamma^\mu e_L)(\overline{e_R}\gamma_\mu e_R)
\Big] + \text{h.c.}
\label{eq:L_mueee}
\end{align}
where the Fierz rearrangement has been used in the four-fermion operators. Here, \(G_F\) is the Fermi constant, and \(A_{L,R}\) and $h_i$ are dimensionless coupling constants for the corresponding operators with the assumption that the momentum dependencies of the form factors are negligible. Several comments on the Lagrangian are warranted:

\begin{figure}[t!]
    \centering
    \begin{subfigure}[b]{0.32\textwidth}
        \centering
        \includegraphics[width=\textwidth]{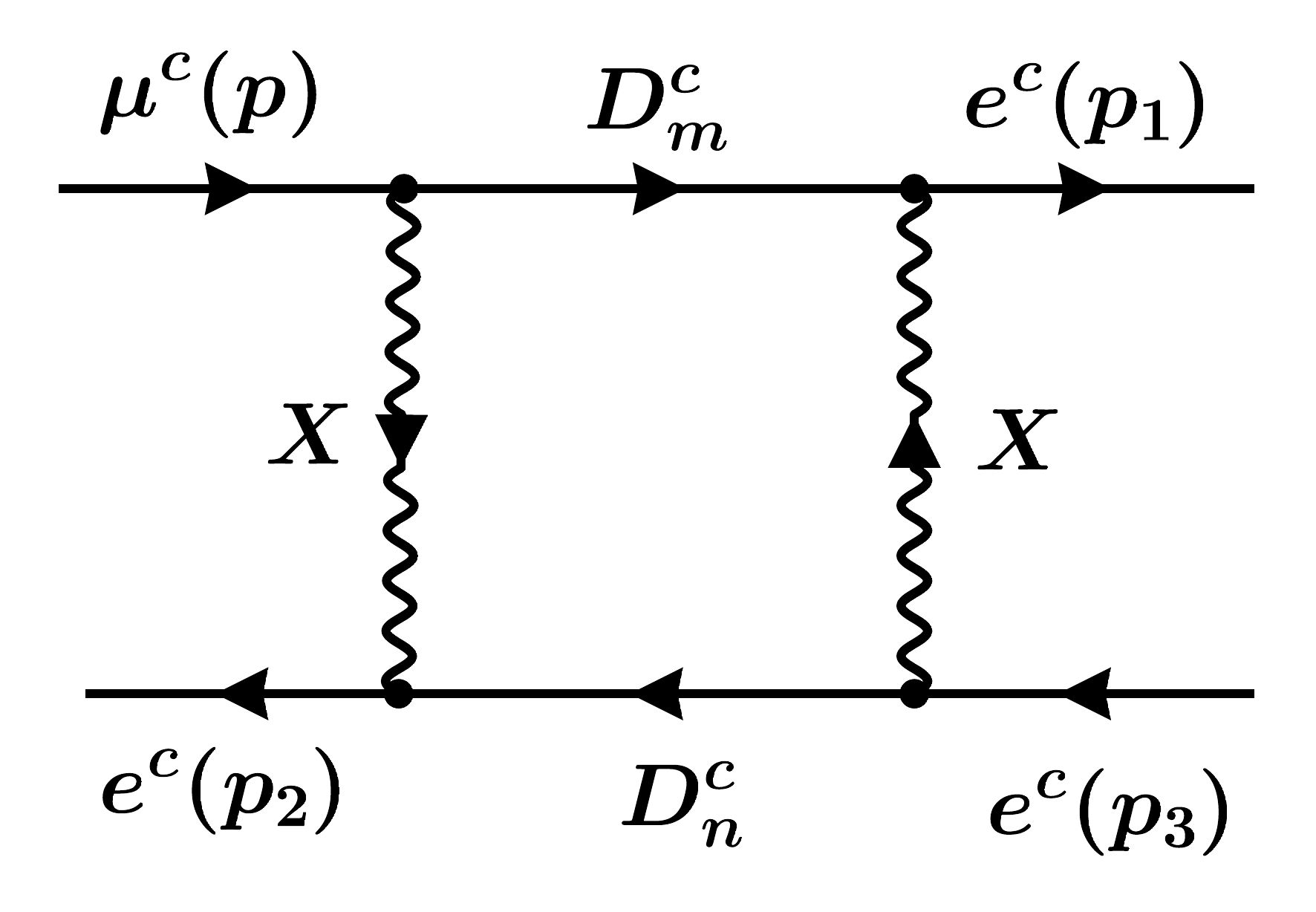}
        \caption{}
    \end{subfigure}
    \hfill
    \begin{subfigure}[b]{0.32\textwidth}
        \centering
        \includegraphics[width=\textwidth]{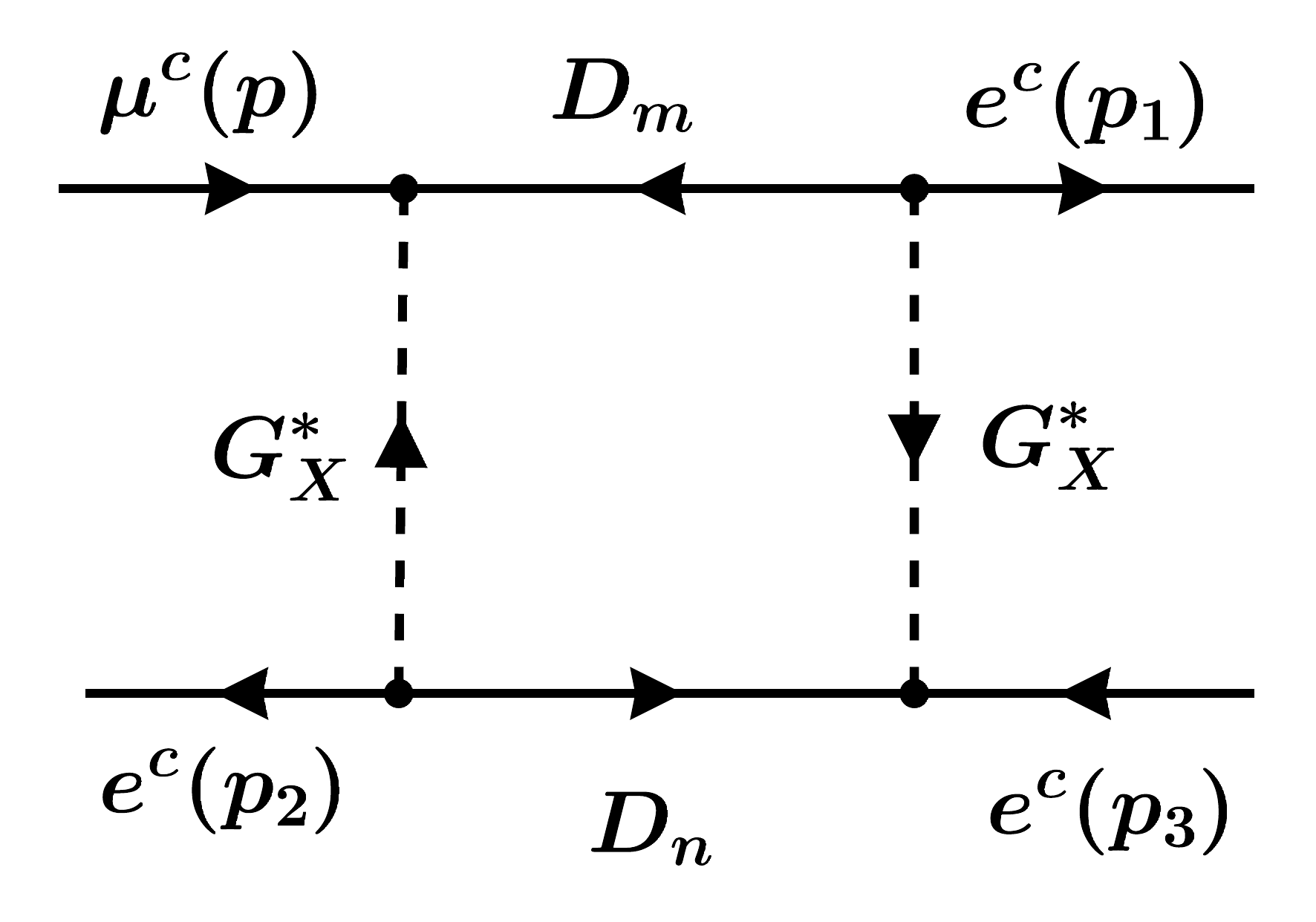}
        \caption{}
    \end{subfigure}
    \begin{subfigure}[b]{0.32\textwidth}
        \centering
        \includegraphics[width=\textwidth]{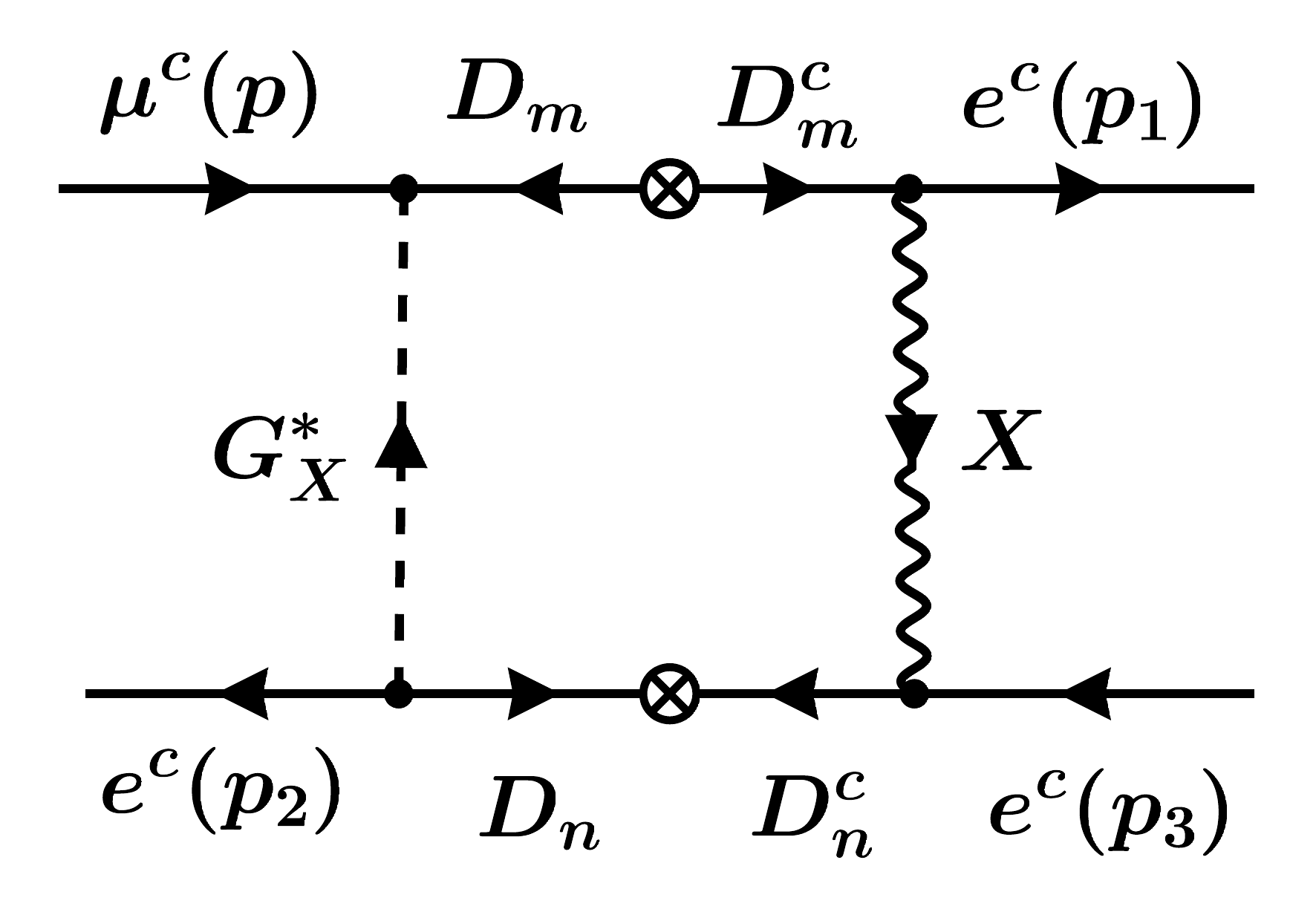}
        \caption{}
    \end{subfigure}

    \vspace{0.4cm} 

    \begin{subfigure}[b]{0.4\textwidth}
        \centering
        \includegraphics[width=\textwidth]{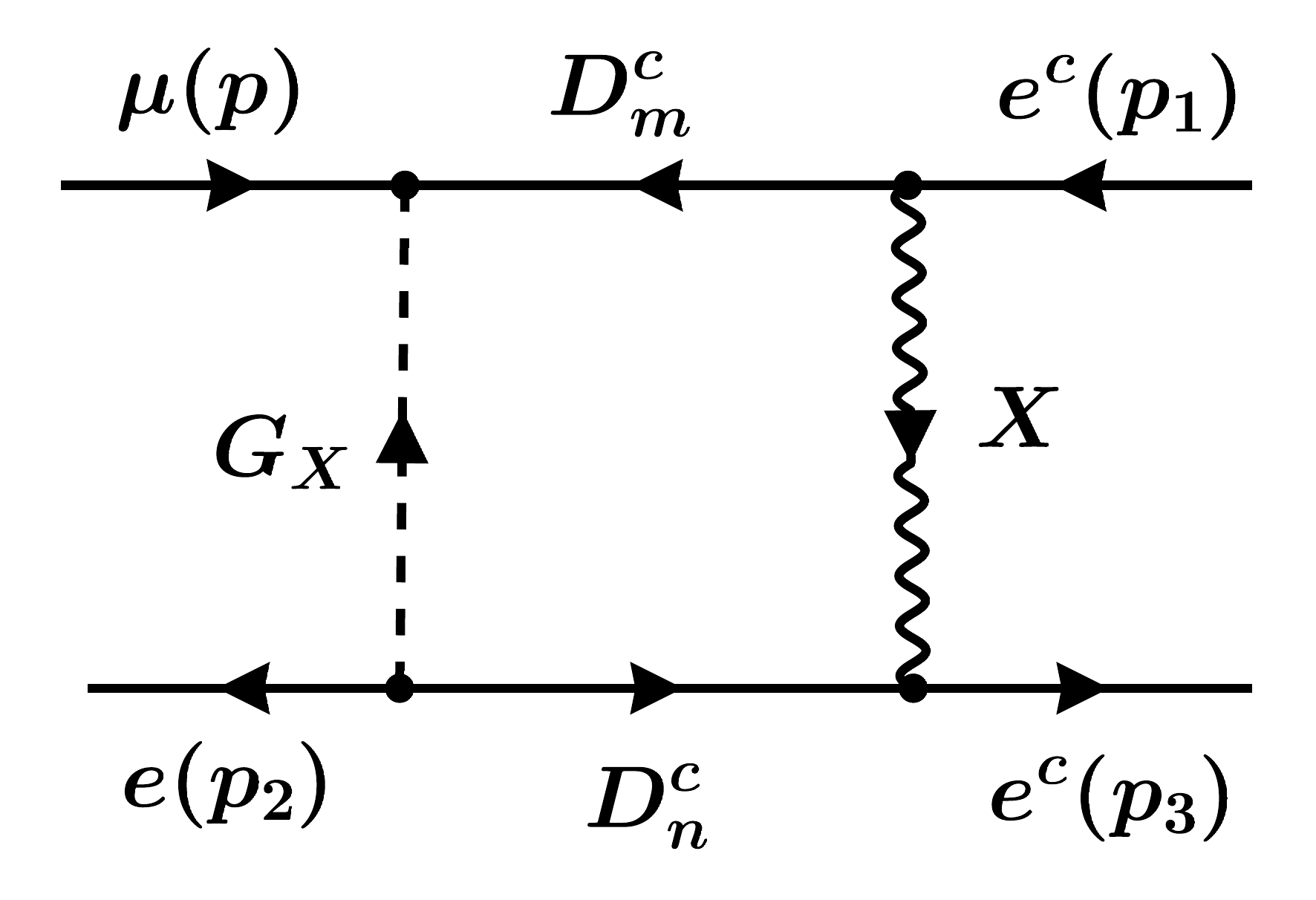}
        \caption{}
    \end{subfigure}
    \hfill
    \begin{subfigure}[b]{0.4\textwidth}
        \centering
        \includegraphics[width=\textwidth]{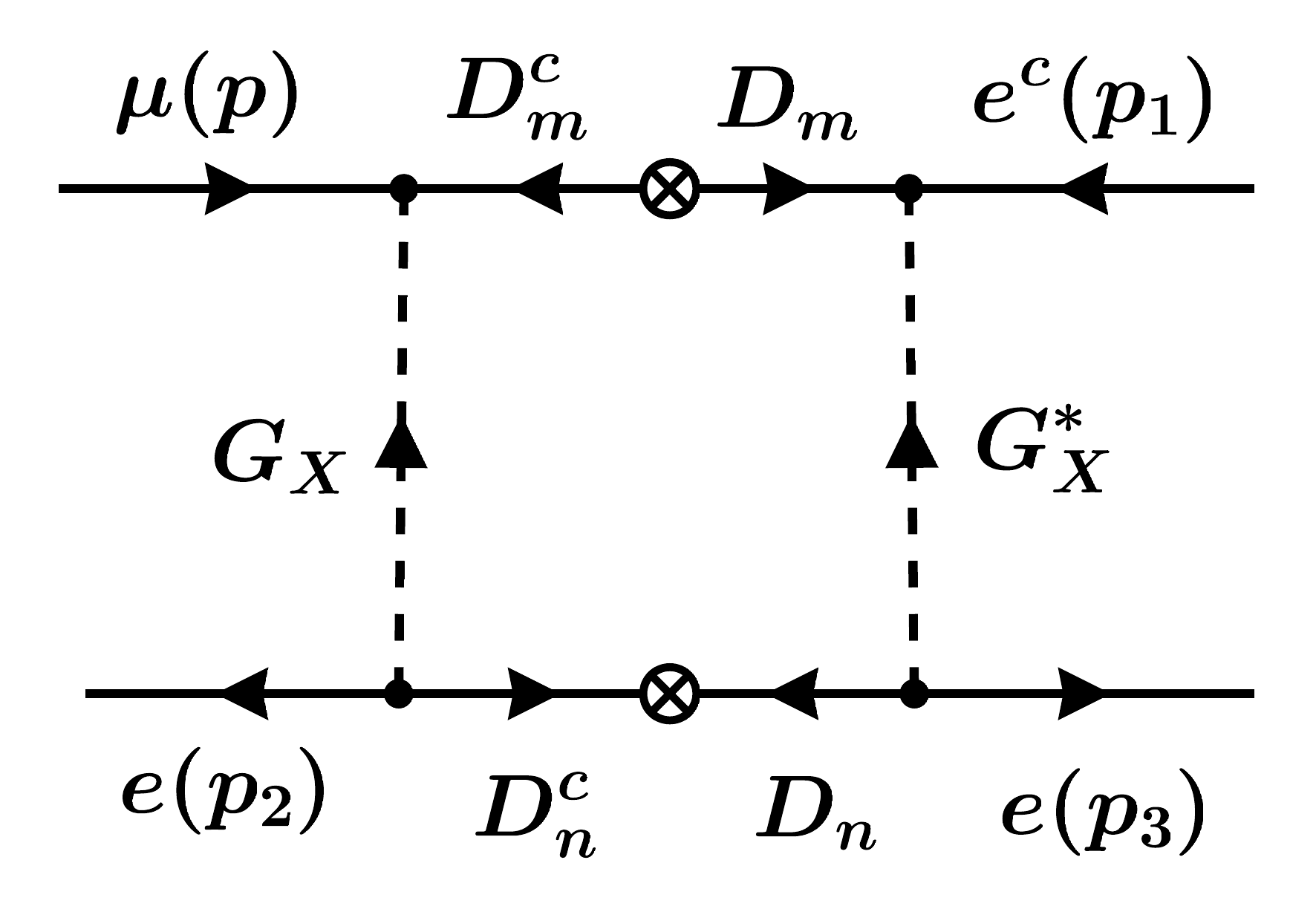}
        \caption{}
    \end{subfigure}
    
    \vspace{.4cm}
        \begin{subfigure}[b]{0.4\textwidth}
        \centering
        \includegraphics[width=\textwidth]{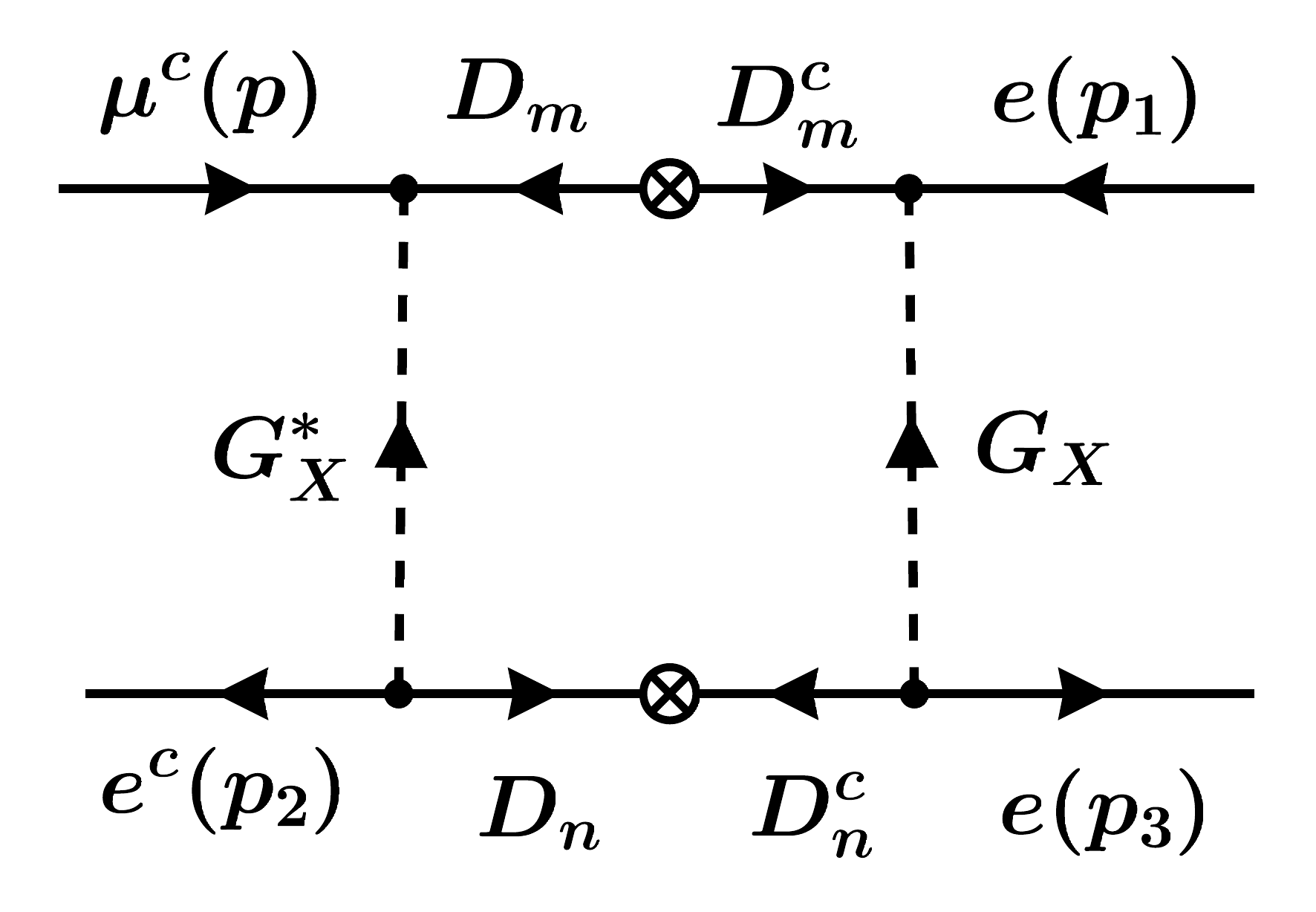}
        \caption{}
        \end{subfigure}
        \hfill
    \begin{subfigure}[b]{0.4\textwidth}
        \centering
        \includegraphics[width=\textwidth]{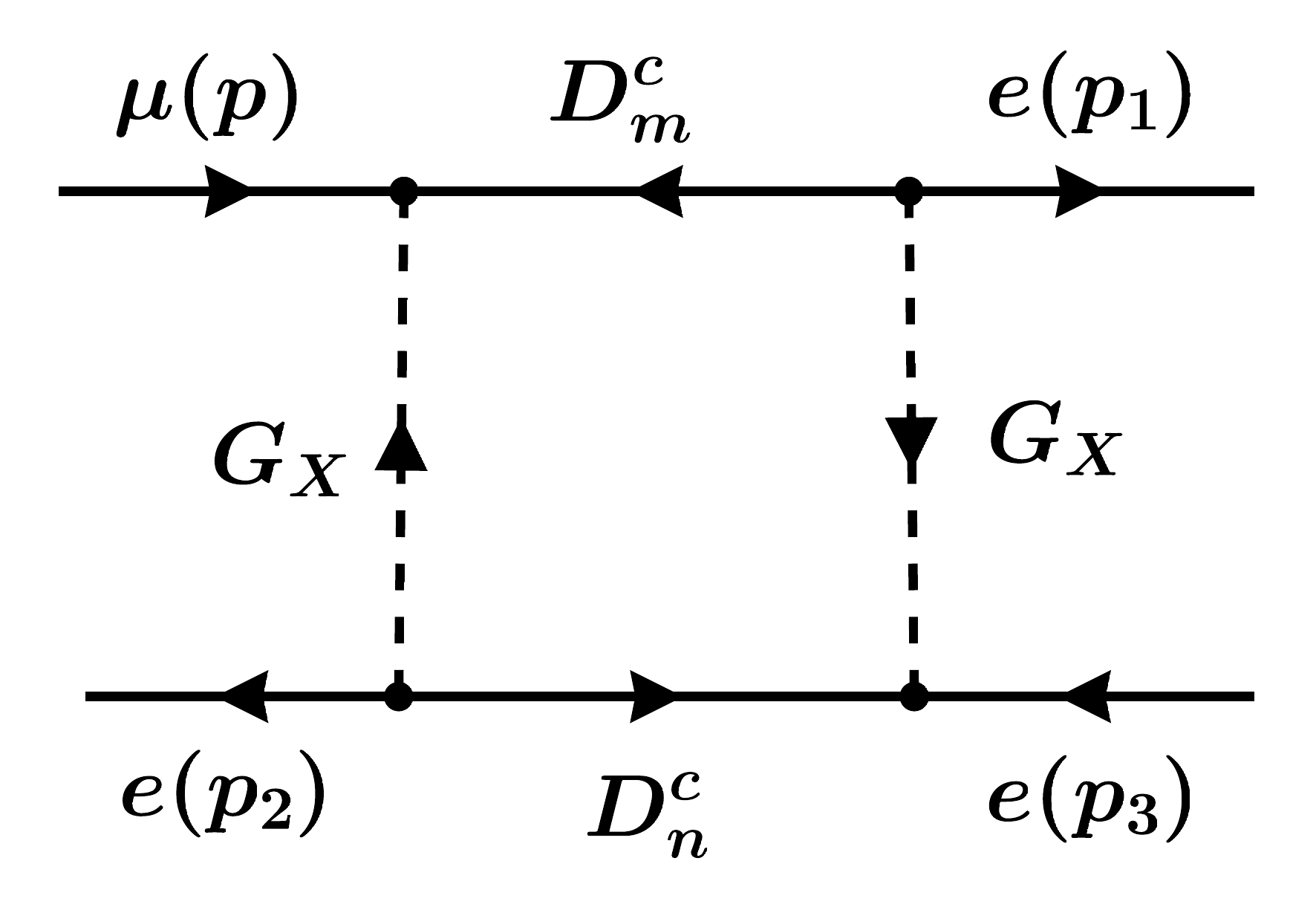}
        \caption{}
    \end{subfigure}
    \caption{Box topologies contributing to the \(\mu \to eee\) decay. \(m,\,n\) denotes the flavor indices of the internal heavy field. The set of diagrams related by $p_1\leftrightarrow p_2$ is implied as well.} 
    \label{fig:mu_eee_boxes}
\end{figure}

The first comment is about the photonic contribution to this decay process. There are two types of amplitudes for photonic transition \(\mu \to e \gamma^*\), the
monopole and dipole transitions. In the Lagrangian, we have converted the monopole contribution to the vector-vector interaction. The smallness of \(A_R\) in our model has already been discussed previously in Sec.~\ref{subsubsec:mueg}, and the matching between Eqs.~\eqref{eqn:Tmuegamma} and \eqref{eq:L_mueee} gives us 
\(A_L=\sigma_L/(4\sqrt{2}\,G_F\,m_{\mu})\).
The extra factor of \(1/2\) comes from the basis change between \(\epsilon_{\mu}q_{\nu}\) and \(F_{\mu \nu}=i(q_{\mu}\epsilon_{\nu}-q_{\nu}\epsilon_{\mu})\).

The second comment is on the vector-vector interactions (terms with \(h_3,\, h_4,\,h_5,\,\text{and } h_6\) coefficients). We are considering lepton flavor violation mediated by the leptoquark gauge boson, \(X_{\mu}\). As we know, in our setup, \(X_{\mu}\) would couple right-handed leptons to the down isospin dominantly. So, only \(h_3\) and \(h_5\) will contribute through penguins and boxes. The other two coefficients, \(h_4\) and \(h_6\), will be induced only through self-energy corrections or Goldstone contributions in boxes which are strongly suppressed. For example, in the topology of Fig.~\ref{fig:mu_eee_boxes}\,(g), \(h_4\) is induced with a suppression of \(m_{\mu}m_e^3/M_X^4\) with external momentum set to zero, which gives essentially zero contribution.

The third comment is about the scalar four-fermion operators, \(h_1\) and \(h_2\). From Fig.~\ref{fig:mu_eee_boxes} (d, e, f), we see that they also suffer suppression from light lepton mass \(\sim m_{\mu}m_e/M_X^2\). As a result, they are also subleading in the parameter space of our model. 

Since we presented the detailed expression for the dipole form factor $\sigma_L$ in the previous section, we now concentrate on the neutral–current Wilson coefficients $h_3$ and $h_5$. These coefficients capture the short–distance neutral–current effects arising from $\gamma/Z$ penguins and box diagrams mediated by the vector leptoquark \( X_\mu \) (see Figs.~\ref{fig:mu_eee_penguins}--\ref{fig:mu_eee_boxes}). They are finite, gauge–independent, and satisfy the Slavnov–Taylor identities when the self–energy subtractions are included, following the procedure of Refs.~\cite{Bishara:2021buy,Brod:2019bro}. These coefficients have the general form:
\begin{align}
h_3 &=
\frac{3}{16\pi^2}
\left(\frac{g_4}{g_L}\right)^{2}
\left(\frac{M_W}{M_X}\right)^{2}
\left[
e^2\left(\mathcal{F}^{(\gamma Z)}_{1} + \mathcal{F}^{(\gamma Z)}_{2}\right)
+ g_4^2\mathcal{F}^{(B)}
\right], \label{eqn:g3_general}\\[3pt]
h_5 &=
\frac{3\,e^2}{16\pi^2}
\left(\frac{g_4}{g_L}\right)^{2}\!
\left(\frac{M_W}{M_X}\right)^{2}
\left(
\mathcal{F}^{(\gamma Z)}_{1} + \mathcal{F}^{(\gamma Z)}_{2}
\right).
\label{eqn:g5_general}
\end{align}
The functions \(\mathcal{F}_{1,2}^{(\gamma Z)}\) and \(\mathcal{F}^{(B)}\) are the gauge-
independent combinations of the photon penguin with the $Z$ penguin and the photon penguin with the box diagrams. For \(\mathcal{F}_{1,2}^{(\gamma Z)}\), we implement the same method used in \(\mu \to e \gamma\) of separating the light $d_m=(d,s,b)$ and heavy $D_m=(D_1,D_2,D_3)$ internal fermions flowing in the loop. We define
\begin{align}
\mathcal{F}^{(\gamma Z)}_{1}
&= \sum_{m=1}^{3} \rho_m^{d} 
\big[ Q_F F_{\gamma Z}^{F}(t_m) + Q_V F_{\gamma Z}^{V}(t_m) \big],
\label{eqn:FgammaZ-1}\\
\mathcal{F}^{(\gamma Z)}_{2}
&= \sum_{m=1}^{3} \rho_m^{D} 
\big[ Q_F F_{\gamma Z}^{F}(T_m) + Q_V F_{\gamma Z}^{V}(T_m) \big],
\label{eqn:FgammaZ-2}
\end{align}
with \( Q_F=-\tfrac{1}{3} \), \( Q_V=+\tfrac{2}{3} \), \( t_i=m_{d_i}^2/m_X^2 \), and \( T_i=m_{D_i}^2/m_X^2 \), and the flavor structure \(\rho_m^{d,D}\) follows the same definition as in Eqs.~\eqref{eqn:rhoL}--\eqref{eqn:rho_H}. The penguin form factors are given as:
\begin{align}
F_{\gamma Z}^{F}(t) &= 
\frac{14t^2 - 21t + 1}{12(t-1)^3}
+ \frac{-9t^2 + 16t - 4}{6(t-1)^4}\ln t, 
\label{eqn:FgammaZ-F}\\
F_{\gamma Z}^{V}(t) &= 
\frac{6t^4 - 18t^3 - 32t^2 + 87t - 37}{12(t-1)^3}
+ \frac{t(8t^3 - 2t^2 - 15t + 6)}{6(t-1)^4}\ln t.
\label{eqn:FgammaZ-V}
 \end{align}
We see that the function $F_{\gamma Z}(t)$ develops an infrared log-singularity in the limit $t = m_f^2/m_X^2 \to 0$, corresponding to a light–fermion loop in the effective theory. In the Standard Model, this logarithm reproduces the leading term associated with the operator mixing $Q_{7\gamma}\to Q_9$~\cite{Buchalla:1995vs}.  In our framework, the same light fermion contribution cancels against the Standard Model part, and we end up having a finite, gauge–independent short-distance contribution in the Wilson coefficients $h_3$. 

As for the box diagram form factor, we can express it as follows:
\begin{equation}
\mathcal{F}^{(B)} =
\sum_{m,n=1}^{3} 
\rho_m^{D} \rho_n^{D} 
\big[ f_{V}^{R,BZ}(T_m,T_n) + f_{V}^{R,B'Z}(T_m,T_n) \big]
\end{equation}
where the symmetric two–mass kernels are given by
\begin{align}
f_{V}^{R,BZ}(x,y) &=
\frac{x y}{4}
\!\left[
\frac{3x(xy(xy+x-6)+4)-(x-1)(xy(xy-2)+4)}{(x-1)^2(xy-1)^2}\ln x
\right. \nonumber\\[-1pt]
&\qquad \left.
+ \frac{3y(-4xy+x+y+2)}{(y-1)^2(xy-1)^2}\ln y+ \frac{12-3y}{(x-1)(y-1)} + \frac{2xy-5}{xy-1}
\right],
\label{eqn:FVRBZ}
\\[2pt]
f_{V}^{R,B'Z}(x,y) &=
\frac{x y}{4}
\!\left[
\frac{3x^2 y(xy+x-2)-(x-1)(xy(xy-2)+4)}{(x-1)^2(xy-1)^2}\ln x
+ \frac{3y(x+y-2)}{(y-1)^2(xy-1)^2}\ln y
\right. \nonumber\\[-1pt]
&\qquad \left.
-\frac{3y^2}{(x-1)(y-1)} + \frac{2xy-5}{xy-1}
\right],
\label{eqn:FVRBpZ}
\end{align}
which are UV–finite and symmetric under $x\leftrightarrow y$ exchange.
\begin{figure}[!t]
    \centering
    \includegraphics[width=0.6\textwidth]{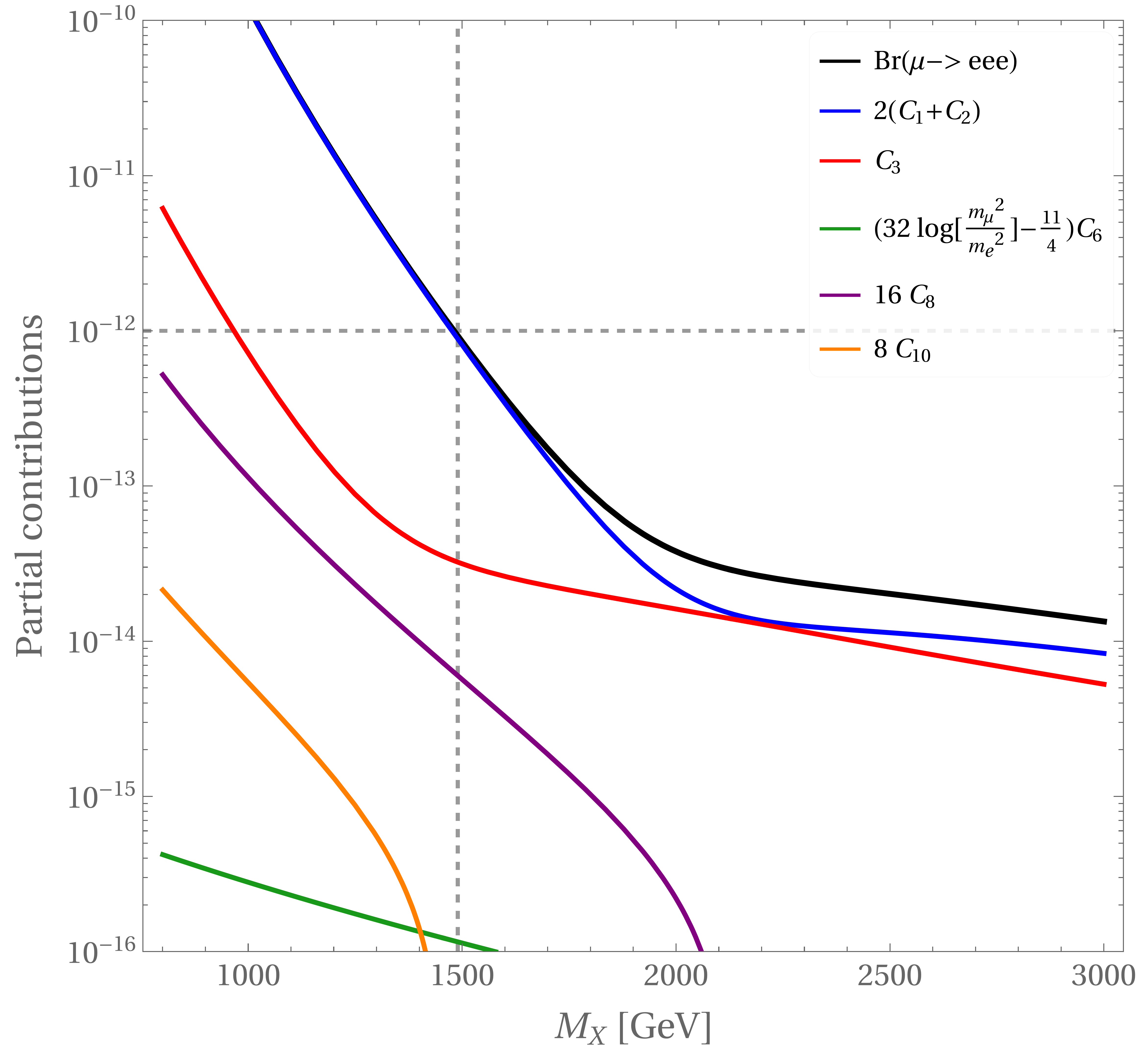}
\caption{Decomposition of $\mathrm{Br}(\mu\to e e e)$ as a function of $M_X$ at fixed mixings and
heavy down–quark masses $M_D=(7640,7600,6700)\,\mathrm{GeV}$. The black curve represents the total rate, while the colored curves show the relevant individual terms in Eq.~\eqref{eq:mutoeee_terms}. When the dipole piece $C_6$ is suppressed, the contact contributions $C_{1,2,3}$ can dominate.} \label{fig:partial_contributions_mutoeee}
\end{figure}

Now that we have finite, gauge-independent contributions to \(\mu \to eee\) in hand, we can write down the
kinematically integrated branching ratio, following Ref.~\cite{Kuno:1999jp}:
\begin{equation}
\mathrm{Br}(\mu \to eee)=2(C_1+C_2)+(C_3+C_4)
+\,32\!\left[\ln\!\left(\frac{m_\mu^2}{m_e^2}\right)-\frac{11}{4}\right](C_5 + C_6)
+16(C_7 + C_8) + 8(C_9 + C_{10})\,.
\label{eq:mutoeee_terms}
\end{equation}
Here the coefficients are given by
\begin{equation}
\begin{aligned}
C_1 &= \tfrac{|h_1|^2}{16} + |h_3|^2, \quad
C_2 = \tfrac{|h_2|^2}{16} + |h_4|^2, \quad
C_3 = |h_5|^2, \quad
C_4 = |h_6|^2, \\[2pt]
C_5 &= |eA_R|^2, \quad
C_6 = |eA_L|^2, \quad
C_7 = \Re(eA_R h_4^\ast), \quad
C_8 = \Re(eA_L h_3^\ast), \\[2pt]
C_9 &= \Re(eA_R h_6^\ast), \quad
C_{10} = \Re(eA_L h_5^\ast).
\end{aligned}
\label{eq:C1toC10}
\end{equation}
For completeness, we have included all the contributions here, but as we discussed, not all will have significant contributions to this decay in our setup. To have a clear picture, we plot in Fig.~\ref{fig:partial_contributions_mutoeee} the major contributors. As expected, \(A_L\,~h_3\), and \(h_5\) play important roles. The contribution coming from the pure dipole term \(C_6\) will always show up in \(\mu \to e \gamma\) process first, because here they will always have a kinematic suppression given by
\begin{equation}
\frac{B(\mu\!\to e e e)}{B(\mu\!\to e\gamma)}
\simeq
\frac{\alpha}{3\pi}
\left[
\ln\!\left(\frac{m_\mu^2}{m_e^2}\right)
- \frac{11}{4}
\right]
= 0.006,
\label{eq:Br_ratio_mueee_muegamma}
\end{equation}
which we can check with any benchmark case in our setup. For example, with \(M_D=(7.64,7.60,6.70)\) TeV, independent of \(M_X\), the ratio is \(\sim 0.006\).
\begin{figure}[!t]
    \centering
    \includegraphics[width=0.65\textwidth]{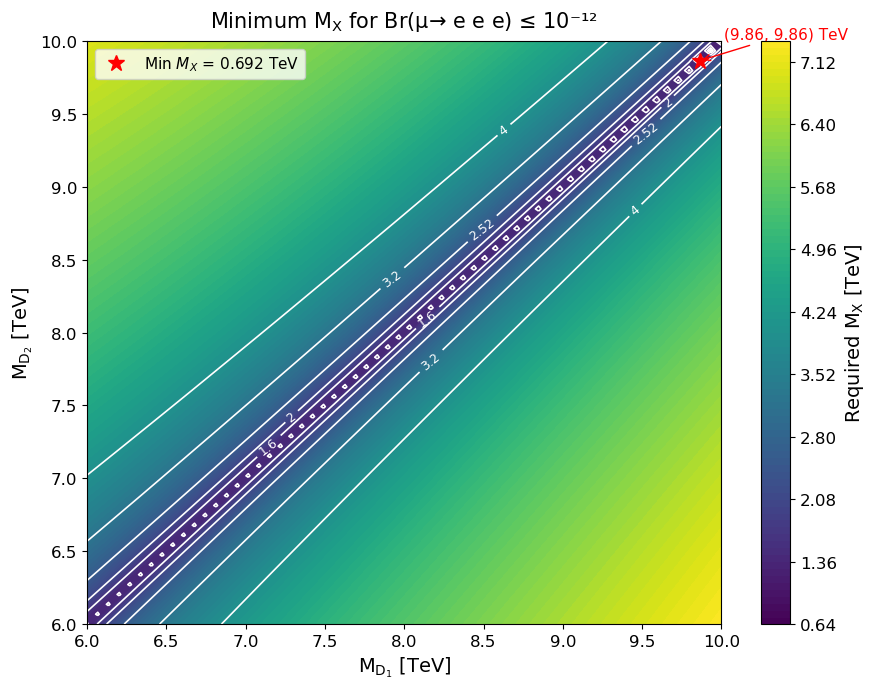}
\caption{Minimum $M_X$ required by the
$\mathrm{BR}(\mu\!\to\!eee)\le 1.0\times10^{-12}$ in the $(M_{D_1},M_{D_2})$ plane (with $M_{D_3}=6.7~\mathrm{TeV}$ fixed). The color bar indicates the required $M_X$ in TeV, with the white lines representing iso–$M_X$ contours. We show the global minimum on the scan with a red star.}
 \label{fig:mu_to_eee}
\end{figure}
For the scan in Fig.~\ref{fig:mu_to_eee}, we use the same benchmark scenario as in \(\mu \to e \gamma\), since the same mixing and couplings enter into the calculation. The plot shows the minimum $M_X$ that is required to satisfy $\mathrm{Br}(\mu\to eee)\leq10^{-12}$~(see Table~\ref{tab:FLAVOR}) in the $(M_{D_1},\,M_{D_2})$ plane, keeping $M_{D_3}=6.7~\mathrm{TeV}$ fixed. The diagonal valley is also visible in the plot, as we were expecting from the discussions of  the previous section. 
\subsubsection{$\mu$-$e$ conversion in nuclei}
\label{subsubsec:mueconv}
\begin{figure}[!]
    \centering
    \begin{subfigure}[b]{0.32\textwidth}
        \centering
        \includegraphics[width=\textwidth]{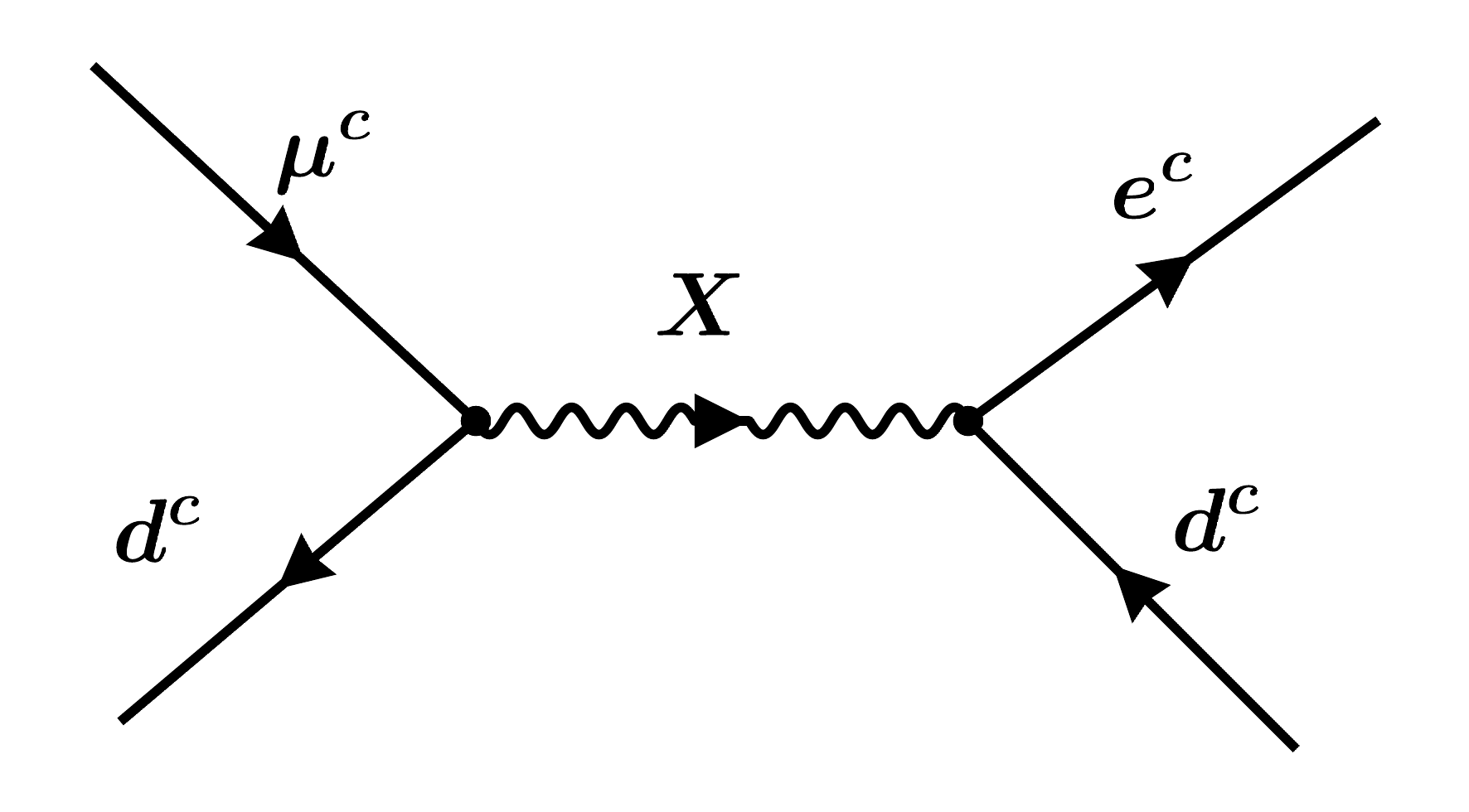}
        \caption{}
    \end{subfigure}
    \hfill
    \begin{subfigure}[b]{0.32\textwidth}
        \centering
        \includegraphics[width=\textwidth]{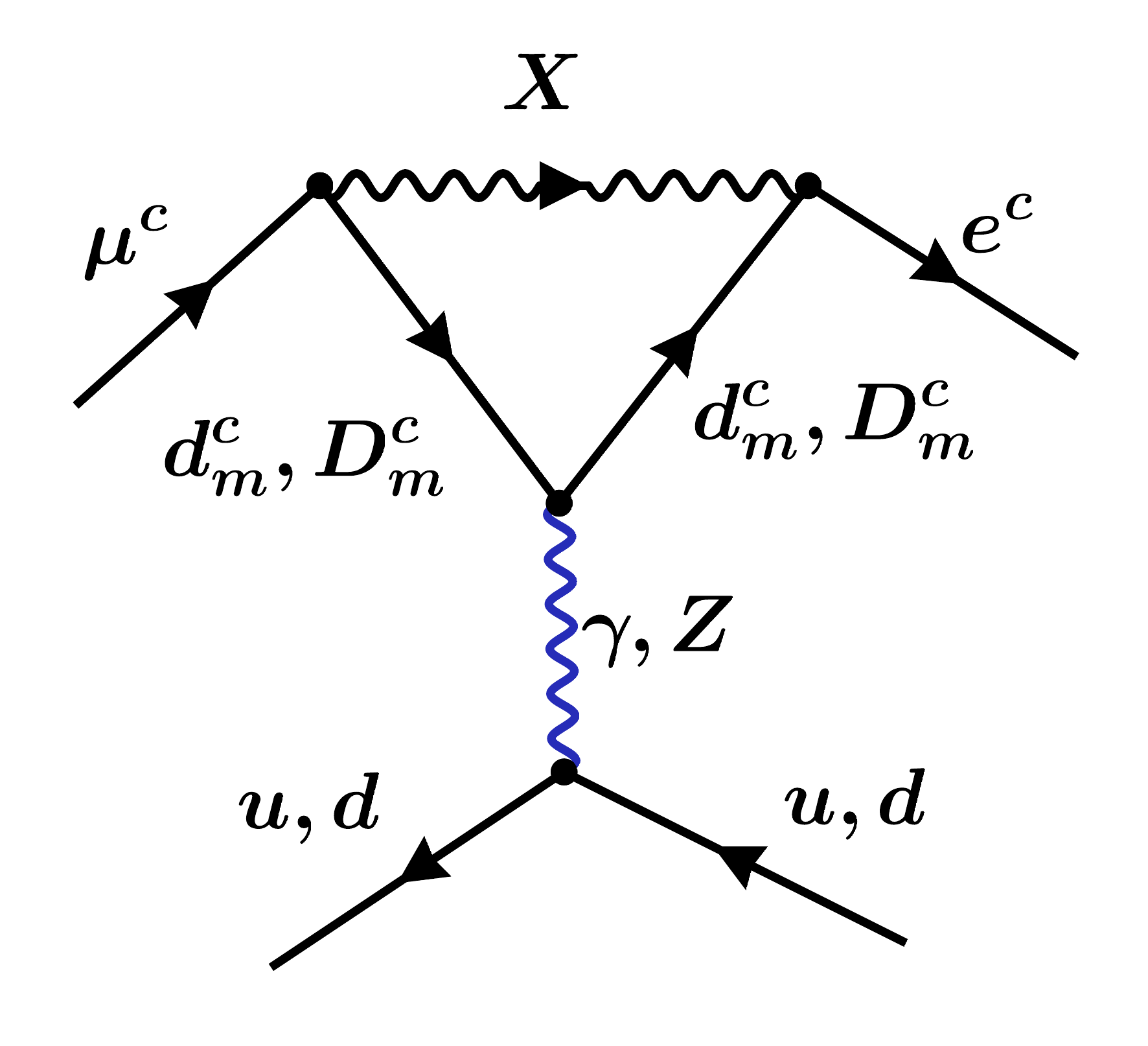}
        \caption{}
    \end{subfigure}
    \hfill
    \begin{subfigure}[b]{0.32\textwidth}
        \centering
        \includegraphics[width=\textwidth]{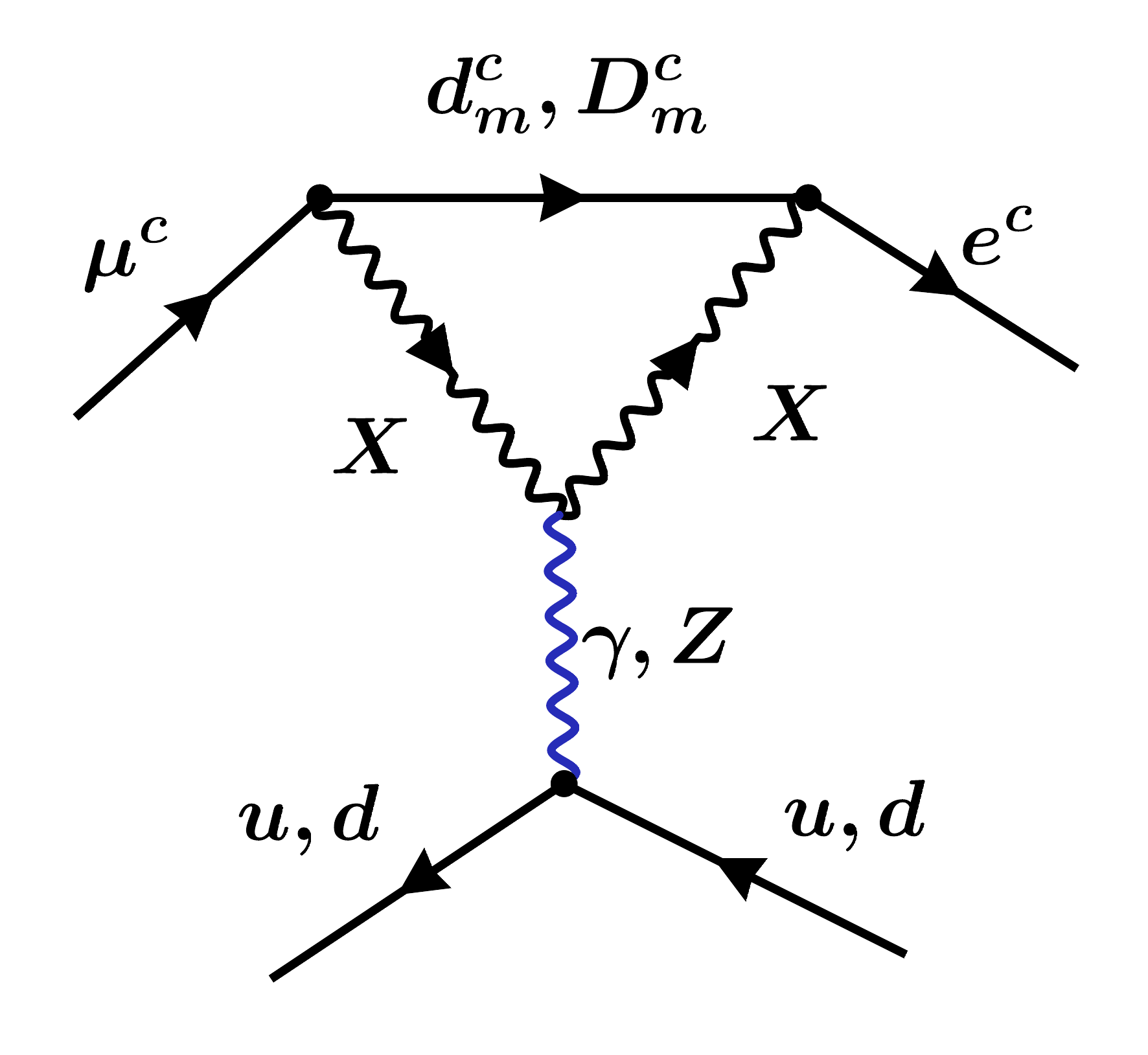}
        \caption{}
    \end{subfigure}

    \vspace{.4cm}
    
        \begin{subfigure}[b]{0.32\textwidth}
        \centering
        \includegraphics[width=\textwidth]{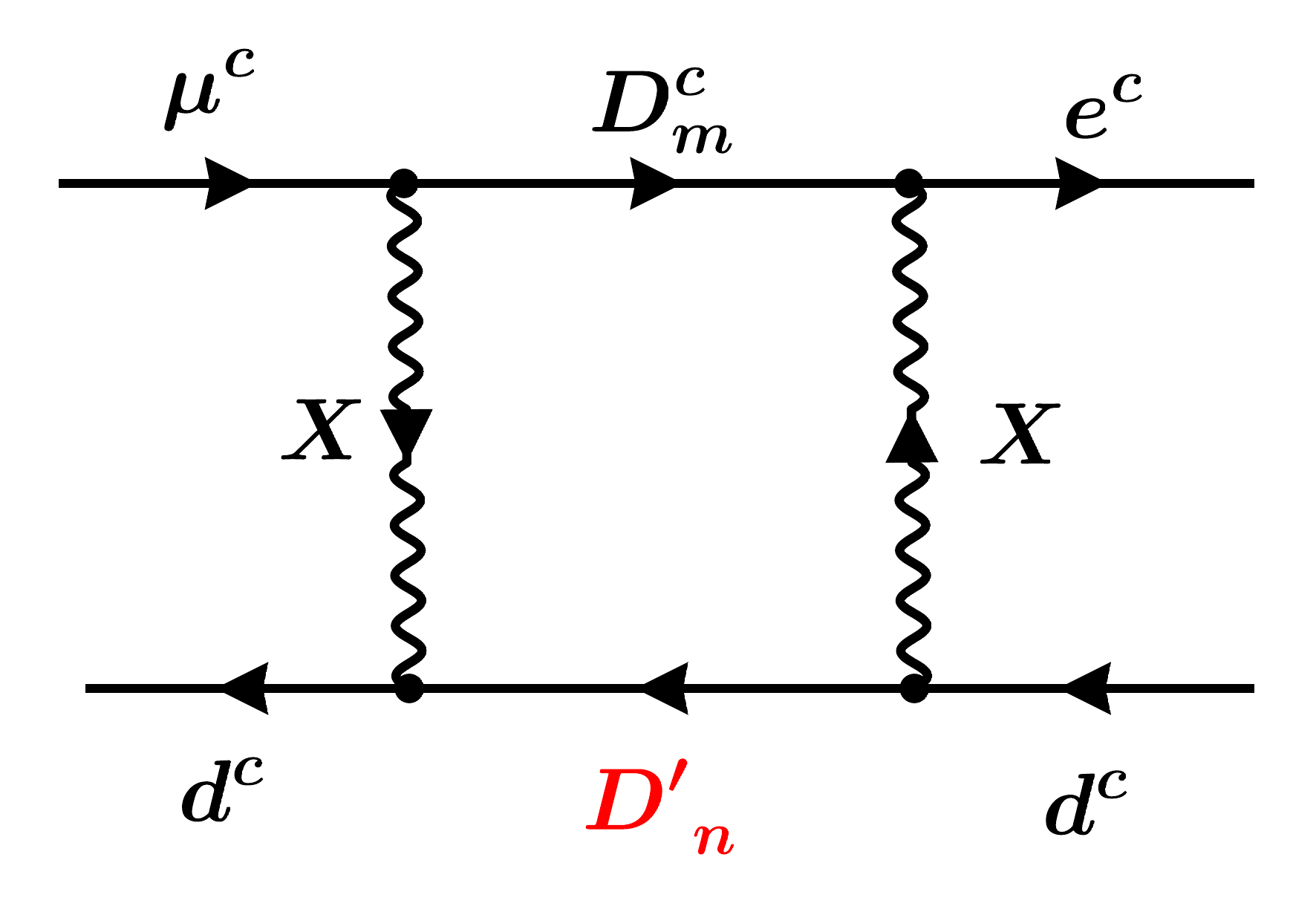}
        \caption{}
    \end{subfigure}
        \hfill
        \begin{subfigure}[b]{0.32\textwidth}
        \centering
        \includegraphics[width=\textwidth]{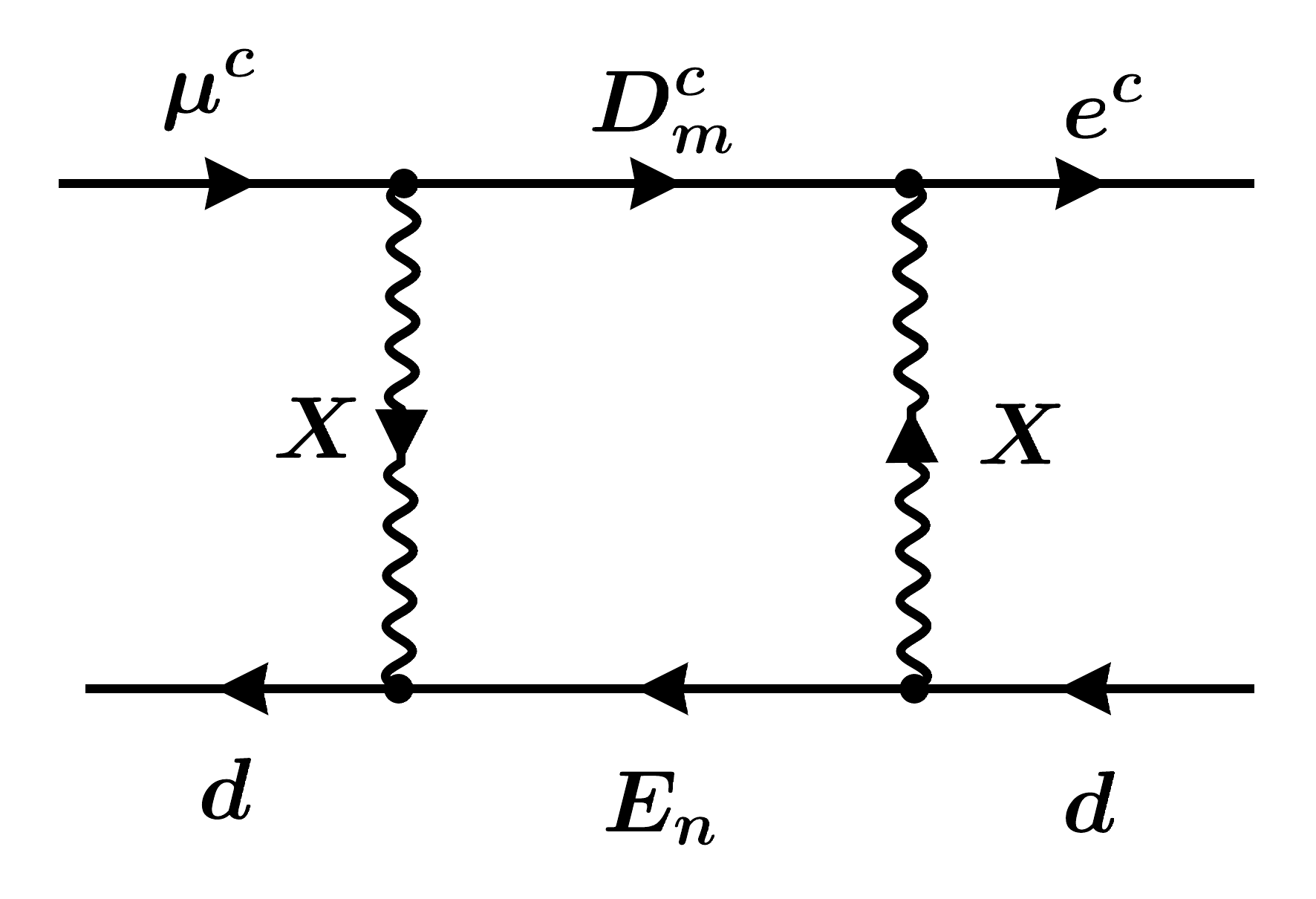}
        \caption{}
    \end{subfigure}
        \hfill
        \begin{subfigure}[b]{0.32\textwidth}
        \centering
        \includegraphics[width=\textwidth]{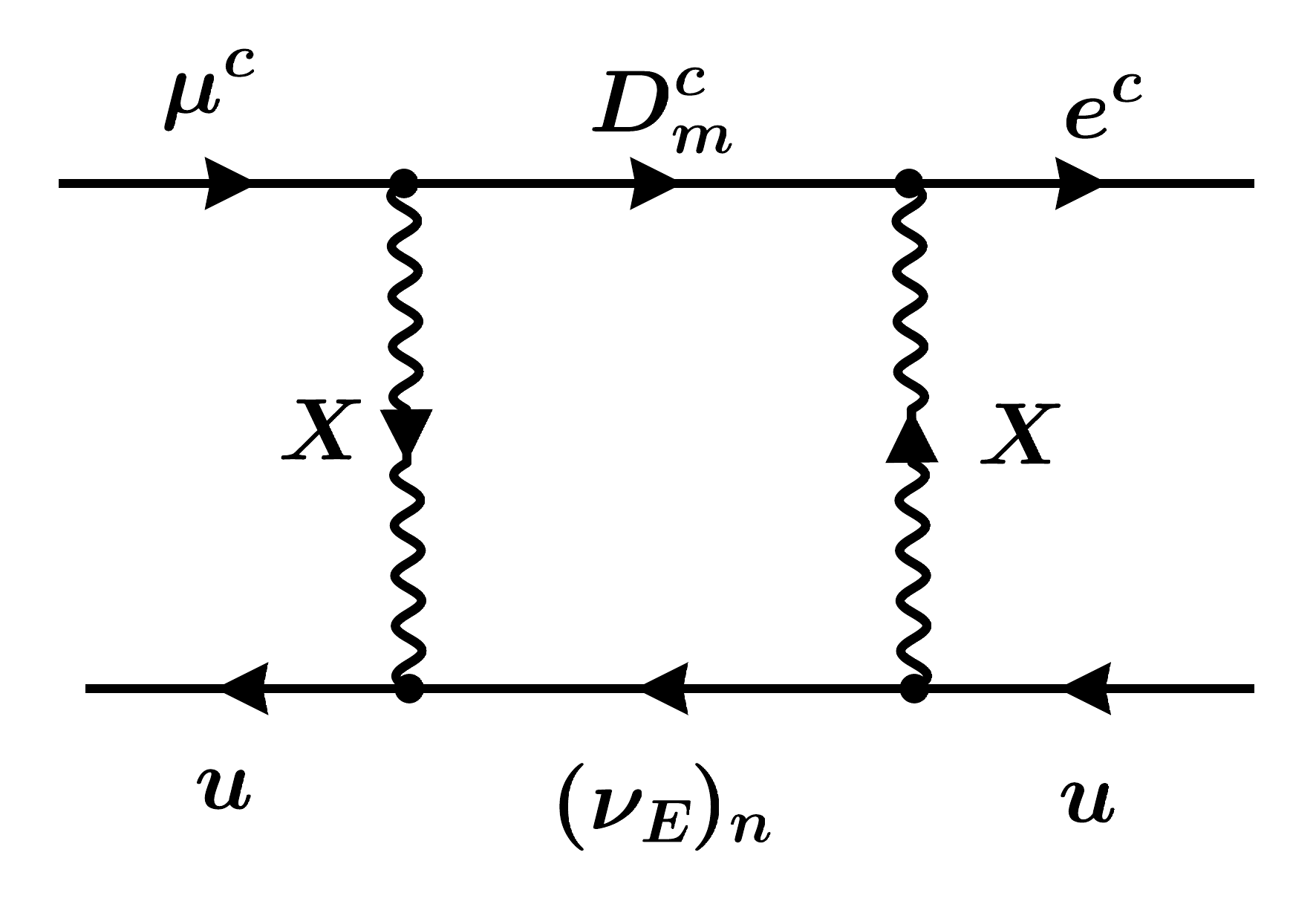}
        \caption{}
    \end{subfigure}
    \caption{Some representative diagrams that contribute to the \(\mu-e\) conversion. (a) shows the tree-level contribution mediated by the leptoquark gauge boson \(X_{\mu}\). (b) and (c) represent $\gamma/Z$ penguin diagrams, other diagrams with the unphysical modes can be inferred by replacing the external lepton legs in \(\mu \to eee\) process (see Fig.~\ref{fig:mu_eee_penguins}) with valence quarks. (d)-(f) are the box diagrams involving two different classes of VLQ where \(\mu \to e\) conversion can happen while interacting with both chiralities of quarks.}
    \label{fig:mueconversion_diagrams}
\end{figure}

Among the well-studied cLFV processes, coherent $\mu-e$ conversion in nuclei has a special place. When an atom captures a negative muon, it can fall to the \(1s\) orbital and form a muonic atom. Instead of the usual decay or nuclear capture, this muon can convert into an electron through a flavor-violating transition, leaving the nucleus intact~\cite{Czarnecki:1998iz}. This neutrino-less process will conserve lepton number but violate flavor by one unit. We are interested in the coherent conversion where all nucleons contribute constructively.  As a result, its rate can be significantly enhanced, especially when flavor-violating gauge boson interactions are present in the theory. This process can be more constraining in some parts of the parameter space where $\mu\to e\gamma$ and/or $\mu\to eee$ rates remain small. Unlike these two processes, tree-level diagrams contribute to $\mu-e$ conversion. It is also induced at the one-loop level as shown in Fig.~\ref{fig:mueconversion_diagrams}.

We can write down the most general LFV interaction Lagrangian that contributes to the \(\mu \to e \) conversion in nuclei following Ref.~\cite{Kitano:2002mt}: 
\begin{align}
\mathcal{L}_{\rm int}
&= -\frac{4G_F}{\sqrt{2}}
\big( m_\mu A_R\,\bar{\mu}\sigma^{\mu\nu}P_L e\,F_{\mu\nu}
     + m_\mu A_L\,\bar{\mu}\sigma^{\mu\nu}P_R e\,F_{\mu\nu}
\big)\notag\\
&\quad -\frac{G_F}{\sqrt{2}} 
\sum_{q}
\Big[
\big(g_{LS(q)}\,\bar{e}P_R\mu + g_{RS(q)}\,\bar{e}P_L\mu\big)\,\bar{q}q
+ \big(g_{LP(q)}\,\bar{e}P_R\mu + g_{RP(q)}\,\bar{e}P_L\mu\big)\,\bar{q}\gamma_5 q \notag\\
&\qquad
+ \big(g_{LV(q)}\,\bar{e}\gamma^\mu P_L\mu + g_{RV(q)}\,\bar{e}\gamma^\mu P_R\mu\big)\,\bar{q}\gamma_\mu q
+ \big(g_{LA(q)}\,\bar{e}\gamma^\mu P_L\mu + g_{RA(q)}\,\bar{e}\gamma^\mu P_R\mu\big)\,\bar{q}\gamma_\mu \gamma_5 q\notag\\
&\qquad
+ \frac{1}{2}\big(
g_{LT(q)}\,\bar{e}\sigma^{\mu\nu}P_R\mu + g_{RT(q)}\,\bar{e}\sigma^{\mu\nu}P_L\mu
\big)\,\bar{q}\sigma_{\mu\nu}q
\Big]+ \mathrm{h.c.} .
\label{eq:L_mu_e_conversion}
\end{align}
where all the \(g\)’s are all dimensionless coupling constants for the corresponding operators, where the momentum dependencies have been neglected. The magnitude of each coupling depends on the type of new physics under consideration that contributes to LFV. Actually, many of the terms won't play a significant role in our setup. First, since we are interested in the coherent processes in which the final state of the nucleus is the same as
the initial one, the matrix elements $\langle N|\bar{q}\gamma_5 q|N\rangle$, $\langle N|\bar{q}\gamma_\mu \gamma_5 q|N\rangle$, and $\langle N|\bar{q}\sigma_{\mu\nu} q|N\rangle$ vanish. And specific to our setup, as discussed in the previous section,  \(A_R\) and the scalar current \(\bar{q}q\) are negligible. Only the dipole term with \(A_L\), vector current \(\bar{q} \gamma_\mu q\) with lepton flavor violation in the right-handed chirality, \(\bar{e^c} \gamma^\mu \mu^c\) and their interference will contribute.

\begin{figure}[!h]
\centering
\includegraphics[scale=0.34]{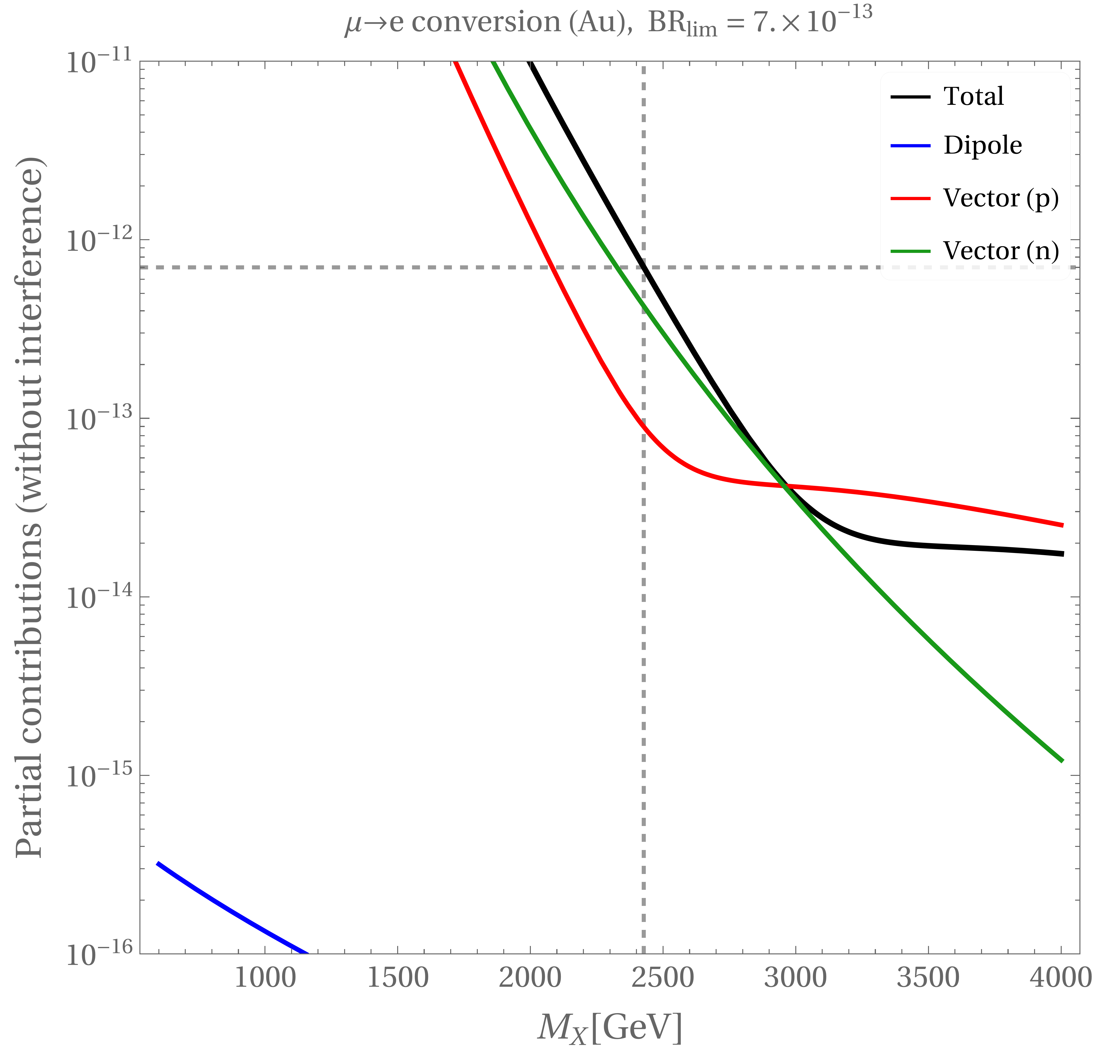}
\hspace{.4cm}
\includegraphics[scale=0.36]{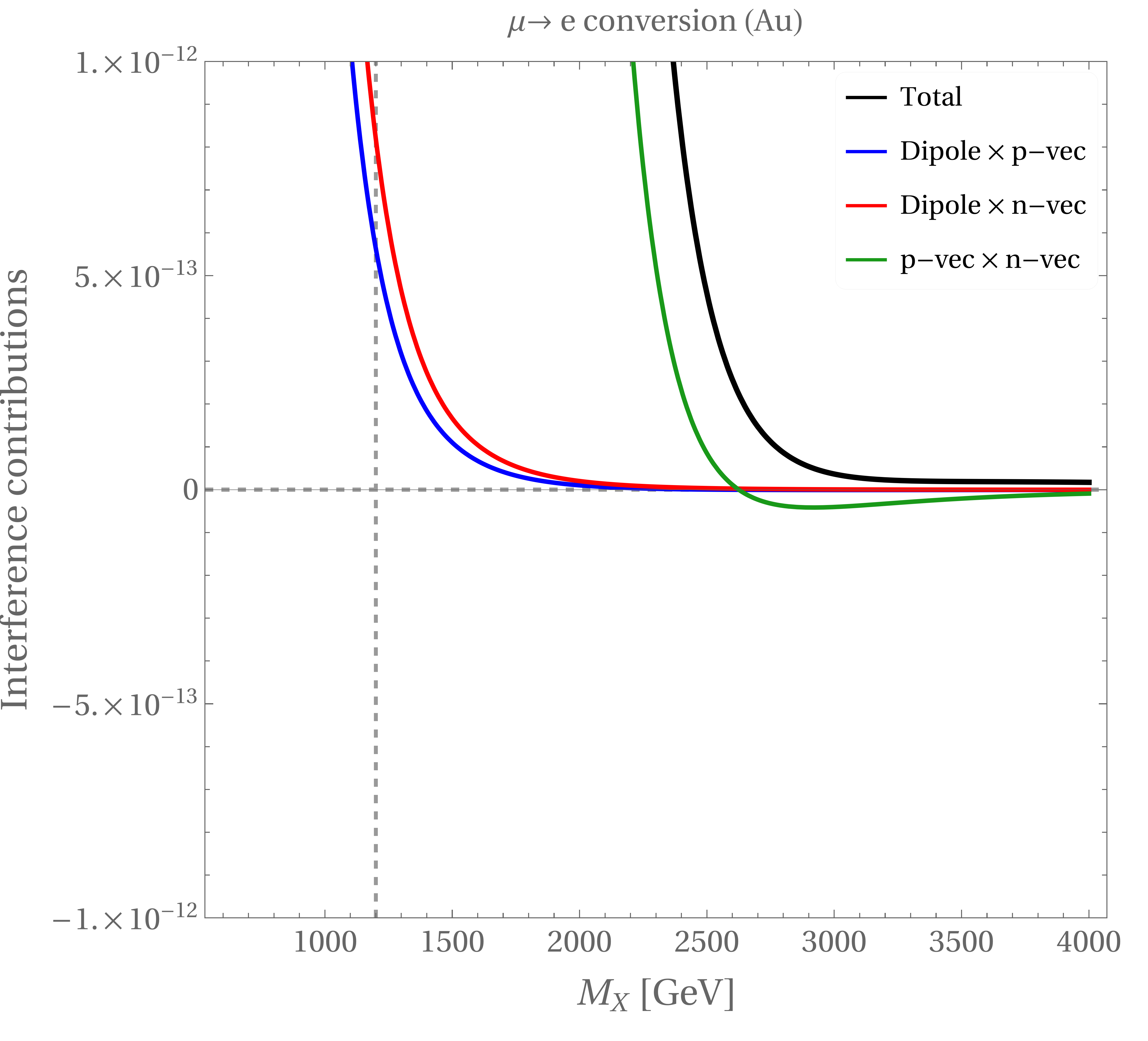}
\caption{Individual and interference contributions to the $\mu \to e$ conversion rate in gold~\cite{SINDRUMII:2006dvw}. The left panel shows the partial branching fractions from dipole and vector (proton and neutron) terms without interference, and the right panel presents their interference terms. The vector contributions dominate over the dipole part, and the total rate surpasses the current experimental limit 
${\rm BR}_{\text{lim}}(\mu \to e) = 7.0\times10^{-13}$ for $M_X \le 2.5~{\rm TeV}$. For this plot, we take the exotic mass benchmark $(M_{D_1}, M_{D_2}, M_{D_3}) = (7640,7600,6700)~{\rm GeV}$, the same combination that was taken to plot Fig.~\ref{fig:partial_contributions_mutoeee}.}
\label{fig:partial_contributions_mutoeconv(Au)}
\label{fig:contributions_mueconv}
\end{figure}
Now there are a few new features in this conversion that are absent in the \(\mu \to eee\) process. First, here we have a tree-level contribution mediated by the leptoquark gauge boson \(X_{\mu}\). Second, in the box diagrams, different classes of vector-like fermions can run in the loop, and \(\bar{\mu^c}\gamma_{\mu}e^c\) current can couple to both chiral currents \(\bar{q^c}\gamma_{\mu}q^c\) and  \(\bar{q}\gamma_{\mu}q\)~(see Fig.~\ref{fig:mueconversion_diagrams}). This simplifies the interaction Lagrangian specific to our setup:
\begin{equation}
\begin{aligned}
\mathcal{L}_{\mu \to e ~\mathrm{conv}} &=
-\frac{4 G_F}{\sqrt{2}}
\Big[m_\mu A_L\,\mu \sigma^{\mu\nu} e^c F_{\mu\nu}
+\frac{1}{4}
\sum g_{RV(q)}\,(\bar{e^c} \gamma^\mu \mu^c)\, (\bar{q} \gamma_\mu q+\bar{q^c}\gamma_{\mu}q^c)\Big]+ \mathrm{h.c.}
\end{aligned}
\label{eq:L_mu_e_conversion_this_setup}
\end{equation}
As a result, the final formula will be simplified as well and take this simple form:
\begin{equation}
\begin{aligned}
\Gamma_{\mathrm{conv}} &= 2 G_F^2 
\Big|A_L^\ast D+\tilde{g}^{(p)}_{RV} V^{(p)}+
\tilde{g}^{(n)}_{RV} V^{(n)}\Big|^2 .
\end{aligned}
\label{eq:omega_conv_final_form}
\end{equation}
Here \(A_L=\sigma_L/(4\sqrt{2}\,G_F\,m_{\mu})\) has already been discussed elaborately in the previous sections and can be used directly from Eqs.~\eqref{eqn:sigma_L}--\eqref{eqn:yX}. The effective couplings $\tilde{g}_{RV}^{(p/n)}$ in Eq.~\eqref{eq:omega_conv_final_form} are given by
\begin{align}
\tilde{g}^{(p)}_{RV} &= 2\,g_{RV(u)} + g_{RV(d)} ,
\label{eq:gtilde_p_vector} \\[3pt]
\tilde{g}^{(n)}_{RV} &= g_{RV(u)} + 2\,g_{RV(d)} .
\label{eq:gtilde_n_vector}
\end{align}
and the overlap integrals \(D,V^{(p)},V^{(n)}\) are listed in the appendix~\ref{Appen:Overlap_integrals} for convenience. We follow closely the discussions of the previous section on \(\mu \to eee\) from Ref.~\cite{Bishara:2021buy,Brod:2019bro} for the calculation of \(g_{RV(q)}\).
\begin{figure}[t!]
    \centering
    \begin{subfigure}[b]{0.47\textwidth}
        \centering
        \includegraphics[width=\textwidth]{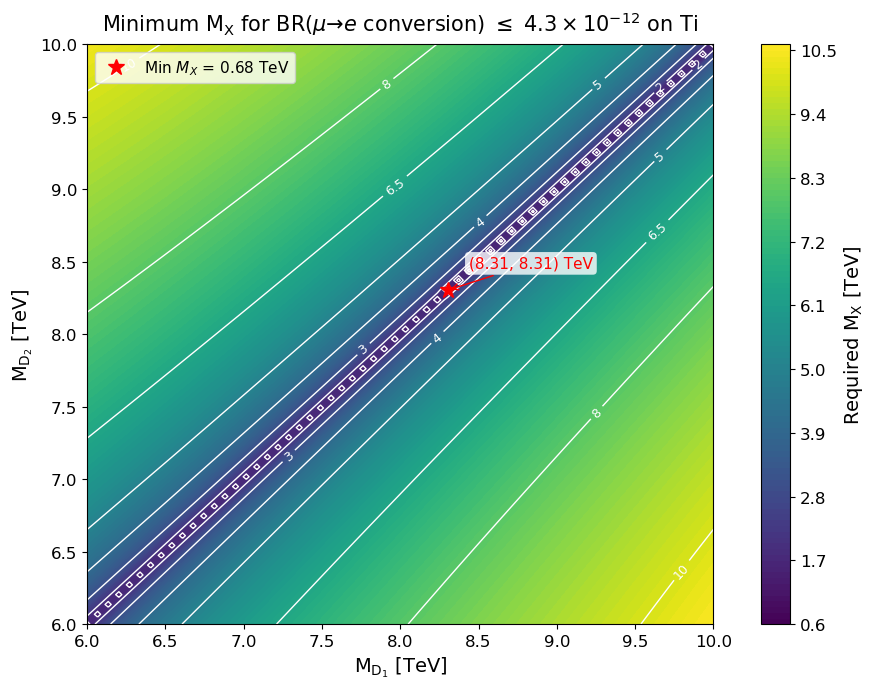}
        \caption{}
    \end{subfigure}
    \hfill
    \begin{subfigure}[b]{0.47\textwidth}
        \centering
        \includegraphics[width=\textwidth]{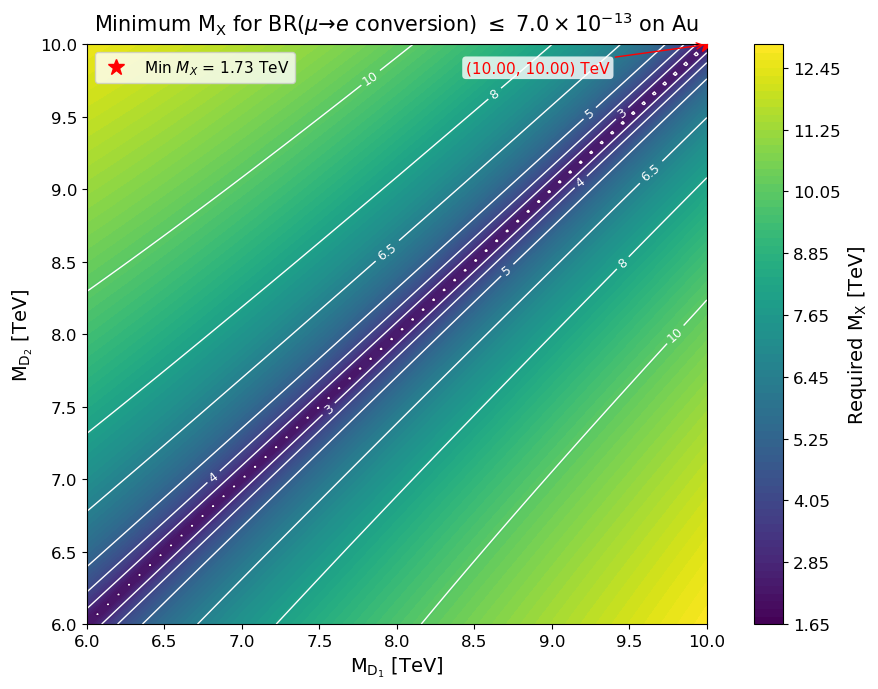}
        \caption{}
    \end{subfigure} 
    \vspace{.4cm}
        \begin{subfigure}[b]{0.47\textwidth}
        \centering
        \includegraphics[width=\textwidth]{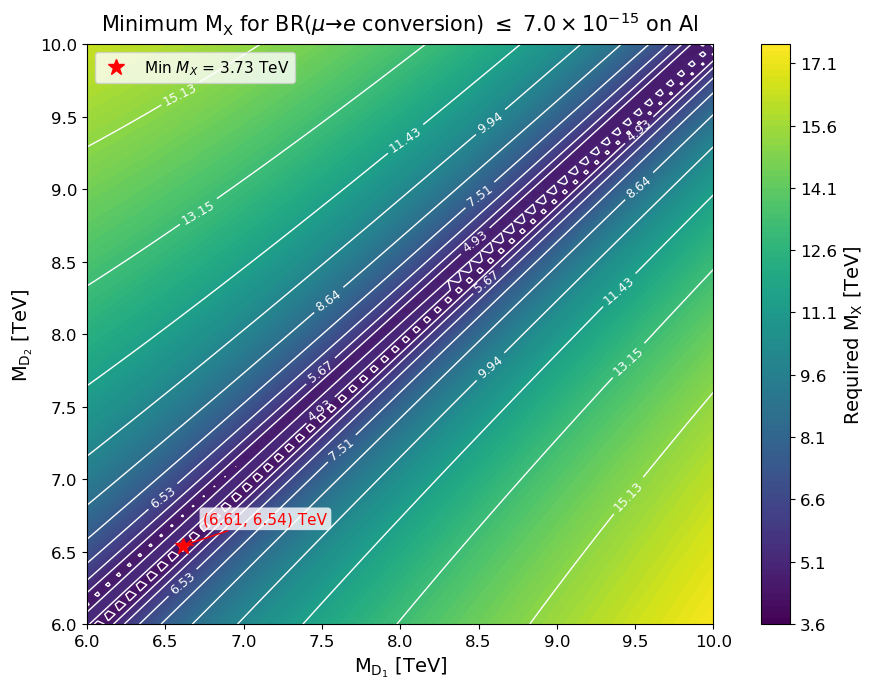}
        \caption{}
        \end{subfigure}
        \hfill
    \begin{subfigure}[b]{0.47\textwidth}
        \centering
        \includegraphics[width=\textwidth]{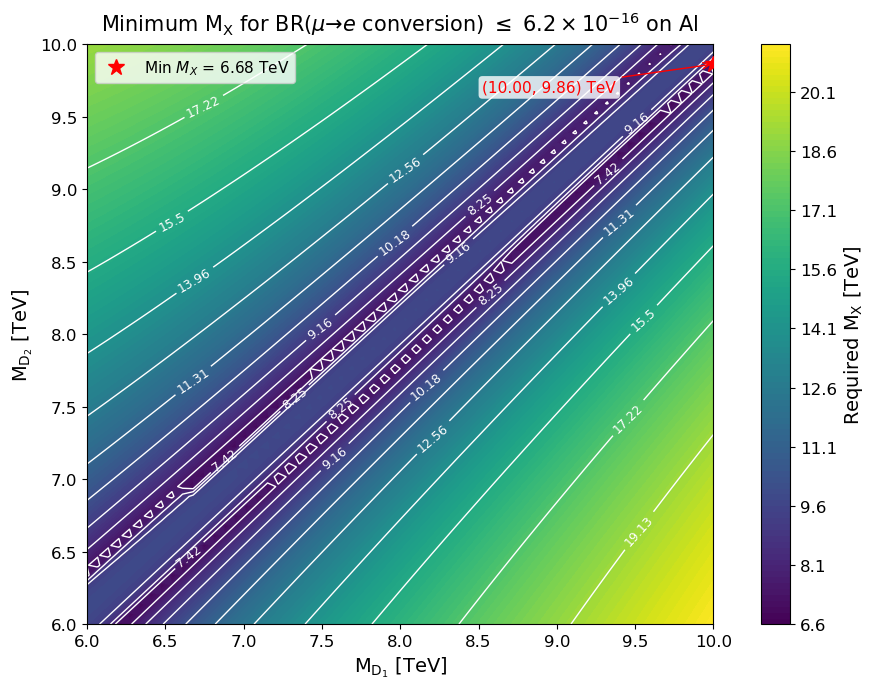}
        \caption{}
    \end{subfigure}
    \caption{Minimum \(M_X\) required to satisfy the experimental and projected bounds on \(\mathrm{Br}(\mu \to e\,\mathrm{conv})\) from various targets: Plots in~(a) and~(b) correspond to the current SINDRUM~II limits on Ti~\cite{SINDRUMII:1993gxf} and Au~\cite{SINDRUMII:2006dvw}, respectively. Plots~(c) and~(d) represent the expected sensitivities of future experiments, COMET Phase~I~\cite{COMET:2018auw} and Mu2e~\cite{Mu2e:2022ggl} respectively, both using aluminum targets. Each plot shows contours of the minimum \(M_X\) required in the \((M_{D_1},M_{D_2})\) plane (with \(M_{D_3}=6.7~\mathrm{TeV}\) fixed).} 
    \label{fig:mueconv_projections}
\end{figure}
Here, we should make one comment on the tree-level contribution to the vector current~(see Fig.~\ref{fig:mueconversion_diagrams} (a)). It is extremely suppressed in our setup due to the structure of the optimized CKM-like matrix (see Eq.~\eqref{eqn:optimized_K_mu_e}), which was needed to eliminate the \(K\)-meson decays. Thus, the same flavor alignment that ensures vanishing \(\mathrm{Br}(K\to \mu e)\) also guarantees that the tree-level \(\mu \to e\) conversion is highly suppressed. For an order of magnitude estimate, we use
\begin{equation}
    \mathrm{Br^{tree}_{\mu \to e~\mathrm con(Ti)}}\sim \dfrac{2 G_F^2}{\Gamma_{\mathrm {capt}}^{\mathrm{Ti}}}\Big|2\left(\dfrac{g_4}{g_L}\right)^2\left(\dfrac{M_W}{M_X}\right)^2\left(V^{(p)}+2V^{(n)}\right)(K_{d^ce^c})_{11}(K_{d^ce^c})_{12}^*\Big|^2
\end{equation}
with the overlap integrals, \( V^{(p)},~V^{(n)}\), in units of \(m_{\mu}^{5/2}\) and the total capture rate, \(\Gamma_{\mathrm{capt}}^{\mathrm{Ti}}\) in units of \(10^6~\mathrm{s}^{-1}\) listed in Table~\ref{tab:overlap}, the choice for \(K_{ij}\) from Eq.~\eqref{eqn:optimized_K_mu_e} and \(M_X\sim 1~\mathrm{TeV}\), we find the \( \mathrm{Br^{tree}_{\mu \to e~\mathrm con(Ti)}}\sim 2\times 10^{-18}\).  The gauge invariant and finite combinations from penguins and box diagrams yields the expression
\begin{equation}
  g_{RV(q)}=\frac{3}{16\pi^2}
\left(\frac{g_4}{g_L}\right)^{2}
\left(\frac{M_W}{M_X}\right)^{2}
\left[
e^2\left(\mathcal{F}^{(\gamma Z)}_{1} + \mathcal{F}^{(\gamma Z)}_{2}\right)
+ g_4^2\mathcal{G}^{(B)}
\right],
\label{eqn: gRVq}
\end{equation}
where the form for \(\mathcal{F}^{(\gamma Z)}_{1},\,\mathcal{F}^{(\gamma Z)}_{2}\) has been discussed in Eqs.~\eqref{eqn:FgammaZ-1}--\eqref{eqn:FgammaZ-V}. For the box contribution \(\mathcal{G}^B\), we have taken the different classes of vector-like fermions \(\textcolor{red}{D'},\,E,\,\nu_{E}\) at the same mass scale and the CKM-like mixing exactly as \(\rho^H\) (see Eq.~\eqref{eqn:benchmark2} in benchmark-2), so that \(\mathcal{G}^B\sim\mathcal{F}^B\) and we use the mass kernels defined in Eqs.~\eqref{eqn:FVRBZ}--\eqref{eqn:FVRBpZ} for the parameter scan. To emphasize the point that \(\mu \to e\) conversion in itself can be more constraining when \(\mu \to e \gamma\) and \(\mu\to eee\) processes are absent, we present Fig.~\ref{fig:contributions_mueconv}. The left panel shows the individual contributions, and the right panel includes the interference terms to keep track of possible cancellations. In our setup, the vector contribution ($p$- or $n$-type, depending on the relevant mass scale) clearly surpasses the dipole part.  Because of the attractiveness from the experimental point of view (the emitted electron has a mono-energetic spectrum at $E_e \simeq 105~\text{MeV}$, well above the Michel endpoint ($52.8~\text{MeV}$), so no need for coincidence measurement), several experiments have already searched for it~\cite{Ahmad:1988ur,SINDRUMII:1993gxf,SINDRUMII:2006dvw}, and future experiments~\cite{COMET:2018auw,Mu2e:2022ggl} are aiming to push the sensitivity much further. Specially, the upcoming Mu2e experiment at Fermilab will improve the sensitivity by nearly four orders of magnitude~\cite{Mu2e:2022ggl}, which will enable us to probe our low-scale leptoquark gauge boson $X_{\mu}$ up to about $20~\text{TeV}$ in the benchmark scenarios considered here (see Fig.~\ref{fig:mueconv_projections} (d)). For all plots in Fig.~\ref{fig:mueconv_projections}, we have made our usual choice for CKM-like mixings from Eq.~\eqref{eqn:benchmark2} and plotted the contours of minimum \(M_X\) required to satisfy the experimental bounds on $\mathrm{Br}(\mu\!\to\! e\,\mathrm{conv})$  in the \((M_{D_1},M_{D_2})\) plane, with \(M_{D_3}=6.7\) TeV fixed.

\subsubsection{Global fit with the three cLFV processes}
\label{subsubsec:global_fit}
\begin{figure}[!t]
    \centering
    \includegraphics[width=0.6\textwidth]{FIG/global_fit.png}
\caption{Global requirement on $M_X$ from the combined constraints $\mathrm{BR}(\mu \to e\gamma)\le 4.2\times10^{-13}$, $\mathrm{Br}(\mu\to eee)\le 1.0\times 10^{-12}$, and $\mathrm{BR}(\mu \to e~$conv.; Au)$\le 7.0\times10^{-13}$. The color at each point in the $(M_{D_1},M_{D_2})$ plane with $M_{D_3}=6.7~\text{TeV}$ fixed gives the information of the smallest mediator mass $M_X$ that satisfies all three limits simultaneously. White curves are contours of constant $M_X$. Throughout the scenario, we have made sure that all three processes share the same CKM-like structure and exotic-fermion content. }
 \label{fig:global_fit}
\end{figure}
Now we proceed to combine the three cLFV channels in a single fit, since they depend on the same parameters in the flavor sector. For loops involving light down-type quarks, we use the CKM matrix with soft–breaking suppression (see Eq.~\eqref{eqn:optimized_K_mu_e}). For loops involving vector-like quarks, we use the same CKM-like matrix (see Eq.~\eqref{eqn:benchmark2}), ensuring that all processes share a common flavor ansatz. At each point in the \((M_{D_1}, M_{D_2})\) plane (where \(M_{D_3}=6.7~\mathrm{TeV}\) is fixed), we scan over the mediator mass \(M_X\) and determine the minimum value that satisfies all current bounds simultaneously:
\[
M_X^{\rm req}(M_{D_1},M_{D_2}) \;=\;
\max\Big\{
M_X^{(\mu\to e\gamma)},\;
M_X^{(\mu\to eee)},\;
M_X^{(\mu-e~\text{conv; Au})}
\Big\},
\]
with \(\mathrm{BR}(\mu\to e\gamma)\le 4.2\times10^{-13}\), \(\mathrm{Br}(\mu\to eee)\le 1.0\times10^{-12}\), and \(\mathrm{BR}(\mu\!\to e~\text{conv.; Au})\le 7.0\times10^{-13}\)~(see Table \ref{tab:FLAVOR}).
The resulting heat map in Fig.~\ref{fig:global_fit} shows the required \(M_X^{\rm req}\)  with iso-lines indicating constant values of \(M_X\). We have also shown the global minimum (\(M_X \sim\) 2.8 TeV with \((M_{D_1}, M_{D_2})\sim (6.00,6.07)\)TeV) within the scan for this benchmark. Now, it is possible that for some combination of this benchmark parameter, destructive interference among dipole, \(\gamma-Z\), and box diagrams is locally relaxing one channel, but the global fit would automatically select the most constraining process at each point, which is the reason for showing this fit. Fig.~\ref{fig:global_fit} represents the resulting \(M_X^{\rm req}\) surface and its iso-contours.

\section{Conclusions}\label{sec:5}
In this work, we have presented a concrete model of low-scale quark-lepton unification based on the Pati-Salam gauge symmetry \(SU(2)_L\times SU(2)_R\times SU(4)_C\) with an \(E_6\)-inspired particle spectrum. In addition to the SM fermions, our model features multi-TeV vector-like down-type quarks and leptons, together with a minimal Higgs sector that achieves two-step breaking of the Pati-Salam gauge group. One of the crucial components of this construction is a softly broken \(Z_2\) symmetry, under which the SM fermions are even, while the exotic states (except $\textcolor{red}{D'^c}$), including \(X_{\mu},\,W'_{\mu}\) are odd, which allows for realizing the low-scale unification. The soft breaking in the right–handed down-quark sector induces small mixings that permit tree–level meson decays mediated by \(X_{\mu}\), but the branching ratios are helicity and mixing suppressed, with  \(Z_2\) unsuppressed contributions arising at one loop. 

We have analyzed the collider signatures, meson decays, neutral meson oscillations, as well as charged lepton flavor violation mediated by \(X_\mu\), and we have shown that this framework can realize quark-lepton unification at accessible energies in a phenomenologically rich way, while remaining compatible with current collider and flavor constraints. While flavor observables by themselves are consistent with \(M_X\) being as low as \(1.1~\text{TeV}\), the LHC limit on the \(Z'_{\mu}\) boson translates into an indirect bound on \(M_X \gtrsim 4.3~\text{TeV}\). We have also discussed the distinctive collider signatures of the TeV-scale vector–like down–type quark with baryon number \(-2/3\),
whose existence is a natural consequence of a consistent baryon–number assignment in the model.

Within this setup, neutrino masses arise from two sources. One is through an effective dimension-7 operator, which is generated after integrating out the heavy neutral leptons of the model. The second source is a radiative contribution of the scotogenic type.
Depending on the region of the parameter space, either mechanism can dominate, or both can participate comparably to explain the observed neutrino masses with TeV-scale new physics.

Looking ahead, our results show that this low–scale realization is a well–defined target for both the energy and intensity frontiers.  The same leptoquark gauge boson \(X_{\mu}\) and vector–like fermions that are essential for low–scale unification also control the rates for \(\mu\to e\gamma\), \(\mu\to eee\), and \(\mu\to e\) conversion, which makes flavor experiments such as MEG~II, Mu3e, Mu2e, and COMET directly sensitive to the relevant parameter space. Parallel searches for leptoquark gauge bosons and vector–like quarks at the LHC and future colliders can probe the same structure from the high–energy side, providing complementary tests of quark–lepton unification at accessible scales.

In summary, the heart of our work is to show that quarks and leptons can be made indistinguishable by a gauge symmetry realized at experimentally accessible energy scales.

\subsection*{Acknowledgments}
The works of KSB and SB are supported in part by US Department of Energy Grant
Number DE-SC 0016013. 
SS acknowledges the financial support from the Slovenian Research Agency (research core funding No. P1-0035 and N1-0321). 
The authors acknowledge the Center for Theoretical Underground Physics and Related Areas (CETUP* 2024) and the Institute for Underground Science at Sanford Underground Research Facility (SURF) for hospitality and for providing a conducive environment during the finalization of this work. The authors thank Jogesh Pati for useful comments and for his enthusiastic support.
\appendix
\section{Coefficients in renormalization group equations}
\label{app:B}
The  beta function coefficients used for the running of the SM gauge couplings of Eq.~\eqref{eqn:beta1}-\eqref{eqn:beta2} are given below:~\cite{Arason:1991ic},
\begin{equation}
    b_i = \left(\frac{41}{10},\; -\frac{19}{6},\; -7\right),\quad  B_{ij}=
  \begin{pmatrix}
    \dfrac{199}{50} & \dfrac{27}{10} & \dfrac{44}{5} \\[6pt]
    \dfrac{9}{10}   & \dfrac{35}{6}  & 12 \\[6pt]
    \dfrac{11}{10}  & \dfrac{9}{2}   & -26
  \end{pmatrix},\quad  C_{if}=
  \begin{pmatrix}
    \dfrac{17}{10} & \dfrac{1}{2} & \dfrac{3}{2} \\[6pt]
    \dfrac{3}{2}   & \dfrac{3}{2}  & \dfrac{1}{2} \\[6pt]
    2  & 2  & 0
  \end{pmatrix},
\end{equation}
with \(f=u,d,e\). For completeness, we also note the Yukawa sector beta functions that enter into the coupled equations with gauge coupling running, 
\begin{equation}
16\pi^2\frac{d Y_{u,d,e}}{dt}=
  Y_{u,d,e}\left(
    \,\beta^{(1)}_{u,d,e}
    + \beta^{(2)}_{u,d,e}
  \right),
\end{equation}
where we considered only the one-loop contributions in our numerical analysis. At one-loop, they can be written as,
\begin{align}
  \beta^{(1)}_{u}&=\frac{3}{2}\left(Y_u^\dagger Y_u - Y_d^\dagger Y_d\right)+Y_2(S)- \left(\frac{17}{20}g_1^2 + \frac{9}{4}g_2^2 + 8g_3^2\right),\\
  \beta^{(1)}_{d} &= \frac{3}{2}\left(Y_d^\dagger Y_d - Y_u^\dagger Y_u\right)+ Y_2(S)- \left(\frac{1}{4}g_1^2 + \frac{9}{4}g_2^2 + 8g_3^2\right),\\
  \beta^{(1)}_{e}&= \frac{3}{2}\,Y_e^\dagger Y_e+ Y_2(S)- \frac{9}{4}\left(g_1^2 + g_2^2\right),
\end{align}
with
\begin{equation}
  Y_2(S)=
  \mathrm{Tr}\!\left(3Y_u^\dagger Y_u + 3Y_d^\dagger Y_d + Y_e^\dagger Y_e\right).
\end{equation}
\section{Input parameters for meson mixing}\label{app:C}
\subsection{$K^0-\bar{K}^0$ oscillation}
\label{subsection: Input parameters for K-K oscillation}
The non-vanishing magic numbers used in Eq. (\ref{eqn:Wilson_running}) for the kaon system are given below~\cite{Ciuchini:1998ix}:
\begin{align}
a_i ~~~&= (0.29,\,-0.69,\,0.79,\,-1.1,\,0.14) \nonumber \\
b_i^{(11)} &= (0.82,\,0,\,0,\,0,\,0), &
c_i^{(11)} &= (-0.016,\,0,\,0,\,0,\,0), \nonumber \\
b_i^{(22)} &= (0,\,2.4,\,0.011,\,0,\,0), &
c_i^{(22)} &= (0,\,-0.23,\,-0.002,\,0,\,0), \nonumber \\
b_i^{(23)} &= (0,\,-0.63,\,0.17,\,0,\,0), &
c_i^{(23)} &= (0,\,-0.018,\,0.0049,\,0,\,0), \nonumber \\
b_i^{(32)} &= (0,\,-0.019,\,0.028,\,0,\,0), &
c_i^{(32)} &= (0,\,0.0028,\,-0.0093,\,0,\,0), \nonumber \\
b_i^{(33)} &= (0,\,0.0049,\,0.43,\,0,\,0), &
c_i^{(33)} &= (0,\,0.00021,\,0.023,\,0,\,0), \nonumber \\
b_i^{(44)} &= (0,\,0,\,0,\,4.4,\,0), &
c_i^{(44)} &= (0,\,0,\,0,\,-0.68,\,0.0055), \nonumber \\
b_i^{(45)} &= (0,\,0,\,0,\,1.5,\,-0.17), &
c_i^{(45)} &= (0,\,0,\,0,\,-0.35,\,-0.0062), \nonumber \\
b_i^{(54)} &= (0,\,0,\,0,\,0.18,\,0), &
c_i^{(54)} &= (0,\,0,\,0,\,-0.026,\,-0.016), \nonumber \\
b_i^{(55)} &= (0,\,0,\,0,\,0.061,\,0.82), &
c_i^{(55)} &= (0,\,0,\,0,\,-0.013,\,0.018). \nonumber
\end{align}
Table~\ref{tab:K_input_param} gives the relevant constants for the $K^0-\bar{K}^0$ phenomenological analysis~\cite{Bauer:2009cf}:
\begin{table}[h!]
\centering
\renewcommand{\arraystretch}{1.3}
\begin{tabular}{|c|c|c|c|c|c|c|c|}
\hline
$\mu$ & $M_K$ & $f_K$ & $(m_d+m_s)$ & $B_1^{sd}$ & $B_4^{sd}$ & $B_5^{sd}$ \\
\hline
$2~\mathrm{GeV}$ & $497.67~\mathrm{MeV}$ & $156.1~\mathrm{MeV}$ & $135\pm 18~\mathrm{MeV}$ & $0.527\pm 0.022$ & $0.9\pm 0.2$ & $0.6\pm0.1$ \\
\hline
\end{tabular}
\caption{Input parameters used for $\Delta M_K$ estimation.}
\label{tab:K_input_param}
\end{table}
\subsection{$D^0-\bar{D}^0$ oscillation}
\label{subsection: Input parameters for D-D oscillation}
The non-vanishing magic numbers are given below~\cite{UTfit:2007eik}:
\begin{align*}
a_i ~~~ &= (0.286,\,-0.692,\,0.787,\,-1.143,\,0.143) \\
b_i^{(11)} &= (0.837,\,0,\,0,\,0,\,0), &
c_i^{(11)} &= (-0.016,\,0,\,0,\,0,\,0), \\
b_i^{(22)} &= (0,\,2.163,\,0.012,\,0,\,0), &
c_i^{(22)} &= (0,\,-0.20,\,-0.002,\,0,\,0), \\
b_i^{(23)} &= (0,\,-0.567,\,0.176,\,0,\,0), &
c_i^{(23)} &= (0,\,-0.016,\,0.006,\,0,\,0), \\
b_i^{(32)} &= (0,\,-0.032,\,0.031,\,0,\,0), &
c_i^{(32)} &= (0,\,-0.004,\,-0.010,\,0,\,0), \\
b_i^{(33)} &= (0,\,0.008,\,0.474,\,0,\,0), &
c_i^{(33)} &= (0,\,0,\,0.025,\,0,\,0), \\
b_i^{(44)} &= (0,\,0,\,0,\,3.63,\,0), &
c_i^{(44)} &= (0,\,0,\,0,\,-0.56,\,0.006), \\
b_i^{(45)} &= (0,\,0,\,0,\,1.21,\,-0.19), &
c_i^{(45)} &= (0,\,0,\,0,\,-0.29,\,-0.006), \\
b_i^{(54)} &= (0,\,0,\,0,\,0.14,\,0), &
c_i^{(54)} &= (0,\,0,\,0,\,-0.019,\,-0.016), \\
b_i^{(55)} &= (0,\,0,\,0,\,0.045,\,0.839), &
c_i^{(55)} &= (0,\,0,\,0,\,-0.009,\,0.018).
\end{align*}
Table~\ref{tab:DD_constants} gives the constants in the $D^0 - \bar{D}^0$ phenomenological analysis~\cite{Bauer:2009cf}:

\begin{table}[h!]
\centering
\begin{tabular}{|c|c|c|c|c|c|c|c|}
\hline
$\mu$ & $M_D$ & $f_D$ & $(m_u+m_c)$ & $B_1^{cu}$ & $B_4^{cu}$ & $B_5^{cu}$ \\
\hline
2.8 GeV & 1.86 GeV & \(212\pm14\)  MeV & \(1.17\pm 0.12\) GeV & \(0.85\pm 0.09\) & \(1.10\pm 0.11\) & \(1.37\pm 0.14\) \\
\hline
\end{tabular}
\caption{Input parameters used for $\Delta M_D$ estimation.}
\label{tab:DD_constants}
\end{table}
\subsection{$B_q^0-\bar{B_q}^0$ oscillation}
\label{subsection: Input parameters for B-B oscillation}
The non-vanishing magic numbers are given below~\cite{Becirevic:2001jj}:
\begin{align*}
a_i~~~&= (0.286,\,-0.692,\,0.787,\,-1.143,\,0.143) \\[1ex]
b_i^{(11)} &= (0.865,\,0,\,0,\,0,\,0), &
c_i^{(11)} &= (-0.017,\,0,\,0,\,0,\,0), \\
b_i^{(22)} &= (0,\,1.879,\,0.012,\,0,\,0), &
c_i^{(22)} &= (0,\,-0.18,\,-0.003,\,0,\,0), \\
b_i^{(23)} &= (0,\,-0.493,\,0.18,\,0,\,0), &
c_i^{(23)} &= (0,\,-0.014,\,0.008,\,0,\,0), \\
b_i^{(32)} &= (0,\,-0.044,\,0.035,\,0,\,0), &
c_i^{(32)} &= (0,\,-0.005,\,-0.012,\,0,\,0), \\
b_i^{(33)} &= (0,\,0.011,\,0.54,\,0,\,0), &
c_i^{(33)} &= (0,\,0,\,0.028,\,0,\,0), \\
b_i^{(44)} &= (0,\,0,\,0,\,2.87,\,0), &
c_i^{(44)} &= (0,\,0,\,0,\,-0.48,\,-0.005), \\
b_i^{(45)} &= (0,\,0,\,0,\,0.961,\,-0.22), &
c_i^{(45)} &= (0,\,0,\,0,\,-0.25,\,-0.006), \\
b_i^{(54)} &= (0,\,0,\,0,\,0.09,\,0), &
c_i^{(54)} &= (0,\,0,\,0,\,-0.013,\,-0.016), \\
b_i^{(55)} &= (0,\,0,\,0,\,0.029,\,0.863), &
c_i^{(55)} &= (0,\,0,\,0,\,-0.007,\,0.019).
\end{align*}
Table~\ref{tab:Bd_input_param} and~\ref{tab:Bs_input_param} gives the relevant constants for the $B_{d,s}^0-\bar{B}_{d,s}^0$ phenomenological analysis respectively~\cite{Bauer:2009cf}:
\begin{table}[h!]
\centering
\renewcommand{\arraystretch}{1.3}
\begin{tabular}{|c|c|c|c|c|c|c|c|}
\hline
$\mu$ & $M_{B_d}$ & $f_{B_d}$ & $(m_b+m_d)$ & $B_1^{bd}$ & $B_4^{bd}$ & $B_5^{bd}$ \\
\hline
$4.2~\mathrm{GeV}$ & $5.2795~\mathrm{GeV}$ & $(200 \pm 20)~\mathrm{MeV}$ & $4.22\pm 0.08~\mathrm{GeV}$ & $0.81\pm 0.08$ & $1.14\pm 0.13$ & $1.72\pm0.19$ \\
\hline
\end{tabular}
\caption{Input parameters used for $\Delta M_{B_d}$ estimation.}
\label{tab:Bd_input_param}
\end{table}
\begin{table}[h!]
\centering
\renewcommand{\arraystretch}{1.3}
\begin{tabular}{|c|c|c|c|c|c|c|c|}
\hline
$\mu$ & $M_{B_s}$ & $f_{B_s}$ & $(m_b+m_s)$ & $B_1^{bs}$ & $B_4^{bs}$ & $B_5^{bs}$ \\
\hline
$4.2~\mathrm{GeV}$ & $497.67~\mathrm{GeV}$ & $245\pm 25~\mathrm{MeV}$ & $4.30\pm0.08~\mathrm{GeV}$ & $0.527\pm 0.022$ & $0.9\pm 0.2$ & $0.6\pm0.1$ \\
\hline
\end{tabular}
\caption{Input parameters used for $\Delta M_{B_s}$ estimation.}
\label{tab:Bs_input_param}
\end{table}
\section{Input parameters for \(\mu \to e\) conversion in nuclei}\label{app:D}
\label{Appen:Overlap_integrals}
\begin{table}[h!]
\centering
\renewcommand{\arraystretch}{1.2}
\begin{tabular}{|c|cccc|}
\hline
Nucleus & \(D\) & 
$V^{(p)}$ & 
$V^{(n)}$ &
$\Gamma_{\mathrm{capt}}$\\
\hline
Al & 0.0362 & 0.0161 & 0.0173 & 0.69 \\
Ti & 0.0864 & 0.0396 & 0.0468 & 2.59 \\
Au & 0.1890 & 0.0974 & 0.1460 & 13.07 \\
\hline
\end{tabular}
\caption{Values of the input parameters for different nuclei used in the calculation.}
\label{tab:overlap}
\end{table}
The overlap integrals, \(D, V^{(p)},~\mathrm{and}~V^{(n)}\), in units of \(m_{\mu}^{5/2}\) are listed in the Table from Ref.~\cite{Kitano:2002mt}. The total capture rates, \(\Gamma_{\mathrm{capt}}\) in units of \(10^6~\mathrm{s}^{-1}\) are used from Ref.~\cite{Suzuki:1987jf}.
\bibliographystyle{style}
\bibliography{reference}

\providecommand{\href}[2]{#2}\begingroup\raggedright\begin{thebibliography}{100}

\bibitem{Pati:1973rp}
J.~C. Pati and A.~Salam, ``{Is Baryon Number Conserved?},'' \href{http://dx.doi.org/10.1103/PhysRevLett.31.661}{{\em Phys. Rev. Lett.} {\bfseries 31} (1973) 661--664}.

\bibitem{Pati:1974yy}
J.~C. Pati and A.~Salam, ``{Lepton Number as the Fourth Color},'' \href{http://dx.doi.org/10.1103/PhysRevD.10.275}{{\em Phys. Rev. D} {\bfseries 10} (1974) 275--289}. [Erratum: Phys.Rev.D 11, 703--703 (1975)].

\bibitem{Kuznetsov:1994tt}
A.~V. Kuznetsov and N.~V. Mikheev, ``{Vector leptoquarks could be rather light?},'' \href{http://dx.doi.org/10.1016/0370-2693(94)90775-7}{{\em Phys. Lett. B} {\bfseries 329} (1994) 295--299}, \href{http://arxiv.org/abs/hep-ph/9406347}{{\ttfamily arXiv:hep-ph/9406347}}.

\bibitem{Valencia:1994cj}
G.~Valencia and S.~Willenbrock, ``{Quark - lepton unification and rare meson decays},'' \href{http://dx.doi.org/10.1103/PhysRevD.50.6843}{{\em Phys. Rev. D} {\bfseries 50} (1994) 6843--6848}, \href{http://arxiv.org/abs/hep-ph/9409201}{{\ttfamily arXiv:hep-ph/9409201}}.

\bibitem{Smirnov:2007hv}
A.~D. Smirnov, ``{Mass limits for scalar and gauge leptoquarks from K(L)0 ---\ensuremath{>} e-+ mu+-, B0 ---\ensuremath{>} e-+ tau+- decays},'' \href{http://dx.doi.org/10.1142/S0217732307024401}{{\em Mod. Phys. Lett. A} {\bfseries 22} (2007) 2353--2363}, \href{http://arxiv.org/abs/0705.0308}{{\ttfamily arXiv:0705.0308 [hep-ph]}}.

\bibitem{Gursey:1975ki}
F.~Gursey, P.~Ramond, and P.~Sikivie, ``{A Universal Gauge Theory Model Based on E6},'' \href{http://dx.doi.org/10.1016/0370-2693(76)90417-2}{{\em Phys. Lett. B} {\bfseries 60} (1976) 177--180}.

\bibitem{Kuznetsov:1994wa}
A.~V. Kuznetsov and N.~V. Mikheev, ``{Could vector leptoquarks be rather light?},'' in {\em {8th International Seminar on High-energy Physics}}.
\newblock 5, 1994.
\newblock \href{http://arxiv.org/abs/hep-ph/9409321}{{\ttfamily arXiv:hep-ph/9409321}}.

\bibitem{Volkas:1995yn}
R.~R. Volkas, ``{Prospects for mass unification at low-energy scales},'' \href{http://dx.doi.org/10.1103/PhysRevD.53.2681}{{\em Phys. Rev. D} {\bfseries 53} (1996) 2681--2698}, \href{http://arxiv.org/abs/hep-ph/9507215}{{\ttfamily arXiv:hep-ph/9507215}}.

\bibitem{Foot:1997pb}
R.~Foot, ``{An Alternative SU(4) x SU(2)L x SU(2)R model},'' \href{http://dx.doi.org/10.1016/S0370-2693(97)01519-0}{{\em Phys. Lett. B} {\bfseries 420} (1998) 333--339}, \href{http://arxiv.org/abs/hep-ph/9708205}{{\ttfamily arXiv:hep-ph/9708205}}.

\bibitem{Foot:1999wv}
R.~Foot and G.~Filewood, ``{Implications of TeV scale SU(4) x SU(2)-L x SU(2)-R quark lepton-lepton unification},'' \href{http://dx.doi.org/10.1103/PhysRevD.60.115002}{{\em Phys. Rev. D} {\bfseries 60} (1999) 115002}, \href{http://arxiv.org/abs/hep-ph/9903374}{{\ttfamily arXiv:hep-ph/9903374}}.

\bibitem{Dolan:2020doe}
M.~J. Dolan, T.~P. Dutka, and R.~R. Volkas, ``{Lowering the scale of Pati-Salam breaking through seesaw mixing},'' \href{http://dx.doi.org/10.1007/JHEP05(2021)199}{{\em JHEP} {\bfseries 05} (2021) 199}, \href{http://arxiv.org/abs/2012.05976}{{\ttfamily arXiv:2012.05976 [hep-ph]}}.

\bibitem{Calibbi:2017qbu}
L.~Calibbi, A.~Crivellin, and T.~Li, ``{Model of vector leptoquarks in view of the $B$-physics anomalies},'' \href{http://dx.doi.org/10.1103/PhysRevD.98.115002}{{\em Phys. Rev. D} {\bfseries 98} no.~11, (2018) 115002}, \href{http://arxiv.org/abs/1709.00692}{{\ttfamily arXiv:1709.00692 [hep-ph]}}.

\bibitem{Helsens:2642473}
C.~Helsens and M.~Selvaggi, ``{Search for high-mass resonances at FCC-hh},'' tech. rep., CERN, Geneva, 2018.
\newblock \url{https://cds.cern.ch/record/2642473}.

\bibitem{Babu:2009aq}
K.~S. Babu, S.~Nandi, and Z.~Tavartkiladze, ``{New Mechanism for Neutrino Mass Generation and Triply Charged Higgs Bosons at the LHC},'' \href{http://dx.doi.org/10.1103/PhysRevD.80.071702}{{\em Phys. Rev.} {\bfseries D80} (2009) 071702},
\href{http://arxiv.org/abs/0905.2710}{{\ttfamily arXiv:0905.2710 [hep-ph]}}.

\bibitem{Bambhaniya:2013yca}
G.~Bambhaniya, J.~Chakrabortty, S.~Goswami, and P.~Konar, ``{Generation of neutrino mass from new physics at TeV scale and multilepton signatures at the LHC},'' \href{http://dx.doi.org/10.1103/PhysRevD.88.075006}{{\em Phys. Rev. D} {\bfseries 88} no.~7, (2013) 075006}, \href{http://arxiv.org/abs/1305.2795}{{\ttfamily arXiv:1305.2795 [hep-ph]}}.

\bibitem{Ma:2006km}
E.~Ma, ``{Verifiable radiative seesaw mechanism of neutrino mass and dark matter},'' \href{http://dx.doi.org/10.1103/PhysRevD.73.077301}{{\em Phys. Rev. D} {\bfseries 73} (2006) 077301}, \href{http://arxiv.org/abs/hep-ph/0601225}{{\ttfamily arXiv:hep-ph/0601225}}.

\bibitem{Tao:1996vb}
Z.-j. Tao, ``{Radiative seesaw mechanism at weak scale},'' \href{http://dx.doi.org/10.1103/PhysRevD.54.5693}{{\em Phys. Rev. D} {\bfseries 54} (1996) 5693--5697}, \href{http://arxiv.org/abs/hep-ph/9603309}{{\ttfamily arXiv:hep-ph/9603309}}.

\bibitem{Preskill:1979zi}
J.~Preskill, ``{Cosmological Production of Superheavy Magnetic Monopoles},'' \href{http://dx.doi.org/10.1103/PhysRevLett.43.1365}{{\em Phys. Rev. Lett.} {\bfseries 43} (1979) 1365}.

\bibitem{Jeannerot:2000sv}
R.~Jeannerot, S.~Khalil, G.~Lazarides, and Q.~Shafi, ``{Inflation and monopoles in supersymmetric SU(4)C x SU(2)(L) x SU(2)(R)},'' \href{http://dx.doi.org/10.1088/1126-6708/2000/10/012}{{\em JHEP} {\bfseries 10} (2000) 012}, \href{http://arxiv.org/abs/hep-ph/0002151}{{\ttfamily arXiv:hep-ph/0002151}}.

\bibitem{Barbieri:2017tuq}
R.~Barbieri and A.~Tesi, ``{$B$-decay anomalies in Pati-Salam SU(4)},'' \href{http://dx.doi.org/10.1140/epjc/s10052-018-5680-9}{{\em Eur. Phys. J. C} {\bfseries 78} no.~3, (2018) 193}, \href{http://arxiv.org/abs/1712.06844}{{\ttfamily arXiv:1712.06844 [hep-ph]}}.

\bibitem{Blanke:2018sro}
M.~Blanke and A.~Crivellin, ``{$B$ Meson Anomalies in a Pati-Salam Model within the Randall-Sundrum Background},'' \href{http://dx.doi.org/10.1103/PhysRevLett.121.011801}{{\em Phys. Rev. Lett.} {\bfseries 121} no.~1, (2018) 011801}, \href{http://arxiv.org/abs/1801.07256}{{\ttfamily arXiv:1801.07256 [hep-ph]}}.

\bibitem{Heeck:2018ntp}
J.~Heeck and D.~Teresi, ``{Pati-Salam explanations of the B-meson anomalies},'' \href{http://dx.doi.org/10.1007/JHEP12(2018)103}{{\em JHEP} {\bfseries 12} (2018) 103}, \href{http://arxiv.org/abs/1808.07492}{{\ttfamily arXiv:1808.07492 [hep-ph]}}.

\bibitem{Fuentes-Martin:2020pww}
J.~Fuentes-Martin, G.~Isidori, J.~Pag{\`e}s, and B.~A. Stefanek, ``{Flavor non-universal Pati-Salam unification and neutrino masses},'' \href{http://dx.doi.org/10.1016/j.physletb.2021.136484}{{\em Phys. Lett. B} {\bfseries 820} (2021) 136484}, \href{http://arxiv.org/abs/2012.10492}{{\ttfamily arXiv:2012.10492 [hep-ph]}}.

\bibitem{Greljo:2018tuh}
A.~Greljo and B.~A. Stefanek, ``{Third family quark{\textendash}lepton unification at the TeV scale},'' \href{http://dx.doi.org/10.1016/j.physletb.2018.05.033}{{\em Phys. Lett. B} {\bfseries 782} (2018) 131--138}, \href{http://arxiv.org/abs/1802.04274}{{\ttfamily arXiv:1802.04274 [hep-ph]}}.

\bibitem{FernandezNavarro:2022gst}
M.~Fern{\'a}ndez~Navarro and S.~F. King, ``{B-anomalies in a twin Pati-Salam theory of flavour including the 2022 LHCb $ {R}_{K^{\left(\ast \right)}} $ analysis},'' \href{http://dx.doi.org/10.1007/JHEP02(2023)188}{{\em JHEP} {\bfseries 02} (2023) 188}, \href{http://arxiv.org/abs/2209.00276}{{\ttfamily arXiv:2209.00276 [hep-ph]}}.

\bibitem{Kobayashi:2004ya}
T.~Kobayashi, S.~Raby, and R.-J. Zhang, ``{Searching for realistic 4d string models with a Pati-Salam symmetry: Orbifold grand unified theories from heterotic string compactification on a Z(6) orbifold},'' \href{http://dx.doi.org/10.1016/j.nuclphysb.2004.10.035}{{\em Nucl. Phys. B} {\bfseries 704} (2005) 3--55}, \href{http://arxiv.org/abs/hep-ph/0409098}{{\ttfamily arXiv:hep-ph/0409098}}.

\bibitem{Cvetic:2004ui}
M.~Cvetic, T.~Li, and T.~Liu, ``{Supersymmetric Pati-Salam models from intersecting D6-branes: A Road to the standard model},'' \href{http://dx.doi.org/10.1016/j.nuclphysb.2004.07.036}{{\em Nucl. Phys. B} {\bfseries 698} (2004) 163--201}, \href{http://arxiv.org/abs/hep-th/0403061}{{\ttfamily arXiv:hep-th/0403061}}.

\bibitem{Assel:2010wj}
B.~Assel, K.~Christodoulides, A.~E. Faraggi, C.~Kounnas, and J.~Rizos, ``{Classification of Heterotic Pati-Salam Models},'' \href{http://dx.doi.org/10.1016/j.nuclphysb.2010.11.011}{{\em Nucl. Phys. B} {\bfseries 844} (2011) 365--396}, \href{http://arxiv.org/abs/1007.2268}{{\ttfamily arXiv:1007.2268 [hep-th]}}.

\bibitem{Chamseddine:2013rta}
A.~H. Chamseddine, A.~Connes, and W.~D. van Suijlekom, ``{Beyond the Spectral Standard Model: Emergence of Pati-Salam Unification},'' \href{http://dx.doi.org/10.1007/JHEP11(2013)132}{{\em JHEP} {\bfseries 11} (2013) 132}, \href{http://arxiv.org/abs/1304.8050}{{\ttfamily arXiv:1304.8050 [hep-th]}}.

\bibitem{Hewett:1988xc}
J.~L. Hewett and T.~G. Rizzo, ``{Low-Energy Phenomenology of Superstring Inspired E(6) Models},'' \href{http://dx.doi.org/10.1016/0370-1573(89)90071-9}{{\em Phys. Rept.} {\bfseries 183} (1989) 193}.

\bibitem{Mohapatra:1980qe}
R.~N. Mohapatra and R.~Marshak, ``{Local B-L Symmetry of Electroweak Interactions, Majorana Neutrinos and Neutron Oscillations},'' \href{http://dx.doi.org/10.1103/PhysRevLett.44.1316}{{\em Phys. Rev. Lett.} {\bfseries 44} (1980) 1316--1319}. [Erratum: Phys.Rev.Lett. 44, 1643 (1980)].

\bibitem{Aulakh:2002zr}
C.~S. Aulakh and A.~Girdhar, ``{SO(10) a la Pati-Salam},'' \href{http://dx.doi.org/10.1142/S0217751X0502001X}{{\em Int. J. Mod. Phys. A} {\bfseries 20} (2005) 865--894}, \href{http://arxiv.org/abs/hep-ph/0204097}{{\ttfamily arXiv:hep-ph/0204097}}.

\bibitem{Arkani-Hamed:1998jmv}
N.~Arkani-Hamed, S.~Dimopoulos, and G.~R. Dvali, ``{The Hierarchy problem and new dimensions at a millimeter},'' \href{http://dx.doi.org/10.1016/S0370-2693(98)00466-3}{{\em Phys. Lett. B} {\bfseries 429} (1998) 263--272}, \href{http://arxiv.org/abs/hep-ph/9803315}{{\ttfamily arXiv:hep-ph/9803315}}.

\bibitem{deAdelhartToorop:2010vtu}
R.~de~Adelhart~Toorop, F.~Bazzocchi, and L.~Merlo, ``{The Interplay Between GUT and Flavour Symmetries in a Pati-Salam x S4 Model},'' \href{http://dx.doi.org/10.1007/JHEP08(2010)001}{{\em JHEP} {\bfseries 08} (2010) 001}, \href{http://arxiv.org/abs/1003.4502}{{\ttfamily arXiv:1003.4502 [hep-ph]}}.

\bibitem{King:2014iia}
S.~F. King, ``{A to Z of Flavour with Pati-Salam},'' \href{http://dx.doi.org/10.1007/JHEP08(2014)130}{{\em JHEP} {\bfseries 08} (2014) 130}, \href{http://arxiv.org/abs/1406.7005}{{\ttfamily arXiv:1406.7005 [hep-ph]}}.

\bibitem{Saad:2017pqj}
S.~Saad, ``{Fermion Masses and Mixings, Leptogenesis and Baryon Number Violation in Pati-Salam Model},'' \href{http://dx.doi.org/10.1016/j.nuclphysb.2019.114630}{{\em Nucl. Phys. B} {\bfseries 943} (2019) 114630}, \href{http://arxiv.org/abs/1712.04880}{{\ttfamily arXiv:1712.04880 [hep-ph]}}.

\bibitem{FileviezPerez:2013zmv}
P.~Fileviez~Perez and M.~B. Wise, ``{Low Scale Quark-Lepton Unification},'' \href{http://dx.doi.org/10.1103/PhysRevD.88.057703}{{\em Phys. Rev. D} {\bfseries 88} (2013) 057703}, \href{http://arxiv.org/abs/1307.6213}{{\ttfamily arXiv:1307.6213 [hep-ph]}}.

\bibitem{Butterworth:2025ttw}
J.~Butterworth, H.~Debnath, P.~Fileviez~Perez, and P.~Wang, ``{Quark-Lepton Unification Signatures},'' \href{http://arxiv.org/abs/2512.00143}{{\ttfamily arXiv:2512.00143 [hep-ph]}}.

\bibitem{Ma:1986we}
E.~Ma, ``{Particle Dichotomy and Left-Right Decomposition of E(6) Superstring Models},'' \href{http://dx.doi.org/10.1103/PhysRevD.36.274}{{\em Phys. Rev. D} {\bfseries 36} (1987) 274}.

\bibitem{Babu:1987kp}
K.~S. Babu, X.-G. He, and E.~Ma, ``{New Supersymmetric Left-Right Gauge Model: Higgs Boson Structure and Neutral Current Analysis},'' \href{http://dx.doi.org/10.1103/PhysRevD.36.878}{{\em Phys. Rev. D} {\bfseries 36} (1987) 878}.

\bibitem{Allwicher:2021rtd}
L.~Allwicher, P.~Arnan, D.~Barducci, and M.~Nardecchia, ``{Perturbative unitarity constraints on generic Yukawa interactions},'' \href{http://dx.doi.org/10.1007/JHEP10(2021)129}{{\em JHEP} {\bfseries 10} (2021) 129}, \href{http://arxiv.org/abs/2108.00013}{{\ttfamily arXiv:2108.00013 [hep-ph]}}.

\bibitem{Mohapatra:1979ia}
R.~N. Mohapatra and G.~Senjanovic, ``{Neutrino Mass and Spontaneous Parity Nonconservation},'' \href{http://dx.doi.org/10.1103/PhysRevLett.44.912}{{\em Phys. Rev. Lett.} {\bfseries 44} (1980) 912}.

\bibitem{Grimus:2000vj}
W.~Grimus and L.~Lavoura, ``{The Seesaw mechanism at arbitrary order: Disentangling the small scale from the large scale},'' \href{http://dx.doi.org/10.1088/1126-6708/2000/11/042}{{\em JHEP} {\bfseries 11} (2000) 042}, \href{http://arxiv.org/abs/hep-ph/0008179}{{\ttfamily arXiv:hep-ph/0008179}}.

\bibitem{lu2002inverses}
T.-T. Lu and S.-H. Shiou, ``Inverses of 2$\times$ 2 block matrices,'' {\em Computers \& Mathematics with Applications} {\bfseries 43} no.~1-2, (2002) 119--129.

\bibitem{Alloul:2013bka}
A.~Alloul, N.~D. Christensen, C.~Degrande, C.~Duhr, and B.~Fuks, ``{FeynRules 2.0 - A complete toolbox for tree-level phenomenology},'' \href{http://dx.doi.org/10.1016/j.cpc.2014.04.012}{{\em Comput. Phys. Commun.} {\bfseries 185} (2014) 2250--2300}, \href{http://arxiv.org/abs/1310.1921}{{\ttfamily arXiv:1310.1921 [hep-ph]}}.

\bibitem{Degrande:2011ua}
C.~Degrande, C.~Duhr, B.~Fuks, D.~Grellscheid, O.~Mattelaer, and T.~Reiter, ``{UFO - The Universal FeynRules Output},'' \href{http://dx.doi.org/10.1016/j.cpc.2012.01.022}{{\em Comput. Phys. Commun.} {\bfseries 183} (2012) 1201--1214}, \href{http://arxiv.org/abs/1108.2040}{{\ttfamily arXiv:1108.2040 [hep-ph]}}.

\bibitem{Alwall:2014hca}
J.~Alwall, R.~Frederix, S.~Frixione, V.~Hirschi, F.~Maltoni, O.~Mattelaer, H.~S. Shao, T.~Stelzer, P.~Torrielli, and M.~Zaro, ``{The automated computation of tree-level and next-to-leading order differential cross sections, and their matching to parton shower simulations},'' \href{http://dx.doi.org/10.1007/JHEP07(2014)079}{{\em JHEP} {\bfseries 07} (2014) 079}, \href{http://arxiv.org/abs/1405.0301}{{\ttfamily arXiv:1405.0301 [hep-ph]}}.

\bibitem{ATLAS:2019erb}
{\bfseries ATLAS} Collaboration, G.~Aad {\em et~al.}, ``{Search for high-mass dilepton resonances using 139 fb$^{-1}$ of $pp$ collision data collected at $\sqrt{s}=$13 TeV with the ATLAS detector},'' \href{http://dx.doi.org/10.1016/j.physletb.2019.07.016}{{\em Phys. Lett. B} {\bfseries 796} (2019) 68--87}, \href{http://arxiv.org/abs/1903.06248}{{\ttfamily arXiv:1903.06248 [hep-ex]}}.

\bibitem{Altarelli:1989ff}
G.~Altarelli, B.~Mele, and M.~Ruiz-Altaba, ``{Searching for New Heavy Vector Bosons in $p \bar{p}$ Colliders},'' \href{http://dx.doi.org/10.1007/BF01556677}{{\em Z. Phys. C} {\bfseries 45} (1989) 109}. [Erratum: Z.Phys.C 47, 676 (1990)].

\bibitem{ATLAS:2018tvr}
{\bfseries ATLAS} Collaboration, ``{Prospects for searches for heavy $Z^\prime$ and $W^\prime$ bosons in fermionic final states with the ATLAS experiment at the HL-LHC},''.

\bibitem{ATLAS:2023ibb}
{\bfseries ATLAS} Collaboration, G.~Aad {\em et~al.}, ``{Search for vector-boson resonances decaying into a top quark and a bottom quark using pp collisions at $ \sqrt{s} $ = 13 TeV with the ATLAS detector},'' \href{http://dx.doi.org/10.1007/JHEP12(2023)073}{{\em JHEP} {\bfseries 12} (2023) 073}, \href{http://arxiv.org/abs/2308.08521}{{\ttfamily arXiv:2308.08521 [hep-ex]}}.

\bibitem{ATLAS:2018umm}
{\bfseries ATLAS} Collaboration, M.~Aaboud {\em et~al.}, ``{Search for R-parity-violating supersymmetric particles in multi-jet final states produced in $p$-$p$ collisions at $\sqrt{s} =13$ TeV using the ATLAS detector at the LHC},'' \href{http://dx.doi.org/10.1016/j.physletb.2018.08.021}{{\em Phys. Lett. B} {\bfseries 785} (2018) 136--158}, \href{http://arxiv.org/abs/1804.03568}{{\ttfamily arXiv:1804.03568 [hep-ex]}}.

\bibitem{ATLAS:2017oes}
{\bfseries ATLAS} Collaboration, M.~Aaboud {\em et~al.}, ``{Search for new phenomena in a lepton plus high jet multiplicity final state with the ATLAS experiment using $ \sqrt{s}=13 $ TeV proton-proton collision data},'' \href{http://dx.doi.org/10.1007/JHEP09(2017)088}{{\em JHEP} {\bfseries 09} (2017) 088}, \href{http://arxiv.org/abs/1704.08493}{{\ttfamily arXiv:1704.08493 [hep-ex]}}.

\bibitem{CMS:2019ybf}
{\bfseries CMS} Collaboration, A.~M. Sirunyan {\em et~al.}, ``{Searches for physics beyond the standard model with the $M_\mathrm{T2}$ variable in hadronic final states with and without disappearing tracks in proton-proton collisions at $\sqrt{s}=$ 13 TeV},'' \href{http://dx.doi.org/10.1140/epjc/s10052-019-7493-x}{{\em Eur. Phys. J. C} {\bfseries 80} no.~1, (2020) 3}, \href{http://arxiv.org/abs/1909.03460}{{\ttfamily arXiv:1909.03460 [hep-ex]}}.

\bibitem{Valencia:1997xe}
G.~Valencia, ``{Long distance contribution to K(L) ---{\ensuremath{>}} lepton+ lepton-},'' \href{http://dx.doi.org/10.1016/S0550-3213(98)00116-3}{{\em Nucl. Phys. B} {\bfseries 517} (1998) 339--352}, \href{http://arxiv.org/abs/hep-ph/9711377}{{\ttfamily arXiv:hep-ph/9711377}}.

\bibitem{GomezDumm:1998gw}
D.~Gomez~Dumm and A.~Pich, ``{Long distance contributions to the K(L) ---{\ensuremath{>}} mu+ mu- decay width},'' \href{http://dx.doi.org/10.1103/PhysRevLett.80.4633}{{\em Phys. Rev. Lett.} {\bfseries 80} (1998) 4633--4636}, \href{http://arxiv.org/abs/hep-ph/9801298}{{\ttfamily arXiv:hep-ph/9801298}}.

\bibitem{E871:2000wvm}
{\bfseries E871} Collaboration, D.~Ambrose {\em et~al.}, ``{Improved branching ratio measurement for the decay K0(L) --{\ensuremath{>}} mu+ mu-},'' \href{http://dx.doi.org/10.1103/PhysRevLett.84.1389}{{\em Phys. Rev. Lett.} {\bfseries 84} (2000) 1389--1392}.

\bibitem{BNL:1998apv}
{\bfseries BNL} Collaboration, D.~Ambrose {\em et~al.}, ``{New limit on muon and electron lepton number violation from K0(L) ---\ensuremath{>} mu+- e-+ decay},'' \href{http://dx.doi.org/10.1103/PhysRevLett.81.5734}{{\em Phys. Rev. Lett.} {\bfseries 81} (1998) 5734--5737}, \href{http://arxiv.org/abs/hep-ex/9811038}{{\ttfamily arXiv:hep-ex/9811038}}.

\bibitem{LHCb:2020pcv}
{\bfseries LHCb} Collaboration, R.~Aaij {\em et~al.}, ``{Search for the Rare Decays $B^0_s\to e^+e^-$ and $B^0\to e^+e^-$},'' \href{http://dx.doi.org/10.1103/PhysRevLett.124.211802}{{\em Phys. Rev. Lett.} {\bfseries 124} no.~21, (2020) 211802}, \href{http://arxiv.org/abs/2003.03999}{{\ttfamily arXiv:2003.03999 [hep-ex]}}.

\bibitem{CMS:2019bbr}
{\bfseries CMS} Collaboration, A.~M. Sirunyan {\em et~al.}, ``{Measurement of properties of B$^0_\mathrm{s}\to\mu^+\mu^-$ decays and search for B$^0\to\mu^+\mu^-$ with the CMS experiment},'' \href{http://dx.doi.org/10.1007/JHEP04(2020)188}{{\em JHEP} {\bfseries 04} (2020) 188}, \href{http://arxiv.org/abs/1910.12127}{{\ttfamily arXiv:1910.12127 [hep-ex]}}.

\bibitem{LHCb:2017myy}
{\bfseries LHCb} Collaboration, R.~Aaij {\em et~al.}, ``{Search for the decays $B_s^0\to\tau^+\tau^-$ and $B^0\to\tau^+\tau^-$},'' \href{http://dx.doi.org/10.1103/PhysRevLett.118.251802}{{\em Phys. Rev. Lett.} {\bfseries 118} no.~25, (2017) 251802}, \href{http://arxiv.org/abs/1703.02508}{{\ttfamily arXiv:1703.02508 [hep-ex]}}.

\bibitem{LHCb:2017hag}
{\bfseries LHCb} Collaboration, R.~Aaij {\em et~al.}, ``{Search for the lepton-flavour violating decays B$_{(s)}^{0} \to e^{\pm}\mu^{\mp}$},'' \href{http://dx.doi.org/10.1007/JHEP03(2018)078}{{\em JHEP} {\bfseries 03} (2018) 078}, \href{http://arxiv.org/abs/1710.04111}{{\ttfamily arXiv:1710.04111 [hep-ex]}}.

\bibitem{Belle:2021rod}
{\bfseries Belle} Collaboration, H.~Atmacan {\em et~al.}, ``{Search for $B^{0} \to \tau^\pm \ell^\mp$ ($\ell=e,\mu$) with a hadronic tagging method at Belle},'' \href{http://dx.doi.org/10.1103/PhysRevD.104.L091105}{{\em Phys. Rev. D} {\bfseries 104} no.~9, (2021) L091105}, \href{http://arxiv.org/abs/2108.11649}{{\ttfamily arXiv:2108.11649 [hep-ex]}}.

\bibitem{ATLAS:2018cur}
{\bfseries ATLAS} Collaboration, M.~Aaboud {\em et~al.}, ``{Study of the rare decays of $B^0_s$ and $B^0$ mesons into muon pairs using data collected during 2015 and 2016 with the ATLAS detector},'' \href{http://dx.doi.org/10.1007/JHEP04(2019)098}{{\em JHEP} {\bfseries 04} (2019) 098}, \href{http://arxiv.org/abs/1812.03017}{{\ttfamily arXiv:1812.03017 [hep-ex]}}.

\bibitem{LHCb:2017rmj}
{\bfseries LHCb} Collaboration, R.~Aaij {\em et~al.}, ``{Measurement of the $B^0_s\to\mu^+\mu^-$ branching fraction and effective lifetime and search for $B^0\to\mu^+\mu^-$ decays},'' \href{http://dx.doi.org/10.1103/PhysRevLett.118.191801}{{\em Phys. Rev. Lett.} {\bfseries 118} no.~19, (2017) 191801}, \href{http://arxiv.org/abs/1703.05747}{{\ttfamily arXiv:1703.05747 [hep-ex]}}.

\bibitem{ParticleDataGroup:2024cfk}
{\bfseries Particle Data Group} Collaboration, S.~Navas {\em et~al.}, ``{Review of particle physics},'' \href{http://dx.doi.org/10.1103/PhysRevD.110.030001}{{\em Phys. Rev. D} {\bfseries 110} no.~3, (2024) 030001}.

\bibitem{MEG:2016leq}
{\bfseries MEG} Collaboration, A.~M. Baldini {\em et~al.}, ``{Search for the lepton flavour violating decay $\mu ^+ \rightarrow \mathrm {e}^+ \gamma $ with the full dataset of the MEG experiment},'' \href{http://dx.doi.org/10.1140/epjc/s10052-016-4271-x}{{\em Eur. Phys. J. C} {\bfseries 76} no.~8, (2016) 434}, \href{http://arxiv.org/abs/1605.05081}{{\ttfamily arXiv:1605.05081 [hep-ex]}}.

\bibitem{BaBar:2009hkt}
{\bfseries BaBar} Collaboration, B.~Aubert {\em et~al.}, ``{Searches for Lepton Flavor Violation in the Decays tau+- ---\ensuremath{>} e+- gamma and tau+- ---\ensuremath{>} mu+- gamma},'' \href{http://dx.doi.org/10.1103/PhysRevLett.104.021802}{{\em Phys. Rev. Lett.} {\bfseries 104} (2010) 021802}, \href{http://arxiv.org/abs/0908.2381}{{\ttfamily arXiv:0908.2381 [hep-ex]}}.

\bibitem{SINDRUM:1987nra}
{\bfseries SINDRUM} Collaboration, U.~Bellgardt {\em et~al.}, ``{Search for the Decay mu+ ---\ensuremath{>} e+ e+ e-},'' \href{http://dx.doi.org/10.1016/0550-3213(88)90462-2}{{\em Nucl. Phys. B} {\bfseries 299} (1988) 1--6}.

\bibitem{Hayasaka:2010np}
K.~Hayasaka {\em et~al.}, ``{Search for Lepton Flavor Violating Tau Decays into Three Leptons with 719 Million Produced Tau+Tau- Pairs},'' \href{http://dx.doi.org/10.1016/j.physletb.2010.03.037}{{\em Phys. Lett. B} {\bfseries 687} (2010) 139--143}, \href{http://arxiv.org/abs/1001.3221}{{\ttfamily arXiv:1001.3221 [hep-ex]}}.

\bibitem{Buras:1998raa}
A.~J. Buras, ``{Weak Hamiltonian, CP violation and rare decays},'' in {\em {Les Houches Summer School in Theoretical Physics, Session 68: Probing the Standard Model of Particle Interactions}}, pp.~281--539.
\newblock 6, 1998.
\newblock \href{http://arxiv.org/abs/hep-ph/9806471}{{\ttfamily arXiv:hep-ph/9806471}}.

\bibitem{Peskin:1995ev}
M.~E. Peskin and D.~V. Schroeder, \href{http://dx.doi.org/10.1201/9780429503559}{{\em {An Introduction to quantum field theory}}}.
\newblock Addison-Wesley, Reading, USA, 1995.

\bibitem{DiLuzio:2019jyq}
L.~Di~Luzio, M.~Kirk, A.~Lenz, and T.~Rauh, ``{$\Delta M_s$ theory precision confronts flavour anomalies},'' \href{http://dx.doi.org/10.1007/JHEP12(2019)009}{{\em JHEP} {\bfseries 12} (2019) 009}, \href{http://arxiv.org/abs/1909.11087}{{\ttfamily arXiv:1909.11087 [hep-ph]}}.

\bibitem{UTfit:2007eik}
{\bfseries UTfit} Collaboration, M.~Bona {\em et~al.}, ``{Model-independent constraints on $\Delta F=2$ operators and the scale of new physics},'' \href{http://dx.doi.org/10.1088/1126-6708/2008/03/049}{{\em JHEP} {\bfseries 03} (2008) 049}, \href{http://arxiv.org/abs/0707.0636}{{\ttfamily arXiv:0707.0636 [hep-ph]}}.

\bibitem{Bauer:2009cf}
M.~Bauer, S.~Casagrande, U.~Haisch, and M.~Neubert, ``{Flavor Physics in the Randall-Sundrum Model: II. Tree-Level Weak-Interaction Processes},'' \href{http://dx.doi.org/10.1007/JHEP09(2010)017}{{\em JHEP} {\bfseries 09} (2010) 017}, \href{http://arxiv.org/abs/0912.1625}{{\ttfamily arXiv:0912.1625 [hep-ph]}}.

\bibitem{Dcruz:2023mvf}
R.~Dcruz, ``{Flavor physics constraints on left-right symmetric models with universal seesaw},'' \href{http://dx.doi.org/10.1016/j.nuclphysb.2024.116519}{{\em Nucl. Phys. B} {\bfseries 1001} (2024) 116519}, \href{http://arxiv.org/abs/2301.10786}{{\ttfamily arXiv:2301.10786 [hep-ph]}}.

\bibitem{Ciuchini:1998ix}
M.~Ciuchini {\em et~al.}, ``{Delta M(K) and epsilon(K) in SUSY at the next-to-leading order},'' \href{http://dx.doi.org/10.1088/1126-6708/1998/10/008}{{\em JHEP} {\bfseries 10} (1998) 008}, \href{http://arxiv.org/abs/hep-ph/9808328}{{\ttfamily arXiv:hep-ph/9808328}}.

\bibitem{Senjanovic:1979cta}
G.~Senjanovic and P.~Senjanovic, ``{Suppression of Higgs Strangeness Changing Neutral Currents in a Class of Gauge Theories},'' \href{http://dx.doi.org/10.1103/PhysRevD.21.3253}{{\em Phys. Rev. D} {\bfseries 21} (1980) 3253}.

\bibitem{Mohapatra:1983ae}
R.~N. Mohapatra, G.~Senjanovic, and M.~D. Tran, ``{Strangeness Changing Processes and the Limit on the Right-handed Gauge Boson Mass},'' \href{http://dx.doi.org/10.1103/PhysRevD.28.546}{{\em Phys. Rev. D} {\bfseries 28} (1983) 546}.

\bibitem{Zhang:2007da}
Y.~Zhang, H.~An, X.~Ji, and R.~N. Mohapatra, ``{General CP Violation in Minimal Left-Right Symmetric Model and Constraints on the Right-Handed Scale},'' \href{http://dx.doi.org/10.1016/j.nuclphysb.2008.05.019}{{\em Nucl. Phys. B} {\bfseries 802} (2008) 247--279}, \href{http://arxiv.org/abs/0712.4218}{{\ttfamily arXiv:0712.4218 [hep-ph]}}.

\bibitem{Petcov:1976ff}
S.~T. Petcov, ``{The Processes $\mu \rightarrow e + \gamma, \mu \rightarrow e + \overline{e}, \nu' \rightarrow \nu + \gamma$ in the Weinberg-Salam Model with Neutrino Mixing},'' {\em Sov. J. Nucl. Phys.} {\bfseries 25} (1977) 340. [Erratum: Sov.J.Nucl.Phys. 25, 698 (1977), Erratum: Yad.Fiz. 25, 1336 (1977)].

\bibitem{MEGII:2025}
{MEG II Collaboration}, ``New limit on the $\mu^+ \to e^+ \gamma$ decay with the meg ii experiment,'' \href{http://dx.doi.org/10.1140/epjc/s10052-025-14906-3}{{\em Eur. Phys. J. C} {\bfseries 85} no.~10, (2025) 1177}, \href{http://arxiv.org/abs/2504.15711}{{\ttfamily arXiv:2504.15711 [hep-ex]}}.

\bibitem{Cheng:2000ct}
T.~P. Cheng and L.~F. Li, \href{http://dx.doi.org/10.1093/oso/9780198506218.001.0001}{{\em {Gauge theory of elementary particle physics: Problems and solutions}}}.
\newblock 2000.

\bibitem{Lavoura:2003xp}
L.~Lavoura, ``{General formulae for f(1) ---\ensuremath{>} f(2) gamma},'' \href{http://dx.doi.org/10.1140/epjc/s2003-01212-7}{{\em Eur. Phys. J. C} {\bfseries 29} (2003) 191--195}, \href{http://arxiv.org/abs/hep-ph/0302221}{{\ttfamily arXiv:hep-ph/0302221}}.

\bibitem{Antusch:2006vwa}
S.~Antusch, C.~Biggio, E.~Fernandez-Martinez, M.~B. Gavela, and J.~Lopez-Pavon, ``{Unitarity of the Leptonic Mixing Matrix},'' \href{http://dx.doi.org/10.1088/1126-6708/2006/10/084}{{\em JHEP} {\bfseries 10} (2006) 084}, \href{http://arxiv.org/abs/hep-ph/0607020}{{\ttfamily arXiv:hep-ph/0607020}}.

\bibitem{Abada:2007ux}
A.~Abada, C.~Biggio, F.~Bonnet, M.~B. Gavela, and T.~Hambye, ``{Low energy effects of neutrino masses},'' \href{http://dx.doi.org/10.1088/1126-6708/2007/12/061}{{\em JHEP} {\bfseries 12} (2007) 061}, \href{http://arxiv.org/abs/0707.4058}{{\ttfamily arXiv:0707.4058 [hep-ph]}}.

\bibitem{Kuno:1999jp}
Y.~Kuno and Y.~Okada, ``{Muon decay and physics beyond the standard model},'' \href{http://dx.doi.org/10.1103/RevModPhys.73.151}{{\em Rev. Mod. Phys.} {\bfseries 73} (2001) 151--202}, \href{http://arxiv.org/abs/hep-ph/9909265}{{\ttfamily arXiv:hep-ph/9909265}}.

\bibitem{Bishara:2021buy}
F.~Bishara, J.~Brod, M.~Gorbahn, and U.~Moldanazarova, ``{Generic one-loop matching conditions for rare meson decays},'' \href{http://dx.doi.org/10.1007/JHEP07(2021)230}{{\em JHEP} {\bfseries 07} (2021) 230}, \href{http://arxiv.org/abs/2104.10930}{{\ttfamily arXiv:2104.10930 [hep-ph]}}.

\bibitem{Brod:2019bro}
J.~Brod and M.~Gorbahn, ``{The $Z$ Penguin in Generic Extensions of the Standard Model},'' \href{http://dx.doi.org/10.1007/JHEP09(2019)027}{{\em JHEP} {\bfseries 09} (2019) 027}, \href{http://arxiv.org/abs/1903.05116}{{\ttfamily arXiv:1903.05116 [hep-ph]}}.

\bibitem{Buchalla:1995vs}
G.~Buchalla, A.~J. Buras, and M.~E. Lautenbacher, ``{Weak Decays beyond Leading Logarithms},'' \href{http://dx.doi.org/10.1103/RevModPhys.68.1125}{{\em Rev. Mod. Phys.} {\bfseries 68} (1996) 1125--1144}, \href{http://arxiv.org/abs/hep-ph/9512380}{{\ttfamily arXiv:hep-ph/9512380}}.

\bibitem{Czarnecki:1998iz}
A.~Czarnecki, W.~J. Marciano, and K.~Melnikov, ``{Coherent muon electron conversion in muonic atoms},'' \href{http://dx.doi.org/10.1063/1.56214}{{\em AIP Conf. Proc.} {\bfseries 435} no.~1, (1998) 409--418}, \href{http://arxiv.org/abs/hep-ph/9801218}{{\ttfamily arXiv:hep-ph/9801218}}.

\bibitem{Kitano:2002mt}
R.~Kitano, M.~Koike, and Y.~Okada, ``{Detailed calculation of lepton flavor violating muon electron conversion rate for various nuclei},'' \href{http://dx.doi.org/10.1103/PhysRevD.76.059902}{{\em Phys. Rev. D} {\bfseries 66} (2002) 096002}, \href{http://arxiv.org/abs/hep-ph/0203110}{{\ttfamily arXiv:hep-ph/0203110}}. [Erratum: Phys.Rev.D 76, 059902 (2007)].

\bibitem{SINDRUMII:2006dvw}
{\bfseries SINDRUM II} Collaboration, W.~H. Bertl {\em et~al.}, ``{A Search for muon to electron conversion in muonic gold},'' \href{http://dx.doi.org/10.1140/epjc/s2006-02582-x}{{\em Eur. Phys. J. C} {\bfseries 47} (2006) 337--346}.

\bibitem{SINDRUMII:1993gxf}
{\bfseries SINDRUM II} Collaboration, C.~Dohmen {\em et~al.}, ``{Test of lepton flavor conservation in mu ---\ensuremath{>} e conversion on titanium},'' \href{http://dx.doi.org/10.1016/0370-2693(93)91383-X}{{\em Phys. Lett. B} {\bfseries 317} (1993) 631--636}.

\bibitem{COMET:2018auw}
{\bfseries COMET} Collaboration, R.~Abramishvili {\em et~al.}, ``{COMET Phase-I Technical Design Report},'' \href{http://dx.doi.org/10.1093/ptep/ptz125}{{\em PTEP} {\bfseries 2020} no.~3, (2020) 033C01}, \href{http://arxiv.org/abs/1812.09018}{{\ttfamily arXiv:1812.09018 [physics.ins-det]}}.

\bibitem{Mu2e:2022ggl}
{\bfseries Mu2e} Collaboration, F.~Abdi {\em et~al.}, ``{Mu2e Run I Sensitivity Projections for the Neutrinoless Conversion Search in Aluminum},'' \href{http://dx.doi.org/10.3390/universe9010054}{{\em Universe} {\bfseries 9} no.~1, (2023) 54}, \href{http://arxiv.org/abs/2210.11380}{{\ttfamily arXiv:2210.11380 [hep-ex]}}.

\bibitem{Ahmad:1988ur}
S.~Ahmad {\em et~al.}, ``{Search for Muon - Electron and Muon - Positron Conversion},'' \href{http://dx.doi.org/10.1103/PhysRevD.38.2102}{{\em Phys. Rev. D} {\bfseries 38} (1988) 2102}.

\bibitem{Arason:1991ic}
H.~Arason, D.~J. Castano, B.~Keszthelyi, S.~Mikaelian, E.~J. Piard, P.~Ramond, and B.~D. Wright, ``{Renormalization group study of the standard model and its extensions. 1. The Standard model},'' \href{http://dx.doi.org/10.1103/PhysRevD.46.3945}{{\em Phys. Rev. D} {\bfseries 46} (1992) 3945--3965}.

\bibitem{Becirevic:2001jj}
D.~Becirevic, M.~Ciuchini, E.~Franco, V.~Gimenez, G.~Martinelli, A.~Masiero, M.~Papinutto, J.~Reyes, and L.~Silvestrini, ``{$B_d - \bar{B}_d$ mixing and the $B_d \to J/\psi K_s$ asymmetry in general SUSY models},'' \href{http://dx.doi.org/10.1016/S0550-3213(02)00291-2}{{\em Nucl. Phys. B} {\bfseries 634} (2002) 105--119}, \href{http://arxiv.org/abs/hep-ph/0112303}{{\ttfamily arXiv:hep-ph/0112303}}.

\bibitem{Suzuki:1987jf}
T.~Suzuki, D.~F. Measday, and J.~P. Roalsvig, ``{Total Nuclear Capture Rates for Negative Muons},'' \href{http://dx.doi.org/10.1103/PhysRevC.35.2212}{{\em Phys. Rev. C} {\bfseries 35} (1987) 2212}.

\end{thebibliography}\endgroup

\end{document}